\documentclass[aps,longbibliography,nofootinbib,showkeys]{revtex4-1}

\usepackage{amssymb}
\usepackage{amsfonts} 
\usepackage{graphicx} 
\usepackage{amsthm} 
\usepackage{amsmath} 
\usepackage{epsfig}
\usepackage{hyperref}
\usepackage{color}
\usepackage{bbold}
\usepackage{young}
\usepackage{youngtab}
\usepackage{braket}

\usepackage{pgf}
\usepackage{bm}
\usepackage{bbold}
\usepackage{bbm}
\usepackage{graphicx}

\usepackage[utf8]{inputenc}

\def\nn{\nonumber}
\def\ic{\mathrm{i}}

\def \bc {\begin{center}}
\def \ec {\end{center}}
\def \bi {\begin{itemize}}
\def \ei {\end{itemize}}

\def \ba {\begin{array}}
\def \ea {\end{array}}

\def \bea {\begin{eqnarray}}

\def \eea {\end{eqnarray}}

\def \be {\begin{equation}}
\def \ee {\end{equation}}

\newcommand{\la}{\langle}
\newcommand{\ra}{\rangle}

\def\bc {\bar{\beta}}

\def\nb{{\vec{n}}}

\def\dcat{{\scriptstyle\mathrm{DCAT}}}
\def\2cat{{\scriptstyle\mathrm{2CAT}}}
\def\3cat{{\scriptstyle\mathrm{3CAT}}}
\newcommand{\cat}[1]{{\scriptstyle\mathrm{#1 CAT}}}

 \theoremstyle{remark}
 
\definecolor{darkgreen}{rgb}{0,0.5,0}
\definecolor{darkblue}{rgb}{0.1,0.1,0.5}
\definecolor{darkred}{rgb}{0.8,0,0}
\definecolor{verylightgray}{rgb}{0.87,0.87,0.87}

\def\rojo#1{\textcolor{red}{\textbf{#1}}}

\def\nn{\nonumber}
\def\ic{\mathrm{i}}

\def\nb{{\vec{n}}}
\def\zb{\bm{z}}

\def\wb{\bm{w}}

\def\alphab{\bm{\alpha}}

\newtheorem{teorema}{Theorem}
\newtheorem{proposicion}{Proposition}
\newtheorem{lema}{Lemma}
\newtheorem{corolario}{Corollary}

\graphicspath{
{./figures/}
}

\def\graphwidth{5.5cm}
\def\graphsep{0.25cm}

\begin{document}

\title{Schmidt decomposition  of parity adapted coherent states for symmetric multi-quDits}


\author{Julio Guerrero}
\email{jguerrer@ujaen.es: corresponding author}
\affiliation{Department of Mathematics, University of Ja\'en, Campus Las Lagunillas s/n, 23071 Ja\'en, Spain}
\affiliation{Institute Carlos I of Theoretical and Computational Physics (iC1), University of  Granada,
Fuentenueva s/n, 18071 Granada, Spain}
\author{Antonio Sojo}
\email{asojo@ujaen.es}
\affiliation{Department of Mathematics, University of Ja\'en, Campus Las Lagunillas s/n, 23071 Ja\'en, Spain}
\author{Alberto Mayorgas}
\email{albmayrey97@ugr.es}
\affiliation{Department of Applied Mathematics, University of  Granada,
Fuentenueva s/n, 18071 Granada, Spain}
\author{Manuel Calixto}
\email{calixto@ugr.es}
\affiliation{Department of Applied Mathematics, University of  Granada,
Fuentenueva s/n, 18071 Granada, Spain}
\affiliation{Institute Carlos I of Theoretical and Computational Physics (iC1), University of  Granada,
Fuentenueva s/n, 18071 Granada, Spain}

\date{\today}


\begin{abstract}
In this paper we  study the entanglement in symmetric $N$-quDit systems. In particular we use generalizations to $U(D)$
of spin $U(2)$ coherent states and their projections on definite parity $\mathbbm{c}\in\mathbb{Z}_2^{D-1}$ (multicomponent Schr\"odinger cat) states  and we analyse their  reduced density matrices when  tracing out $M<N$ quDits. The eigenvalues (or Schmidt coefficients) of these reduced density matrices are completely characterized, allowing to proof a theorem for the decomposition of a $N$-quDit Schr\"odinger cat state with a given parity $\mathbbm{c}$ into a sum over all possible parities of tensor products  of Schr\"odinger cat states of $N-M$ and $M$ particles.
Diverse asymptotic properties of the Schmidt eigenvalues are studied and, in particular, for the (rescaled) double thermodynamic limit ($N,M\rightarrow\infty,\,M/N$ fixed), we reproduce and generalize to quDits known results for photon
loss of parity adapted coherent states of the harmonic oscillator, thus providing an unified Schmidt decomposition for both multi-quDits and (multi-mode) photons.
These results allow to determine the entanglement properties of these states and also their decoherence properties under quDit loss, where we demonstrate the robustness of these states.

\end{abstract}

\maketitle

\section{Introduction}

Coherent states (CS), either of the harmonic oscillator (HO) or for a spin system, have many applications in Quantum Mechanics and Quantum Optics (among many other fields), and in particular parity adapted
CS (a particular instance of a  Schr\"odinger cat state) for the HO or for $U(2)$\footnote{We shall consider in this paper unitary groups $U(D)$ instead of special unitary groups $SU(D)$ since the difference between them is an irrelevant global phase and it is easier to write down a basis of the Lie algebra in the case of unitary groups.}
are interesting since they are used in many protocols in Quantum Information Processing
or appear  as the lower energy states   in some nuclear or molecular  models like the Dicke \cite{Dicke} or Lipkin-Meshkov-Glick (LMG) models \cite{lipkin1}.

We shall consider in this paper ``spin'' CS for $U(D)$ in its {symmetric representation} (i.e. made of a fixed number $N$ of indistinguishable quDits) and we shall construct from them  parity adapted CS, i.e states which are invariant under the parity group. The parity group in the case of $U(D)$ is  $\mathbb{Z}_2^{D-1}$, generalizing the case of $SU(2)$, where the parity group is simply  $\mathbb{Z}_2=\{-1,+1\}$.

For applications, it is important to characterize the {entanglement} properties of these states, therefore we shall consider {entropic} entanglement measures
in terms of the entropy of $M$-particle {reduced density matrices} (RDM), i.e. in terms of a bipartition of the $N$ quDits in $N-M$ and $M$ quDits and tracing out the $N-M$  subsystem.

There is an intense debate in the literature concerning the notion of entanglement in systems of identical and indistinguishable particles (like the case of symmetric multi-quDits). Some authors \cite{BENATTI20201,e23040479} consider that particle entanglement, obtained through RDM by tracing out a number of particles \cite{Molmer,LoFranco2016,Sciara2017}, cannot be used as a quantum resource in quantum information tasks for indistinguishable particles. Other authors consider that the entanglement due  to exchange symmetry can indeed be useful in those tasks, providing some examples of it \cite{PhysRevLett.112.150501_Plenio,PhysRevX.10.041012}. The authors of \cite{BENATTI20201,e23040479} propose mode entanglement as the only meaningful way of defining entanglement for indistinguishable particles.

In this paper we shall consider the notion of particle entanglement in symmetric multi-quDit systems since it is mathematically consistent and  physically justified in the case of parity adapted coherent states. The case of mode entanglement will be considered in a future work.

To characterize the entanglement of  parity adapted $U(D)$ CS, we need to compute the Schmidt decomposition of these states when a bipartition in $N-M$ and $M$ particles is considered. The main result of this paper is that, under this decomposition, parity adapted CS decompose as the convolution over all possible parities, of tensor products
of  parity adapted CS of $N-M$ and $M$ particles. The coefficients of this decomposition (Schmidt coefficients) and their squares (Schmidt eigenvalues) are determined for all $N$ and $M$, and the main features of them and their behaviour under various limits are estudied. For a detailed account of these features, a variety of graphical tools are used, like contour plots, angular plots, etc., but, in the case of a large $D$, information diagrams (see \cite{Guerrero22} and references therein) prove to be a valuable tool when appropriate colormaps are used (see the Supplementary Material).

The content of the paper is the following. In Sec. \ref{SymmetricRep} the symmetric representation of $U(D)$ is reviewed in order to fix notation (second quantization approach), and parity transformations and parity projectors are introduced. In Sec. \ref{CS-U(D)} CS for the symmetric representation of $U(D)$ are reviewed, and some results for their relation with identical tensor product states are given. In Sec. \ref{ParityCS} parity adapted CS are defined and some of their properties are given, in particular their
behaviour under certain limits. In Sec. \ref{EntanglementMeasures} we specify the entanglement measure used in this paper to study the entanglement properties of parity adapted CS. In Sec. \ref{SchmidtDec} the main result of this paper is proved, namely the Schmidt decomposition of parity adapted CS of $N$ particles into $N-M$ and $M$ particle subsystems, with the determination of the Schmidt coefficients and eigenvalues. In Sec. \ref{Limits} some limits for the Schmidt eigenvalues are studied.
In Sec. \ref{PhysApp} some physical appliations and possible methods to generate these states are given, in particular the interesting subject of quDit loss, where the Schmidt decomposition here provided can be of crucial importance.
 The paper ends with a conclusing section \ref{Conclusions}. In the Supplementary Material, a reminder
 of the subject of information diagrams is included, and a exhaustive set of figures shown the entanglement properties of parity adapted CS for $D>2$ is provided.

\section{Symmetric representation of $\mathbf{U(D)}$}
\label{SymmetricRep}


The fully {symmetric representation} of dimension $\tbinom{N+D-1}{N}$
 of $U(D)$ can be realized as a system of $N$ {identical and indistinguishable} particles (atoms)  with $D$ levels (internal states), that will be referred to as quDits, with levels labelled by $|0\ra,\,|1\ra,\,\ldots,|D-1\ra$ (for $D=2$ we have the standard qubit usually  labelled by $|0\ra$,$|1\ra$ or $|\uparrow\ra$,$|\downarrow\ra$).

Introducing the boson operators $a_i,a_i^\dag$ ($a_i^\dag$ creates a quDit in the $i$-th level and $a_i$ annihilates it), the Lie algebra of $U(D)$ can be realized (in the Schwinger representation \cite{Schwinger}) as:
\be
S_{ij}=a^\dag_i a_j, \; i,j=0,\dots,D-1\,.\label{UDgen}
\ee

The operator $S_{ii}$ is the {number operator} for the population of the $i$-th level, and  $S_{ij}$ ($i\neq j$) {creates} a quDit in the level $i$ and {annihilates} another one in the level $j$. Therefore $S_{ij}$ {preserves} the total number $N$ of quDits.

It is important to note that the $U(D)$ operators $S_{ij}$ are collective operators, in the sense that they do not act on the individual states of the particles (which can be an ill-defined concept due to symmetrization). They have in fact been built up using the second quantization formalism \cite{Dirac,Fock1932}.
They fulfill the {commutation relations}:
\be
\left[S_{{ij}},S_{{kl}}\right]=\delta_{{jk}} S_{{il}} -\delta_{{il}} S_{{kj}}.\label{commurel}
\ee

The carrier {Hilbert space} ${\cal H}_N$ of our symmetric  $N$-quDit system  is spanned by the Bose-Einstein-Fock basis states ($|\vec{0}\ra$ denotes the Fock vacuum):
\be
|\vec{n}\ra=|n_0,\dots, n_{D-1}\ra=
\frac{(a_0^\dag)^{n_0}\dots(a_{D-1}^\dag)^{n_{D-1}}}{\sqrt{n_0!\dots n_{D-1}!}}|\vec{0}\ra\,,\qquad
n_0+\dots+n_{D-1}=N\,,
\label{symmetricbasis}
\ee
where $n_i\geq 0$ denotes the {occupancy number} of the $i$th-level (the eigenvalue of $S_{ii}$), with the {restriction}   given by the linear Casimir of $U(D)$, $C_1=\sum_{i=0}^{D-1} S_{ii}$, the total number of quDits.
The expansion of a general symmetric $N$-quDit state $\psi$ in the Fock basis will be written as
\be
|\psi\ra=\sum_{\|\vec{n}\|_1=N}\,c_{\vec{n}}|\vec{n}\ra,\label{psisym}
\ee
where the sum is restricted to those $\vec{n}$ such that  $\|\vec{n}\|_1=n_0+\dots+n_{D-1}=N$.

In order to define the notion of entanglement, we shall consider in ${\cal H}_N$ various families of states. The simplest states that one usually introduces, in order to define separable (non-entangled) states, are Tensor Product States (TPS) $|\vec{\psi}\ra^{(N)}= |\psi_{1}\ra\otimes \cdots \otimes |\psi_{N}\ra$. However, due to the exchange symmetry, in  ${\cal H}_N$ the only TPS are the subset of \textit{identical} tensor product states (ITPS):
%
%
\be
{\cal A}_N^{ITPS} = \{|\psi\;\ra^{\otimes N}\in {\cal H}_N\,:\, |\psi\ra^{\otimes N}=|\psi\ra\otimes \stackrel{N}{\cdots} \otimes |\psi\ra\}\subset {\cal H}_N
\,.
\ee
with $|\psi\ra\in {\cal H}_1$.
If the one-quDit state $|\psi\ra\in {\cal H}_1$ is expressed as:
\be
 |\psi\ra = w_0 |0\ra + w_1|1\ra + \cdots + w_{D-1} |D-1\ra
 \equiv |w_0,w_1,\ldots,w_{D-1}\ra \equiv
 |\wb\ra\,,  \qquad |w_0|^2+|w_1|^2\cdots +|w_{D-1}|^2=1 \,,
 \label{GenState}
\ee
then it can be shown (see \cite{Sugita2003}) that the expression of the $N$-quDit state
$|\psi\ra^{\otimes N}$ in the Fock basis is:
\begin{equation}
 |\psi\ra^{\otimes N} \equiv|\wb\ra^{\otimes N}= \sum_{\|\vec{n}\|_1=N} \sqrt{\binom{N}{\vec{n}}}\left(\prod_{j=0}^{D-1} w_j^{n_j}\right)|\vec{n}\ra\,,
 \label{ITPS-Fock}
\end{equation}
where $\tbinom{N}{\vec{n}}=\frac{N!}{\vec{n}\:!} $ is a multinomial and $\vec{n}\:!$ stands for
$\prod_{i=0}^{D-1}n_i!$.

As an important example of ITPS we have the coherent states discussed below.


In place of TPS, in ${\cal H}_N$ we can consider the subset of  Symmetrized TPS (STPS), obtained from a TPS   under symmetrization:

\be
{\cal A}_N^{STPS} = \left\{|\vec{\psi}\;\ra^{(\Sigma N)}\in {\cal H}_N\,:\, |\vec{\psi}\ra^{(\Sigma N)}=\frac{1}{{\cal N}(\vec{\psi})^{(\Sigma N)}}\Sigma|\vec{\psi}\ra  =\frac{1}{{\cal N}(\vec{\psi})^{(\Sigma N)}}\frac{1}{N!}\sum_{\sigma\in S_N}|\psi_{\sigma(1)}\ra\otimes \stackrel{N}{\cdots} \otimes |\psi_{\sigma(N)}\ra\right\}
\subset {\cal H}_N
\,,
\ee

with $\left({\cal N}(\vec{\psi})^{(\Sigma N)}\right)^2={}^{(N)}\la\vec{\psi}|\Sigma|\vec{\psi}\ra^{(N)}$, $S_N$ is the symmetric group of permutations of $N$ elements, and where $\Sigma$ is the symmetrization operator, i.e. the projector operator onto the symmetric subspace under $S_N$.

Another important family of states is the Hilbert subspace spanned by
${\cal A}_N^{ITPS}$, i.e. ${\cal H}_N^{ITPS}={\rm span}({\cal A}_N^{ITPS})$, containing all possible finite linear combinations of ITPS. We shall focus on this paper  on states obtained as finite linear combinations of coherent states (see Sec. \ref{CS-U(D)}), which belong to ${\cal H}_N^{ITPS}$.


\subsection{Parity operators for $\mathbf{U(D)}$}

Parity operators play an important role in the representation theory of the group $U(D)$, and also in its physical applications. They are given by (Roman $\ic$ denotes the imaginary unit):
\be
\Pi_j=\exp(\ic \pi S_{jj}) \quad , \quad j=0,1,\ldots,D-1\,,
\label{parityop}
\ee
with the action on Fock states:
\be
\Pi_j|\vec{n}\ra=(-1)^{n_j}|\vec{n}\ra\,,
\label{parityop-fock}
\ee
indicating the {even} (+) or {odd} ($-$) character  of the {population} $n_{j}$ of each level $j=0,\dots,D-1$.
Note that parity operators, like the $U(D)$ generators $S_{ij}$, are collective operators, in the sense that they only depend on the populations of each level, and do not depend on the individual particle states (which could be ill-defined due to symmetrization).

The action of parity operators on ITPS states is given by:
\be
\Pi_j|\wb\ra^{\otimes N} =  \sum_{\|\vec{n}\|_1=N}\sqrt{\binom{N}{\vec{n}}}
\left( \prod_{i=0}^{D-1}w_i^{n_i}\right)(-1)^{n_j} |\vec{n}\ra
= |w_0,\ldots,-w_j,\ldots,w_{D-1}\ra^{\otimes N}\,.
\ee

Due to the constraint of the fixed number of particles equating to $N$, we have the relation $\Pi_0\dots\Pi_{D-1}|\vec{n}\ra=(-1)^N|\vec{n}\ra$.
Hence, {discarding} for instance $\Pi_0$, the true discrete parity symmetry group corresponds to the finite Abelian group $\mathbb{Z}_2^{D-1}=\mathbb{Z}_2\times\stackrel{D-1}{\dots}\times\mathbb{Z}_2$.

Taking this into account, let us denote by $\Pi^\mathbbm{b}=\Pi_1^{b_1}\dots \Pi_{D-1}^{b_{D-1}}$, where $\Pi_1^{b_i}=(\Pi_i)^{b_i}$ and the {binary string} $\mathbbm{b}=[b_1,\dots,b_{D-1}]\in\{0,1\}^{D-1}$ denotes one of the $2^{D-1}$ elements of the parity group $\mathbb{Z}_2^{D-1}$. There are $2^{D-1}$ {parity invariant} subspaces labelled by  the {inequivalent} group {characters} $\chi_\mathbbm{c}$, with  $\mathbbm{c}=[c_1,\dots,c_{D-1}]\in\{0,1\}^{D-1}$ denoting elements of the Pontryagin {dual group} $\widehat{\mathbb{Z}_2^{D-1}}\sim \mathbb{Z}_2^{D-1}$.

The {projectors} onto these {invariant} subspaces of {definite parity} $\mathbbm{c}$ are given by the {Fourier Transform} (FT) between $\mathbb{Z}_2^{D-1}$ and its dual $\widehat{\mathbb{Z}_2^{D-1}}$ (which in this case is a multidimensional Discrete Fourier Transform of dimension $2\times\stackrel{D-1}{\dots}\times 2$, whose matrix realization is the Walsh-Hadamard transform \cite{Kunz-1979}):
\be
\Pi_\mathbbm{c}=2^{1-D}\sum_{\mathbbm{b}} \chi_\mathbbm{c}( \mathbbm{b}) \Pi^\mathbbm{b}\,,\label{projpar}
\ee
with group characters $\chi_\mathbbm{c}( \mathbbm{b})=(-1)^{\mathbbm{c}\cdot \mathbbm{b}}=(-1)^{c_1b_1+\dots+c_{D-1}b_{D-1}}$. The sum in $\mathbbm{b}$ is on the whole parity group $\mathbb{Z}_2^{D-1}$, but we shall omit it for notational convenience (the same applies to the sums in $\mathbbm{c}$ that run on the dual group, which is isomorphic to $\mathbb{Z}_2^{D-1}$).

Using the properties of the characters $\chi_\mathbbm{c}$ of the parity group $\mathbb{Z}_2^{D-1}$, the projectors satisfy:
\be
\Pi_\mathbbm{c}\Pi_{\mathbbm{c}'}=\delta_{\mathbbm{c},\mathbbm{c}'}\Pi_\mathbbm{c}
\ee
and since they are self-adjoint, they are orthogonal projectors.

By the {Fourier inversion formula}  between $\widehat{\mathbb{Z}_2^{D-1}}$ and $\mathbb{Z}_2^{D-1}$, the parity operators can be recovered from the parity projectors through the inverse FT:
\be
\Pi^\mathbbm{b}=\sum_{\mathbbm{c}} \chi_\mathbbm{c}(\mathbbm{b})\Pi_\mathbbm{c} \,,
\ee
where we have used that, in this case, the caracters are real and therefore
$\overline{\chi_\mathbbm{c}(\mathbbm{b})}=\chi_\mathbbm{c}(\mathbbm{b})$.
Denoting by $\mathbb{0}$ and $\mathbb{1}$ the binary strings $[0,0,\ldots,0]$ and $[1,1,\ldots,1]$, respectively, we obtain:
\begin{eqnarray}
 \Pi^\mathbb{0}&=&\sum_{\mathbbm{c}} \Pi_\mathbbm{c} = I\,, \label{ParityResolutionIdentity} \\
 \Pi^\mathbb{1}&=&\sum_{\mathbbm{c}} (-1)^{\mathbbm{c}\cdot\mathbbm{1}} \Pi_\mathbbm{c} \label{TotalParity}\,,
\end{eqnarray}
with $I$ the identity operator on ${\cal H}_N$.
Note that $\Pi^\mathbb{1}=\Pi_1\dots\Pi_{D-1}$ represents the {total parity} of all the states $1,2,\ldots D-1$, and that
$\Pi^\mathbb{0}=I$ and $\Pi^\mathbbm{1}$ generate a $\mathbb{Z}_2$ subgroup (the ``total parity'' subgroup) of the parity group.

By Eq. (\ref{ParityResolutionIdentity}) and the orthogonality of the projectors, the Hilbert space ${\cal H}_N$ decomposes into a direct sum of parity adapted (or projected) subspaces:
\be
{\cal H}_N= \bigoplus_{\mathbb{c}} {\cal H}_N^{(\mathbb{c})} \,.
\ee

In order to clarify the  meaning of the different objects appearing in the remaining sections, we shall make explicit
the number of particles  (say $M$) of the  involved representations in the notation of $U(D)$ generators, parity, projection and identity operators:
 \be
 S_{ij}\longrightarrow S_{ij}^{(M)}\quad \,,\quad \Pi^{\mathbbm{b}} \longrightarrow \Pi^{\mathbbm{b}(M)}\,, \quad  \Pi_\mathbbm{c} \longrightarrow \Pi_\mathbbm{c}^{(M)}
 \quad \,, \quad I \longrightarrow I^{(M)} \,.
 \ee

 Let us state the following Lemma, which will be helpful in the proof of the rest of results of this paper.

 \begin{lema}
  $M$-particle parity operators $\Pi_i^{(M)}$  can be factorized into one-particle parity operators:
  \be
 \Pi_i^{(M)}=\prod_{k=1}^M I^{(k-1)}\otimes \Pi_i^{(1)}\otimes I^{(M-k)}\,.
 \label{oneparticleparityop}
  \ee
  \label{lemafactorizationparityop}
 \end{lema}

 \noindent \textit{Proof:} The proof is a consequence of the fact that $M$-particle collective $U(D)$ generators $S_{ij}^{(M)}$ can be decomposed as a sum of one-particle $U(D)$ generators $S_{ij}^{(1)}$ (see \cite{Sugita2003}):
 \be
 S_{ij}^{(M)}=\sum_{k=1}^M I^{(k-1)}\otimes S_{ij}^{(1)}\otimes I^{(M-k)} \,.
 \ee

 Then eq. (\ref{oneparticleparityop}) follows from the definition of parity operators (\ref{parityop}) and the commutativity of the diagonal operators $S_{ii}^{(M)}$.$\blacksquare$

 According to this Lemma the parity operators are \textit{identical tensor product operators} (ITPO). These operators preserve the subset ${\cal A}_N^{ITPS}$ and therefore leave invariant the Hilbert subspace ${\cal H}_N^{ITPS}$. They also commute with the symmetrization operator $\Sigma$, as $\Pi^\mathbbm{b}$ and $\Pi_\mathbbm{c}$ also do.

\section{Coherent states for the symmetric representation of $\mathbf{U(D)}$}
\label{CS-U(D)}

$U(D)$-spin {coherent states} are defined as \cite{Perelomov}:
\be
	|\zb\ra=\frac{1}{\sqrt{N!}}\left(
	\frac{a_0^\dag+z_1a_1^\dag+\cdots+z_{D-1} a_{D-1}^\dag}{\sqrt{1+|z_1|^2+\cdots+|z_{D-1}|^2}}\right)^{N}|\vec{0}\ra \,. \label{cohD}
\ee

They are labeled by $D-1$ complex numbers $z_j\in\mathbb{C}$ arranged in the column vector $\zb=(z_1,z_2,\dots,z_{D-1})^t\in \mathbb{C}^{D-1}$.
They are in fact labelled by the points in the complex projective space $\mathbb{CP}^{D-1}$, but for simplicity we have considered the chart in
$\mathbb{CP}^{D-1}$ where $z_0\neq 0$ and divided all coefficients by $z_0$ (see, for instance, \cite{nuestroPRE}).

Note that the state for $\zb=\mathbf{0}$, $|\mathbf{0}\ra=\frac{a_0^{\dag\,N}}{\sqrt{N!}}|\vec{0}\ra$ (usually denoted as highest-weight state), should not be confused with the Fock vacuum $|\vec{0}\ra$.

The {coefficients}  $c_\nb(\zb)$
of $|\psi\ra=|\zb\ra$ in the {Fock basis} are  (see eq. (\ref{psisym}))
\be
c_\nb(\zb)=\sqrt{\frac{N!}{ \vec{n}\:!}}\frac{\prod_{i=1}^{D-1} z_i^{n_i}}{(1+\zb^\dag\zb)^{N/2}},\label{coefCS}\qquad
\ee
where $\zb^\dag\zb=|z_1|^2+\dots+|z_{D-1}|^2$ denotes the standard scalar product in $\mathbb{C}^{D-1}$ and $\vec{n}\:!$ stands for $\prod_{i=0}^{D-1}n_i!$.

From the previous expression, and comparing with eq. (\ref{ITPS-Fock}), it is clear that coherent states are
ITPS, cf. eq. (\ref{GenState}), with coefficents $w_j=\frac{z_j}{\sqrt{1+\zb^\dag\zb}}\,,j=0,1,\ldots D-1$ (where $z_0=1$).

In general, these CS are {non-orthogonal}  since the scalar product 
\be \la \zb|\zb'\ra=\frac{(1+\zb^\dag \zb')^N}{(1+\zb^\dag \zb)^{N/2}(1+\zb'^\dag \zb')^{N/2}}\label{scprod}
\ee
is not necessarily zero.  However, they constitute an {overcomplete} set of states closing a resolution of the identity \cite{Perelomov}
\begin{align}\label{resounity}
	I=&\,\int_{\mathbb{C}^{D-1}}|\zb\ra\la\zb|d\mu(\zb),\\
	d\mu(\zb)=&\,\frac{(D-1)!}{\pi^{D-1}}\binom{N+D-1}{N}\frac{d^2z_1\dots d^2z_{D-1}}{(1+\zb^\dag \zb)^{D}}\nonumber\,,
\end{align}
with $d^2z_i=d{\rm Re}(z_i) d{\rm Im}(z_i)$ the usual (Lebesgue) measure on $\mathbb{C}\sim\mathbb{R}^2$.

It turns out  that a CS for a representation with $N$  particles is the tensor product of $N$  copies of one-particle CS states, all of them with the same value of  $\zb$ (see, for instance, \cite{Sugita2003}). Therefore coherent states belong to the subset ${\cal A}_N^{ITPS}$, and will be denoted as $|\zb\ra^{\otimes N}$ when we need to specify the number of quDits in the representation.

In fact, the only ITPS are coherent states, as it is proven in the following Lemma and Proposition.

\begin{lema}
 There is a bijective map between ${\cal H}_1$ and the set of one-particle coherent states.
 \label{lemabijection}
\end{lema}

\textit{Proof:} An arbitrary state in ${\cal H}_1$ is given by eq. (\ref{GenState}),
%
%
with $|w_0|^2+|w_1|^2+\cdots+|w_{D-1}|^2=1$. By the proyective character of ${\cal H}_1$ (invariance under a global phase),
the topology of this space is that of the product of the first hyper-octant of $\mathbb{S}^{D-1}$ times the hypertorus $\mathbb{T}^{D-1}$ (the relative phases), see \cite{bengtsson_zyczkowski_2006}.
Note that for $U(2)$ this reduces to $\mathbb{S}^2\equiv \mathbb{CP}^{1}$, usually referred to as the Bloch sphere (for pure states). In general, it corresponds to the projective space $\mathbb{CP}^{D-1}$, which we shall call the \textit{Bloch projective space}.

To construct the bijection with one-particle coherent states, let us consider a particular chart and  suppose $w_0\neq 0$, then invariance under a global phase allows to choose  $w_0>0$.

One-particle coherent states are of the form
\begin{equation}
 |\zb\ra = \frac{|0\ra + z_1|1\ra + z_2|2\ra+\cdots +z_{D-1}|D-1\ra}{\sqrt{1+|z_1|^2+\cdots+|z_{D-1}|^2}}\,,
\end{equation}
and the geometry is also that of the projective space $\mathbb{CP}^{D-1}$, where again we have used the chart where $z_0\neq 0$ (see, for instance, \cite{QIP-2021-Entanglement}).

Thus, to a one-particle coherent state we can associate a unique element of the Bloch projective space given by:
\be
w_0=\frac{1}{\sqrt{1+|z_1|^2+\cdots+|z_{D-1}|^2}}>0\,, \qquad
w_i=\frac{z_i}{\sqrt{1+|z_1|^2+\cdots+|z_{D-1}|^2}}\,,
\label{CS2Bloch}
\ee
for $i=1,\ldots,D-1$.
 Conversely, to an element of the Bloch projective space (with $w_0>0$) we can associate the coherent state with parameters:
\begin{equation}
 z_i=\frac{w_i}{w_0}\,. \label{Bloch2CS}
\end{equation}
The case $w_0=0$ is handled considering a different chart
in the projective space of coherent states and in the Bloch projective space.
$\blacksquare$

\begin{proposicion}
There is a biyection between ${\cal A}_N^{ITPS}$ and the set of $N$-particle coherent states.
\label{Prop-ITPS}
\end{proposicion}

\noindent \textit{Proof:} Consider tensor products of the same one-particle state and apply the previous Lemma for each particle.$\blacksquare$

\vspace{0.3cm}

As a corollary of this result, it turns out that ${\cal H}_N^{ITPS}$ coincides with ${\rm span}(\{|\zb\ra^{\otimes N}:\,\zb\in\mathbb{C}^{D-1}\}$.

From Eq. (\ref{Bloch2CS}) it is clear that the parameterization $\zb$ for coherent states is the projective version of the parametrization $\wb$ for the Bloch projective space. Thus, we can identify the general one-particle state (\ref{GenState}) and its corresponding $N$-particle ITPS as:
\be
|\psi\ra = |\wb\ra = |\zb\ra \quad ,\quad |\psi\ra^{\otimes N} = |\wb\ra^{\otimes N} = |\zb\ra^{\otimes N}\,.
\ee

However, the parametrization $\zb$ has important properties from the analytical and geometrical point of view, making them more suitable in the applications.
According to these results, the familiy of coherent states, due to their useful geometric and analytic properties, is a convenient way of parametrizing ${\cal H}_N^{ITPS}$.

We can go one step further and write any state in ${\cal H}_N$ as a finite linear combination of coherent states (or ITPS, as you like).

\begin{teorema}
Any state in ${\cal H}_N$ can be written as a finite linear combination of ITPS, i.e. ${\cal H}_N^{ITPS}= {\cal H}_N$.
\label{TeoremaSampling}
\end{teorema}

\noindent\textit{Proof:} We provide here an sketch of the proof, leaving a detailed proof for a future publication \cite{GeneralDcatDecomposition}.
Although coherent states are not orthogonal, they generate the whole space ${\cal H}_N$ since they are overcomplete by eq. (\ref{resounity}). Since ${\cal H}_N$ is finite-dimensional, it is possible to find a finite subset of coherent states generating the whole ${\cal H}_N$ (see for instance \cite{Calixto08} for the case of $SU(2)$), and therefore
${\cal H}_N^{ITPS}= {\cal H}_N$.$\blacksquare$

Note that this theorem does not hold in the case of non-compact groups, where the irreducible unitary representations are realized in an infinite dimensional Hilbert space ${\cal H}_\infty$, and therefore finite linear combinations of coherent states are only  dense in the Hilbert space, i.e. ${\cal H}_N^{ITPS} \neq {\cal H}_\infty$ but $\overline{{\cal H}_N^{ITPS} }= {\cal H}_\infty$ (see \cite{Calixto11b}
for the case of $SU(1,1)$).

For the case of $SU(2)$ an stronger result can be given: ${\cal H}_N$ is made of symmetrized tensor product states.
\begin{teorema}
 For $D=2$, ${\cal H}_N={\cal A}_N^{STPS}$.
 \label{MajoranaSU2}
\end{teorema}

\noindent\textit{Proof:} The usual proof is given in terms of the Majorana representation and Majorana constellation \cite{Majorana1932,Devi2012}. The Majorana representation for the SU(2) state $|\psi\ra$ of spin $s$, with $N=2s$,  is given in terms of the Husimi amplitude $M(z)=\la z|\psi\ra$. Ignoring a global factor depending on $|z|$, the Majorana function $M(z)$ is a polynomial in $\bar{z}$ which is  characterized by its zeros, having up to $2s$ zeros. Completing this zeros with the zeros at infinity  of $M(z)$, the Majorana representation of the state $|\psi\ra$ is characterized by $2s$ points on the sphere (once the complex zeros are mapped to the sphere by inverse stereographic projection and the zeros at infinity are associated with the North pole of the sphere), denoted as Majorana constellation.

On the other hand, the Majorana representation of a STPS $|\vec{\psi}\ra$ is given by:
\bea
M_{\vec{\psi}}(z)&=&\frac{1}{{\cal N}(\vec{\psi})^{(\Sigma N)}}\frac{1}{N!}\sum_{\sigma\in S_N}\la z|\psi_{\sigma(1)}\ra\otimes \cdots \otimes \la z|\psi_{\sigma(N)}\ra \nn \\
&=& \frac{1}{{\cal N}(\vec{\psi})^{(\Sigma N)}} \la z|\psi_1\ra \cdots \la z|\psi_N\ra\,,
\eea
which coincides, up to a global normalization factor independent of $z$, with the Majorana representation of a (non-symmetrical) TPS state. By Lemma \ref{lemabijection} every $|\psi_i\ra$ coincides with a coherent state $|z_i\ra$, thus
\be
M_{\vec{\psi}}(z)=\frac{1}{{\cal N}(\vec{\psi})^{(\Sigma N)}} \la z|z_1\ra \cdots \la z|z_N\ra\,.
\ee

The zeros of $M_{\vec{\psi}}(z)$ are clearly those values of $z$ that make zero any of the overlaps $\la z|z_i\ra$, for some $i=1,\ldots,N$, and, for $SU(2)$ coherent states, the unique zero values of the overlaps are the antipodal points $-\frac{1}{\bar{z}_i}$. By the unicity of the representation of the Majorana function in terms of its zeros, we conclude that any state of spin $s$ corresponds to a STPS.$\blacksquare$.

Unfortunately, although we can define a Majorana representation for $D\geq 3$ in terms of the Husimi amplitudes $M(\zb)=\la \zb|\Psi\ra$, defined on the projective spaces $\mathbb{CP}^{D-1}$, they cannot be characterized in terms of their zeros (which are no longer isolated points), and a result like Theorem \ref{MajoranaSU2} is not available. However, Theorem \ref{TeoremaSampling} is still valid and it can be used to compute the Schmidt decomposition of arbitrary parity adapted  multi-quDit states \cite{GeneralDcatDecomposition}.

\section{Parity adapted $\mathbf{U(D)}$-spin  coherent states}
\label{ParityCS}

Parity operators $\Pi_j=\exp(\ic \pi S_{jj})$ act on CS as
\be\label{parityCS}
    \Pi_j|\zb\ra=|z_1,\dots,-z_j,\dots,z_{D-1}\ra \,, \qquad j=0,1,\ldots,D-1\,,
\ee
thus $\Pi_j$ just {changes the sign} of $z_j$ in $|\zb\ra$. Let us denote $\zb^\mathbbm{b}\equiv((-1)^{b_1}z_1,\dots,(-1)^{b_{D-1}}z_{D-1})^t$,
then
\be
|\zb^\mathbbm{b}\ra \equiv \Pi^\mathbbm{b}|\zb\ra= \prod_{i=1}^{D-1}\Pi_i^{b_i}|\zb\ra \,.
\ee
Define also parity $\mathbbm{c}$ adapted $U(D)$-spin CS as
\be\label{DCAT}
|\zb\ra_\mathbbm{c}\equiv \frac{\Pi_{\mathbbm{c}}|\zb\ra}{\mathcal{N}_\mathbbm{c}(\zb)}=\frac{2^{1-D}}{\mathcal{N}_\mathbbm{c}(\zb)}\sum_{\mathbbm{b}}\chi_\mathbbm{c}(\mathbbm{b})|\zb^\mathbbm{b}\ra \,,
\ee
as the {normalized projection} of $|\zb\ra$ onto the  {invariant subspace} of parity $\mathbbm{c}$ (or normalized FT of $|\zb^\mathbbm{b}\ra$),
where the normalization factor is given by:
\be
\mathcal{N}_\mathbbm{c}(\zb)^2= 2^{1-D} \sum_{\mathbbm{b}} \chi_\mathbbm{c}(\mathbbm{b}) \la \zb|\zb^\mathbbm{b}\ra = 2^{1-D} \frac{ \sum_{\mathbbm{b}} \chi_\mathbbm{c}(\mathbbm{b}) (1+\zb^\dag\zb^\mathbbm{b})^N}{(1+\zb^\dag\zb)^N}
  \,,\label{normcat}
\ee
i.e., $\mathcal{N}_\mathbbm{c}(\zb)^2$ is the FT of the overlap between a coherent state and its parity transformed versions. Note that $\mathcal{N}_\mathbbm{c}(\zb)$ is a function only of the absolute values $|z_i|\,,\,i=1,\ldots,D-1$, and does not depend on the relative phases.

Note that  $\mathcal{N}_\mathbbm{c}(\zb)$ can be zero (as well as $\Pi_{\mathbbm{c}}|\zb\ra$)
for some particular values of $\zb$ and $\mathbbm{c}$. We shall discuss these cases in the next subsections.

The rest of the paper will be devoted to study the entanglement properties of these parity adapted CS, which can be seen as particular instances of Entangled CS \cite{PhysRevA.45.6811} (see also the review \cite{Sanders_2012}).

\subsection{Limit values for normalization factors}

It is worth showing some particular values of the normalization factor, that will be usefull in computing  some limit values of the Schmidt eigenvalues in Sec. \ref{Limits}.

At $\zb=\mathbf{0}$ we have
\be
\mathcal{N}_\mathbbm{c}(\mathbf{0})^2=\delta_{\mathbbm{c},\mathbb{0}}\,,
\label{Nz0}
\ee
implying that the normalization at $\zb=\mathbf{0}$ is zero except for the completely even unnormalized parity adapted CS state (see Sec. \ref{Dcatz0} for a detailed study of the limit $\zb\rightarrow \mathbf{0}$ of these states).
 
 The $\|\zb\|\rightarrow \infty$ limit {does not exist} (except for the case $D=2$, see below), since its value depends on the {direction}.
 Using {hyper-spherical coordinates} in the first hyper-octant, $|z_i|=r y_i(\vec{\theta})$, $\vec{\theta}\in [0,\frac{\pi}{2}]^{D-2}\,,i=1,2,\ldots D-1$, we can compute the directional limits:
 \be
 \lim_{r\rightarrow \infty}  \mathcal{N}_\mathbbm{c}(\zb)^2 =
 2^{1-D}  \sum_{\mathbbm{b}} \chi_\mathbbm{c}( \mathbbm{b}) Y_\mathbbm{b}(\vec{\theta})^{N}\,, \label{N-zinf-angdep}
\ee
 with
 $Y_{\mathbbm{b}}(\vec{\theta})=\sum_{i=1}^{D-1} (-1)^{b_i}y_i(\vec{\theta})^2$.
 For instance, for  $D=3$ we have polar coordinates, and in this case $Y_{\mathbbm{b}}(\theta)=(-1)^{b_1}\cos^2\theta + (-1)^{b_2}\sin^2\theta$\,.

 For $D=2$ the limit $|z|\rightarrow \infty$ exists, ant it is given by:
 \be
 \lim_{|z|\rightarrow \infty}  \mathcal{N}_c(z)^2 =\delta_{N_c,0}\,.
 \label{N-zinf}
\ee
with $N_c=(N-c)\; {\rm mod}\; 2$.

Another interestig limit is the thermodynamic limit, i.e. when the number of particles $N$ grows to infinity:
\be
 \lim_{N\rightarrow \infty}  \mathcal{N}_\mathbbm{c}(\zb)^2 =2^{-k}\delta_{\mathbbm{c}_0,\mathbb{0}_0}\,,
\label{N-Ninf}
 \ee
where $k=\|\zb\|_0$, with the $0$-norm being the number of nonzero components of the vector $\zb$, and $\mathbbm{c}_0$ indicates the subset of components of $\mathbbm{c}$ whose indices coincide with the indices of the zero components of $\zb$.

Finally, let us consider the rescaled thermodynamic limit, when  $N$ grows to infinity but at the same time $\zb$ approach $\mathbf{0}$ such that $\sqrt{N}\zb =\alphab$ is finite. This process is equivalent to the group contraction from $U(D)$  to the harmonic oscillator (HO) group in $D-1$ dimensions\footnote{The harmonic oscillator group in $D-1$ dimensions is the Lie  group generated by the Lie algebra of the canonical annihilation and creation operators in $D-1$ dimensions, including
the corresponding number operators and the rotations. Alternatively, this contraction procedure can be seen as the large $N$ limit of the generalized Holstein-Primakoff realization of $SU(D)$, see \cite{Providencia2006,PhysRevB.48.3190}.}, HO$_{D-1}$:
\be
 \lim_{N\rightarrow \infty}  \mathcal{N}_\mathbbm{c}\left(\frac{\alphab}{\sqrt{N}}\right)^2 = \prod_{i=1}^{D-1}
 \left(\mathcal{N}_{c_i}^{HO}(\alpha_i)\right)^2\,,
\label{N-NinfR}
 \ee
where $\mathcal{N}_{c}^{HO}(\alpha)$ are the normalization factors of the even ($c=0$) and odd ($c=1$)
Schr\"odinger cat states of the one-dimensional harmonic oscillator:
\be
\left(\mathcal{N}_{c}^{HO}(\alpha)\right)^2 = \frac{1}{2} \sum_{b=0,1} (-1)^{bc} \la \alpha|(-1)^b\alpha\ra
= \frac{1}{2} \sum_{b=0,1} (-1)^{bc} e^{((-1)^b-1)|\alpha|^2}
= e^{-|\alpha|^2} {\rm exp}_c(|\alpha|^2)\,, \label{NormHW}
\ee
with $|\alpha\ra$ the coherent states  of the harmonic oscillator, and where:
\be
{\rm exp}_c(x)=\left\{\ba{lcl} \cosh(x) &,& c=0 \quad\hbox{(even)}\,,\\
                              \sinh(x) &,& c=1  \quad\hbox{(odd)}\,.\ea \right.
\ee

\begin{center}
\begin{figure}[h!]
\includegraphics[width=\graphwidth]{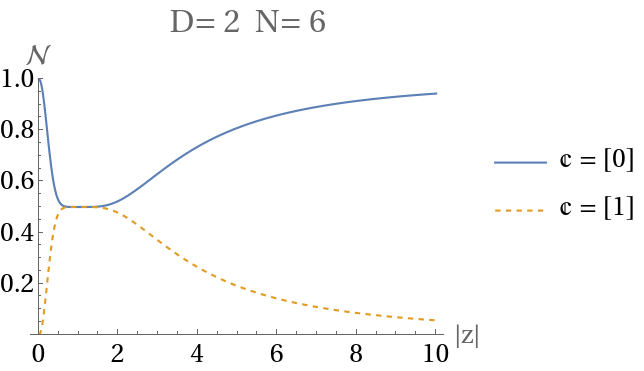}\hspace{\graphsep}\includegraphics[width=\graphwidth]{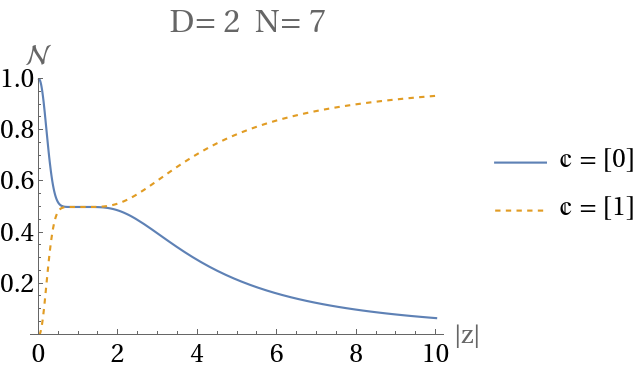}\hspace{\graphsep}
\includegraphics[width=\graphwidth]{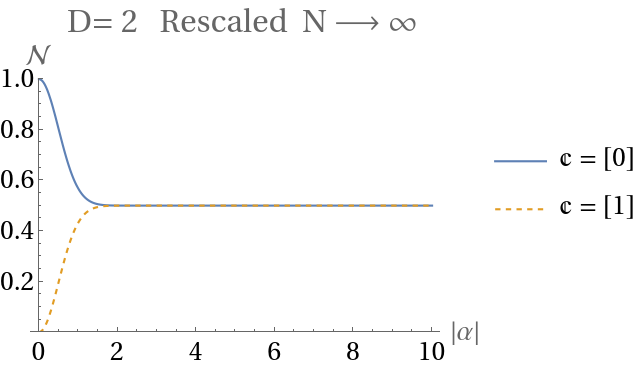}\hspace{\graphsep}
\caption{Plots of the normalization factor $\mathcal{N}_\mathbbm{c}(\zb)^2$ for $D=2$, \, $N=6$ (left), $N=7$ (center) and in the restaled thermodynamic limit (right), for $|z|,|\alpha|\in [0,10]$.}
\label{FigNorms}
\end{figure}
\end{center}

The behaviour of the normalization factors $\mathcal{N}_\mathbbm{c}(\zb)^2$ is summarized in Fig. \ref{FigNorms} for the case $D=2$.

\subsection{Fock coefficients}

The coefficients $c_\nb(\zb)_\mathbbm{c}$ of $|\zb\ra_\mathbbm{c}$  in the Fock basis  are:
\be
c_\nb(\zb)_\mathbbm{c}=\frac{2^{1-D}}{\mathcal{N}_\mathbbm{c}(\zb)}\sum_{\mathbbm{b}}\chi_{\mathbbm{c}+\mathbbm{n}}(\mathbbm{b})c_\nb(\zb)
= \frac{c_\nb(\zb)}{\mathcal{N}_\mathbbm{c}(\zb)}\delta_{\vec{n}_0\; {\rm mod}\; 2,\mathbbm{c}}\,,\label{coefDCAT}
\ee
where $\vec{n}_0=(n_1,\ldots,n_{D-1})$ is retrieved from $\nb$ by removing $n_0$. With this expression of the Fock coefficients for $|\zb\ra_\mathbbm{c}$, the squared norm can be rewritten as
\be
\mathcal{N}_\mathbbm{c}(\zb)^2=\sum_{\|\vec{n}\|_1=N}\delta_{\vec{n}_0\; {\rm mod}\; 2,\mathbbm{c}}
|c_{\vec{n}}(\zb)|^2
\ee

Thus, parity adapted $U(D)$-spin CS  $|\zb\ra_\mathbbm{c}$ in Eq. (\ref{DCAT}) {contain only} basis Fock states $|\nb\ra$ with the {same parity} as $\mathbbm{c}$ (in the indices $i=1,\ldots,D-1$).

Due to these properties, parity adapted CS can be considered multicomponent  Schr\"odinger cat states,
being an extension to $D$ levels and parity $\mathbbm{c}\in\mathbb{Z}_2^{D-1}$ of $SU(2)$
Schr\"odinger cat states \cite{TwoModeSU2-SU11-CAT}, which in turn are the $SU(2)$ version of the traditional even and odd Schr\"odinger cat states for one-mode harmonic oscillator \cite{Dodonovcat}. They are also the $U(D)$ version (extended to all possible parities) of Schr\"odinger cat states of the multimode  harmonic oscillator \cite{Manko-MultimodeCats}. The $SU(2)$ version of these states are related to spin cat states \cite{Huang2015,PhysRevA.105.062456,PhysRevA.104.053721,Maleki:20}, with interesting metrological properties. We shall call them ``$\mathbbm{c}$-$\dcat$s'', or ``$\dcat$s'' for short when the parity $\mathbbm{c}$ is not relevant.

It is interestig to note that if we consider parity adapted CS but restricted to the total parity subgroup, i.e. the one generated by $\Pi^\mathbb{0}=I$ and $\Pi^\mathbbm{1}$,
the resulting states are the $U(D)$ version of the \textit{even} and \textit{odd} multimode (or {po}{ly}{chro}{mat}ic) Schr\"odinger cat states introduced in \cite{Manko-MultimodeCats}. In this case, the even or odd parity refers to the total parity, i.e, that of the sum $n_1+n_2+\ldots + n_{D-1}=N-n_0$.

Finally, note that due to the equivalence between the set of coherent states and the Bloch projective space, given by Lemma \ref{lemabijection}  and Proposition \ref{Prop-ITPS}, we can extend the definitions in this section to the states parametrized by the Bloch projective space and the ITPS obtained from them, i.e. we can define in the obvious way the ($N$-particle states):
\be
|\wb^\mathbbm{b}\ra\quad ,\quad  |\wb\ra_\mathbbm{c} \quad ,\quad |\psi^\mathbbm{b}\ra \quad ,\quad  |\psi\ra_\mathbbm{c}\,,
\ee
where $|\psi\ra$ here stands for $N$ identical copies of the state given in Eq. (\ref{GenState}).
Note that the parity transformed states $|\wb^\mathbbm{b}\ra$ and $|\psi^\mathbbm{b}\ra$ are $N$-particle ITPS, whereas the parity adapted states $|\wb\ra_\mathbbm{c}$ and $|\psi\ra_\mathbbm{c}$ are finite sums of ITPS.

Also, by Theorem \ref{TeoremaSampling}, any state $|\Psi\ra\in {\cal H}_N$ (not necessarily an ITPS) can be writen as a finite sum of coherent states. Therefore we can define $|\Psi\ra^\mathbbm{b}$ and
$|\Psi\ra_\mathbbm{c}$ as the corresponding finite sums of parity transformed or parity adapted coherent states. Although  parity transformations are well defined for any state through its action on Fock states, see eq. (\ref{parityop-fock}), for the purpose of the Schmidt decomposition the expansion in terms of coherent states  will be useful in order to use the factorization property of parity transformations given in Lemma \ref{lemafactorizationparityop}.

\subsection{Limit values for $\dcat$s at $\mathbf{z}=\mathbf{0}$: \textit{Fock-cat} states}
\label{Dcatz0}

It could seem that $\mathbbm{c}$-$\dcat$s for parities $\mathbbm{c}\neq \mathbb{0}$ do not exist at $\zb=\mathbf{0}$, in view of the zero vale of the norm of the unnormalized parity adapted states
$\Pi_{\mathbbm{c}}|\zb\ra$ (see Eq. (\ref{Nz0}) and Fig. \ref{FigNorms} for $D=2$). However, if we consider the normalized $\mathbbm{c}$-$\dcat$s states $|\zb\ra_\mathbbm{c}$ in Eq. (\ref{DCAT}), they are well-behaved at $\zb=\mathbf{0}$, and their Fock coefficients have the expression:
\be
\lim_{\zb\rightarrow \mathbf{0}}c_\nb(\zb)_\mathbbm{c}=\delta_{\vec{n}_0,\mathbbm{c}}\,.\label{coefDCATz0}
\ee

This means that $\mathbbm{c}$-$\dcat$s are Fock states at $z=0$, in particular the ones given by:
\be
\lim_{\zb\rightarrow \mathbf{0}}|\zb\ra_\mathbbm{c} = |n_0=N-k,\nb_0=\mathbbm{c}\ra
=|N-k,c_1,\ldots,c_{D-1}\ra\,,\label{DCATz0}
\ee
with $k=\|\mathbbm{c}\|_0$  the number of non-zero components of $\mathbbm{c}$. We shall call these states  \textit{Fock-cat} states, since they are Fock states but sharing many properties with $\dcat$s, since they are limits of $\dcat$s when $\zb\rightarrow\mathbf{0}$.

In the thermodynamic limit ($N\rightarrow\infty$), the same result still applies, but now $n_0\rightarrow\infty$ in all cases.

In the rescaled thermodynamic limit (the contraction to $D-1$ harmonic oscillators), the result is also similar, but in this case the $\zb\rightarrow\mathbf{0}$ limit is the $(D-1)$-dimensional harmonic oscillator Fock state:
\be
\lim_{\zb\rightarrow \mathbf{0}}|\zb\ra_\mathbbm{c}^{\rm HO} = |\mathbbm{c}\ra=|c_1,\ldots,c_{D-1}\ra\,.
\ee

\subsection{Limit values for $\dcat$s when $\|\mathbf{z}\|\rightarrow \infty$}
\label{DcatzInf}

As it can be seen for the case $D=2$ at Fig. \ref{FigNorms} (the two leftmost graphics), the norm approaches either zero or one when $\|z\|\rightarrow \infty$.
A similar behaviour can be observed for higher values of $D$ at the coordinate axes.
This suggests that we should
study with detail these limits in order to properly identify these states.

Define the vector $\mathbbm{z}\in \{0,1\}^{D-1}$ such that $\|\mathbbm{z}\|_0=1$, i.e. $\mathbbm{z}$ is a unitary vector pointing in the positive direction of one of the coordinate axes in $\mathbb{R}^{D-1}$ (that is, $\mathbbm{z}$ is a particular element of the canonical basis of  $\mathbb{R}^{D-1}$). Then the Fock coefficients of the $\mathbbm{c}$-$\dcat$s in the limit along the coordinate axes are given by:
\be
\lim_{r\rightarrow \infty} c_{\vec{n}}(r \mathbbm{z})_\mathbbm{c}=\delta_{\vec{n}_0,\mathbbm{c}+(N-\|\mathbbm{c}\|_0-N_\mathbbm{c})\mathbbm{z}}\,,
\label{coefDCATinf}
\ee
with $N_\mathbbm{c}=(N-\|\mathbbm{c}\|_0)\; {\rm mod}\; 2$.
This means again that $\mathbbm{c}$-$\dcat$s approach Fock states in these limits, in particular the ones given by:
\be
\lim_{r\rightarrow \infty} |r \mathbbm{z}\ra_\mathbbm{c}= |N_\mathbbm{c},\mathbbm{c}+(N-\|\mathbbm{c}\|_0-N_\mathbbm{c})\mathbbm{z}\ra\,,
\label{DCATinf}
\ee
and thus they are also \textit{Fock-cat} states.

In some particular cases, depending on the parity of $N$ and on $\mathbb{c}$ and $\mathbb{z}$, the resulting Fock state has all the particles in the same level (i.e. they are ITPS, with no entanglement). For instance, for $D=2$ the resulting Fock state in the limit $|z|\rightarrow \infty$ is $|0,N\ra$ if $N$ and $\mathbb{c}$ have the same parity.
For $D=3$, $\mathbb{c}=[0,0]$,
and even $N$, in the limit $\zb\rightarrow (\infty,0)$ the resulting state is $|0,N,0\ra$ and in the limit
$\zb\rightarrow (0,\infty)$ the resulting state is $|0,0,N\ra$. For $\mathbb{c}=[1,0]$ and odd $N$, in the limit $\zb\rightarrow (\infty,0)$ the resulting state is $|0,N,0\ra$. For $\mathbb{c}=[0,1]$ and odd $N$, in the limit $\zb\rightarrow (0,\infty)$ the resulting state is $|0,0,N\ra$. For $\mathbb{c}=[1,1]$, in none of the cases we obtain a Fock state with all the particles in the same level.

It should be noted that $z_i\rightarrow\infty$ for some $i=1,\ldots,D-1$ means that the chosen chart (the one for which $z_0\neq 0$) is no longer valid since $z_0$ turns out to be zero. In this case the chart for which $z_i\neq 0$ should be used instead.

Taking into account this fact, the cases $\|\zb\|\rightarrow 0$ and $z_i\rightarrow\infty$ for some $i=1,\ldots,D-1$ should stand on the same foot. For instance,
the Fock state
$|0,\ldots,0,N,0,\ldots,0\ra$, with the $N$ particles at level $i$, corresponds (in the notation used in Eq. (\ref{GenState})) to $w_i=1,w_j=0$ for $j\neq i$ and $\mathbbm{c}=\mathbb{0}$ if $N$ is even and   $\mathbbm{c}=[0,\ldots,0,1,0,\ldots,0]$ (with a one at position $i$) if $N$ is odd. The other cases can be treated in a similar fashion.

\section{Entanglement measures}
\label{EntanglementMeasures}


Entanglement implies quantum correlations among the different parts of a multipartite system. We shall restrict to bipartite entanglement of pure states, where different measures of entanglement exist, depending on how non-entangled (separable) states are defined. In \cite{BENATTI20201} some definitions of entanglement are introduced,  and in the most basic one (Entanglement-I) separable states are identical TPS states (ITPS), thus the only separable states are coherent states, as shown in Prop. \ref{Prop-ITPS}. In particlar, symmetrized TPS are entangled according to this definition, due  to the exchange symmetry.

For other definitions of entanglement, like (Entanglement-V) in \cite{BENATTI20201}, also known as mode entanglement, separable states
are obtained, in the second quantization formalism, by acting on the Fock vacuum with a product of creation operators which act on a different set of modes (or levels). An example of separable states are basis Fock states $|\vec{n}\ra$, which however are not separable (except in the case when all the particles are in the same level) with respect to Entanglement-I. Also, with respect to Entanglement-V, coherent states are not separable (see, for instance, \cite{QIP-2021-Entanglement} for U(3) coherent states).

There is an intense debate in the literature about which of the different notions of entanglement is the correct one for indistinguishable particles, in the sense that entanglement could be used as a resource in quantum information, computation, communication or metrology tasks, see \cite{BENATTI20201,PhysRevX.10.041012,e23040479} and references therein.
 
 Without entering in this debate, we shall stick in this paper to the notion of Entanglement-I, which is very common in the literature, and which provides a mathematically consistent notion of entanglement in terms of $M$-particle RDMs (when $N-M$  quDits are traced out) and their corresponding Schmidt coefficients (see \cite{RevModPhys.81.865-Horodecki} and references therein).
We shall use entropic measures on the RDMs to quantify the entanglement (see \cite{Molmer} for the case of qubits). From the physical (and also the mathematical) point of view, this is justified since we shall work only with CS and finite sums of them. Also, in certain situations, like quDit loss (see Sec. \ref{quDitLoss}), the process of losing one or more quDits is modelled by tracing out by these quDits, and our decomposition fits in perfectly in this scheme.

 Since the computation of many entropies (like von Neumann entropy) requires the {knowledge of the eigenvalues} of the RDM, we shall mainly focus on computing the eigenvalues of these $M$-particle RDMs.

 By the {Schmidt decomposition theorem} (see \cite{RevModPhys.81.865-Horodecki}) the nonzero eigenvalues (the squares of the so called {Schmidt coefficients}) of the $M$-particle RDM coincide with those of the $(N-M)$-particle RDM, thus we shall restrict to $1\leq M\leq \lfloor \frac{N}{2}\rfloor$. Only in the study of robustness (see Sec. \ref{quDitLoss}) we will be interested in the $(N-M)$-particle RDM.
 
 Starting with a pure state $|\psi\ra^{(N)}$ in the symmetric irreducible representation of $U(D)$ with $N$ particles/quDits, we define the $M$-particle RDM as:
 \be
 \rho^{(M)} = {\rm Tr}_{N-M} |\psi\ra\la \psi |^{(N)}\,.
 \ee
 Due to the symmetry of the original state $|\psi\ra^{(N)}$, the resulting density matrix  $\rho^{(M)}$ lies in the symmetric irreducible representation of $U(D)$ with $M$ particles.

\section{Schmidt decomposition of DCATs}
\label{SchmidtDec}

In this section we provide the main results of the paper on Schmidt decomposition of $\dcat$s under the bipartition in $M$ and $N-M$ particles, with $1\leq M\leq \lfloor \frac{N}{2}\rfloor$.
 
 \subsection{Decomposition of definite parity projection operators}

The following Lemma will be of extreme importance in the following results about Schmidt coefficients of reduced density matrices for states with definite parity, stating that projectors $\Pi_\mathbbm{c}$ onto subspaces of definite parity $\mathbbm{c}$ for a given number of particles can be decomposed as a sum over all possible parities of tensor products of projectors on subspaces of definite parity with smaller number of particles.
 
 \begin{lema}
 Let $N,M$ integers with $N>1$  and $1\leq M\leq \lfloor \frac{N}{2}\rfloor$. Then:
 \be
 \Pi_\mathbbm{c}^{(N)} = \sum_{\mathbbm{c}'} \Pi_{\mathbbm{c}-\mathbbm{c}'}^{(N-M)}\otimes \Pi_{\mathbbm{c}'}^{(M)}
  = \sum_{\mathbbm{c}'} \Pi_{\mathbbm{c}-\mathbbm{c}'}^{(M)}\otimes \Pi_{\mathbbm{c}'}^{(N-M)}
  =\sum_{\mathbbm{c}'} \Pi_{\mathbbm{c}'}^{(N-M)}\otimes \Pi_{\mathbbm{c}-\mathbbm{c}'}^{(M)}\,.
\ee
 \end{lema}

\noindent Proof:  The proof is obvious using {Convolution Theorem} for $\mathbb{Z}_2^{D-1}$:

 \begin{eqnarray}
  \Pi_\mathbbm{c}^{(N)}&=& 2^{1-D}\sum_{\mathbbm{b}} \chi_\mathbbm{c}( \mathbbm{b}) \Pi^{\mathbbm{b}(N)}
 = 2^{1-D}\sum_{\mathbbm{b}} \chi_\mathbbm{c}( \mathbbm{b}) \left( \Pi^{\mathbbm{b}(N-M)}\otimes I^{(M)}\right)\left( I^{(N-M)}\otimes \Pi^{\mathbbm{b}(M)}\right)\\
 &=&  \sum_{\mathbbm{c}'}  \left(\Pi_{\mathbbm{c}-\mathbbm{c}'}^{(N-M)}\otimes I^{M}\right)\left(I^{(N-M)}\otimes\Pi_{\mathbbm{c}'}^{(M)}\right)   =   \sum_{\mathbbm{c}'} \Pi_{\mathbbm{c}-\mathbbm{c}'}^{(N-M)}\otimes \Pi_{\mathbbm{c}'}^{(M)} \,,\nn
 \end{eqnarray}

where we have used Lemma \ref{lemafactorizationparityop} to factorize $\Pi^{\mathbbm{b}(N)}$:
\bea
\Pi^{\mathbbm{b}(N)}&=& \Pi_1^{b_1}\dots \Pi_{D-1}^{b_{D-1}} = \exp(\ic \pi \sum_{j=1}^{D-1} b_j S_{jj}^{(N)})= \exp(\ic \pi \sum_{j=1}^{D-1} b_j( S_{jj}^{(N-M)}\otimes I^{(M)}+I^{(N-M)}\otimes S_{jj}^{(M)} )) \\
&=&\exp(\ic \pi \sum_{j=1}^{D-1} b_j S_{jj}^{(N-M)}\otimes I^{(M)}) \exp(\ic \pi b_j I^{(N-M)}\otimes S_{jj}^{(M)} ) =
\left( \Pi^{\mathbbm{b}(N-M)}\otimes I^{(M)}\right)\left( I^{(N-M)}\otimes \Pi^{\mathbbm{b}(M)}\right)\,,\nn
\eea

together with the fact that the number operators $S_{jj}$ commute among them for all values of $j$.$\blacksquare$
 
Note the symmetry under the interchange $(N-M) \leftrightarrow M$, due to the Schmidt Theorem, and the symmetry under the interchange $(\mathbbm{c}-\mathbbm{c}')\leftrightarrow\mathbbm{c}'$, due to the Convolution Theorem.

\subsection{Decomposition of parity adapted CS}
 
 Using the previous result applied to a coherent state we obtain the main result of this paper:

 \begin{teorema}
 Let $N,M$ integers with $N>1$  and $1\leq M\leq \lfloor \frac{N}{2}\rfloor$. Then the $\mathbbm{c}$-$\dcat$ of $N$ particles can be decomposed in terms of superpositions of tensor products of $\dcat$s of $M$ and $N-M$ particles as:
 \be
 |\zb\ra_\mathbbm{c}^{(N)} = \sum_{\mathbbm{c}'} l_{\mathbbm{c},\mathbbm{c}'}^{N,M}(\zb) |\zb\ra_{\mathbbm{c}-\mathbbm{c}'}^{(N-M)}\otimes |\zb\ra_{\mathbbm{c}'}^{(M)}\,, \label{SchmidtDecomp}
\ee
with {Schmidt coefficients}
\be
l_{\mathbbm{c},\mathbbm{c}'}^{N,M}(\zb) =\frac{\mathcal{N}_{\mathbbm{c}-\mathbbm{c}'}^{(N-M)}(\zb) \mathcal{N}_{\mathbbm{c}'}^{(M)}(\zb) }{ \mathcal{N}_\mathbbm{c}^{(N)}(\zb) }\,.
\ee
\end{teorema}

\noindent Proof: The proof is obvious by applying the previous Lemma and restoring the normalization of the diverse $\dcat$s appearing in the equation.$\blacksquare$

The {Schmidt eigenvalues} of the RDM obtained after tracing out $N-M$ particles are given by the squares of the Schmidt coefficients:
\be
\lambda_{\mathbbm{c},\mathbbm{c}'}^{N,M}(\zb) =\left(l_{\mathbbm{c},\mathbbm{c}'}^{N,M}(\zb) \right)^2\,.
\ee

Note that the  Schmidt eigenvalues depend only on the absolute values $|z_i|\,,\,i=1,\ldots,D-1$, and do not depend on the relative phases. They are also invariant under the simultaneous interchange of $(N-M) \leftrightarrow M$ and $(\mathbbm{c}-\mathbbm{c}')\leftrightarrow\mathbbm{c}'$.

The $M$-particle RDMs are well-defined density matrices, since the eigenvalues are positive and their sum is one.
 \begin{proposicion} The trace of the $M$-particle RDM  $\rho_\mathbbm{c}^{(M)}(\zb)$ of a $\mathbbm{c}$-$\dcat$ is:
 \be
{\rm Tr}\; \rho_\mathbbm{c}^{(M)}(\zb)={\rm Tr}\;\left({\rm Tr}_{N-M} {}^{(N)}|\zb\ra_{\mathbb{c}\;\mathbb{c}}\la\zb|^{(N)}\right)=
 \sum_{\mathbbm{c}'} \lambda_{\mathbbm{c},\mathbbm{c}'}^{N,M}(\zb)=1 \,.
 \ee
 \end{proposicion}

\noindent Proof:  Using the definition of the Schmidt numbers, it is easily proven that

 \bea
 \sum_{\mathbbm{c}'} \lambda_{\mathbbm{c},\mathbbm{c}'}^{N,M}(\zb) &=& \frac{1}{\left( \mathcal{N}_\mathbbm{c}^{(N)}(\zb)\right)^2 } \sum_{\mathbbm{c}'}\left(\mathcal{N}_{\mathbbm{c}-\mathbbm{c}'}^{(N-M)}(\zb) \mathcal{N}_{\mathbbm{c}'}^{(M)}(\zb)\right)^2
 =\frac{1}{\left( \mathcal{N}_\mathbbm{c}^{(N)}(\zb)\right)^2 } 2^{1-D} \sum_{\mathbbm{b}} \chi_\mathbbm{c}(\mathbbm{b})  \la \zb|\zb^\mathbbm{b}\ra^{(N-M)} \la \zb|\zb^\mathbbm{b}\ra^{(M)}\nn \\
 &=& \frac{1}{ \left(\mathcal{N}_\mathbbm{c}^{(N)}(\zb)\right)^2 }2^{1-D} \sum_{\mathbbm{b}} \chi_\mathbbm{c}(\mathbbm{b})  \la \zb|\zb^\mathbbm{b}\ra^{(N)}=1\,,
 \eea
 where Convolution Theorem  for $\mathbb{Z}_2^{D-1}$ has been used again in the second line.$\blacksquare$

It could seem that the Schmidt decomposition provided by Eq. (\ref{SchmidtDecomp}) is ill-defined since the right-hand side of the equation is not invariant under particle permutations. However, it is
easy to show that, due to the group-theoretical properties of parity adapted CS, it is in fact symmetric, guaranteeing that the standard Schmidt decomposition for distinguishable bipartite systems works in our case without the need of modification (see for instance the modification suggested in \cite{Sciara2017} for the case of two indistinguishable qubits).

\subsection{Schmidt rank of DCATs}

 Although there are $2^{D-1}$ terms in the decomposition of a parity adapted CS, not all the eigenvalues
 $\lambda_{\mathbbm{c},\mathbbm{c}'}^{N,M}$ are in general different from zero. There are some general bounds on the rank that we should take into account:
 \begin{itemize}
  \item If some $z_i=0$ in eq. (\ref{cohD}) $ \Rightarrow $ the number of eigenvalues is reduced in a factor 2 if $c_i=0$ (see Sec. \ref{Limitsz})\\ Thus, the minimum Schmidt number (attained when $\zb\rightarrow\bm{0}$) is $2^{k}$, with $k=\|\mathbbm{c}\|_0$ the number of non-zero
  components of the vector $\mathbbm{c}$.

  \item For a given $M$, the dimension of the symmetric representation with $M$ particles is $\tbinom{M+D-1}{M}$.
 \end{itemize}

 From this considerations, the next Corollary follows.

\begin{corolario}

The Schmidt number of the $M$-particle RDM, defined as the rank of $\rho_\mathbbm{c}^{(M)}(\zb)$,  for a $\mathbbm{c}$-$\dcat$ satisfy the bounds:
 \be
{\rm rank}(\rho_\mathbbm{c}^{(M)}(\zb))
=  \min\{2^{\|\zb\|_0+\|\mathbbm{c}_{\mathbb{0}}\|_0},\tbinom{M+D-1}{M}\}\,,
 \ee
with $\|\zb\|_0\leq D-1$ the number of non-zero entries of the vector $\zb$, and where
$\|\mathbbm{c}_{\mathbb{0}}\|_0$ is the  number of nonzero entries of the vector
$\mathbbm{c}_{\mathbb{0}}$, defined as the subset of components of $\mathbbm{c}$ whose indices coincide with the indices of the zero components of $\zb$.
 
\end{corolario}

 In Table \ref{TableDimensions} some examples for the different dimensions are shown, indicating in red the cases where the dimension of the symmetric representation with $M$ particles is smaller than the maximum number of $\dcat$s.

\begin{table}[h]
 \begin{tabular}{|c|c|c|c|}
 \hline
    &  Full tensor product $D^M$ & Symmetric irrep $\tbinom{M+D-1}{M}$  & Maximum number of $\dcat$s $2^{D-1}$ \\
    \hline
  $D=2,M=1$   & 2  & 2 & 2 \\ \hline
  $D=3,M=1$   & 3  & \rojo{3} & 4 \\ \hline
  $D=3,M=2$   & 9  & 6 & 4 \\ \hline
  $D=4,M=1$   & 4  & \rojo{4} & 8 \\ \hline
  $D=4,M=2$   & 16  & 10 & 8 \\ \hline
  $D=5,M=1$   & 5  & \rojo{5} & 16 \\ \hline
  $D=5,M=2$   & 25  & \rojo{15} & 16 \\ \hline
  $D=5,M=3$   & 125  & 35 & 16 \\ \hline
 \end{tabular}
 \caption{Dimensions of the full tensor product space ${\cal H}_1^{\otimes M}$ and the symmetric representation space ${\cal H}_M$ compared with the maximun number of $\dcat$s, for different values of $D$ and $M$. Red colors refer to cases where the maximum number of $\dcat$s exceeds the dimension of ${\cal H}_M$.}
\label{TableDimensions}
 \end{table}

 As it can be seen, in general, for $D=2$ (qubits) it is enough to consider 1-particle RDMs to account for the two possible $\2cat$s (even and odd). For $D=3$ (qutrits) we need 2-particle RDMs to have enough room to accomodate the four $\3cat$s, and for $D=5$ it is necessary to use 3-particle RDMs to have room for the sixteen $\cat{5}$s. Since $2^{D-1}$ grows exponentially with $D$, whereas the binomial coefficient  $\tbinom{M+D-1}{M}$  grows as a polynomial of degree $M$ in $D$, if we fix $M$, clearly the number of $\dcat$s will
 exceed the maximun rank of the RDMs as $D$ grows. We need to increase also $M$ to have enough room for all possible $\dcat$s in the RDMs.

From the previous discussion,  we can compute the Schmidt rank, i.e. the maximun  Schmidt number for all possible $M$-RDM.
 
 \begin{corolario}
 The Schmidt rank of a $\mathbbm{c}$-$\dcat$  is:
 \be
\hbox{Schmidt rank of } |\zb\ra_\mathbbm{c} = \max_{1\leq M\leq \lfloor \frac{N}{2}\rfloor} {\rm rank}(\rho_\mathbbm{c}^{(M)}(\zb))
= 2^{\|\zb\|_0+\|\mathbbm{c}_{\mathbb{0}}\|_0}  \leq 2^{D-1}\,.
 \ee
 \end{corolario}

 \subsection{$M$-wise Entanglement entropy of parity adapted $U(D)$-spin coherent states}


 The knowledge of the Schmidt coefficients and eigenvalues allows to easily compute the Linear and von Neumann entropies
 \be
 \mathcal{L}(\rho)=\frac{d}{d-1}(1-Tr(\rho^2))\,,\qquad \mathcal{S}(\rho)=- Tr(\rho \log_{d} \rho )
 \label{Entropies}
 \ee
 of the $M$-wise RDM matrices, with $d$ a suitable dimension to normalize the entropies between 0 and 1.

 In our case the chosen dimensions (according to the dimension of the symmetric representation for $M$ particles, when $M$ is finite, and to the maximum rank of the RDM for parity adapted CS) are:
 \be
 d=\left\{\begin{array}{lcr} \tbinom{M+D-1}{M} &,& \qquad \hbox{finite }\, M\,, \\  \\ 2^{D-1} &,& \hbox{ infinite } \, M\,. \end{array} \right.
 \ee

 In Sec. \ref{FigD2} (and in the Supplementary Material)
 we shall show different plots of von Neumann entropy  (line plots for $D=2$, contour plots for $D=3$, angular plots with $\|\zb\|_2=R$ for $D=4$ and information diagrams for arbitrary values of $D$). In these plots we shall observe the main features of the entropy of the $M$-wise RDM of  parity adapted CS, and therefore of the entanglement of these states.

\section{Some interesting limits of the Schmidt eigenvalues}
\label{Limits}

In this section we shall analyse with detail the behaviour of the Schmidt eigenvalues under certain limits ($N\rightarrow\infty$, $M\rightarrow\infty$, $\|\zb\|\rightarrow 0$, $z_i\rightarrow 1$, $\|\zb\|\rightarrow \infty$, etc.). We shall make use of the limit values of the normalization factors
${\cal N}_\mathbbm{c}(\zb)$ computed in Sec. \ref{ParityCS}.

\subsection{Single thermodynamic limit}
\label{sec-RSTL}

In many-body systems, the thermodynamic limit is the limit where the number of particles $N$ grows to infinity. The limit $N\rightarrow\infty$ of the Schmidt eigenvalues has the expression:
 
 \be
 \lim_{N\rightarrow\infty} \lambda_{\mathbbm{c},\mathbbm{c}'}^{N,M}(\zb) = \left(\mathcal{N}_{\mathbbm{c}'}^{(M)}(\zb)\right)^2\,. \label{lambdaTL}
 \ee

It turns out that, in this limit, the Schmidt eigenvalues do not depend on the {original parity}
$\mathbbm{c}$, therefore all $\mathbbm{c}$-$\dcat$s  have the same Schmidt decomposition in the thermodynamic limit. Thus, looking at the RDMs, we cannot infer the parity of the original state in this limit. This confers an universal character to the thermodynamic limit of the Schmidt decomposition, erasing all information about the parity $\mathbbm{c}$ of the original state.

\subsection{Double thermodynamic limit}
\label{RTL}

Another interesting limit is the double thermodynamic limit, when both $N$ and $M$ go to infinity independently. However, this limit is not well-defined. In fact only one of the iterated limits makes sense (since $M\leq N$).
The only possible iterated double thermodynamic limit is:
\be
 \lim_{M\rightarrow\infty}\lim_{N\rightarrow\infty} \lambda_{\mathbbm{c},\mathbbm{c}'}^{N,M}(\zb) =
  2^{-k} \delta_{\mathbbm{c}_0,\mathbbm{c}'_0}\,,
 \ee
where $k=\|\zb\|_0$ is the number of nonzero components of the vector $\zb$, and $\mathbbm{c}_0$ indicates the subset of components of $\mathbbm{c}$ whose indices coincide with the indices of the zero components of $\zb$.

This shows that the $\infty$-RDM of an infinite number of quDits corresponds to a  maximally  mixed state of dimension $2^{k}$.
 
We can also consider the
{rescaled}  {directional} double thermodynamic limit, where both $N$ and $M$ go to infinity but with $M/N=1-\eta$ fixed, and simultaneously the variable $\zb$ is rescaled by $\sqrt{N}$:
\be
 \lim_{\stackrel{N\rightarrow\infty}{M=(1-\eta) N}} \lambda_{\mathbbm{c},\mathbbm{c}'}^{N,M}(\alphab/\sqrt{N}) =
 \prod_{i=1}^{D-1} \lambda^{HO,\eta}_{c_i,c'_i}(\alpha_i) \label{RSTLim}
 \ee
 with
 \be
 \lambda^{HO,\eta}_{c,c'}(\alpha)= \left(\frac{\mathcal{N}_{c-c'}^{HO}(\sqrt{\eta}\alpha) \mathcal{N}_{c'}^{HO}(\sqrt{1-\eta}\alpha) }{ \mathcal{N}_{c}^{HO}(\alpha) }\right)^2
 \ee
with  $\eta\in[\frac{1}{2},1)$. Using the expression of $\mathcal{N}_{c}^{HO}(\alpha)$ given in Sec. \ref{ParityCS}, we arrive at:
 \be
 \lambda^{HO,\eta}_{c,c'}(\alpha)= \frac{{\rm exp}_{c-c'}(\eta |\alpha|^2){\rm exp}_{c'}((1-\eta) |\alpha|^2)}{{\rm exp}_{c}(|\alpha|^2)}
 = \frac{1}{2}+(-1)^{c-c'}\frac{{\rm exp}_{c}((1-2\eta) |\alpha|^2)}{2{\rm exp}_{c}( |\alpha|^2)}  \,.
 \ee

 Note that for $\eta=\frac{1}{2}$ we have $\lambda^{HO,\frac{1}{2}}_{1,c'}(\alpha)=\frac{1}{2}$ (the odd cat state is maximally entangled for all non-zero values of $\alpha$), and for $\eta\rightarrow 1$ we have
 \be
 \lim_{\eta\rightarrow 1}\lambda^{HO,\eta}_{c,c'}(\alpha)=\delta_{c',0}\,,
 \ee
 indicating that the fidelity with respect to the original state approaches 1 as $\eta\rightarrow 1$ (see Sec. \ref{quDitLoss}).

This represents a generalization to higher dimensions of the results of
\cite{Glancy:08} (see also \cite{Enk2000EntangledCS} for the case $\eta=\frac{1}{2}$),
 for {decoherence} of a Schr\"odinger cat state  of the harmonic oscillator (realized with a laser beam) by {photon absorption} modelled by the  passage through
a {beam splitter of transmisivity} $\eta$.

This suggests that in the case of $\dcat$s, the Schmidt decomposition we have obtained can be physically interpreted as a
decoherence process under the loss of $M$-quDits. In this sense, our results  can help in designing quantum systems robust under quDit loss, for instance in quantum error correction protocols (see \cite{Stricker2020,ZANGI2021127322} for the case of qubit loss). We shall further discuss this point in Section \ref{quDitLoss}.

\subsection{Limits when $\|\mathbf{z}\|\rightarrow 0$, $|z_i |\rightarrow 1$, $\|\mathbf{z}\|\rightarrow \infty$}
\label{Limitsz}

In this subsection we shall consider diverse limits of the Schmidt eigenvalues in the variable $\zb$.

%
%
%
 
 The limit at $\zb=\bm{0}=(0,0,\ldots,0)$ is important since it will provide the minimum rank of the RDM. It general its expression is cumbersome and we will only give the cases $D=2$ and $D=3$.

 For $D=2$ we have:
 \be
 \lim_{z\rightarrow 0} \lambda_{c,c'}^{N,M}(z) =
 \frac{M}{N} \delta_{c,c'}+\frac{N-M}{N} \delta_{c',0}\,.
 \ee
 

 Thus at $z=0$ the {Schmidt rank} is 1 (pure state) for the completely {even} ($c=0$) $\2cat$ and 2 for the even case ($c=1$) $\2cat$. Note that this last statement should be understood in the limit sense since for $c\neq 0$ the action of the parity projector onto the highest state (which lies in the completely even subspace) is zero, $\Pi_c|\bm{0}\ra=0$.

 For $D=3$ we have:
 \bea
 \lim_{(z_1,z_2)\rightarrow (0,0)} \lambda_{[0,0],\mathbbm{c}'}^{N,M}(z_1,z_2) &=&
 \left(1,0,0,0\right) \nn\\
 \lim_{(z_1,z_2)\rightarrow (0,0)} \lambda_{[0,1],\mathbbm{c}'}^{N,M}(z_1,z_2) &=&
 \left(\frac{M}{N},\frac{N-M}{N},0,0\right) \nn\\
 \lim_{(z_1,z_2)\rightarrow (0,0)} \lambda_{[1,0],\mathbbm{c}'}^{N,M}(z_1,z_2) &=&
 \left(\frac{M}{N},0,\frac{N-M}{N},0\right) \\
\lim_{(z_1,z_2)\rightarrow (0,0)} \lambda_{[1,1],\mathbbm{c}'}^{N,M}(z_1,z_2) &=&
 \left(\frac{M(M-1)}{N(N-1)},\frac{M(N-M)}{N(N-1)},\frac{M(N-M)}{N(N-1)},\frac{(N-M)(N-M-1)}{N(N-1)}\right) \nn\,,
\eea
where at the right-hand side the vector of Schmidt eigenvalues is shown, ordered according to the decimal expression of $\mathbb{c}'$. From this expression the rank of the RDM at $\zb=\bm{0}$ is easily obtained and generalized to arbitrary $D$, resulting in a rank equal to $2^{\|\mathbb{c}\|_0}$. Note that if some $c_i=0$, in the limit $\zb\rightarrow\bm{0}$ the $\dcat$s with $c_i'=1$ are absent in the Schmidt decomposition.

The expression we have obtained for the Schmidt eigenvalues in the limit $\zb\rightarrow\bm{0}$ provide the Schmidt decomposition of \text{Fock-cat} states appearing in Secs. \ref{Dcatz0} and \ref {DcatzInf}.

 
 The limit when $|z_i|\rightarrow 1\,,\,i=1,\ldots,D-1$ also deserves attention, since at this point the entropy of the RDM takes its maximum value, as can be checked in the graphs shown in the Supplementary Material.
 However, the general analytic expression of the limit  is cumbersome, therefore we shall consider only some special cases.

 For $D=2$ we have:
 \be
 \lim_{|z|\rightarrow 1} \lambda_{c,c'}^{N,M}(z) = \frac{1}{2}\,,\qquad \forall c,c',\forall N,M,\quad 1\leq M\leq \lfloor \frac{N}{2}\rfloor\,.
 \label{z1limitD2}
 \ee
 Then we conclude that for $D=2$ the RDM is maximally mixed at $|z|=1$.

For $D=3$ we have:

\bea
 \lim_{(|z_1|,|z_2|)\rightarrow (1,1)}& &\lambda_{\mathbbm{c},\mathbbm{c}'}^{N,M}(\zb) =  \label{z1limitD3}\\
& & \frac{ (3^{N-M}+(-1)^{c_1 - c_1'} + (-1)^{c_2 - c_2'} + (-1)^{c_1 + c_2 - c_1' - c_2'  + N-M})
 (3^M+(-1)^{c_1'} + (-1)^{c_2'} + (-1)^{c_1' + c_2' + M} )  }{4 (3^N+(-1)^{c_1} + (-1)^{c_2} + (-1)^{c_1 + c_2 + N})  }
 \,, \nonumber
 \eea

$\forall c,c',\forall N,M1\quad 1\leq M\leq \lfloor \frac{N}{2}\rfloor$.
In particular we have for $M=1$:
\be
 \lim_{(|z_1|,|z_2|)\rightarrow (1,1)} \lambda_{\mathbbm{c},[1,1]}^{N,1}(\zb) =0
 \,,\qquad \forall \mathbb{c},\forall N>1\,,
 \label{z1limitD3M1-0}
 \ee
which is due to the fact that ${\cal N}_{[1,1]}(\zb)^{(1)}=0\,,\forall \zb\in\mathbb{C}^2$.
Also, in this case for the other values of $\mathbbm{c}'\neq [1,1]$ and $k>1$:
\bea
 \lim_{(|z_1|,|z_2|)\rightarrow (1,1)} \lambda_{[0,0],\mathbbm{c}'}^{2k,1}(\zb) &=&\frac{1}{3}\,, \\
 \lim_{(|z_1|,|z_2|)\rightarrow (1,1)} \lambda_{[1,1],\mathbbm{c}'}^{2k+1,1}(\zb) &=&\frac{1}{3}\,.
 \label{z1limitD3M1}
 \eea

In all other cases the nonzero eigenvalues are practically  $\frac{1}{3}$, approaching $\frac{1}{3}$ for large odd $N$. Then we conclude that for $M=1$ the RDM is (approximatelly) maximally mixed when $(|z_1|,|z_2|)=(1,1)$.

 Similar conclusions can be obtained for larger values of $D$, although the expressions are cumbersome. Then we can conclude that at the point $|z_i|=1,\,i=1,\ldots,D-1$, the rank of the $M$-wise RDM is
${\rm min}\{\tbinom{M+D-1}{M},2^{D-1}\}$.



 The $\|\zb\|\rightarrow \infty$ limit {does not exist} (except for the case $D=2$, see below), since its value depends on the {direction}.
 Using {hyper-spherical coordinates} in the first hyper-octant, $|z_i|=r y_i(\vec{\theta})$, $\vec{\theta}\in [0,\frac{\pi}{2}]^{D-2}\,,i=1,2,\ldots D-1$, we can compute the directional limits:

 \be
 \lim_{r\rightarrow \infty}  \lambda_{\mathbbm{c},\mathbbm{c}'}^{N,M}(r,\vec{\theta}) =
 2^{1-D} \frac{ \sum_{\mathbbm{b}} \chi_{\mathbbm{c}-\mathbbm{c}'}( \mathbbm{b}) Y_{\mathbbm{b}}(\vec{\theta})^{N-M}  \sum_{\mathbbm{b}} \chi_{\mathbbm{c}'}( \mathbbm{b}) Y_{\mathbbm{b}}(\vec{\theta})^{M}}{\sum_{\mathbbm{b}} \chi_{\mathbbm{c}}( \mathbbm{b}) Y_{\mathbbm{b}}(\vec{\theta})^{N}}\,, \label{zinf-angdep}
\ee

 with
 $Y_{\mathbbm{b}}(\vec{\theta})=\sum_{i=1}^{D-1} (-1)^{b_i}y_i(\vec{\theta})^2$.

 \vspace{0.5cm}
 For instance, for  $D=3$ we have polar coordinates $Y_{\mathbbm{b}}(\vec{\theta})=(-1)^{b_1}\cos^2\theta + (-1)^{b_2}\sin^2\theta$\,.
 
The case $D=2$ deserves special attention, since the limit exists, but care should be taken since in some particular cases undetermined limits can appear. The result is:
 \be
 \lim_{|z|\rightarrow \infty}  \lambda_{c,c'}^{N,M}(z) =
 \frac{1}{2}\left[  1+ \frac{N-M}{N} (-1)^{c'+M} + \frac{M}{N} (-1)^{c-c'+N-M}\right]\,.
 \label{zinf-D2}
\ee

\section{Physical applications}
\label{PhysApp}

Quantum superpositions of macroscopically distinct quasi-classical states (the so-called Schr\"odinger cat states) are an important resource for quantum metrology, quantum communication and quantum computation. In particular, superpositions $\frac{1}{2{\cal N}(\alpha)}\left(|\alpha\ra\pm |e^{{\rm i}\phi}\alpha\ra\right)$ of harmonic  oscillator CS $|\alpha\ra$ with the same $|\alpha|$ but with different phases (like the even $\phi=0$ and odd $\phi=\pi$ parity adapted CS discussed here) are a common resource in a large variety of experiments (see for instance the encode of a logical qubit in the subspace generated by this kind
of superpositions, which is protected against phase-flip errors  \cite{Grimm2020,PhysRevX.9.041009}).

In this section we shall discuss how the $\dcat$s introduced in this paper can be generated using different Hamiltonians, and how the Schmidt decomposition found here can be usefull to study the interesting problem of quDit loss.

\subsection{$\dcat$s generation }
\label{DcatGeneration}

In \cite{Glancy:08} different methods of producing optical Schr\"odinger cats for the harmonic oscillator were discussed, and some of them have been realized experimentally \cite{KittenGeneration}.
The experimental creation of optical Schr\"odinger
cat states in cavity QED is discussed in \cite{Haroche1995}.
A Schr\"odinger cat state of an ion in a trap has been generated expermentally \cite{SchrodingerCatAtom}, and a Schr\"odinger cat state formed by two interacting Bose condensates of atoms in different internal
states (two-well) has been proposed \cite{PhysRevA.57.1208}.
Another proposal is the generation of  optomechanical Schr\"odinger cat states in a cavity Bose-Einstein condensate \cite{LI2022}, with a considerable enhancement in the size of the mechanical Schr\"odinger cat state.

An important tecnique to produce Schr\"odinger cat states is the use of Kerr o Kerr-like media, like
in \cite{Grimm2020,PhysRevX.9.041009}, which allows to create, control and measure a qubit in the subspace
generated by various Schr\"odinger cat states, which is protected against phase-flip errors.

Another Hamiltonian where Schr\"odinger cat states appear is the Lipkin-Meshkov-Glick (LMG) nuclear model \cite{lipkin1}. In this case they appear as the ground state solution in the thermodynamic limit,
or as approximate solutions for the lowest eigenstates of the Hamiltonian for finite $N$.
The LMG model has been also realized in circuit QED scheme  \cite{Larson_2010}, and proposed for
optimal state prepration with collective spins \cite{StatePreparationCollectiveSpins}.

\subsubsection{LMG D-Level model}
\label{LMG}

It is surprising that in the literature, when  multimode (with $D>2$ modes) systems are considered, the only studied cat states are the ones associated to the total parity subgroup $\mathbb{Z}_2 \subset\mathbb{Z}_2^{D-1}$ (the one formed by the parity transformations in Eqns. (\ref{ParityResolutionIdentity})-(\ref{TotalParity}), see for instance \cite{Manko-MultimodeCats1995,Fastovets2021}.
However, there are models, like the $D$-level LMG model, where the  Hamiltonian is invariant under parity transformations, where the lowest energy eigenstate and some of the first excited states are parity adapted CS.
More precisely
in the limit where the interaction parameter $\lambda\rightarrow\infty$ in the LMG $D$-level model for a finite number of particles, the lowest energy state is approximatelly (with a high fidelity) a  completelly even $\dcat$ with $\zb=(1,1,\ldots,1)$ \cite{nuestroPRE}, which corresponds to a  maximally entangled state among all parity adapted CS, as was shown in Eqns. (\ref{z1limitD3})-(\ref{z1limitD3M1}). Also, some of the lower excited energy eigenstates are also $\mathbbm{c}$-$\dcat$s with different parities $\mathbbm{c}$.

\subsubsection{Generation by Kerr-like effect}

The optical Kerr effect is a universal technique to generate non-classical states in quantum optics.
In \cite{kirchmair2013} multicomponent Schr\"odinger cat states were generated in a circuit QED where an intense artifical Kerr effect is created, allowing for single-photon Kerr regime.

In \cite{Mirrahimi_2014}, a logical qubit is encoded in two or four harmonic oscillator Schr\"odinger cat states of a microwave cavity, of the form $\frac{1}{2{\cal N}(\alpha)}(|\alpha\ra \pm |-\alpha\ra)$ and $\frac{1}{2{\cal N}(\alpha)}(|{\rm i}\alpha\ra \pm |-{\rm i}\alpha\ra)$, realizing the $\pi/2$ rotation around the $z$-axis by means of Kerr effect, and exploiting multi-photon driven dissipative processes. A two-photon driven dissipative process is used to
stabilize a logical qubit basis of two-component Schrödinger cat states against photon dephasing errors, while a  four-photon driven dissipative process stabilizes a logical qubit in
four-component Schr\"odinger cat states, which is protected against single-photon loss.

In \cite{Grimm2020}  a logical qubit is created in the subspace of four Schr\"odinger cat states, $\frac{1}{2{\cal N}(\alpha)}(|\alpha\ra \pm |-\alpha\ra)$ and $\frac{1}{2{\cal N}(\alpha)}(|\alpha\ra \pm {\rm i} |-\alpha\ra)$, using a Hamiltonian that encompasses both Kerr effect and squeezing in a superconducting microwave resonator. The created qubit is protected under phase-flip errors.

Kerr-like effect in $U(2)$ and $U(D)$ can also be exploited to create Schr\"odinger cat states, using Hamiltonians of the type $H_{\rm kerr} =\chi J_z^2$ ($J_z=\frac{1}{2}\left(S_{22}-S_{11}\right)$ is the third component of the angular momentum operator) for $SU(2)$ or
\be
H_{\rm kerr} = \chi\sum_{j=1}^{D-1} S_{jj}^2
\ee
for the case of $U(D)$. Note that this Hamiltonian is a particular case of the interacting term of the   LMG Hamiltonian for general $D$
\cite{nuestroPRE},
which, in the rescaled double thermodinamic limit, approaches the (multimode) Kerr Hamiltonian plus a squeezing term (see \cite{PhysRevB.48.3190}).

Denoting the revival time as $T_{\rm rev}=\frac{2\pi}{\chi}$, for times $t=T_{\rm rev}/q$,
with $q=2,3,\ldots$,  this kind of Hamiltonians produce  multicomponent Schr\"odinger cat states, where
the number of components is $q$ for odd $q$ and $\frac{q}{2}$ for even $q$ (see \cite{Arjika} for the case of $SU(2)$ and the harmonic oscillator).

\subsection{QuDit loss}
\label{quDitLoss}

A commented in Sec. \ref{sec-RSTL}, when the rescaled directional double thermodynamic limit was discussed and the results in the literature for the study of photon loss were recovered (and generalized to a larger number or harmonic oscillators, or polycromatic lasers), we can infer that the Schmidt decomposition of parity adapted CS in terms of a sum for all parities of tensor products of parity adapted CS with smaller number of particles, really corresponds to the physical process of quDit loss, when some quDits of a symmetric multi-quDit state are lost by some irreversible process (like decoherence by interaction with the environment or other similar process), and that this process is mathematically described by the partial trace (see, for instance \cite{PhysRevA.98.062335}).

This suggests that in the case of $\mathbbm{c}$-$\dcat$s, the Schmidt decomposition we have obtained  can be used to describe a
decoherence process under the loss of $M$-quDits. In this sense, our results  can help in designing quantum error correction protocols (see \cite{Stricker2020,ZANGI2021127322} for the case of qubit loss).

Our Schmidt decomposition can also be useful in the generalization to quDits of the study of \textit{robustness of entanglement under qubit loss} \cite{PhysRevA.98.062335}. In this context, robustness is defined
as the survival of entanglement after the loss of $M$ qubits, i.e. the RDM after tracing out $M$ qubits, $\rho^{(N-M)}_{\,\mathbbm{c}}(\zb)$ (which describes in general a mixed state), is entangled.

In our case, as it can be checked in the limit values discussed in Sec. \ref{Limits} and in the Figures in Sec. \ref{FigD2} (and in the Supplementary Material), the rank of the RDM is larger than one except possibly in the cases
$\|\zb\|\rightarrow 0$ and $|z_i|\rightarrow \infty$ (i.e, along some of the axes), and the entropy of the RDM is lower than the NEMSs (Not Entangled Mixed States) limit except in the cases marked in red in Table \ref{TableDimensions} (changing $M$ by $N-M$), when maximally mixed (not entangled) RDM can appear.

The limits of the $\dcat$s when $\|\zb\|\rightarrow 0$ and $|z_i|\rightarrow \infty$ were studied in Secs. \ref{Dcatz0} and \ref{DcatzInf}, and they turn to be Fock states and in some particular cases they are ITPS. Thus, except for the cases where the original state is an ITPS (and therefore it is not entangled), the rank of the resulting RDM is larger than one and $\rho^{(N-M)}_{\,\mathbbm{c}}(\zb)$ is entangled (except in the cases marked in red in Table \ref{TableDimensions}).

Therefore, we can guarantee that, except in the mentioned cases, the entanglement of the original $\dcat$ is robust under quDit loss. This should be compared with the case of GHZ or NOON states, which are maximally entangled but they are fragile under qubit loss \cite{PhysRevA.98.062335}. In fact, spin cat states \cite{Huang2015} with moderate entanglement can reach the standard quantum limit even in the presence of a relative large amount of qubit loss, whereas GHZ states lose this capability with an small fraction of qubit loss.

Motivated by the example discussed in  \cite{Enk2000EntangledCS,Glancy:08}, where in the case of photon loss when $\eta\rightarrow 1$ the fidelity with the original state approaches one, we would like to introduce the concept of fidelity also in our case. Strictly speaking, the fidelity
\be
F_\mathbbm{c}^{N,M}(\zb)\equiv {}^{(N)}{\!\!\!\!}_\mathbbm{c}\la\zb|\rho^{(N-M)}_{\,\mathbbm{c}}(\zb)|\zb\ra{}^{(N)}_\mathbbm{c} \,,
\ee
makes sense only in the case $N\rightarrow\infty$, since otherwise the number of particles do not match and the expectation value makes no sense. Alternatively, we can define the fidelity for finite $N$ as:
\be
F_\mathbbm{c}^{N,M}(\zb)\equiv \lambda_{\mathbbm{c},\mathbb{0}}^{N,M}(\zb)\,,\label{fidelity}
\ee
i.e., as the component in the Schmidt decompostion where the $(N-M)$ subsystem has the same parity as the original system, and this, according to our definition, corresponds to $\mathbbm{c}'=\mathbb{0}$ in Eqn. (\ref{SchmidtDecomp}).

This definition of fidelity (and its maximization with respect to $\zb$) could be interesting in some protocols where it is
important that the parity $\mathbbm{c}$ of the state should be robust under quDit loss. In other situations, however, it could be interesting the fact that the state has a (quasi) definite parity under quDit loss, without worrying about its value. In this case, $\lambda_{\mathbbm{c},\mathbbm{c}'}^{N,M}(\zb)$ should be maximized for all possible values of $\mathbbm{c}'$, too.
And in other situations, it could be interesting to maximize a balanced combination of robustness $+$ fidelity.


\section{Figures for $D=2$ atom levels (qubits)}
\label{FigD2}

\subsection{Finite number $N$ of qubits}

\begin{center}
\begin{figure}[h!]
\includegraphics[width=\graphwidth]{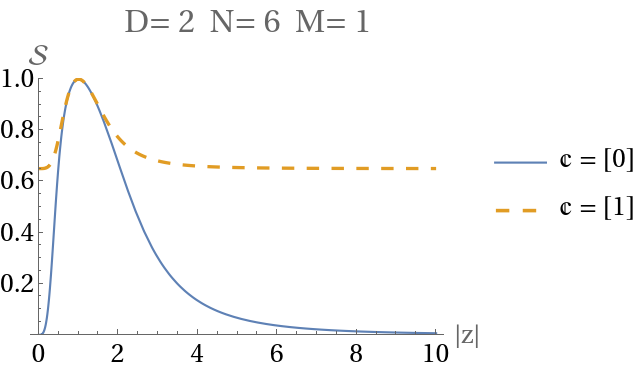}\hspace{\graphsep}\includegraphics[width=\graphwidth]{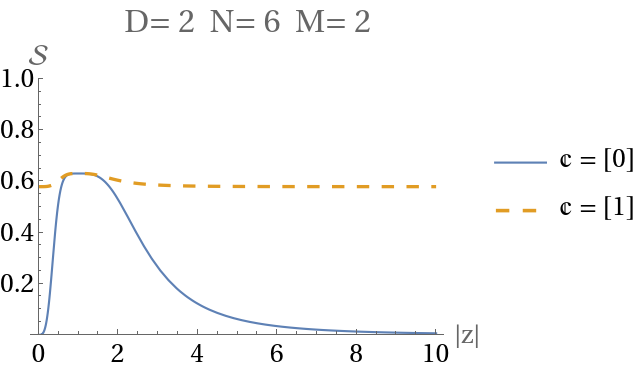}\hspace{\graphsep}
\includegraphics[width=\graphwidth]{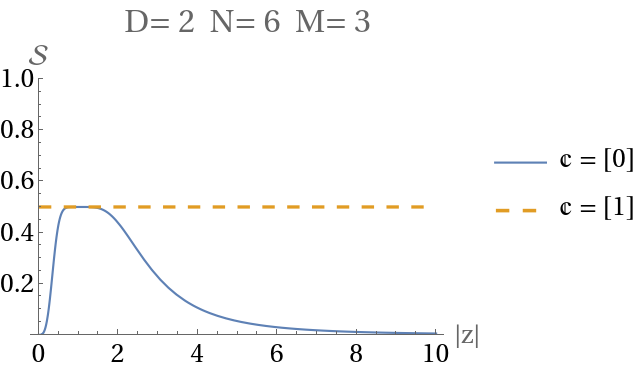}\hspace{\graphsep}
\caption{Plots of von Neumann entropy for $D=2, \ N=6$ and $M=1,2,3$, for $|z|\in [0,10]$.}
\label{NTL-2-6}
\end{figure}
\end{center}

\begin{center}
\begin{figure}[h!]
\includegraphics[width=\graphwidth]{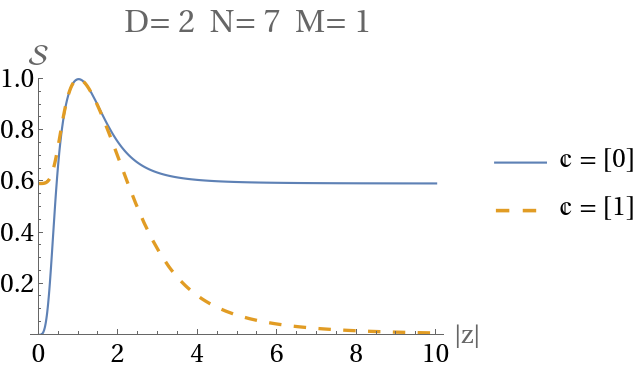}\hspace{\graphsep}\includegraphics[width=\graphwidth]{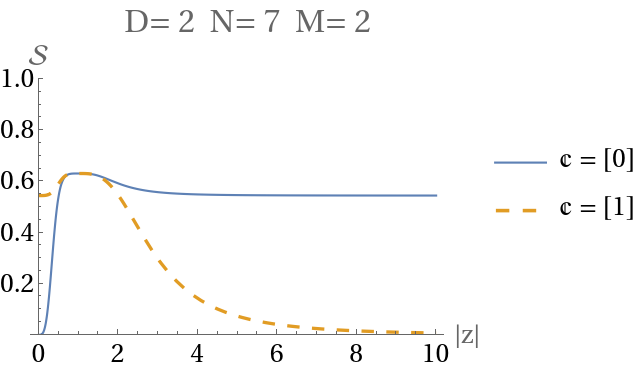}\hspace{\graphsep}
\includegraphics[width=\graphwidth]{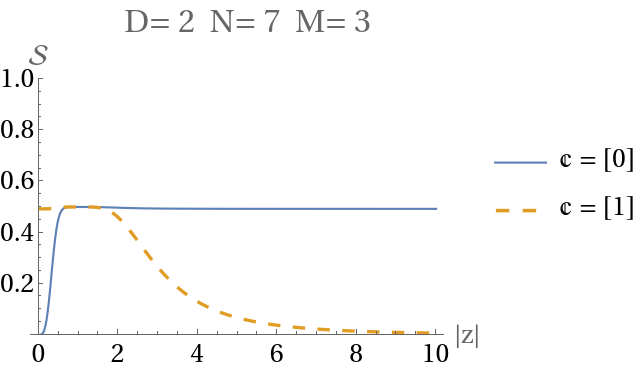}\hspace{\graphsep}
\caption{Plots of von Neumann entropy for $D=2, \,N=7$ and $M=1,2,3$, for $|z|\in [0,10]$.}
\label{NTL-2-7}
\end{figure}
\end{center}

In this section, in Figures \ref{NTL-2-6} ($N=6$) and \ref{NTL-2-7} ($N=7$)   we show  plots of the normalized von Neumann entropy (see Eq. (\ref{Entropies})) of $M$-wise RDMs of the even ($\mathbbm{c}=[0]$)  and odd ($\mathbbm{c}=[1]$) $\mathbbm{c}$-$\2cat$ as a function of $|z|$, for values in the range $[0,10]$ . We can observe that all normalized entropies  reach the maximum for $|z|=1$, with a value 1, i.e. the $\2cat$ is maximally entangled, for $M=1$.

In the case of an even number of particles $N$, for the even parity ($\mathbbm{c}=[0]$) $\2cat$ the entropy is zero (i.e. the $\2cat$ is
a separable pure state) at $z=0$ and for large $|z|$, whereas for the odd ($\mathbbm{c}=[1]$) $\2cat$ both at $z=0$ and for large $|z|$ the entropy takes the same non-zero value, approaching $\frac{1}{2}$ when $\frac{M}{N}$ approaches $\frac{1}{2}$.

For $N$ odd, the behaviour is similar but the even and odd $\mathbb{c}$ cases get interchanged for large $|z|$.

\subsection{Single thermodynamic limit}

\begin{center}
\begin{figure}[h!]
\includegraphics[width=\graphwidth]{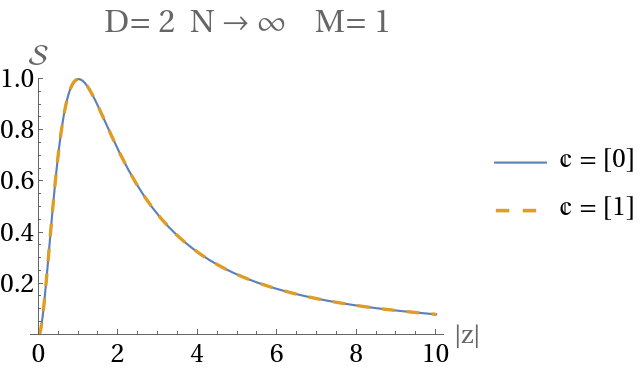}\hspace{\graphsep}\includegraphics[width=\graphwidth]{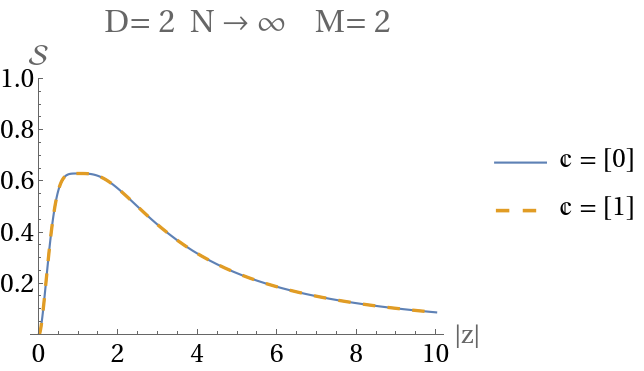}\hspace{\graphsep}
\includegraphics[width=\graphwidth]{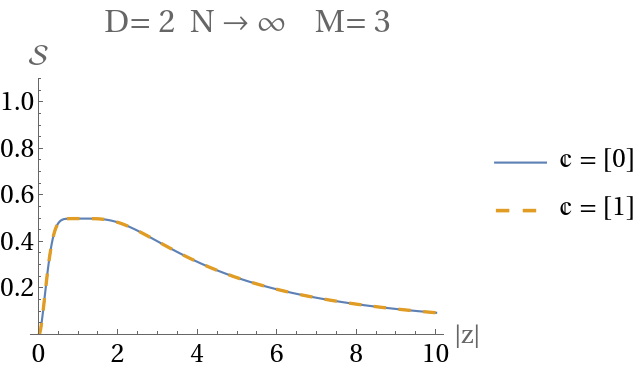}\hspace{\graphsep}
\caption{Plots of von Neumann entropy for $D=2, \ N=6$ and $M=1,2,3$, for $|z|\in [0,10]$.}
\label{TL-2}
\end{figure}
\end{center}

In this section, in Figure \ref{TL-2}    we show  plots of the normalized von Neumann entropy of RDMs of the even ($\mathbbm{c}=[0]$)  and odd ($\mathbbm{c}=[1]$)  $\mathbbm{c}$-$\2cat$ as a function of $|z|$, for values in the range $[0,10]$, in the case $N\rightarrow\infty$. We can observe that all normalized entropies  reach the maximum for $|z|=1$, with a value  of 1 (i.e. the 2cats are maximally entangled) for $M=1$, and approach zero when $|z|$ approach zero or grows to infinity. In this case the entropies coincide for both parities, agreeing with Eq. (\ref{lambdaTL}).

\subsection{Rescaled double thermodynamic limit}

\begin{center}
\begin{figure}[h!]
\includegraphics[width=\graphwidth]{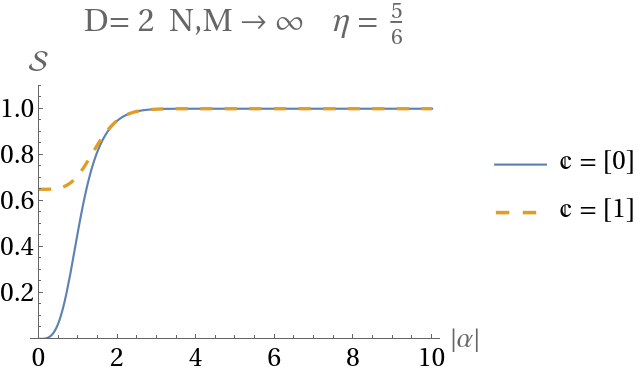}\hspace{\graphsep}\includegraphics[width=\graphwidth]{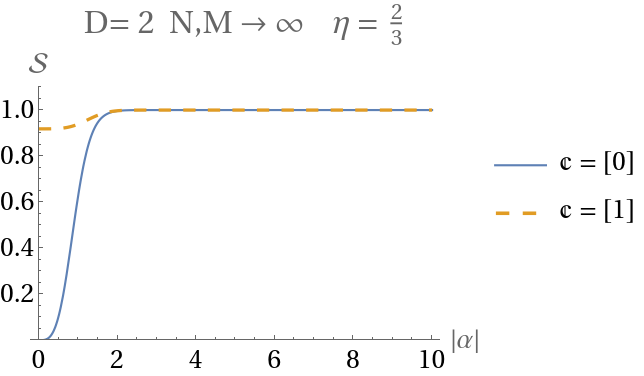}\hspace{\graphsep}
\includegraphics[width=\graphwidth]{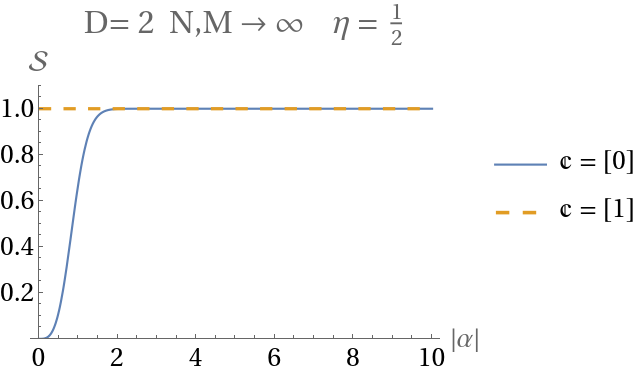}\hspace{\graphsep}
\caption{Plots of von Neumann entropy for $D=2, \ N=6$ and $M=1,2,3$, for $|z|\in [0,10]$.}
\label{RTL-2}
\end{figure}
\end{center}

Finally, in Figure \ref{RTL-2}    we show  plots of the normalized von Neumann entropy of RDM of the even ($\mathbbm{c}=[0]$)  and odd ($\mathbbm{c}=[1]$)  $\mathbbm{c}$-$\2cat$ as a function of $|z|$, for values in the range $[0,10]$, in the case $N,M\rightarrow\infty$ with $M=(1-\eta)N$. We can observe that for the even $\2cat$ the normalized entropy  reach the maximum of 1 when  $|z|$ grows to infinity, and approach zero when $z$ approach zero. For the odd $\2cat$, the normalized entropy  reach the maximum of 1 when  $|z|$ grows to infinity, but approach a non-zero value when $|z|\rightarrow 0$, and this value approach 1
when $\eta$ approach $\frac{1}{2}$, agreeing with the results of Sec. \ref{RTL}.

\section{Conclusions}
\label{Conclusions}

In this paper we provide a thorough discussion of the entanglement properties of symmetric $N$-quDit systems described by parity adapted CS for $U(D)$ ($\mathbbm{c}$-$\dcat$s), in terms of the entropy of the $M$-wise RDM and proving a Schmidt decomposition theorem under a bipartition of the system in terms of $M<N$ and $N-M$ particles (quDits).

We show that the Schmidt decomposition turns out to be a sum over all possible parities of tensor products of parity adapted CS with smaller number of particles. This Schmidt decomposition is well-defined even though we are treating with indistinguishable particles, the reason being the constraints
imposed by the group-theoretical properties of the parity adapted states.

The properties of the Schmidt eigenvalues have been studied for different limit values and different thermodynamic limits, reproducing, in the case of the rescaled double termodynamic limit, known results in the literature for photon loss.
This suggests that the obtained Schmidt decomposition and entanglement properties  could be useful in designing quantum information and computation  protocols with parity adapted CS for quDits of arbitrary $D$, and studying their decoherence properties under quDit loss.

Possible generalizations of this work in different directions are under study. One of them is considering different transformation groups generalizing the parity group $\mathbb{Z}_2^{D-1}$, for instance $\mathbb{Z}_n^{D-1}$ with $n>2$ (an anisotropic version $\mathbb{Z}_{n_1}\times \ldots\mathbb{Z}_{n_{D-1}}$ could also be considered). See \cite{PhysRevA.93.062323} for a the particular case of the one-mode harmonic oscillator.

Another possible generalization is to consider mode entanglement instead of particle entanglement, i.e. considering a bipartition of different modes or levels, for instance $D-K$ and $K$, with $0<K<\lfloor \frac{D}{2}\rfloor$. In this case, it is expected that a similar result for the Schmidt decomposition should hold but the decomposition  involving parity adapted CS of $U(D-K)$ and $U(K)$. Interlevel entanglement for the case $K=1$ has already been discussed in \cite{QIP-2021-Entanglement} for $K=1$.

 \section*{Acknowledgments}
We thank the support of the Spanish MICINN  through the project PGC2018-097831-B-I00 and  Junta de Andaluc\'\i a through the projects  UHU-1262561, FEDER-UJA-1381026 and FQM-381.
AM thanks the Spanish MIU for the FPU19/06376 predoctoral fellowship and AS thanks Junta de Andaluc\'\i a
for a contract under the project FEDER-UJA-1381026.

\bibliography{/home/guerrero/MEGA/Genfimat/Bibliografia/bibliografia.bib}


%

%
%
%
%
%
%
%
%
%
%

\end{document}


\title{Supplementary Material for Entanglement measures  in parity adapted coherent states for symmetric multi-quDits}

\author{Julio Guerrero}
\email{jguerrer@ujaen.es: corresponding author}
\affiliation{Department of Mathematics, University of Ja\'en, Campus Las Lagunillas s/n, 23071 Ja\'en, Spain}
\affiliation{Institute Carlos I of Theoretical and Computational Physics (iC1), University of  Granada,
Fuentenueva s/n, 18071 Granada, Spain}
\author{Antonio Sojo}
\email{aslgta4@gmail.com}
\affiliation{Department of Mathematics, University of Ja\'en, Campus Las Lagunillas s/n, 23071 Ja\'en, Spain}
\author{Alberto Mayorgas}
\email{albmayrey97@ugr.es}
\affiliation{Department of Applied Mathematics, University of  Granada,
Fuentenueva s/n, 18071 Granada, Spain}
\author{Manuel Calixto}
\email{calixto@ugr.es}
\affiliation{Department of Applied Mathematics, University of  Granada,
Fuentenueva s/n, 18071 Granada, Spain}
\affiliation{Institute Carlos I of Theoretical and Computational Physics (iC1), University of  Granada,
Fuentenueva s/n, 18071 Granada, Spain}

\begin{abstract}
 In this Supplementary Material we provide graphics showing the entanglement properties of the $\dcat$s for higher values of $D$, in particular for qutrits, ququatrits and qupentits. For this purpose, we use, in addition to contour plots of von Neumann entropy of RDMs for $D=3$, or angular plots for the large $\|z\|$ behaviour in the case $D=4$, the important tool of information diagrams, consisting in plots of pairs of entropies, like linear and von Neumann entropies.
\end{abstract}

\maketitle

\section{Information Diagrams}
\label{InformationDiagrams}

Information diagrams are an information theoretical tool to study entropic properties of probability distributions, or density matrices\footnote{Since the entropies we shall use here depend on the eigenvalues of the density matrices, whose trace is equal to one, we can identify probability distributions with density matrices from the point of view of the entropies.}.
They were introduced to classify probability distributions according to the relation between two different
information measures, like von Neumann entropy and error probability \cite{IEEETransactions-1994}, or von
Neumann and linear entropies \cite{ie3trans2001-HarremoesTopsoe}. We shall restrict to this last type of information diagrams, although any two pair of entropies or information measures can be used \cite{JPhysA.36.12255}. In this last paper, a detailed study of the boundaries of information diagram is performed, and it is shown that for any pairs of entropies satisfying certain properties (satisfied by most entropies or information measures used in the literature) the probability distributions laying at the boundaries of the information diagram are always the same, assigning an universal character to these probability distributions. In a recent paper \cite{Guerrero22}, a detailed study of information diagrams were performed, and new intraboundaries were found that divided regions with different behaviour with respecto to the rank of the density matrices (or non-zero components of the probability distributions). These properties were used to
study the entropic properties of phase transitions in the 3-level Lipkin-Meshkov-Glick model (see \cite{nuestroPRE,QIP-2021-Entanglement} and references therein).

%

\begin{figure}[h]
\begin{center}
\includegraphics[width=10cm]{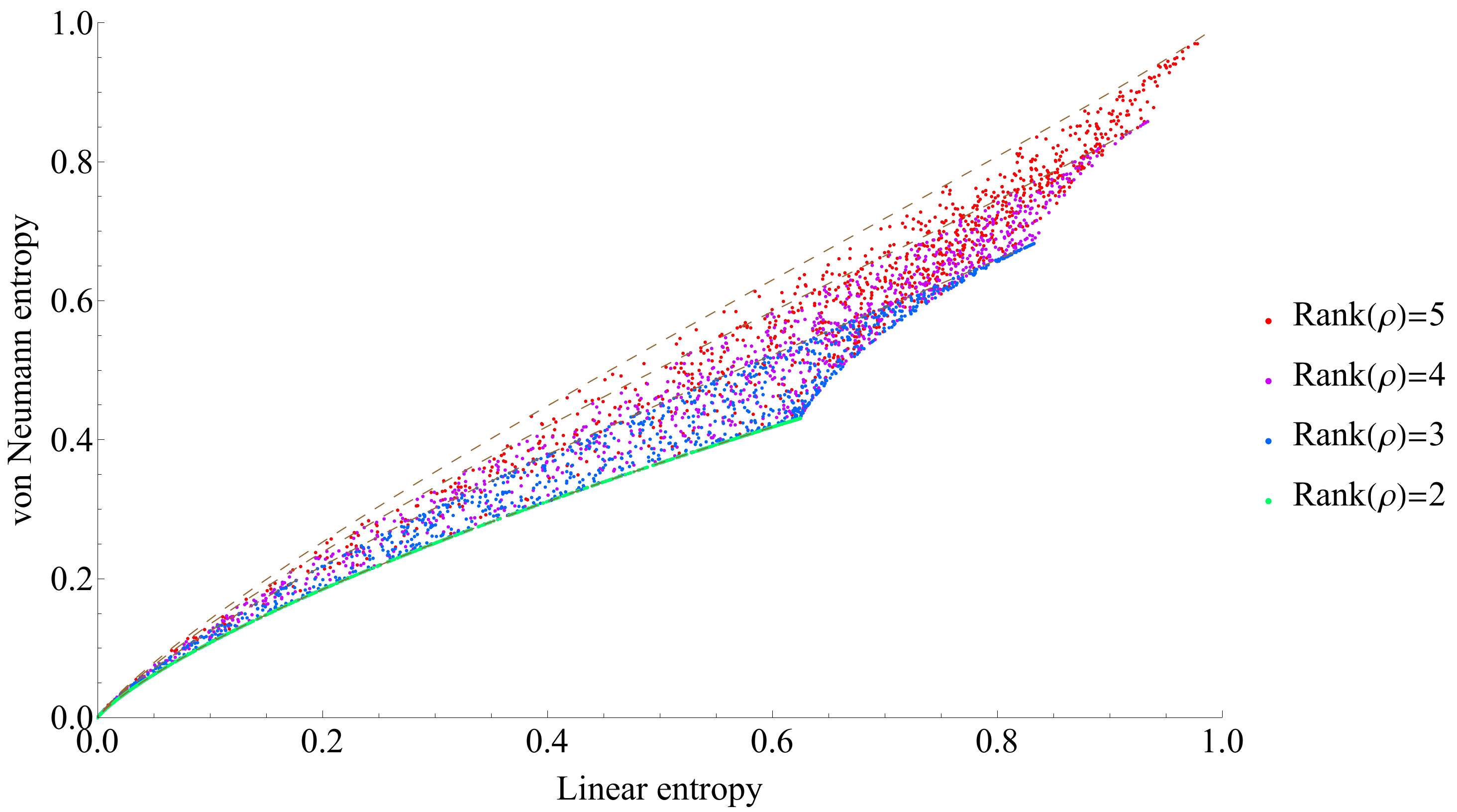}
\end{center}
\caption{Colored plot of  a sample of 20000 density matrices
of dimension $d=5$ randomly generated following a $\chi^2$ distribution for the eigenvalues in an information diagram where the different colors represent the rank of the density matrix (warmer colors represent higher ranks). Repreduced
from \cite{Guerrero22}. 
}
\label{Rank}
\end{figure}

In an information diagram we plot, for each density matrix, a point in the plane $(\cL,\cS)$. All points then lie in a
region bounded by some curves, as can be seen in Figure \ref{Rank}, where random density matrices of dimension $d=6$ have been plot with different colors according to their rank.

\section{Figures for $D=3$ atom levels (qutrits)}
\label{FigD3}

\subsection{Finite number $N$ of qutrits}

\begin{center}
\begin{figure}[h!]
\includegraphics[width=\graphwidth]{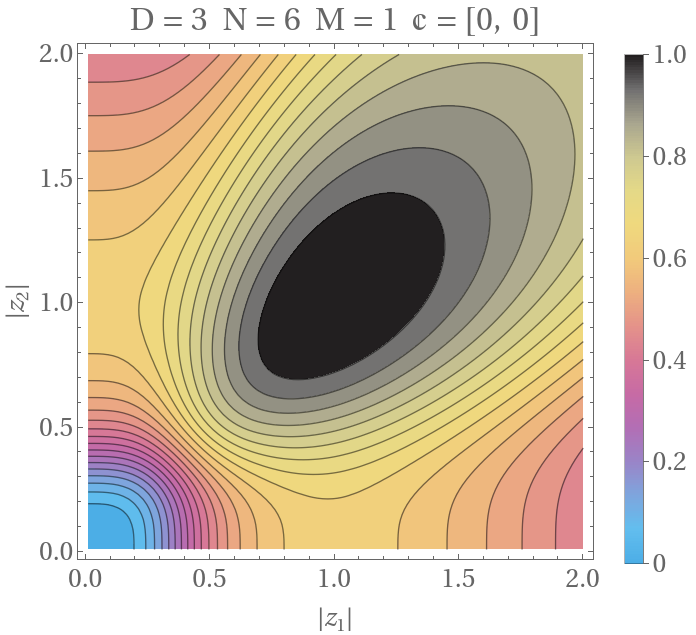}\hspace{\graphsep}\includegraphics[width=\graphwidth]{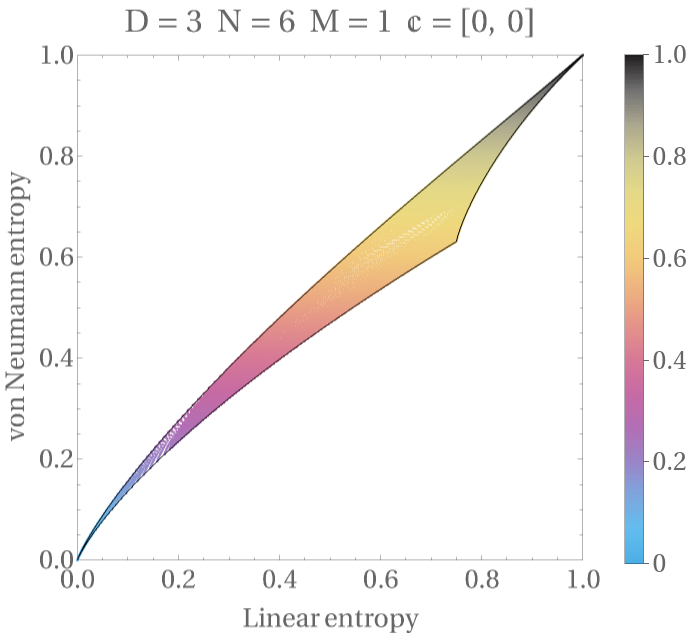}\hspace{\graphsep}
\includegraphics[width=\graphwidth]{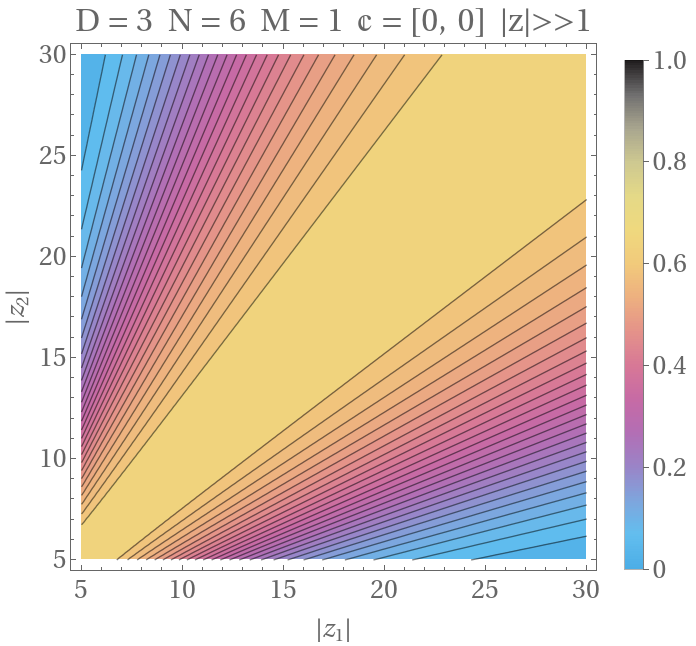}\hspace{\graphsep}\includegraphics[width=\graphwidth]{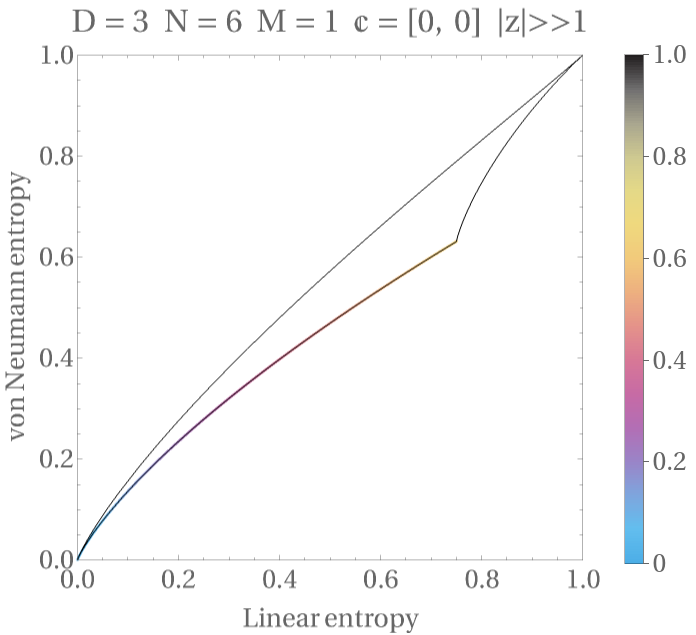}
\caption{Contour plots of von Neumann entropy and information diagrams for $D=3, \quad N=6, \quad  M=1, \quad \hbox{and}\quad \mathbbm{c}=[0,0]$, for $|z_i|\in [0,2]$ (left) and $|z_i|\in [5,30]$ (right).}
\label{NTL-3-6-1-00}
\end{figure}
\end{center}
\begin{center}
\begin{figure}[h!]
\includegraphics[width=\graphwidth]{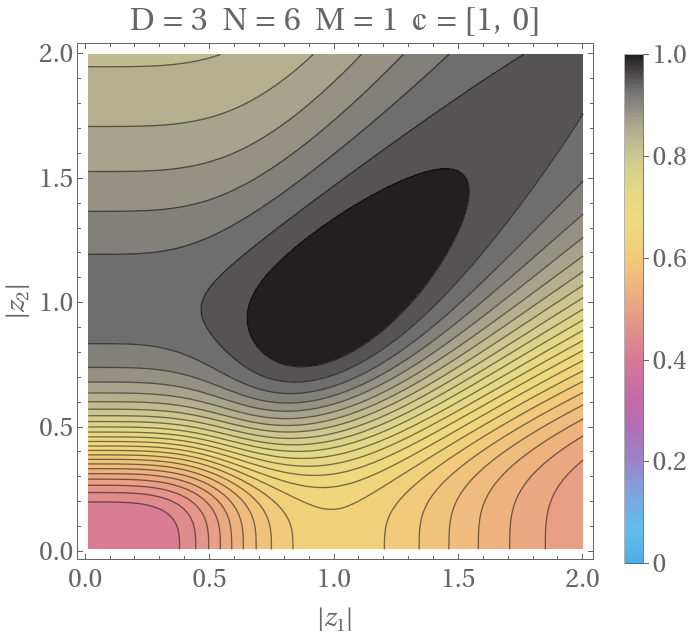}\hspace{\graphsep}\includegraphics[width=\graphwidth]{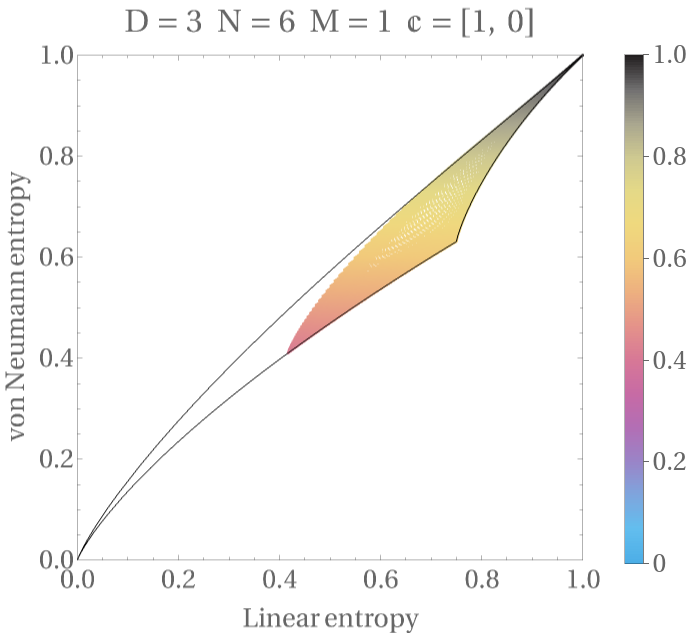}\hspace{\graphsep}
\includegraphics[width=\graphwidth]{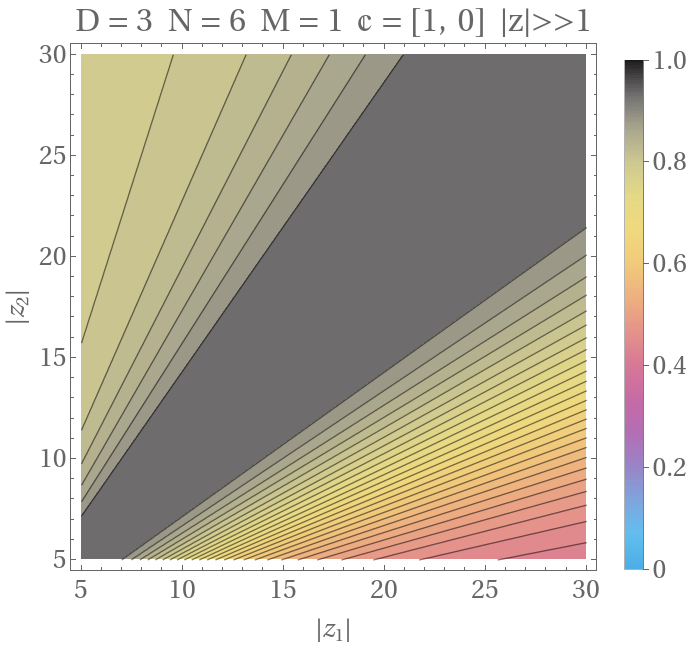}\hspace{\graphsep}
\includegraphics[width=\graphwidth]{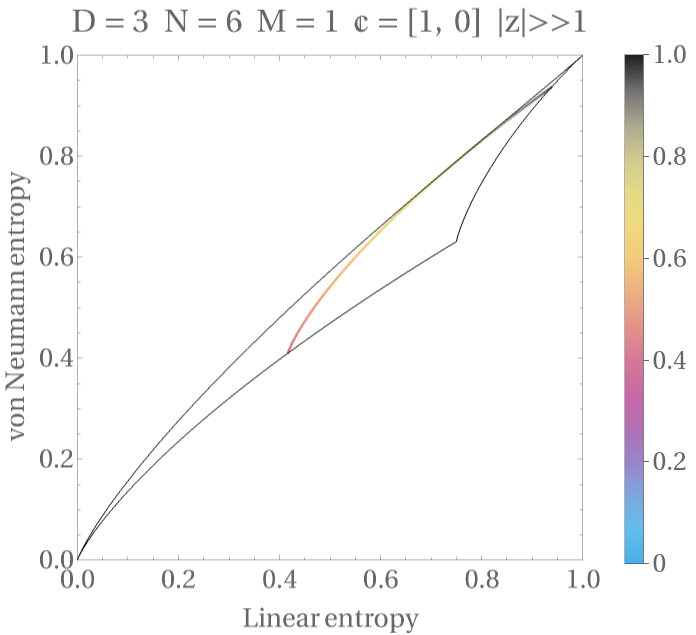}
\caption{Contour plots of von Neumann entropy and information diagrams for $D=3, \quad N=6, \quad  M=1, \quad \hbox{and}\quad \mathbbm{c}=[1,0]$, for $|z_i|\in [0,2]$ (left) and $|z_i|\in [5,30]$ (right).}\label{NTL-3-6-1-10}
\end{figure}
\end{center}
\begin{center}
\begin{figure}[h!]
\includegraphics[width=\graphwidth]{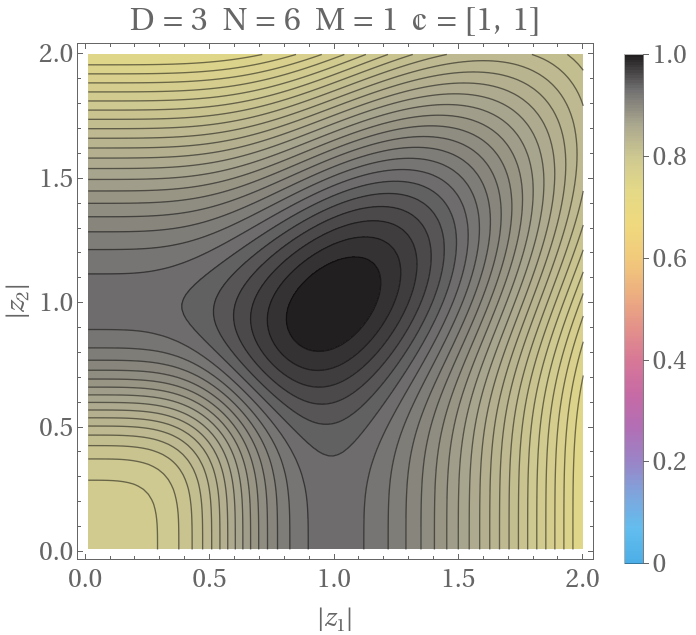}\hspace{\graphsep}\includegraphics[width=\graphwidth]{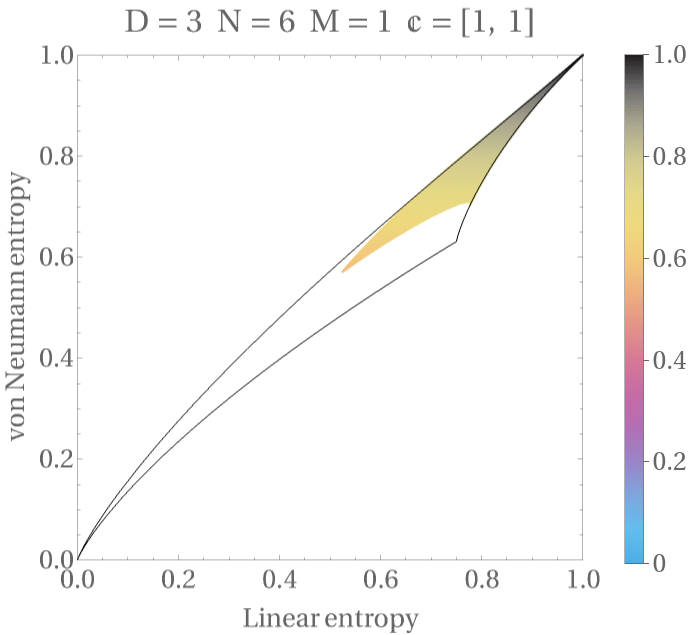}\hspace{\graphsep}
\includegraphics[width=\graphwidth]{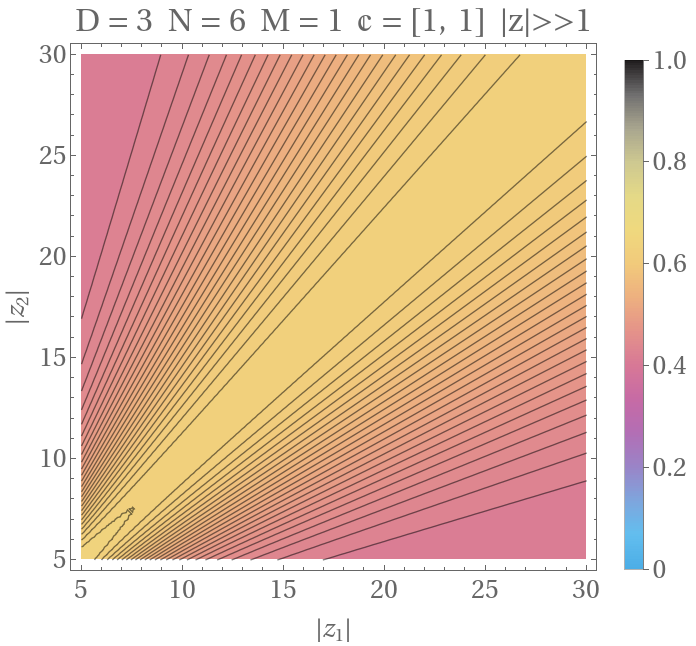}\hspace{\graphsep}\includegraphics[width=\graphwidth]{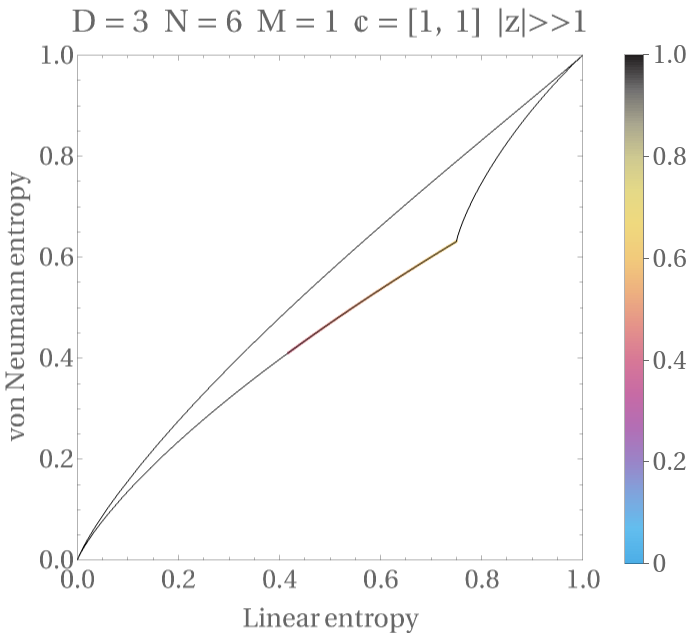}
\caption{Contour plots of von Neumann entropy and information diagrams for $D=3, \quad N=6, \quad  M=1, \quad \hbox{and}\quad \mathbbm{c}=[1,1]$, for $|z_i|\in [0,2]$ (left) and $|z_i|\in [5,30]$ (right).}
\label{NTL-3-6-1-11}
\end{figure}
\end{center}
\begin{center}
\begin{figure}[h!]
\includegraphics[width=\graphwidth]{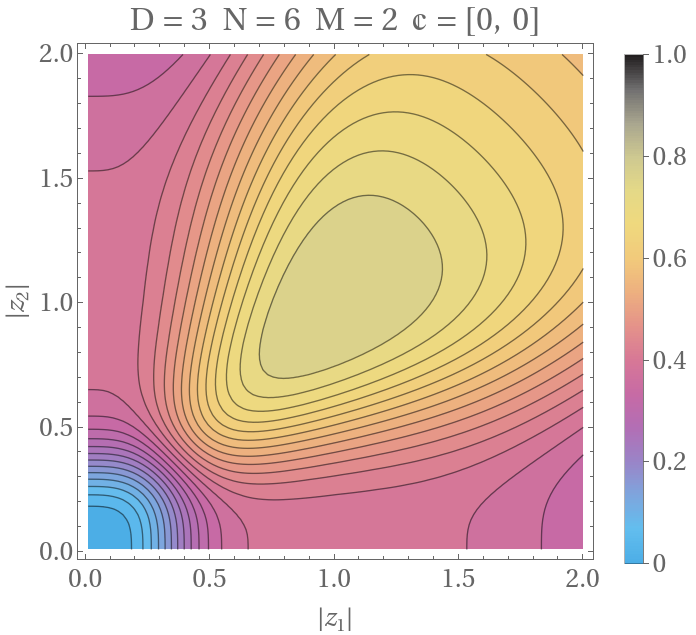}\hspace{\graphsep}\includegraphics[width=\graphwidth]{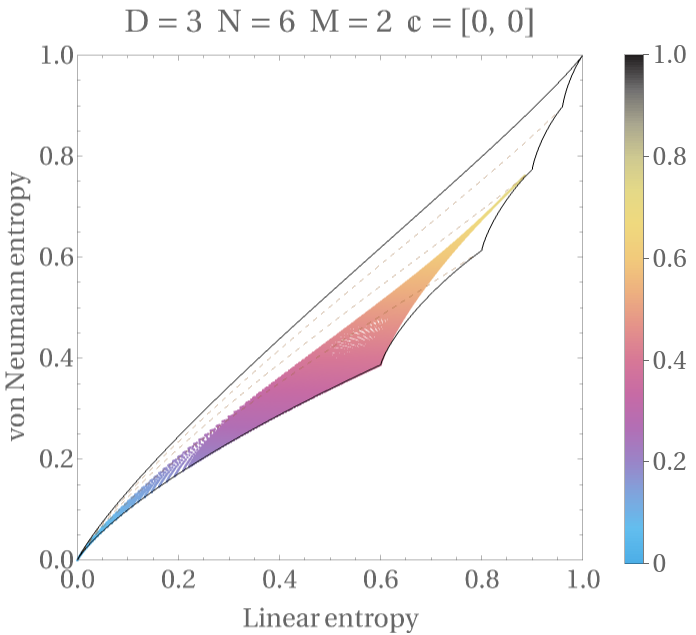}\hspace{\graphsep}
\includegraphics[width=\graphwidth]{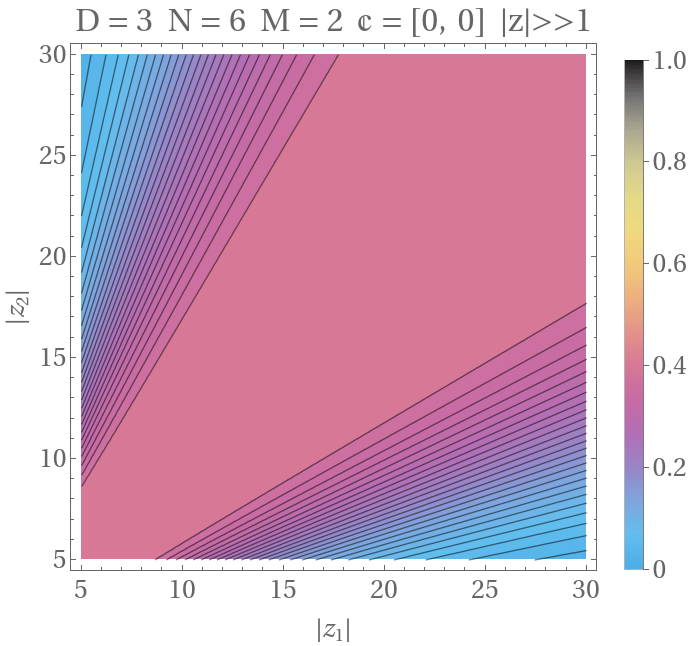}\hspace{\graphsep}\includegraphics[width=\graphwidth]{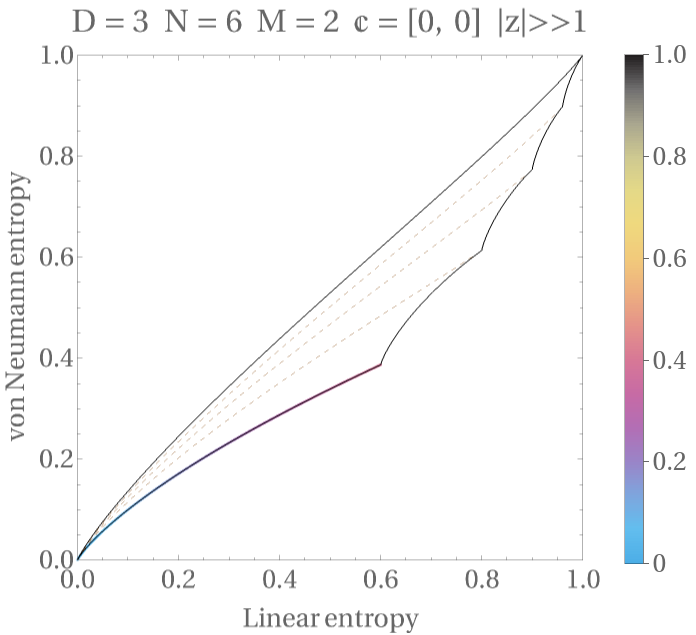}
\caption{Contour plots of von Neumann entropy and information diagrams for $D=3, \quad N=6, \quad  M=2, \quad \hbox{and}\quad \mathbbm{c}=[0,0]$, for $|z_i|\in [0,2]$ (left) and $|z_i|\in [5,30]$ (right).}
\label{NTL-3-6-2-00}
\end{figure}
\end{center}
\begin{center}
\begin{figure}[h!]
\includegraphics[width=\graphwidth]{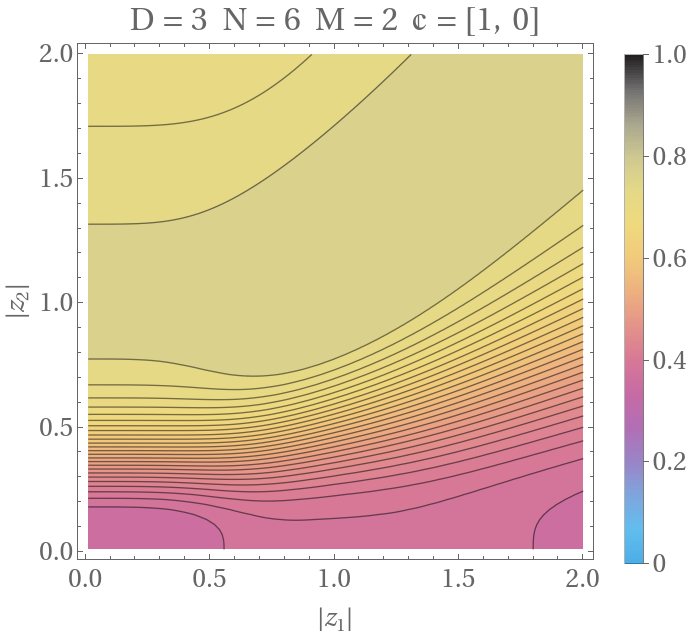}\hspace{\graphsep}\includegraphics[width=\graphwidth]{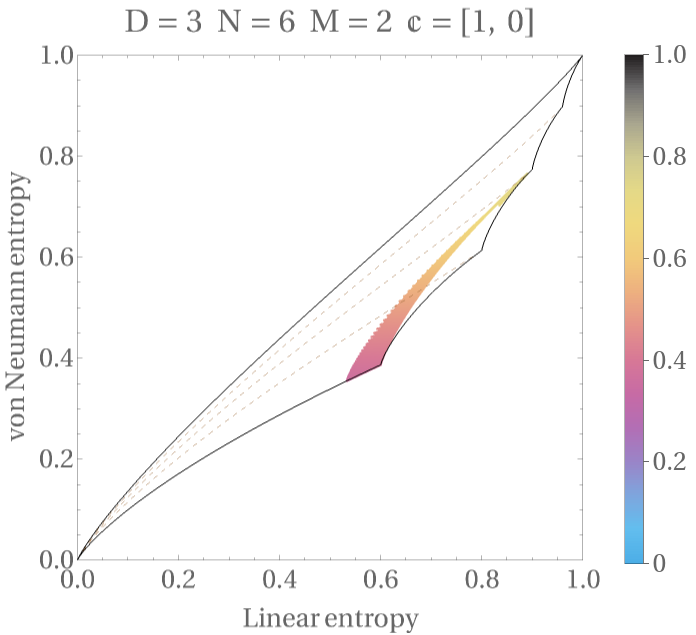}\hspace{\graphsep}
\includegraphics[width=\graphwidth]{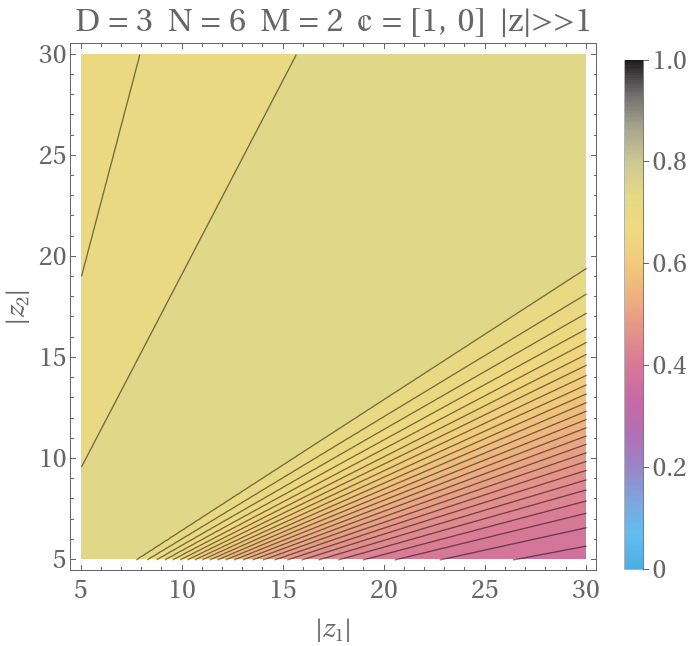}\hspace{\graphsep}\includegraphics[width=\graphwidth]{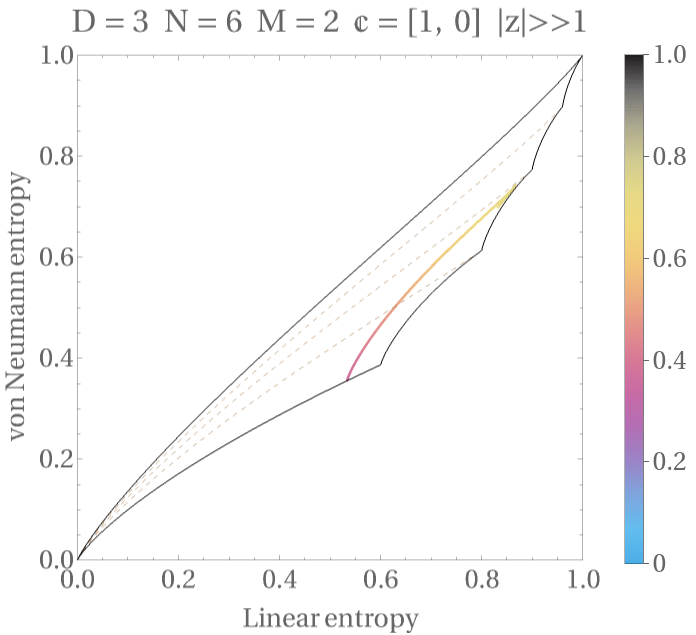}
\caption{Contour plots of von Neumann entropy and information diagrams for $D=3, \quad N=6, \quad  M=2, \quad \hbox{and}\quad \mathbbm{c}=[1,0]$, for $|z_i|\in [0,2]$ (left) and $|z_i|\in [5,30]$ (right).}\label{NTL-3-6-2-10}
\end{figure}
\end{center}
\begin{center}
\begin{figure}[h!]
\includegraphics[width=\graphwidth]{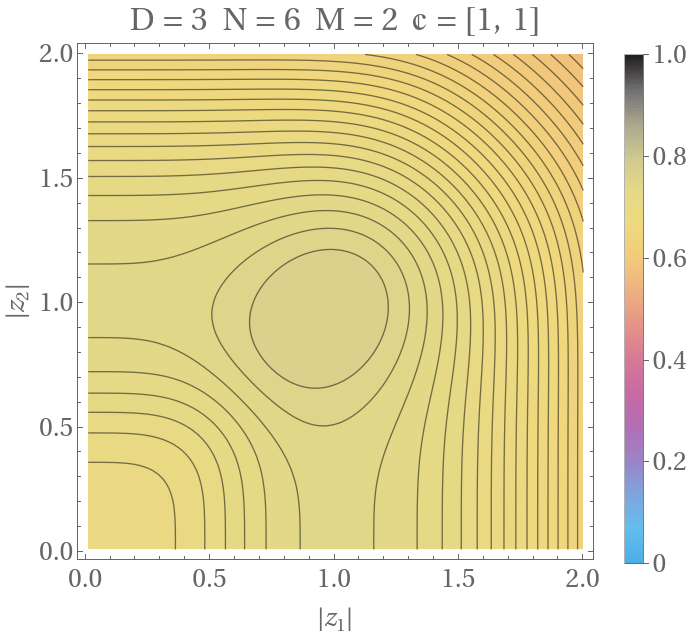}\hspace{\graphsep}\includegraphics[width=\graphwidth]{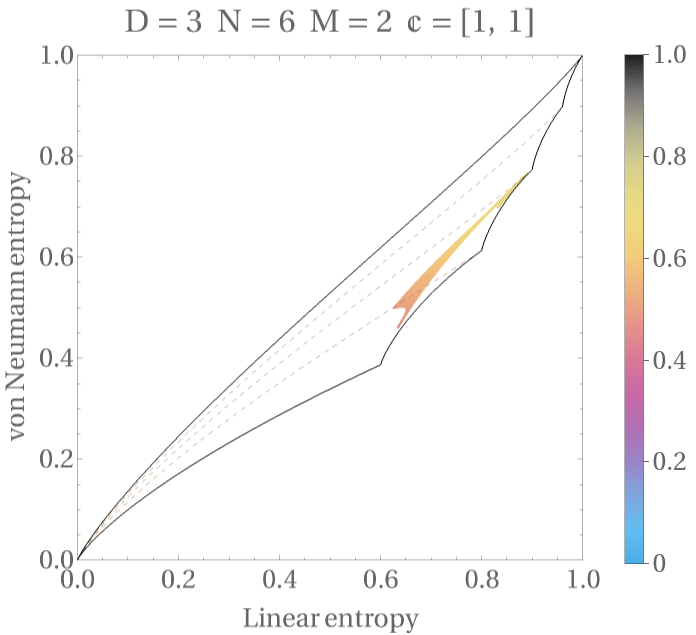}\hspace{\graphsep}
\includegraphics[width=\graphwidth]{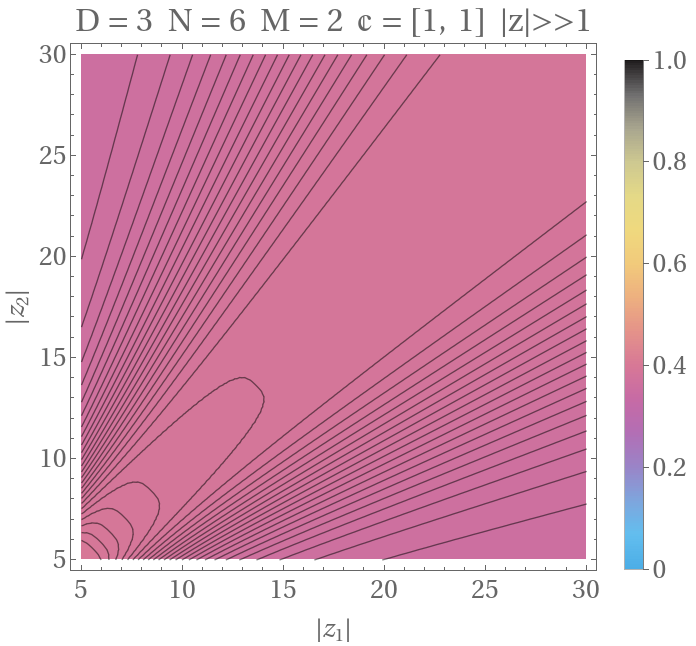}\hspace{\graphsep}\includegraphics[width=\graphwidth]{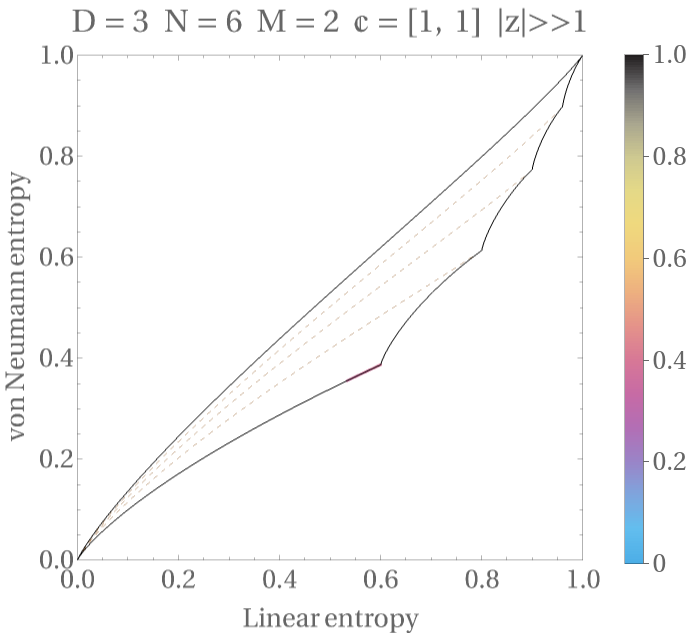}
\caption{Contour plots of von Neumann entropy and information diagrams for $D=3, \quad N=6, \quad  M=2, \quad \hbox{and}\quad \mathbbm{c}=[1,1]$, for $|z_i|\in [0,2]$ (left) and $|z_i|\in [5,30]$ (right).}\label{NTL-3-6-2-11}
\end{figure}
\end{center}
\begin{center}
\begin{figure}[h!]
\includegraphics[width=\graphwidth]{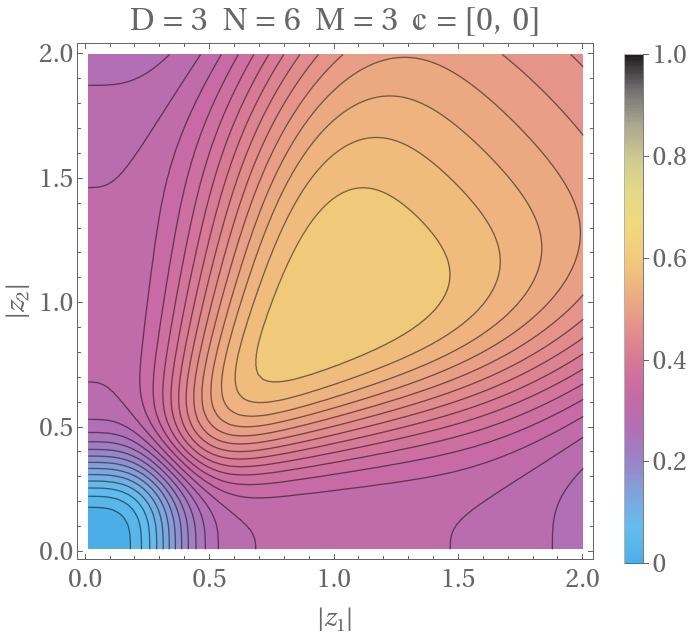}\hspace{\graphsep}\includegraphics[width=\graphwidth]{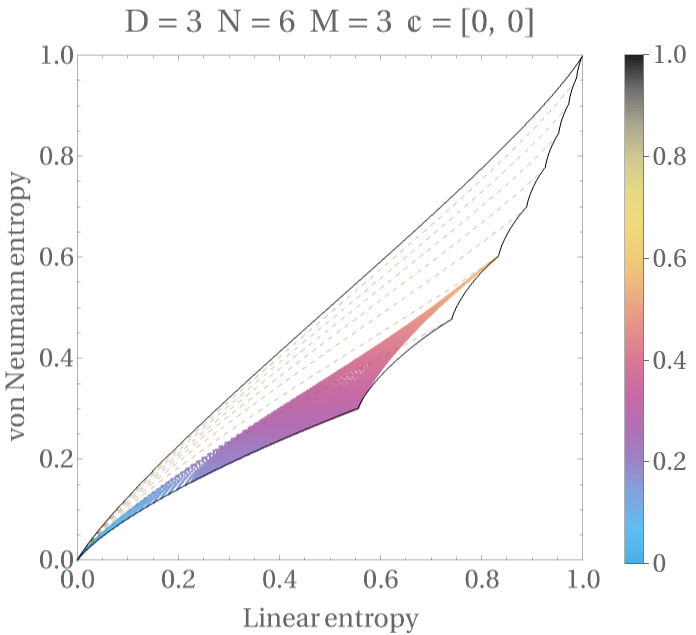}\hspace{\graphsep}
\includegraphics[width=\graphwidth]{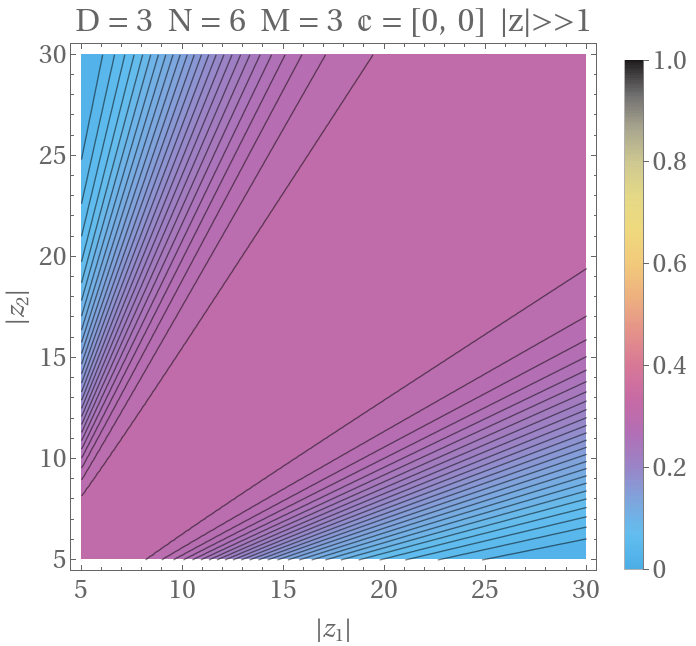}\hspace{\graphsep}\includegraphics[width=\graphwidth]{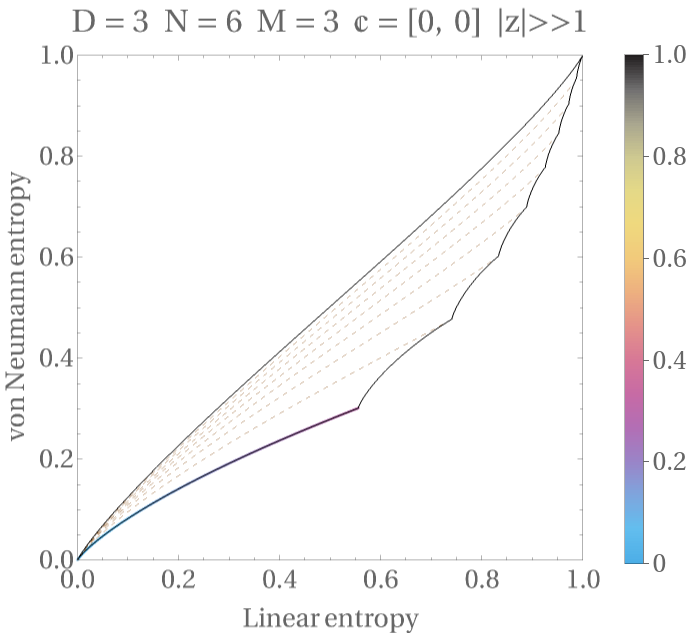}
\caption{Contour plots of von Neumann entropy and information diagrams for $D=3, \quad N=6, \quad  M=3, \quad \hbox{and}\quad \mathbbm{c}=[0,0]$, for $|z_i|\in [0,2]$ (left) and $|z_i|\in [5,30]$ (right).}\label{NTL-3-6-3-00}
\end{figure}
\end{center}
\begin{center}
\begin{figure}[h!]
\includegraphics[width=\graphwidth]{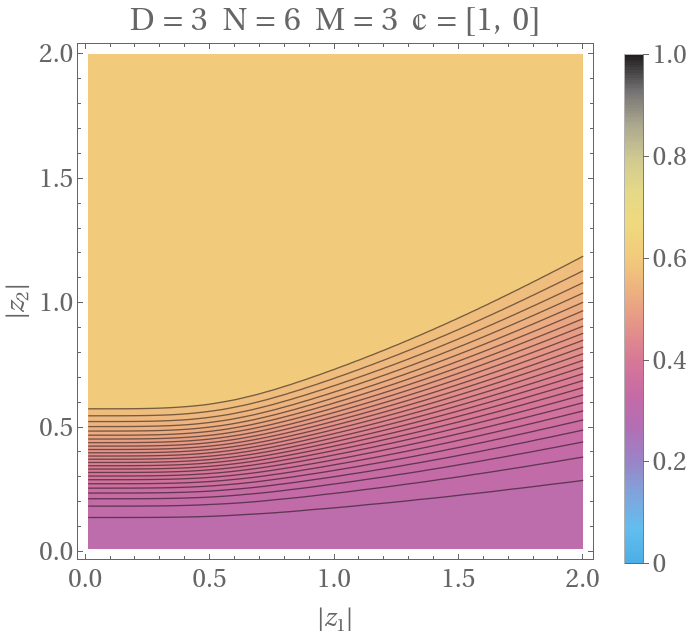}\hspace{\graphsep}\includegraphics[width=\graphwidth]{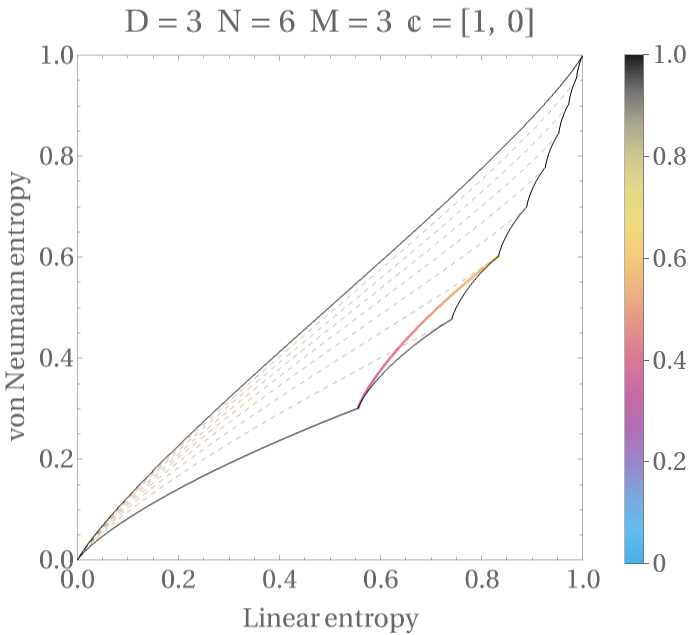}\hspace{\graphsep}
\includegraphics[width=\graphwidth]{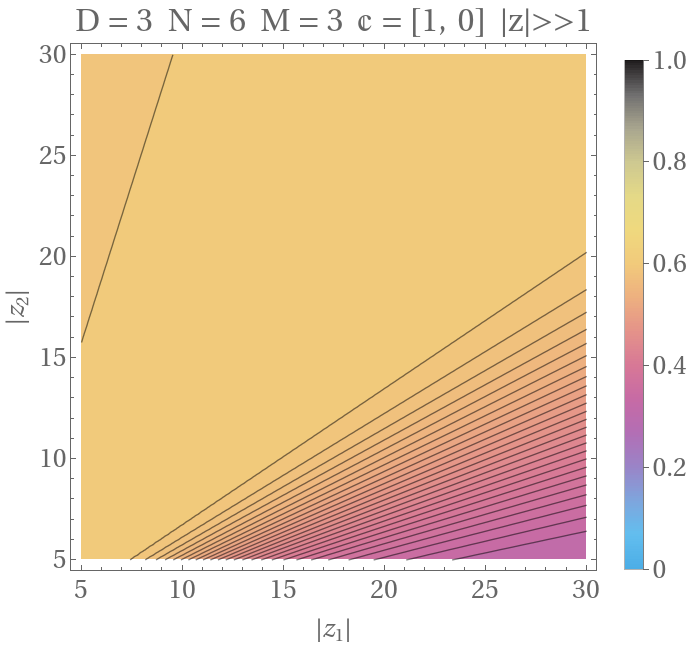}\hspace{\graphsep}\includegraphics[width=\graphwidth]{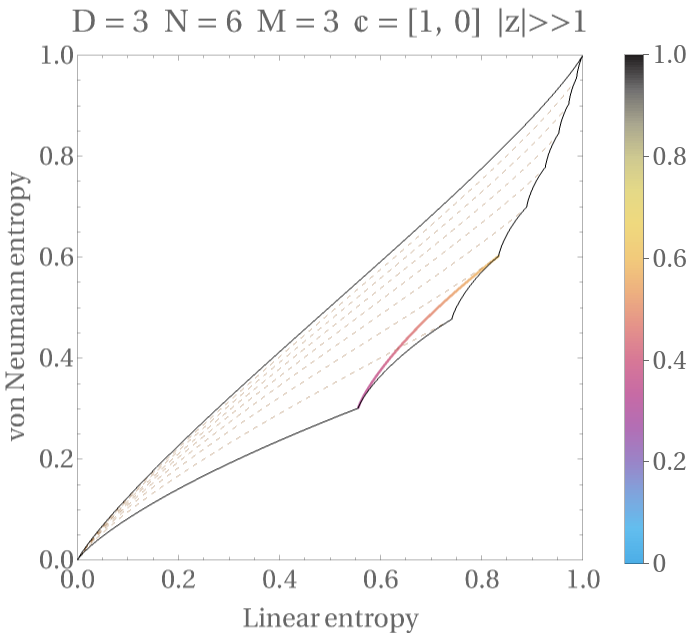}
\caption{Contour plots of von Neumann entropy and information diagrams for $D=3, \quad N=6, \quad  M=3, \quad \hbox{and}\quad \mathbbm{c}=[1,0]$, for $|z_i|\in [0,2]$ (left) and $|z_i|\in [5,30]$ (right).}\label{NTL-3-6-3-10}
\end{figure}
\end{center}
\begin{center}
\begin{figure}[h!]
\includegraphics[width=\graphwidth]{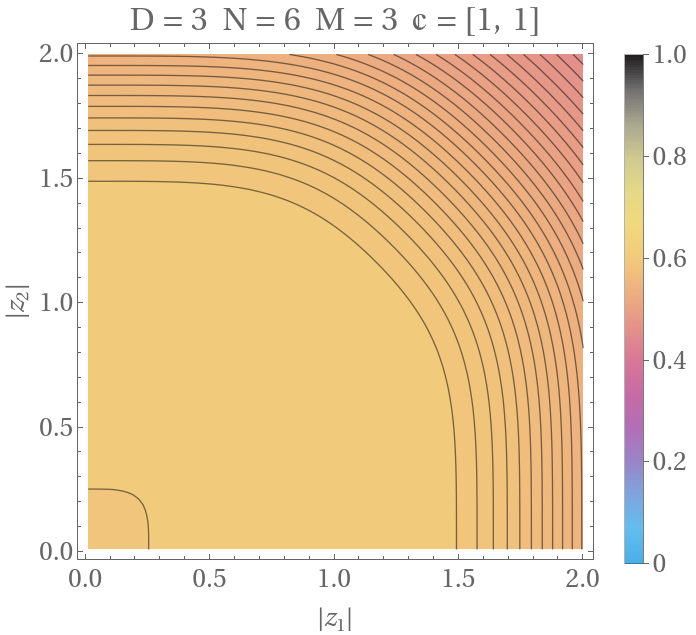}\hspace{\graphsep}\includegraphics[width=\graphwidth]{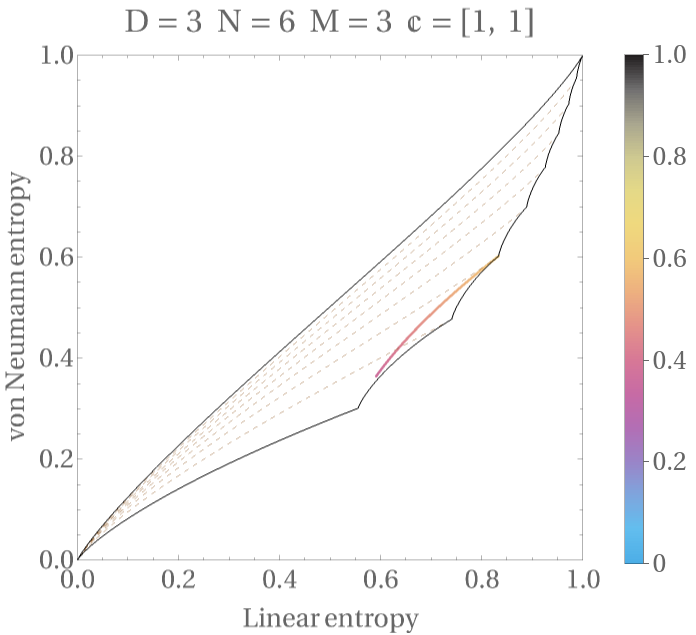}\hspace{\graphsep}
\includegraphics[width=\graphwidth]{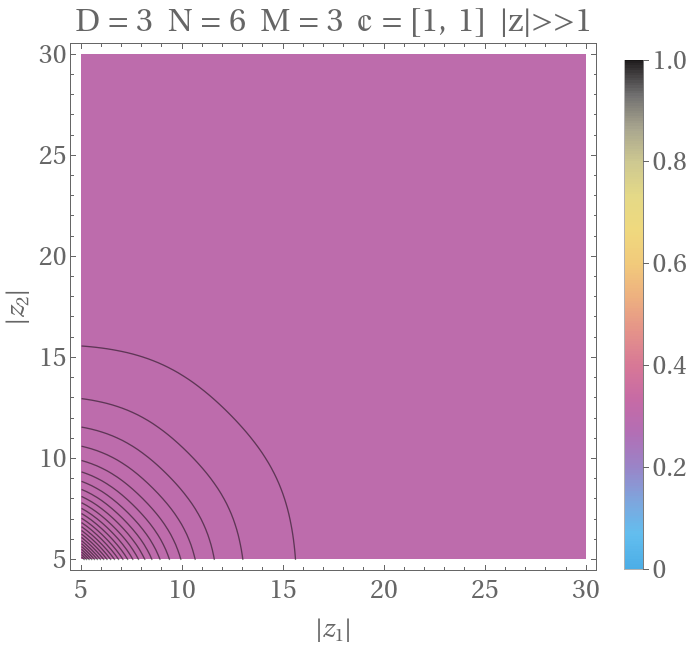}\hspace{\graphsep}\includegraphics[width=\graphwidth]{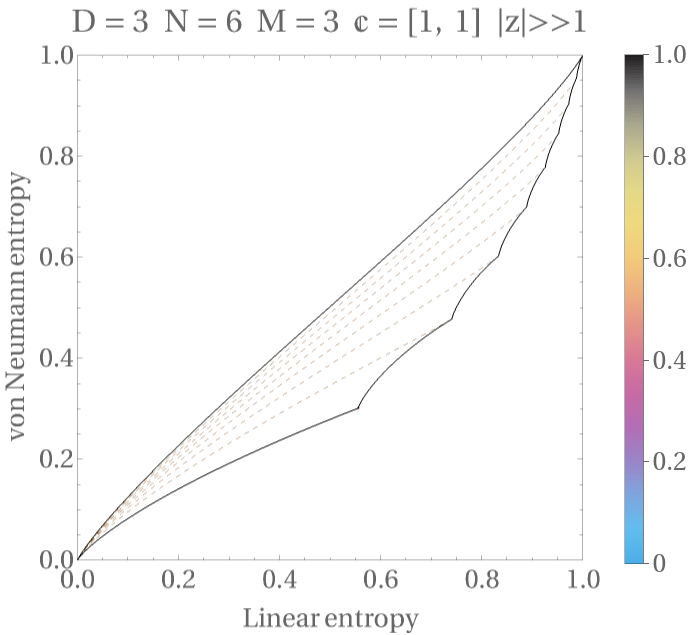}
\caption{Contour plots of von Neumann entropy and information diagrams for $D=3, \quad N=6, \quad  M=3, \quad \hbox{and}\quad \mathbbm{c}=[1,1]$, for $|z_i|\in [0,2]$ (left) and $|z_i|\in [5,30]$ (right).}\label{NTL-3-6-3-11}
\end{figure}
\end{center}

In this section, in Figures \ref{NTL-3-6-1-00} to \ref{NTL-3-6-3-11}, we show contour plots of the normalized von Neumann entropy (i.e. isentropic curves) of RDM for $N=6$ as a function of $|z_1|$ and $|z_2|$, for values in the range $[0,2]$ to appreciate the behaviour near the origin, and the maximum at the point  $(1,1)$, and in the range $[5,30]$ to apreciate the behaviuour in the asympotic limit $\|\zb\|\rightarrow\infty$, where the isentropic curves are lines passing through the origin. Values of $M$ from 1 to 3 and parities $\mathbb{c}=[0,0]$, $[1,0]$ and $[1,1]$ are shown. The case $\mathbb{c}=[0,1]$ is obtained in the contour plots simply by reflection with respect to the diagonal of the case $\mathbb{c}=[1,0]$, and it is not shown.

Next to each countor plot, to the right, an information diagram   with the same parameters is shown, representing, for each RDM matrix, a point in the plane (${\cal L},{\cal S})$, with the same color assigments as in the contour plot, i.e. the value of the von Neuman entropy, which in this case coincides with the vertical axis. Note that the information diagram for $\mathbb{c}=[0,1]$ is exactly the same
as that for $\mathbb{c}=[1,0]$ and it is neither shown.

The common features for all contour plots in the range $|z_i|\in[0,2]$ is a maximum of entanglement entropy at the point $(1,1)$, with a  value of $1$ of the (normalized) von Neumann entropy  and no ent(and corresponding to a maximally mixed RDM) in the case of $M=1$ (and all parities $\mathbb{c}$), but with smaller values of the maximum for larger values of $M$ (this is due to the fact that we are using the dimension of the symmetric representation with $M$ particles as the normalization factor, and this increases with $M$). At the origin $\zb=(0,0)$ the von Neumann entropy reaches the minimum, which is zero (corresponding to a pure state RDM  and no entanglement for the original state) for even parity $\mathbb{c}=[0,0]$, but has a larger value for the other parities, increasing the value with the number on non-zero entries of $\mathbb{c}$. Consequently, the odder states are more entangled that the evener ones.
This is a common feature in all examples considered.

In the information diagram next to the contour plot for $[0,2]$, a sample of 10.000 points  have been plotted, and
we can appreciate that in the case $M=1$ and $\mathbb{c}=(0,0)$ they  completelly fill the information diagram,  whereas for other parities they cover an smaller region concentrated in the upper-right part of the diagram (corresponding to larger values of both entropies, and therefore to a larger amount of entanglement).

For  $|z_i|$ in the range $[5,30]$, the asymptotic behaviour for $\|\zb\|\rightarrow\infty$ is shown. In the contour plot we observe that the isentropic lines are straight lines passing through the origin (as can be deduced from the $r\rightarrow\infty$ limit for the Schmidt eigenvalues given in Eq.
(76) of the paper, which for $D=3$ gives a dependence on $\cos(2\theta)$ for the eigenvalues, and therefore for the entropy), with the maximum value at the diagonal and decreasing when approaching to the abscissa and ordinate axes (simetrically for even and completely odd cases, and non-simmetrically for the other cases). The range of colours in this case is smaller (moving to smaller values of the entropy), and this is reflected in the information diagram in that all points are concentrated in a curve, which lies in the lower boundary for the even and completely odd cases (corresponding to density matrices of rank 2), and having larger ranks for the other cases.


 \subsection{Single thermodynamic limit}
 
\begin{center}
\begin{figure}[h!]
\includegraphics[width=\graphwidth]{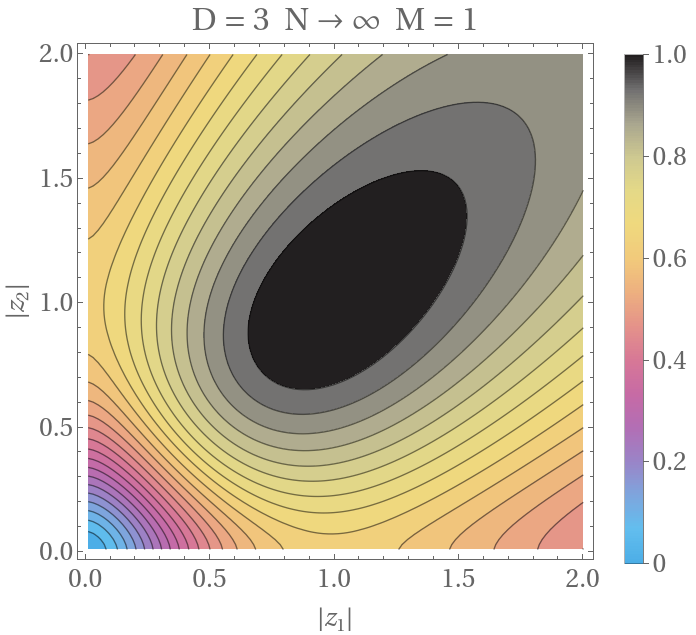}\hspace{\graphsep}
\includegraphics[width=\graphwidth]{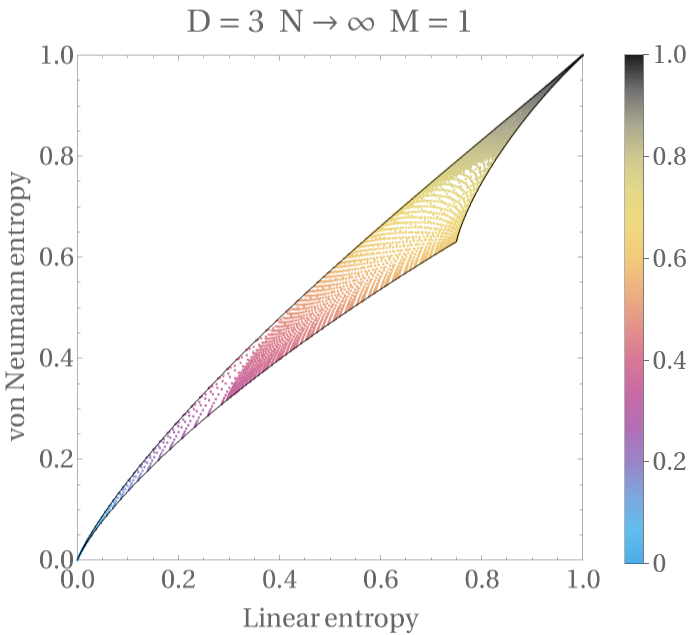}\hspace{\graphsep}
\includegraphics[width=\graphwidth]{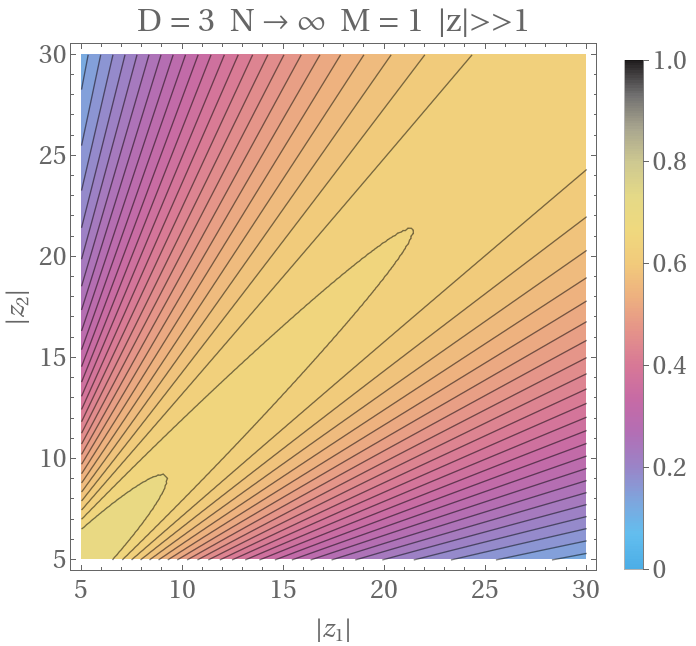}\hspace{\graphsep}
\includegraphics[width=\graphwidth]{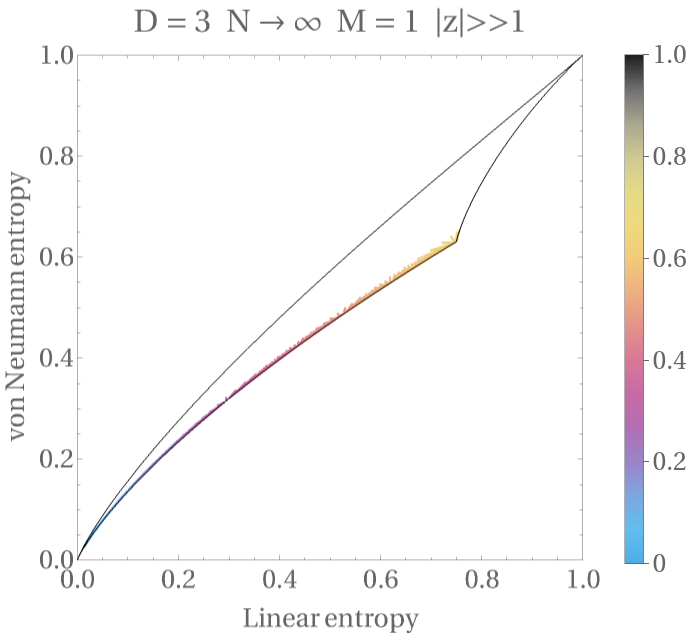}
\caption{Contour plots of von Neumann entropy and information diagrams for $D=3, \quad N\rightarrow\infty, \quad \hbox{and}\quad M=1$, for $|z_i|\in [0,2]$ (left) and $|z_i|\in [5,30]$ (right).}
\label{D3-M1-TL}
\end{figure}
\end{center}

\begin{center}
\begin{figure}[h!]
\includegraphics[width=\graphwidth]{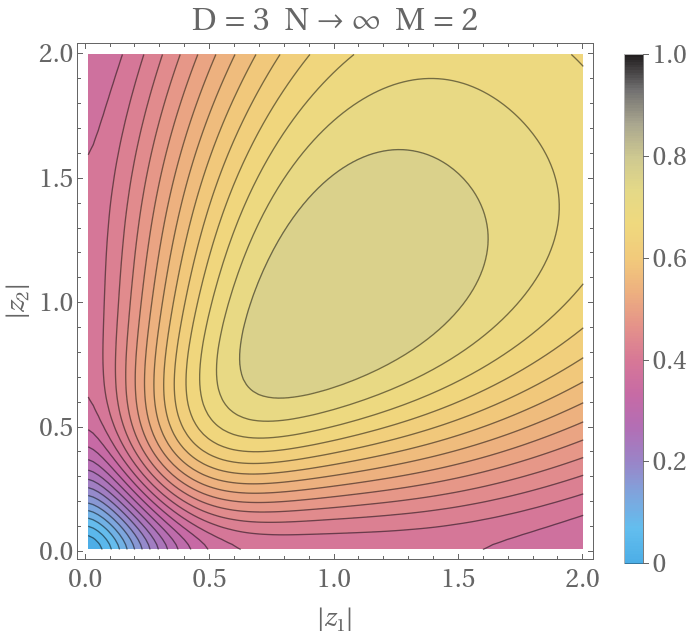}\hspace{\graphsep}
\includegraphics[width=\graphwidth]{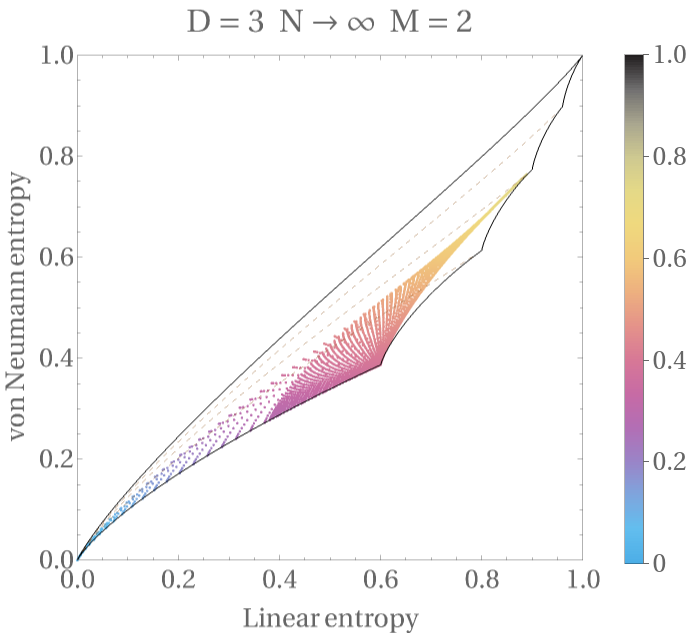}\hspace{\graphsep}
\includegraphics[width=\graphwidth]{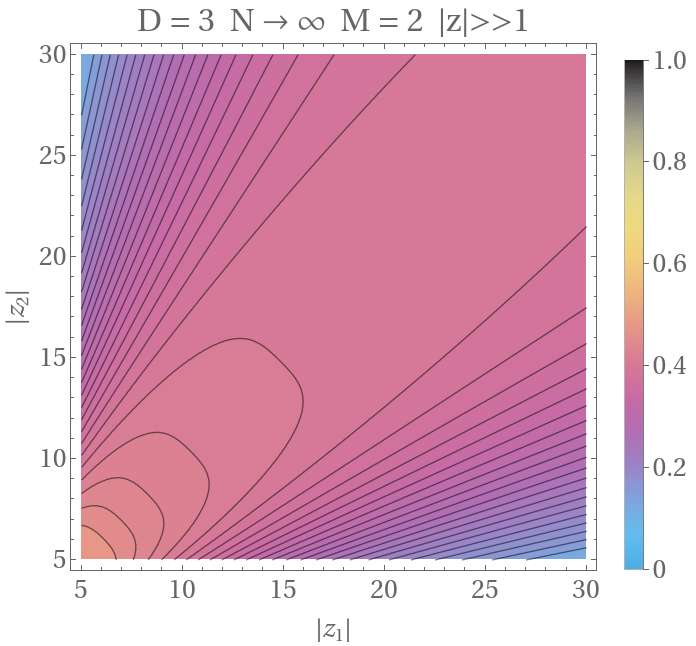}\hspace{\graphsep}
\includegraphics[width=\graphwidth]{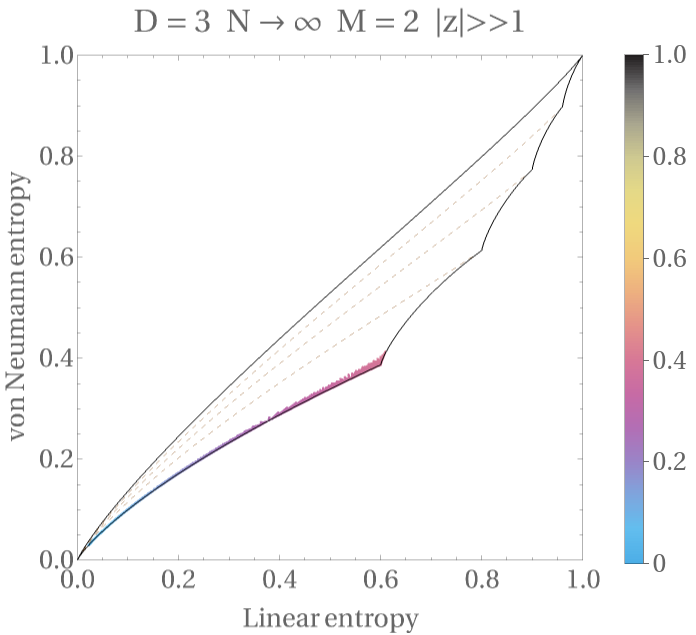}
\caption{Contour plots of von Neumann entropy and information diagrams for $D=3, \quad N\rightarrow\infty, \quad \hbox{and}\quad M=2$, for $|z_i|\in [0,2]$ (left) and $|z_i|\in [5,30]$ (right).}
\label{D3-M2-TL}
\end{figure}
\end{center}

\begin{center}
\begin{figure}[h!]
\includegraphics[width=\graphwidth]{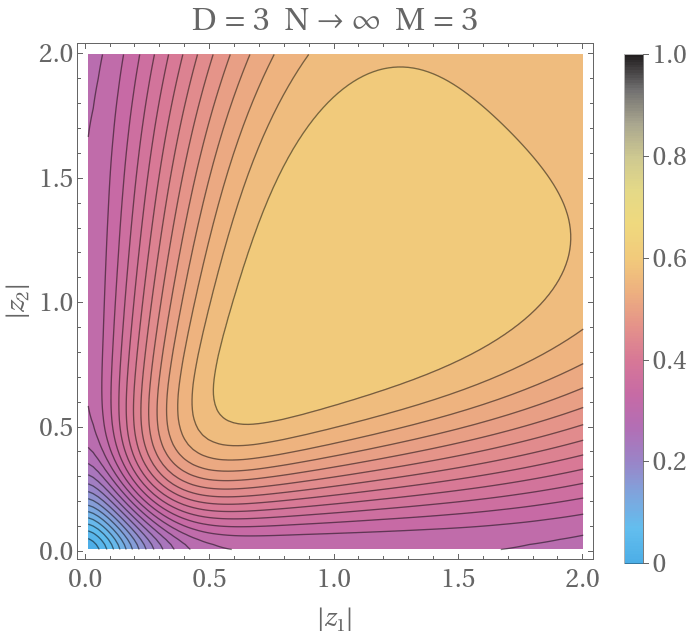}\hspace{\graphsep}
\includegraphics[width=\graphwidth]{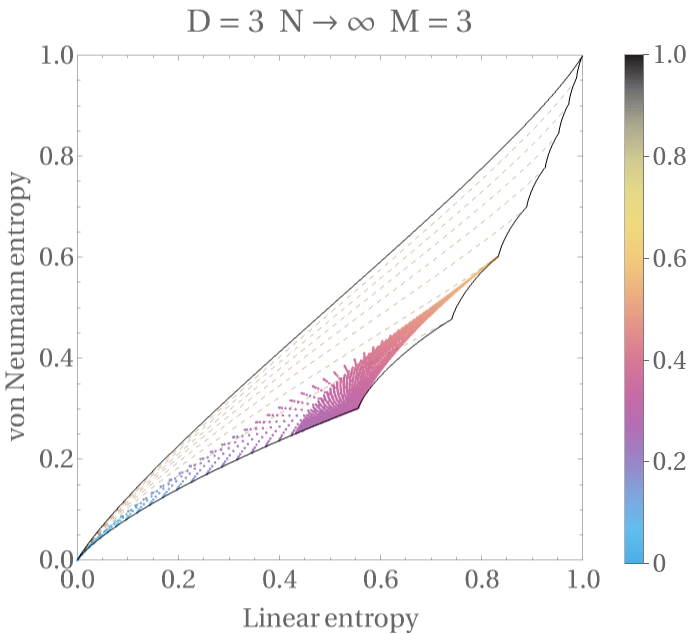}\hspace{\graphsep}
\includegraphics[width=\graphwidth]{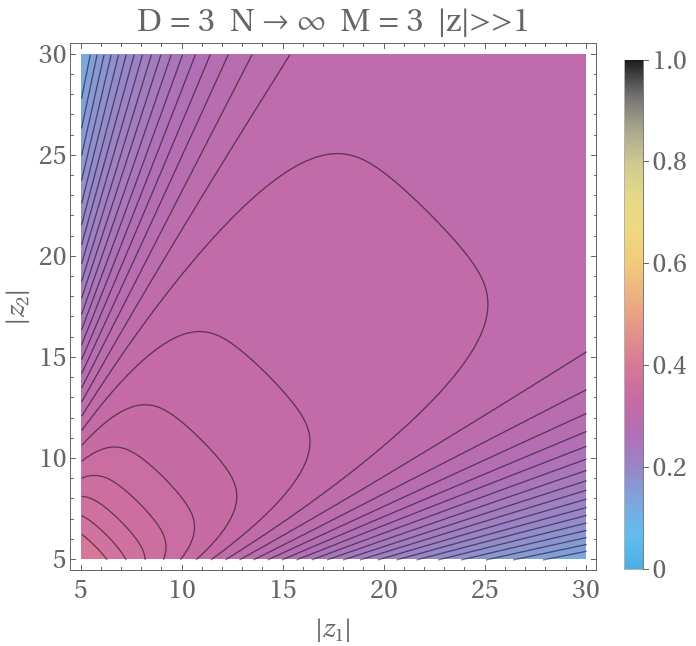}\hspace{\graphsep}
\includegraphics[width=\graphwidth]{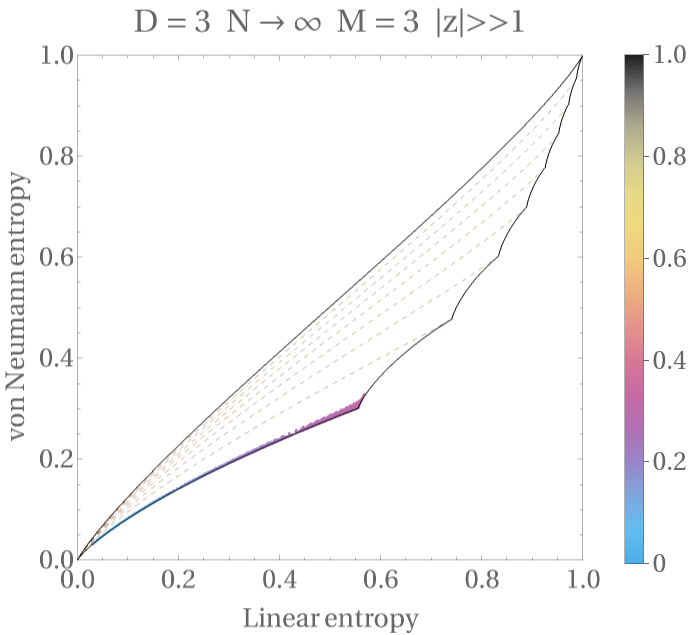}
\caption{Contour plots of von Neumann entropy and information diagrams for $D=3, \quad N\rightarrow\infty, \quad \hbox{and}\quad M=3$, for $|z_i|\in [0,2]$ (left) and $|z_i|\in [5,30]$ (right).}
\label{D3-M3-TL}
\end{figure}
\end{center}

In the case $N\rightarrow\infty$ (the thermodynamic limit), we plot the contour plots and information diagrams for $|z_i|\in [0,2]$ and $|z_i|\in [5,30], i=1,2,3$ . They are qualitatively the same as in the finite $N$ case, with the difference that they do not depend on the original parity $\mathbb{c}$ of the state, and the appearance is that of the even case. See Figures \ref{D3-M1-TL}, \ref{D3-M2-TL}, \ref{D3-M3-TL}.


 \subsection{Rescaled Double thermodynamic limit}

 The case of both $N,M\rightarrow\infty$ (Double thermodynamic limit) will not be considered since in this case the RDM correspond to that of a maximally mixed RDM of dimension $2^k$, with $k=\|\zb\|_0$.
 
 Instead we shall concentrate on the \textit{directional} limit 
 $N,M\rightarrow\infty$ with $M=(1-\eta)N$ and $\eta\in[\frac{1}{2},1)$ , and rescaling $\zb=\frac{\alphab}{\sqrt{N}}$. The case  $\eta\in(0,\frac{1}{2}]$ is obtained interchanging $\eta \leftrightarrow 1-\eta$, i.e. by interchanging $M\leftrightarrow N-M$ before taking the limit.

 In this case, as shown in eq. (65) of the paper, the Schmitd eigenvalues factorize as a product of Schmitd eigenvalues for the decomposition of one-dimensional cat Schr\"odinger states of the harmonic oscillator (with Heisenberg-Weyl symmetry) when a fraction $1-\eta$ of the intensity of the light has been absorbed (i.e the transmisivity of the medium is $\eta$).

\begin{center}
\begin{figure}[h!]
\includegraphics[width=\graphwidth]{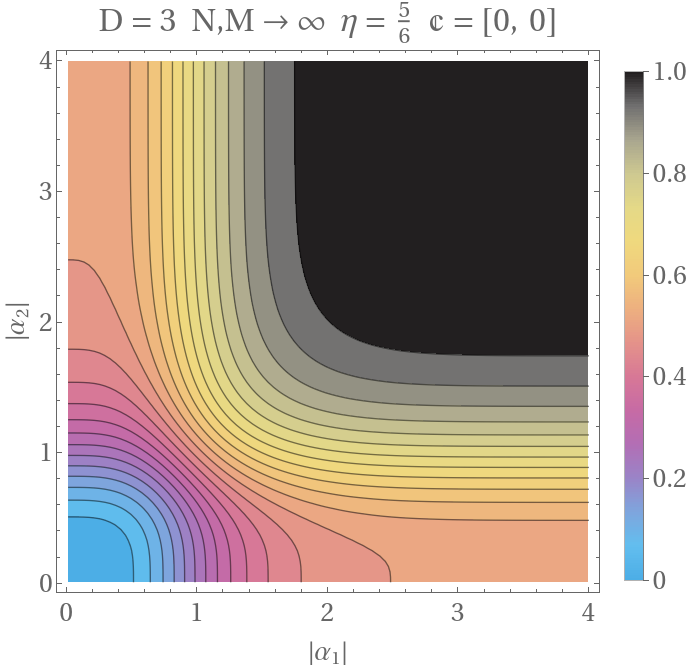}\hspace{\graphsep}\includegraphics[width=\graphwidth]{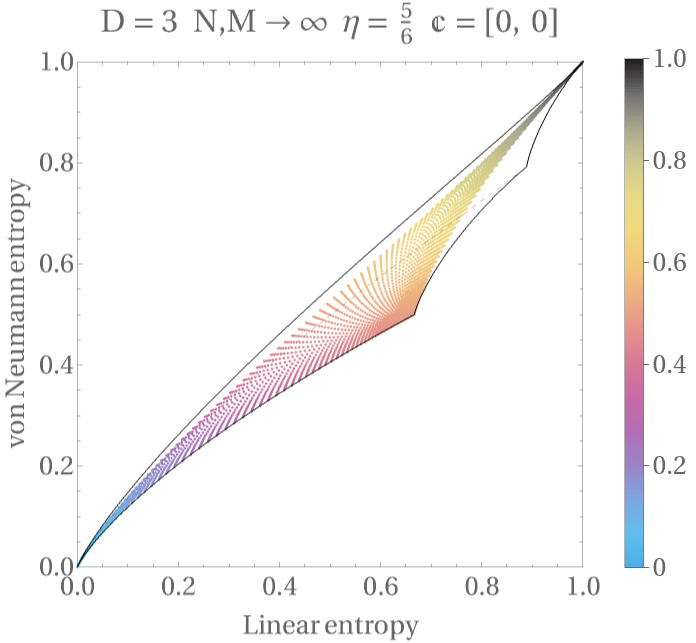}

\caption{Contour plot of von Neumann entropy and information diagram for $D=3, \quad N,M\rightarrow \infty,\quad \eta=\frac{5}{6}, \quad \hbox{and} \quad\mathbbm{c}=[0,0]$, for $|\alpha_i|\in [0,4]$.}
\label{RTL-3-5d6-00}
\end{figure}
\end{center}

\begin{center}
\begin{figure}[h!]
\includegraphics[width=\graphwidth]{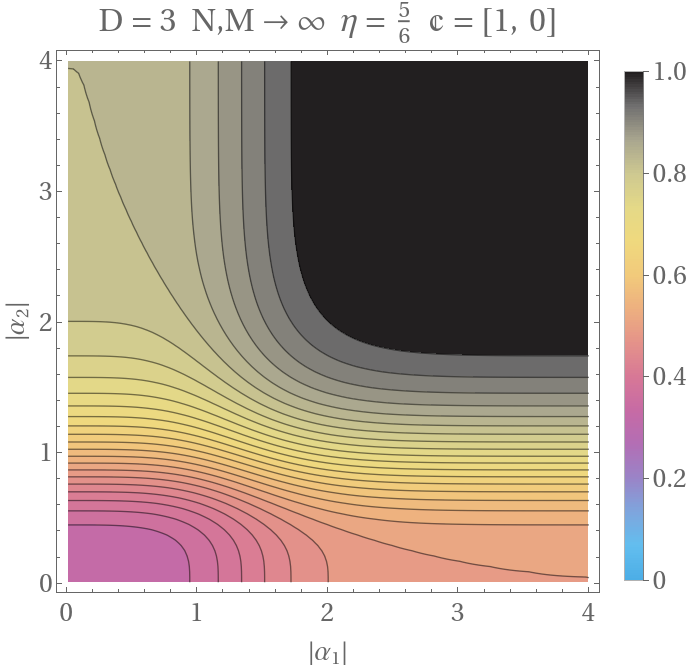}\hspace{\graphsep}\includegraphics[width=\graphwidth]{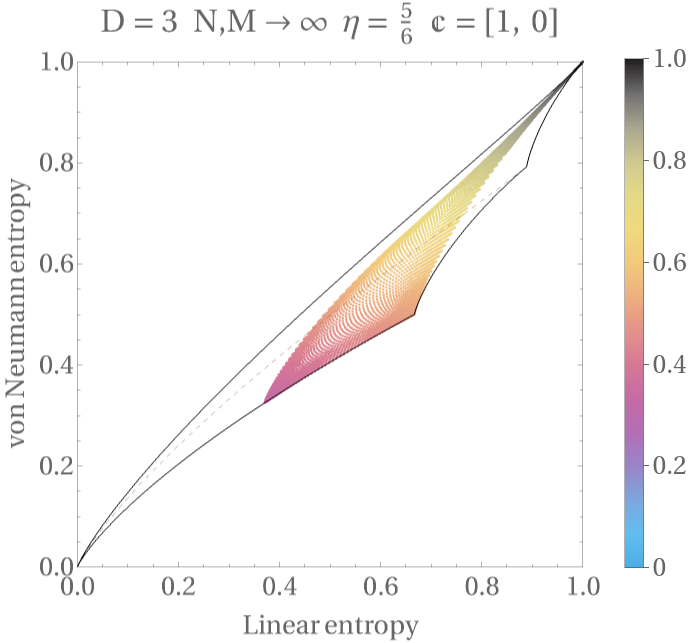}
\caption{Contour plot of von Neumann entropy and information diagram for $D=3, \quad N,M\rightarrow \infty,\quad \eta=\frac{5}{6}, \quad \hbox{and} \quad\mathbbm{c}=[1,0]$, for $|\alpha_i|\in [0,4]$.}\label{NTL-3-5d6-10}
\end{figure}
\end{center}

\begin{center}
\begin{figure}[h!]
\includegraphics[width=\graphwidth]{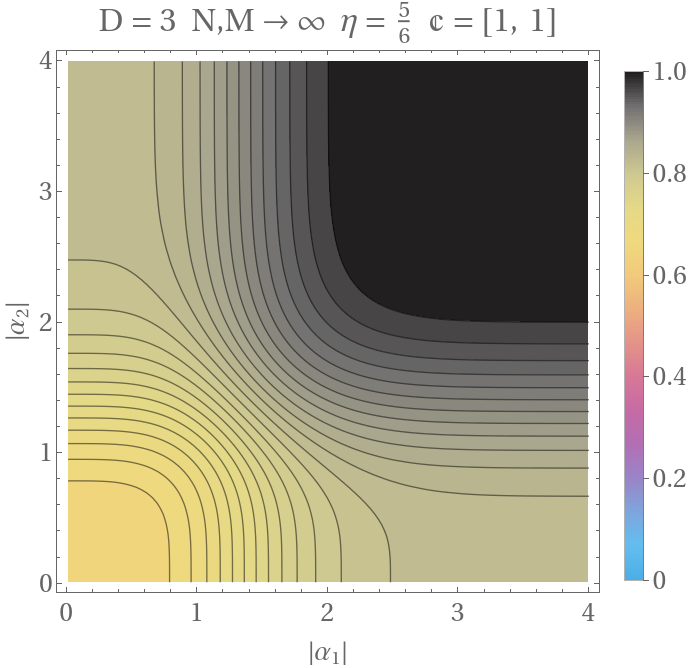}\hspace{\graphsep}\includegraphics[width=\graphwidth]{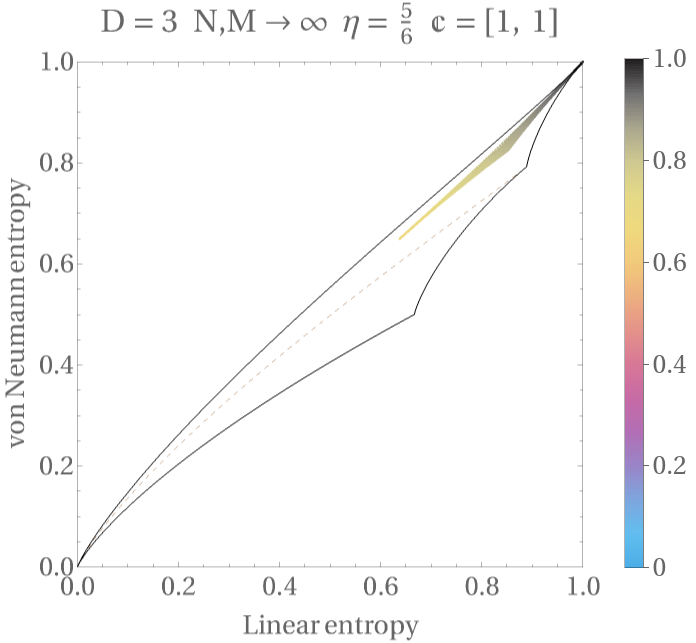}
\caption{Contour plot of von Neumann entropy and information diagram for $D=3, \quad N,M\rightarrow \infty,\quad \eta=\frac{5}{6}, \quad \hbox{and} \quad\mathbbm{c}=[1,1]$, for $|\alpha_i|\in [0,4]$.}\label{NTL-3-5d6-11}
\end{figure}
\end{center}

In  Figures \ref{RTL-3-5d6-00}-\ref{NTL-3-5d6-11} we show Contour plots for von Neumann entropy for
$|\alpha_i|\in [0,4]$, and information diagrams in the same range of $|\alpha_i|$, for the cases $\eta=\frac{5}{6},\frac{2}{3},\frac{1}{2}$ (which correspond to the cases $M=1$, $M=2$ and $M=3$, respectively, when $N=6$, considered in all examples with finite $N$). The factorization of the Schmidt coefficients results in that isentropic lines for large values $|z_i|$ are either horizontal or vertical. Also, due to the exponential behaviour of the Schmidt eigenvalues, the $\|\alphab\|\rightarrow\infty$ limit is reached for relatively small values of $|\alpha_i|$.

It is interestig to note that in the case $\eta=\frac{1}{2}$ (half of the intensity has been absorbed), for $\mathbbm{c}=[1,0]$ one of the factors in the Schmidt eigenvalues is $\frac{1}{2}$ and therefore there are only horizontal isentropic lines (vertical for the case $\mathbbm{c}=[0,1]$, not shown), and for $\mathbbm{c}=[1,1]$ both factors are $\frac{1}{2}$ and therefore the state is maximally entangled for all $\alphab\neq (0,0)$.


\begin{center}
\begin{figure}[h!]
\includegraphics[width=\graphwidth]{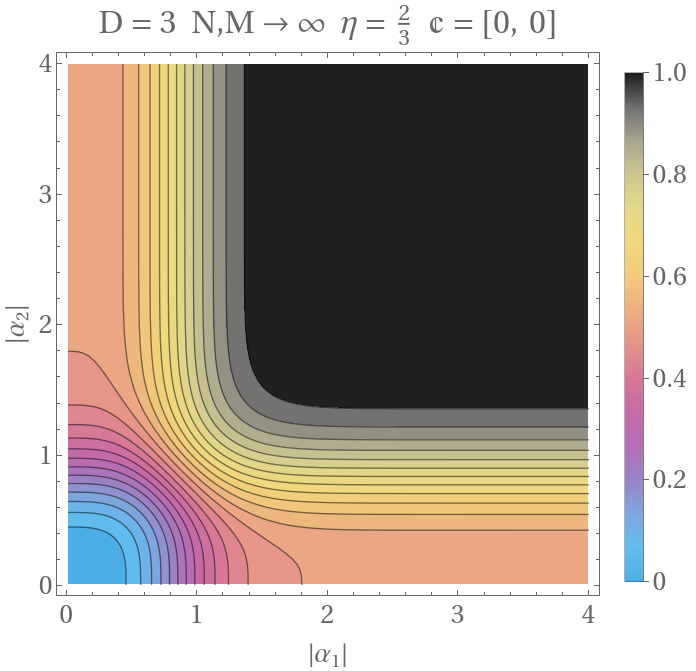}\hspace{\graphsep}\includegraphics[width=\graphwidth]{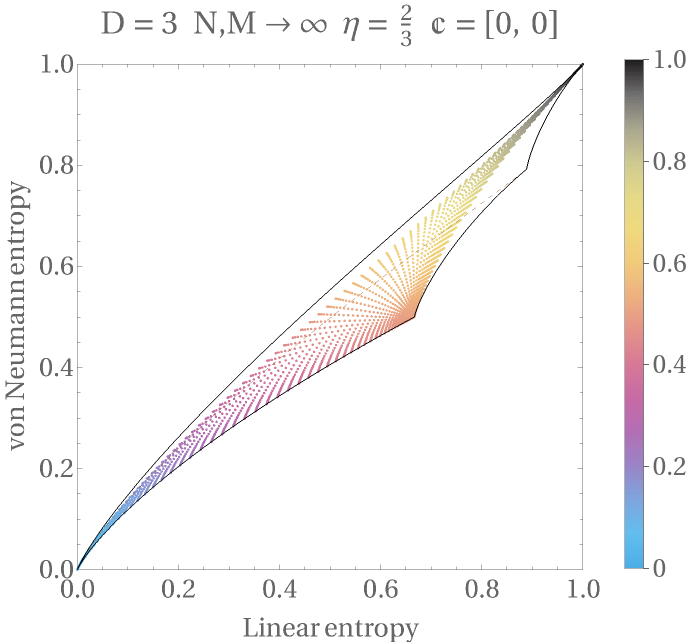}

\caption{Contour plot of von Neumann entropy and information diagram for $D=3, \quad N,M\rightarrow \infty,\quad \eta=\frac{2}{3}, \quad \hbox{and} \quad\mathbbm{c}=[0,0]$, for $|\alpha_i|\in [0,4]$.}
\label{RTL-3-4d6-00}
\end{figure}
\end{center}

\begin{center}
\begin{figure}[h!]
\includegraphics[width=\graphwidth]{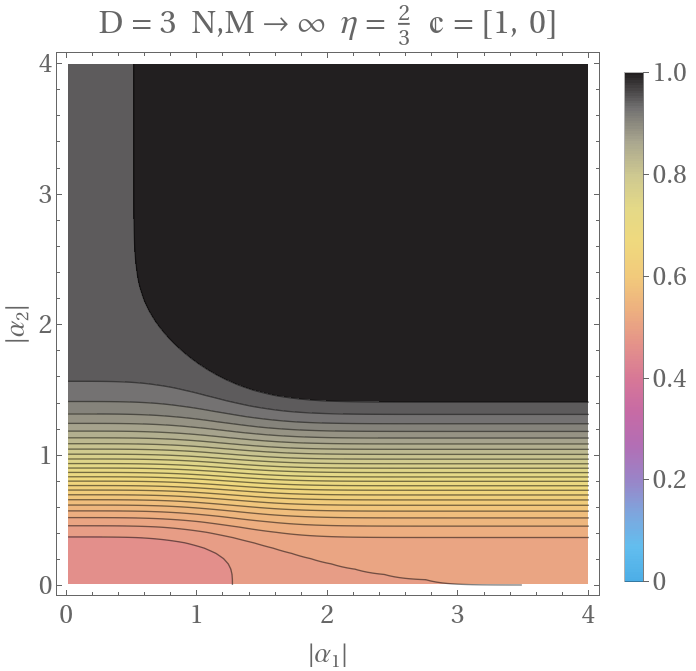}\hspace{\graphsep}\includegraphics[width=\graphwidth]{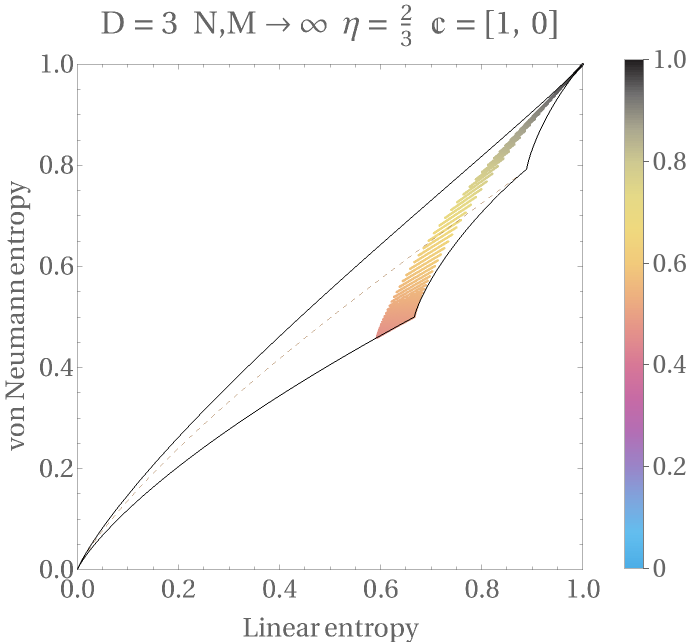}
\caption{Contour plot of von Neumann entropy and information diagram for $D=3, \quad N,M\rightarrow \infty,\quad \eta=\frac{2}{3}, \quad \hbox{and} \quad\mathbbm{c}=[1,0]$, for $|\alpha_i|\in [0,4]$.}
\label{NTL-3-4d6-10}
\end{figure}
\end{center}

 \begin{center}
\begin{figure}[h!]
\includegraphics[width=\graphwidth]{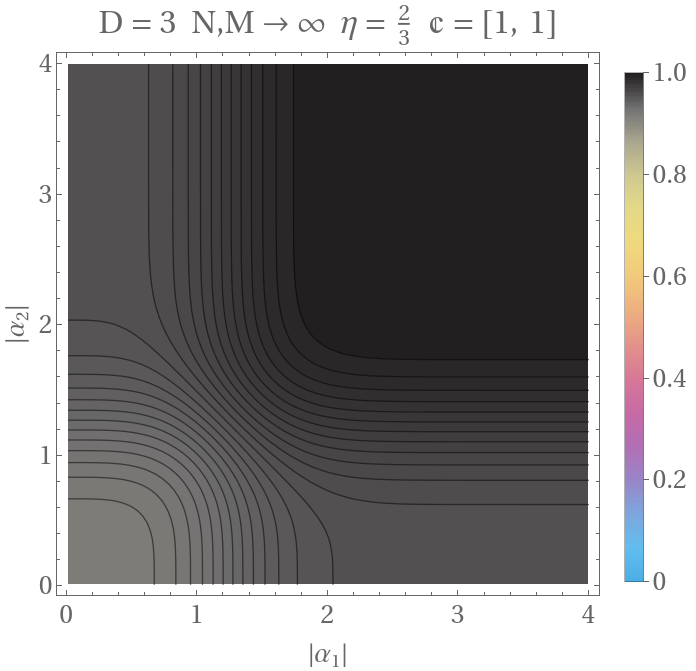}\hspace{\graphsep}\includegraphics[width=\graphwidth]{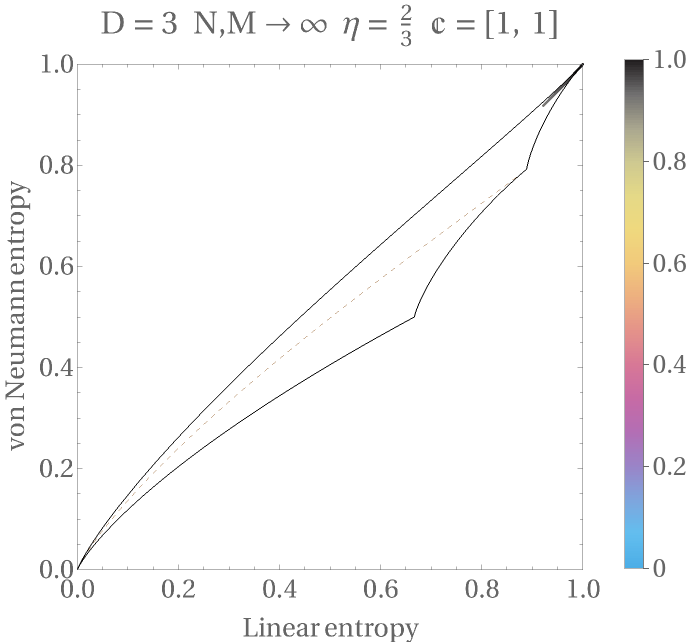}
\caption{Contour plot of von Neumann entropy and information diagram for $D=3, \quad N,M\rightarrow \infty,\quad \eta=\frac{2}{3}, \quad \hbox{and} \quad\mathbbm{c}=[1,1]$, for $|\alpha_i|\in [0,4]$.}
\label{NTL-3-4d6-11}
\end{figure}
\end{center}


\begin{center}
\begin{figure}[h!]
\includegraphics[width=\graphwidth]{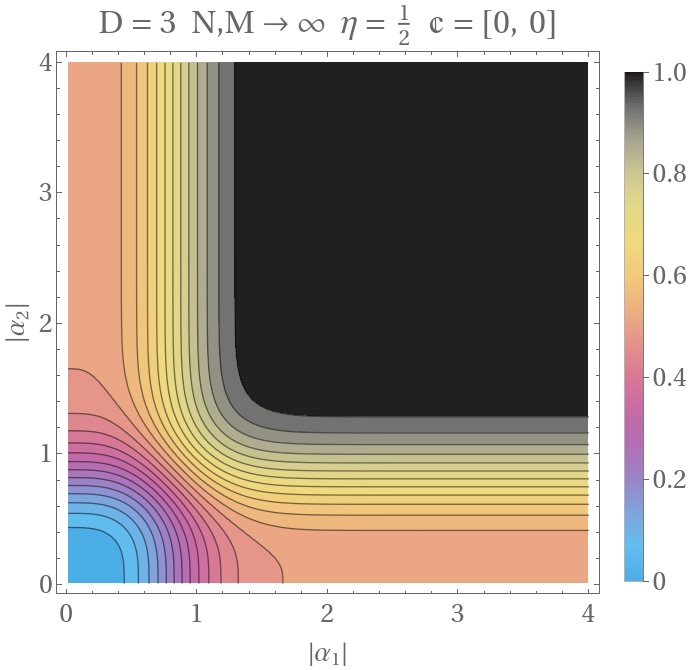}\hspace{\graphsep}\includegraphics[width=\graphwidth]{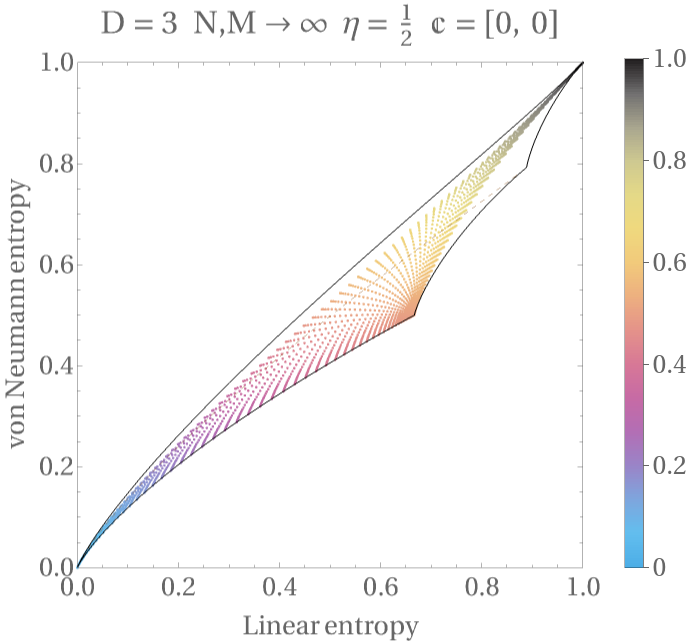}

\caption{Contour plot of von Neumann entropy and information diagram for $D=3, \quad N,M\rightarrow \infty,\quad \eta=\frac{1}{2}, \quad \hbox{and} \quad\mathbbm{c}=[0,0]$, for $|\alpha_i|\in [0,4]$.}
\label{RTL-3-3d6-00}
\end{figure}
\end{center}

\begin{center}
\begin{figure}[h!]
\includegraphics[width=\graphwidth]{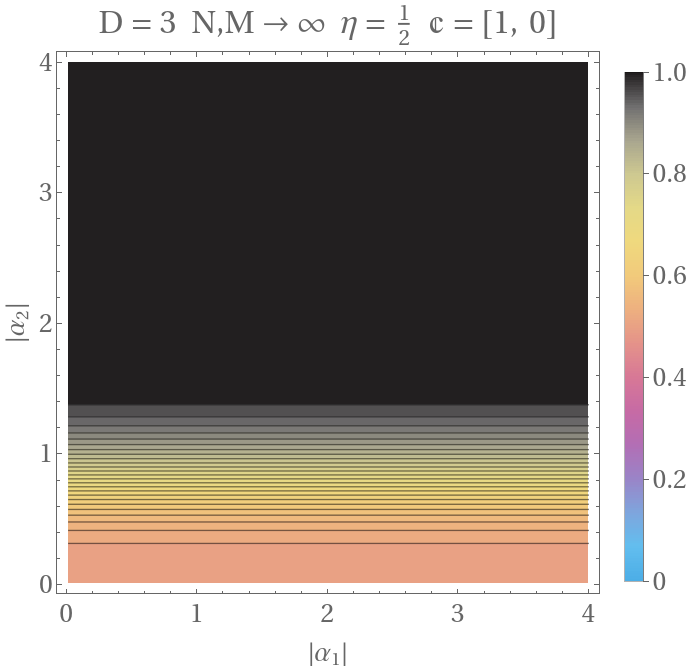}\hspace{\graphsep}\includegraphics[width=\graphwidth]{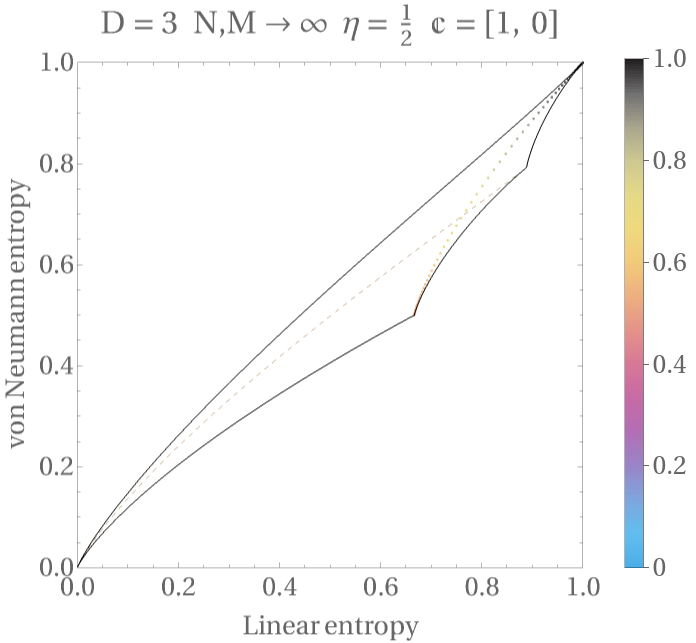}
\caption{Contour plot of von Neumann entropy and information diagram for $D=3, \quad N,M\rightarrow \infty,\quad \eta=\frac{1}{2}, \quad \hbox{and} \quad\mathbbm{c}=[1,0]$, for $|\alpha_i|\in [0,4]$.}\label{NTL-3-3d6-10}
\end{figure}
\end{center}

 \begin{center}
\begin{figure}[h!]
\includegraphics[width=\graphwidth]{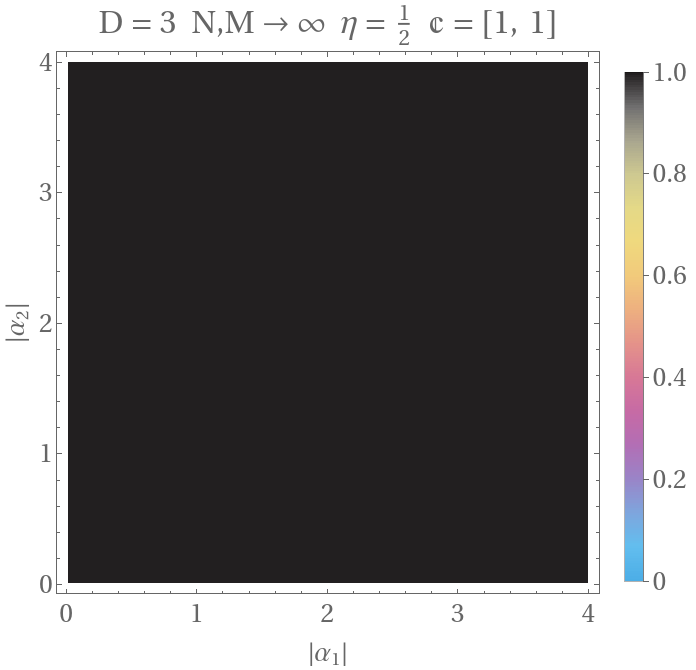}\hspace{\graphsep}\includegraphics[width=\graphwidth]{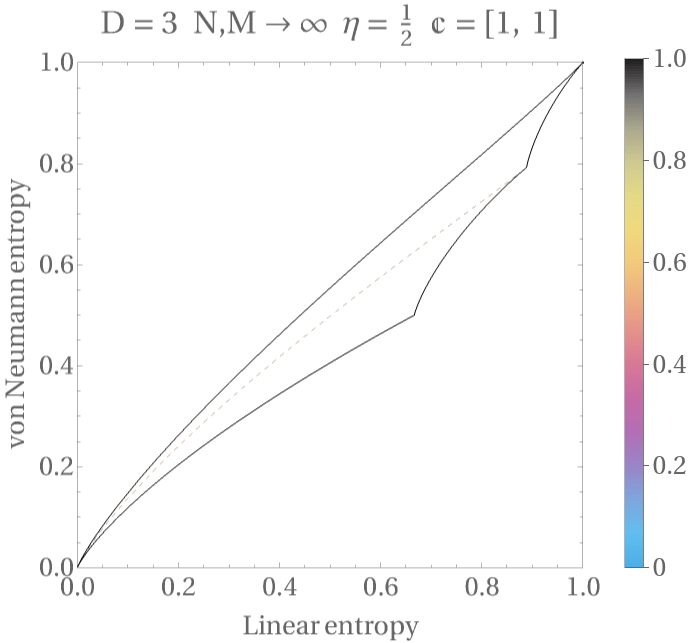}
\caption{Contour plot of von Neumann entropy and information diagram for $D=3, \quad N,M\rightarrow \infty,\quad \eta=\frac{1}{2}, \quad \hbox{and} \quad\mathbbm{c}=[1,1]$, for $|\alpha_i|\in [0,4]$.}\label{NTL-3-3d6-11}
\end{figure}
\end{center}

\section{Figures for $D=4$ atom levels (ququatrit)}
\label{FigD4}

If we increase the number of level/states to $D=4$, the coherent states are parametrized by 3 complex numbers $\zb=(z_1,z_2,z_3)$. In this case contour plots of von Neumann entropy cannot be shown, and they will be replaced by information diagrams (see Sec. \ref{InformationDiagrams}), where points are coloured according to the value of the von Neumann entropy (following the case of $D=3$). The behaviour for large $\|\zb\|$ of the von Neumann entropy, which according to Eq. (76) depend only on the spherical angles $(\theta,\phi)$, can be visualized in the octant of the sphere $\|\zb\|=R>>1$, using contour lines of the von Neumann entropy.

\subsection{Finite number $N$ of ququatrits}

\begin{center}
\begin{figure}[h!]
\includegraphics[width=\graphwidth]{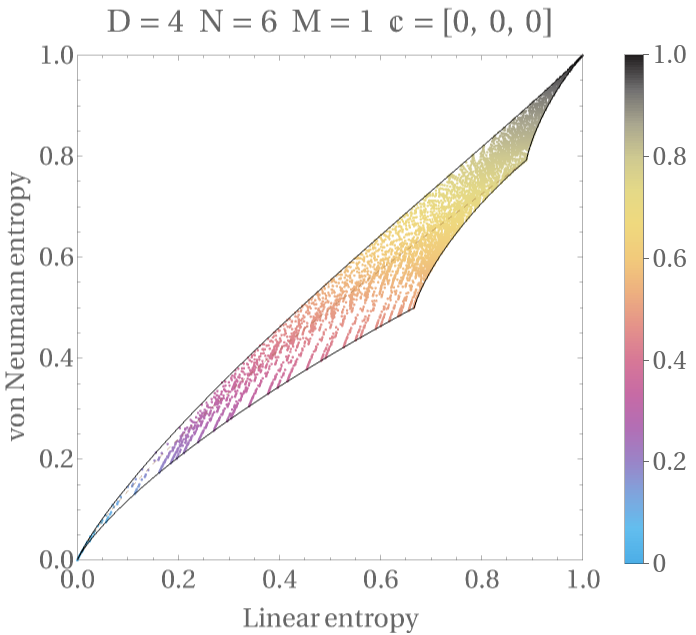}\hspace{\graphsep}
\includegraphics[width=\graphwidth]{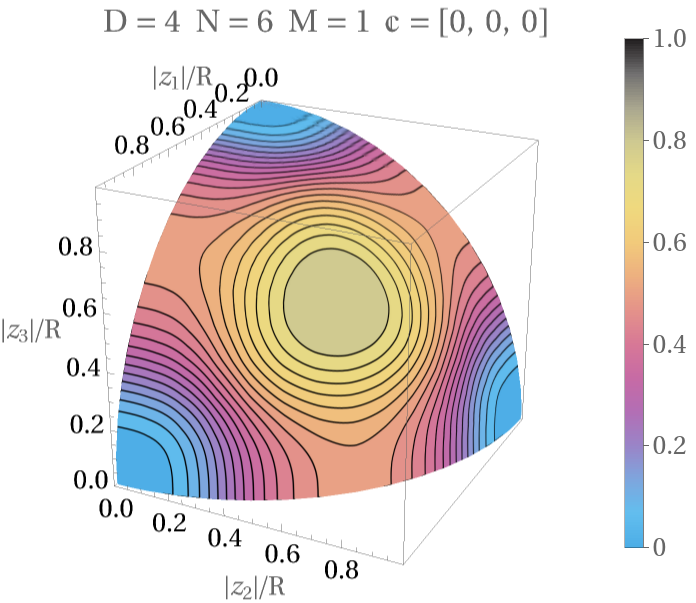}
\caption{Information diagram (left) for $|z_i|\in[0,2]$ and contour plot for the angular dependence (right) for $\|\zb\|=R=10$,  for $D=4, \quad N=6, \quad  M=1, \quad \mathbbm{c}=[0,0,0]$.}
\label{NTL-4-6-1-000}
\end{figure}
\end{center}

\begin{center}
\begin{figure}[h!]
\includegraphics[width=\graphwidth]{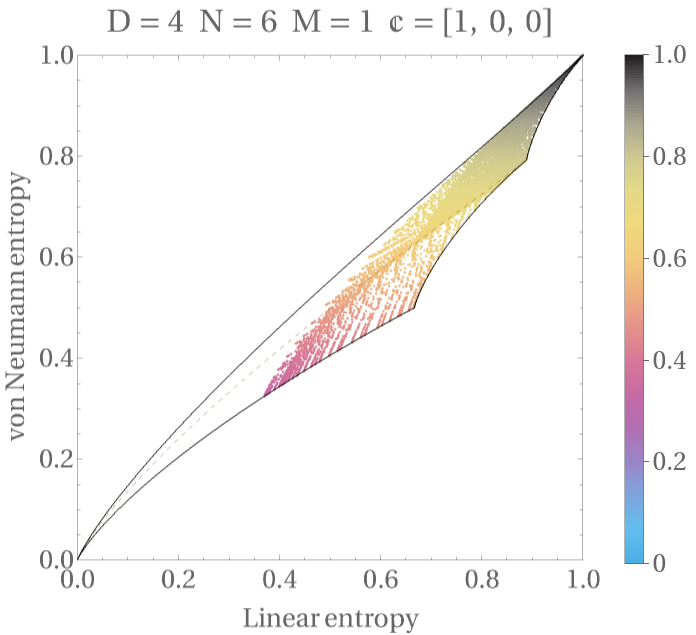}\hspace{\graphsep}
\includegraphics[width=\graphwidth]{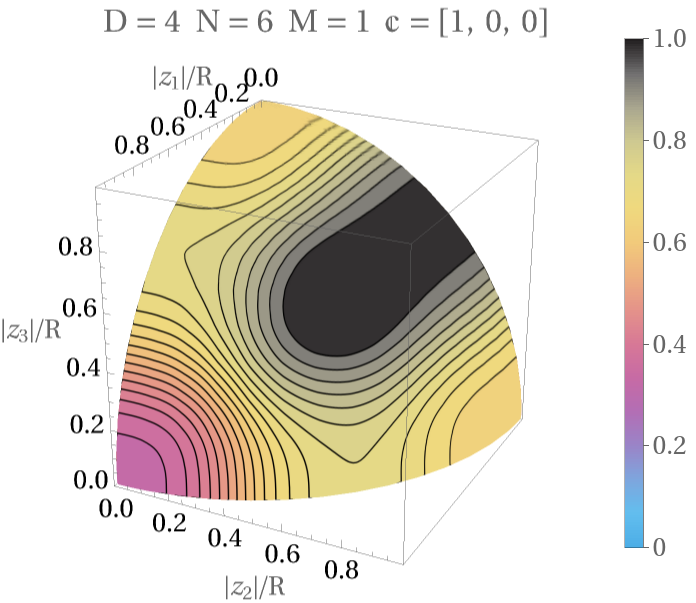}
\caption{Information diagram (left) for $|z_i|\in[0,2]$ and contour plot for the angular dependence (right) for $\|\zb\|=R=10$,  for $D=4, \quad N=6, \quad  M=1, \quad \mathbbm{c}=[1,0,0]$.}
\label{NTL-4-6-1-100}
\end{figure}
\end{center}

\begin{center}
\begin{figure}[h!]
\includegraphics[width=\graphwidth]{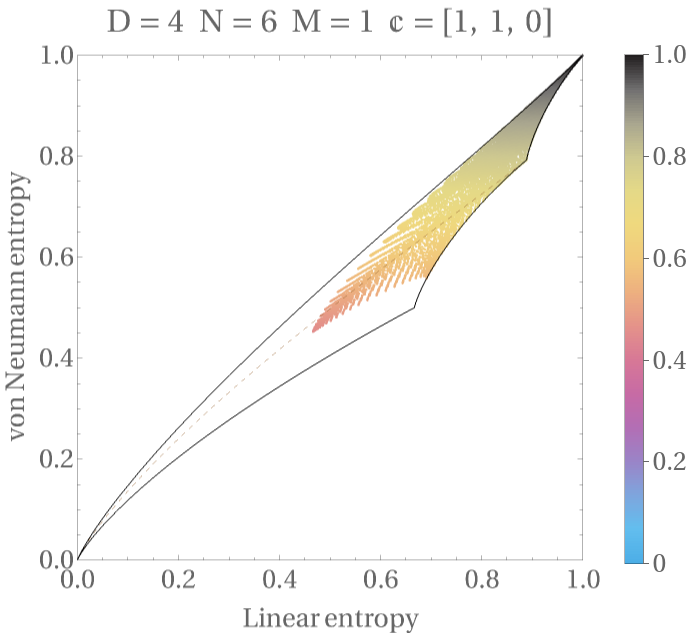}\hspace{\graphsep}
\includegraphics[width=\graphwidth]{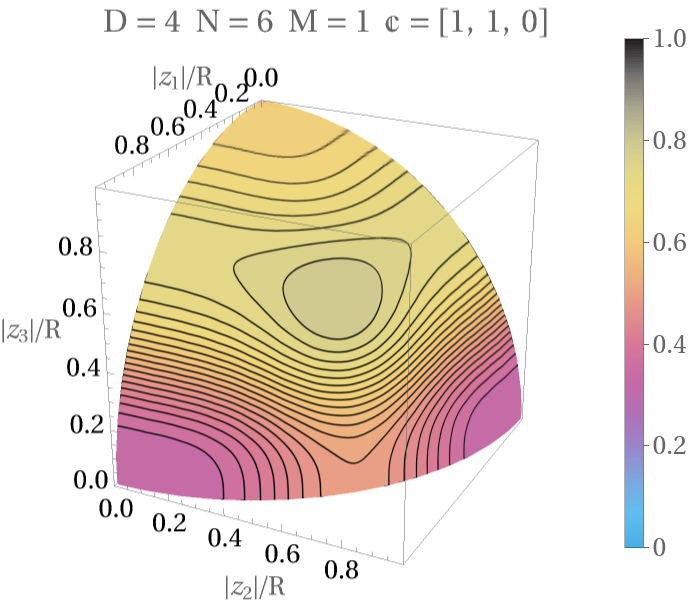}
\caption{Information diagram (left) for $|z_i|\in[0,2]$ and contour plot for the angular dependence (right) for $\|\zb\|=R=10$,  for $D=4, \quad N=6, \quad  M=1, \quad \mathbbm{c}=[1,1,0]$.}
\label{NTL-4-6-1-110}
\end{figure}
\end{center}

\begin{center}
\begin{figure}[h!]
\includegraphics[width=\graphwidth]{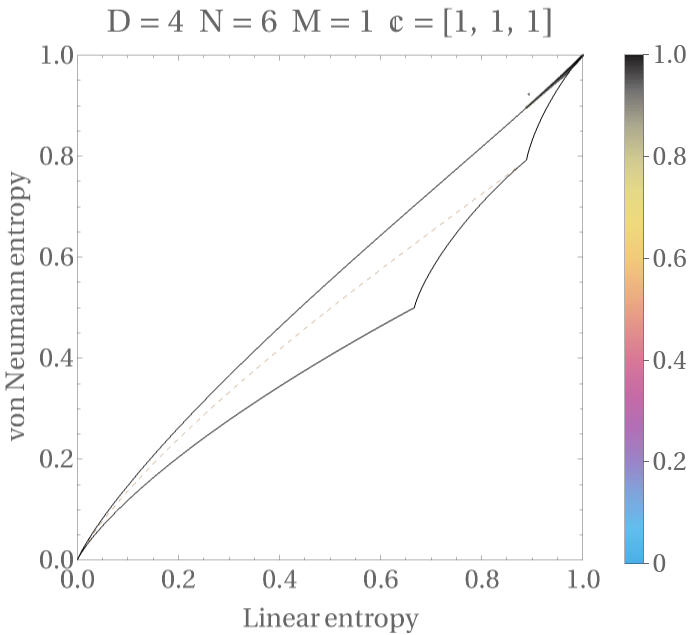}\hspace{\graphsep}
\includegraphics[width=\graphwidth]{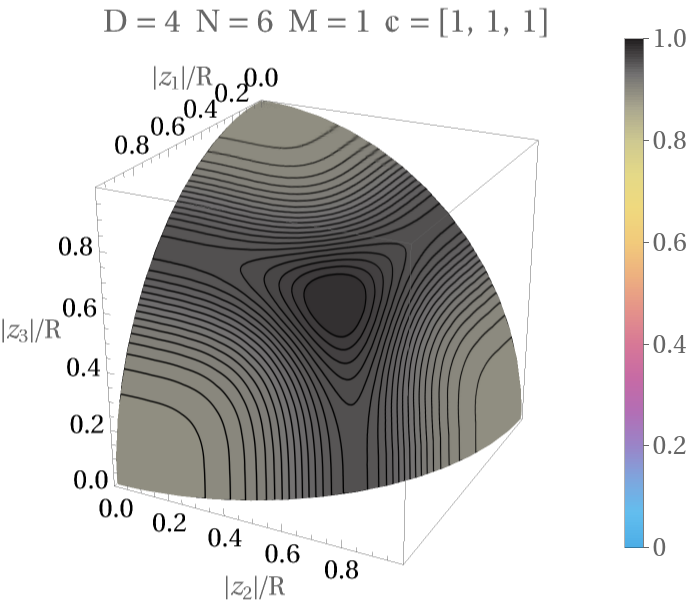}
\caption{Information diagram (left) for $|z_i|\in[0,2]$ and contour plot for the angular dependence (right) for $\|\zb\|=R=10$,  for $D=4, \quad N=6, \quad  M=1, \quad \mathbbm{c}=[1,1,1]$.}
\label{NTL-4-6-1-111}
\end{figure}
\end{center}

\begin{center}
\begin{figure}[h!]
\includegraphics[width=\graphwidth]{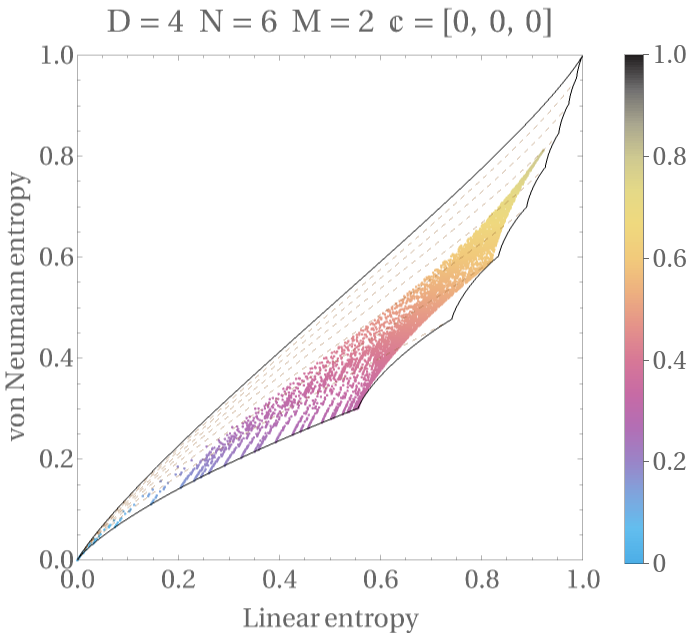}\hspace{\graphsep}
\includegraphics[width=\graphwidth]{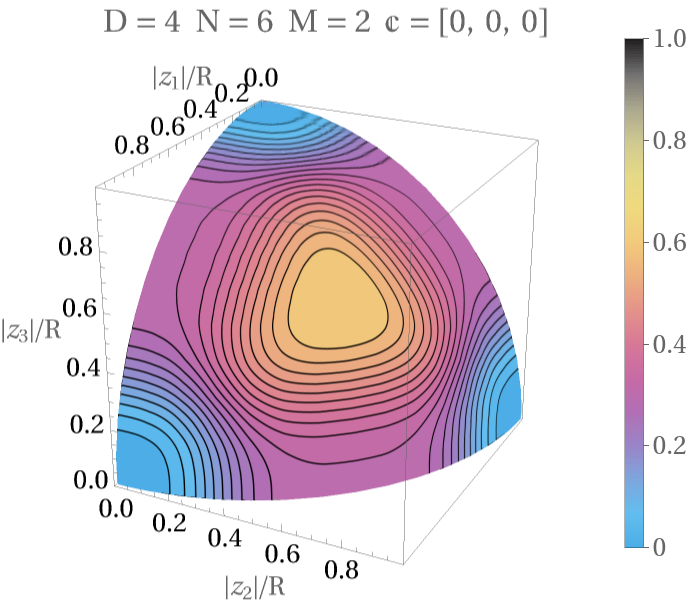}
\caption{Information diagram (left) for $|z_i|\in[0,2]$ and contour plot for the angular dependence (right) for $\|\zb\|=R=10$,  for $D=4, \quad N=6, \quad  M=2, \quad \mathbbm{c}=[0,0,0]$.}
\label{NTL-4-6-2-000}
\end{figure}
\end{center}

\begin{center}
\begin{figure}[h!]
\includegraphics[width=\graphwidth]{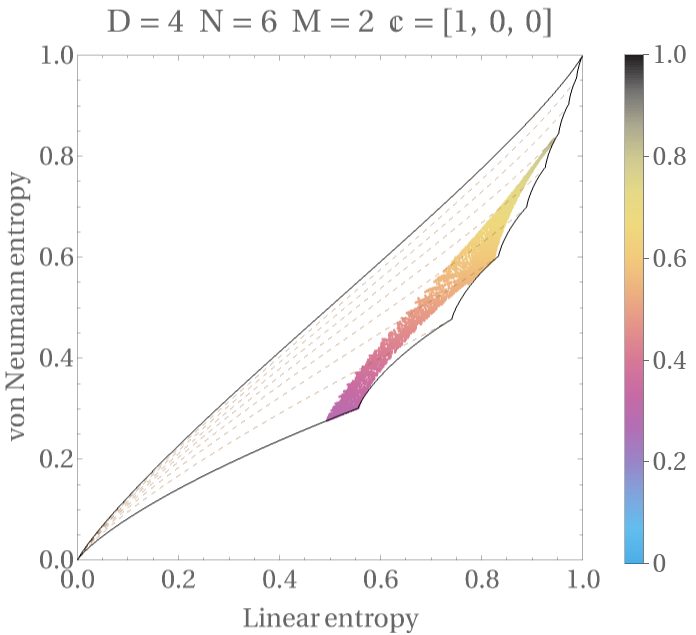}\hspace{\graphsep}
\includegraphics[width=\graphwidth]{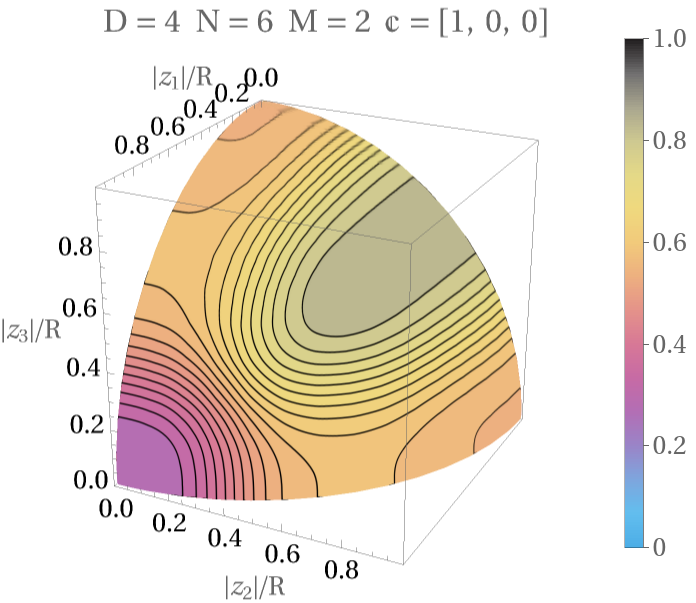}
\caption{Information diagram (left) for $|z_i|\in[0,2]$ and contour plot for the angular dependence (right) for $\|\zb\|=R=10$,  for $D=4, \quad N=6, \quad  M=2, \quad \mathbbm{c}=[1,0,0]$.}
\label{NTL-4-6-2-100}
\end{figure}
\end{center}

\begin{center}
\begin{figure}[h!]
\includegraphics[width=\graphwidth]{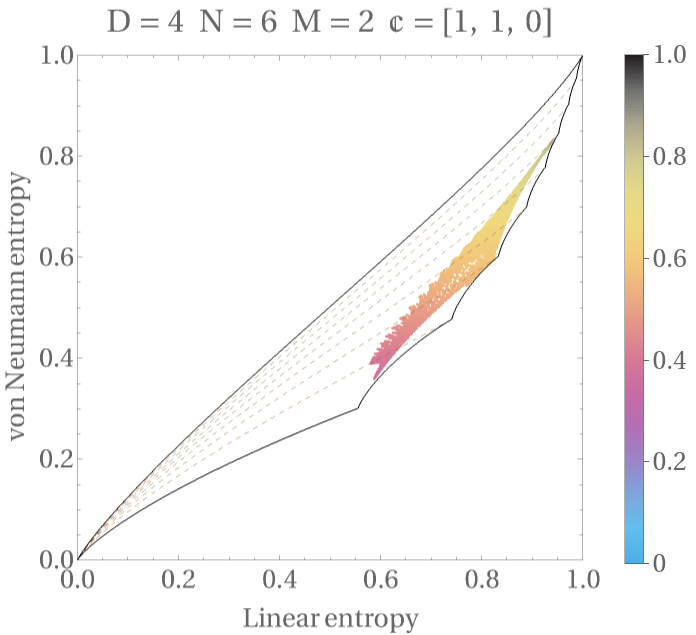}\hspace{\graphsep}
\includegraphics[width=\graphwidth]{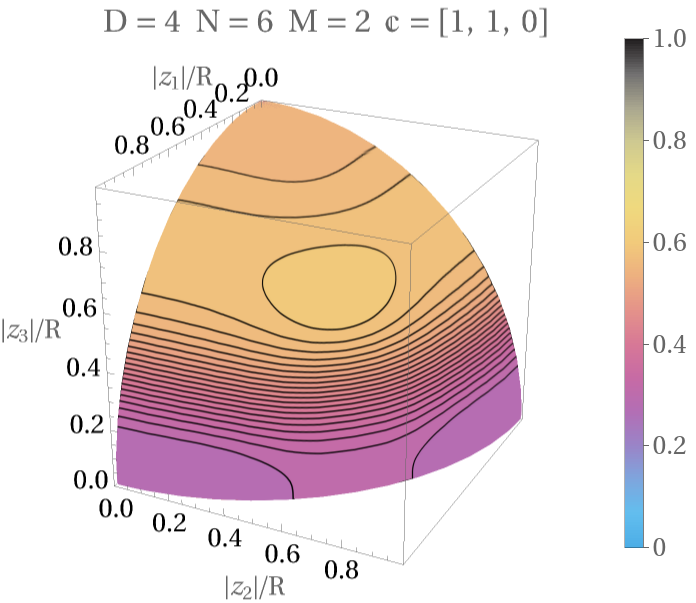}
\caption{Information diagram (left) for $|z_i|\in[0,2]$ and contour plot for the angular dependence (right) for $\|\zb\|=R=10$,  for $D=4, \quad N=6, \quad  M=2, \quad \mathbbm{c}=[1,1,0]$.}
\label{NTL-4-6-2-110}
\end{figure}
\end{center}

\begin{center}
\begin{figure}[h!]
\includegraphics[width=\graphwidth]{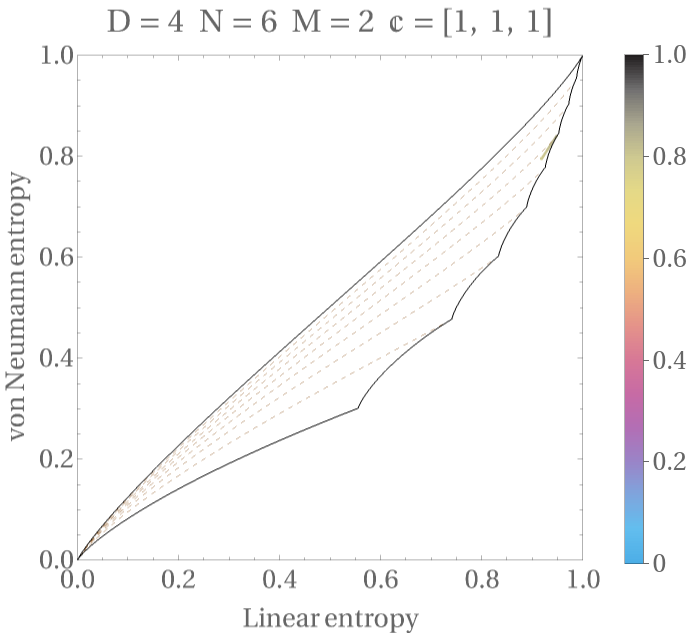}\hspace{\graphsep}
\includegraphics[width=\graphwidth]{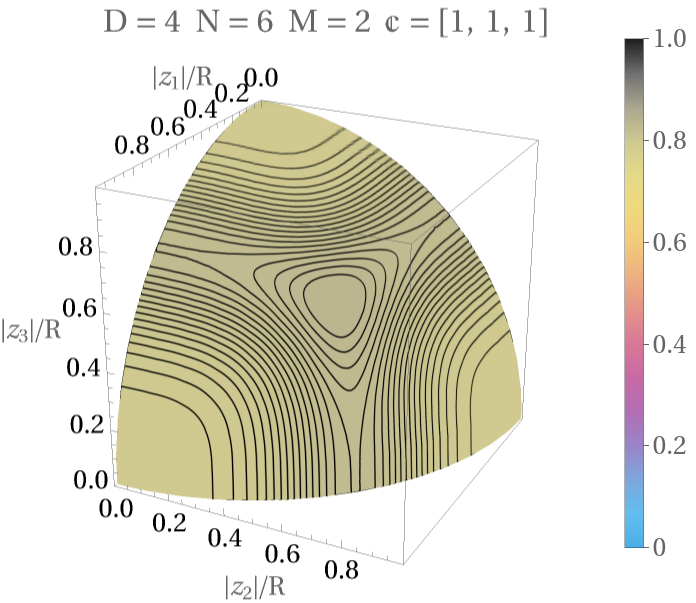}
\caption{Information diagram (left) for $|z_i|\in[0,2]$ and contour plot for the angular dependence (right) for $\|\zb\|=R=10$,  for $D=4, \quad N=6, \quad  M=2, \quad \mathbbm{c}=[1,1,1]$.}
\label{NTL-4-6-2-111}
\end{figure}
\end{center}

\begin{center}
\begin{figure}[h!]
\includegraphics[width=\graphwidth]{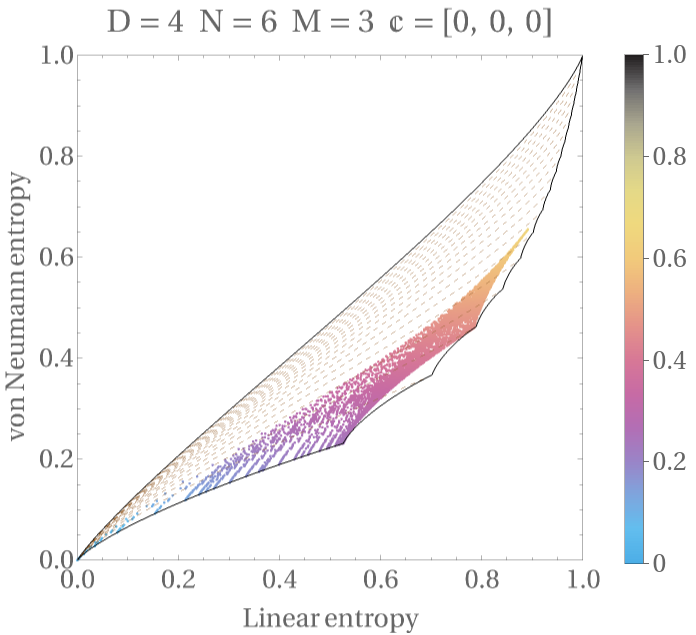}\hspace{\graphsep}
\includegraphics[width=\graphwidth]{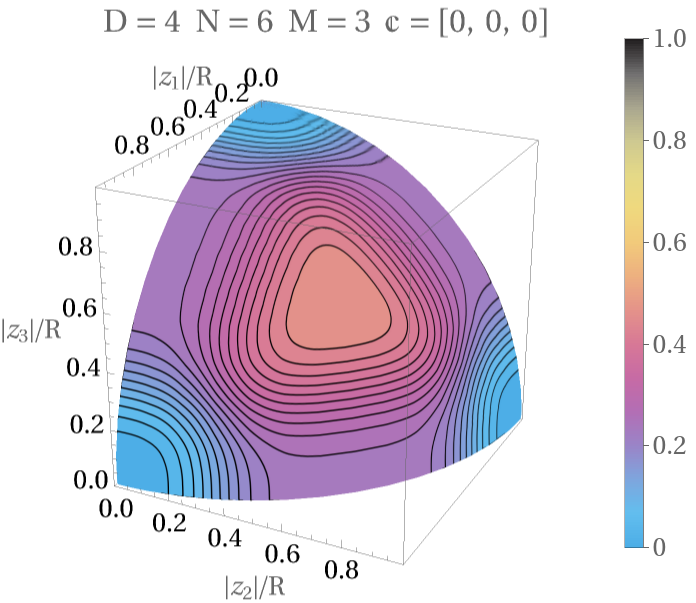}
\caption{Information diagram (left) for $|z_i|\in[0,2]$ and contour plot for the angular dependence (right) for $\|\zb\|=R=10$,  for $D=4, \quad N=6, \quad  M=3, \quad \mathbbm{c}=[0,0,0]$.}
\label{NTL-4-6-3-000}
\end{figure}
\end{center}

\begin{center}
\begin{figure}[h!]
\includegraphics[width=\graphwidth]{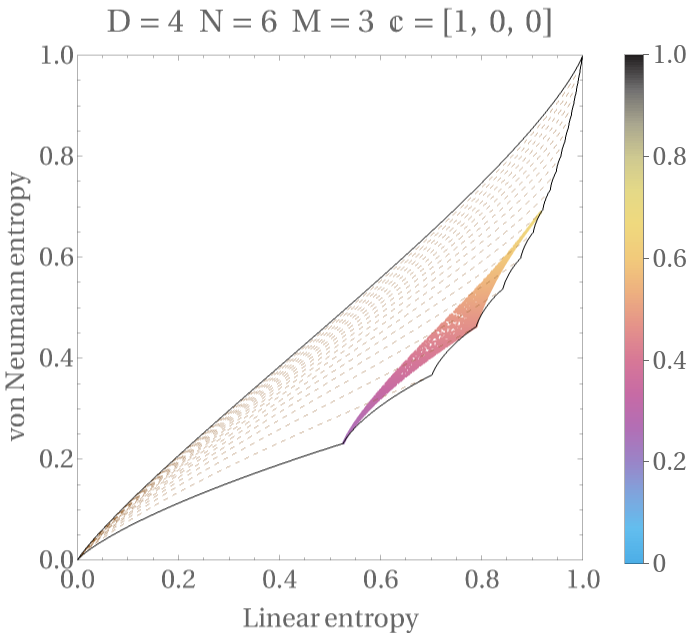}\hspace{\graphsep}
\includegraphics[width=\graphwidth]{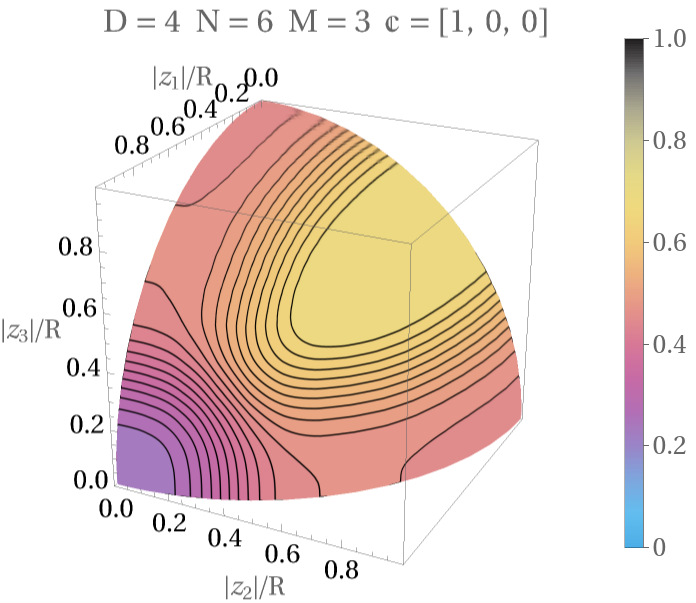}
\caption{Information diagram (left) for $|z_i|\in[0,2]$ and contour plot for the angular dependence (right) for $\|\zb\|=R=10$,  for $D=4, \quad N=6, \quad  M=3, \quad \mathbbm{c}=[1,0,0]$.}
\label{NTL-4-6-3-100}
\end{figure}
\end{center}

\begin{center}
\begin{figure}[h!]
\includegraphics[width=\graphwidth]{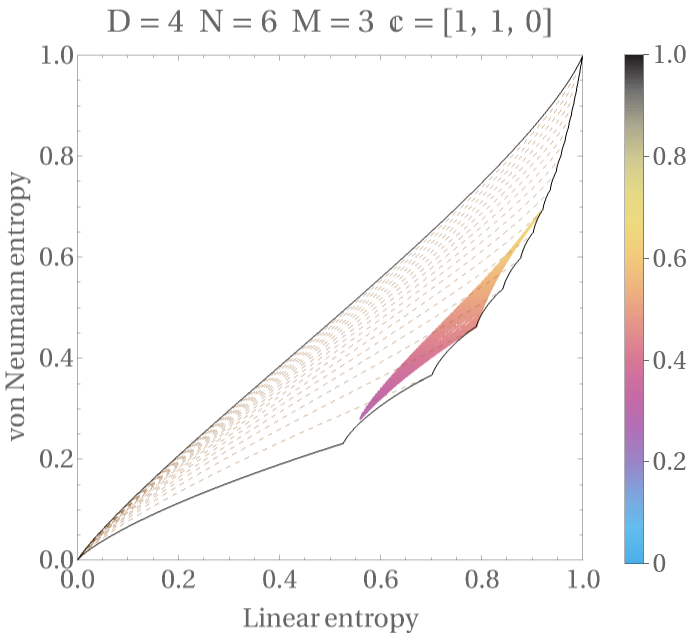}\hspace{\graphsep}
\includegraphics[width=\graphwidth]{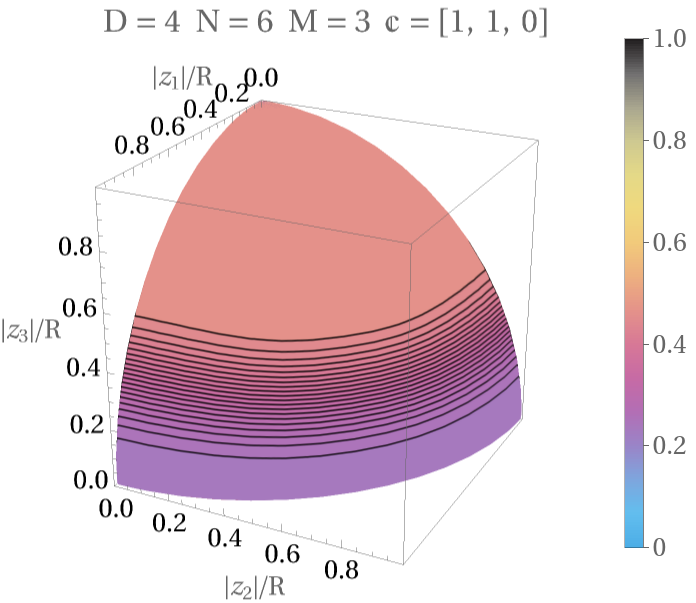}
\caption{Information diagram (left) for $|z_i|\in[0,2]$ and contour plot for the angular dependence (right) for $\|\zb\|=R=10$,  for $D=4, \quad N=6, \quad  M=3, \quad \mathbbm{c}=[1,1,0]$.}
\label{NTL-4-6-3-110}
\end{figure}
\end{center}

\begin{center}
\begin{figure}[h!]
\includegraphics[width=\graphwidth]{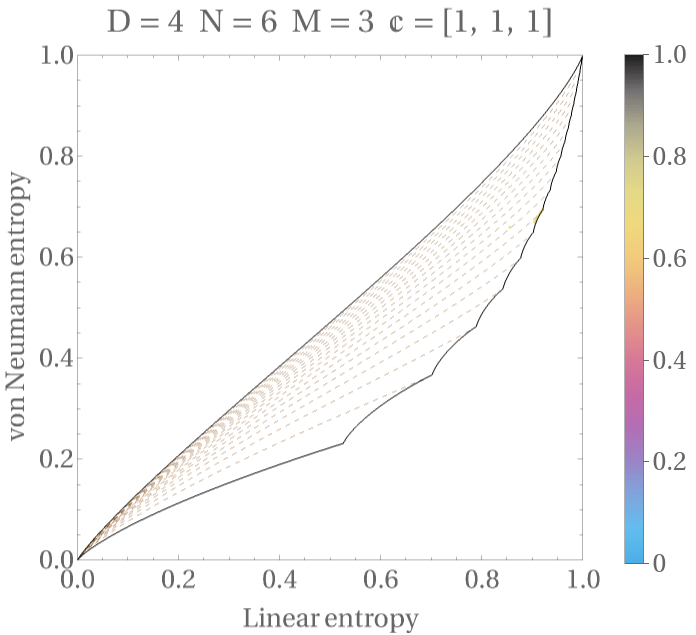}\hspace{\graphsep}
\includegraphics[width=\graphwidth]{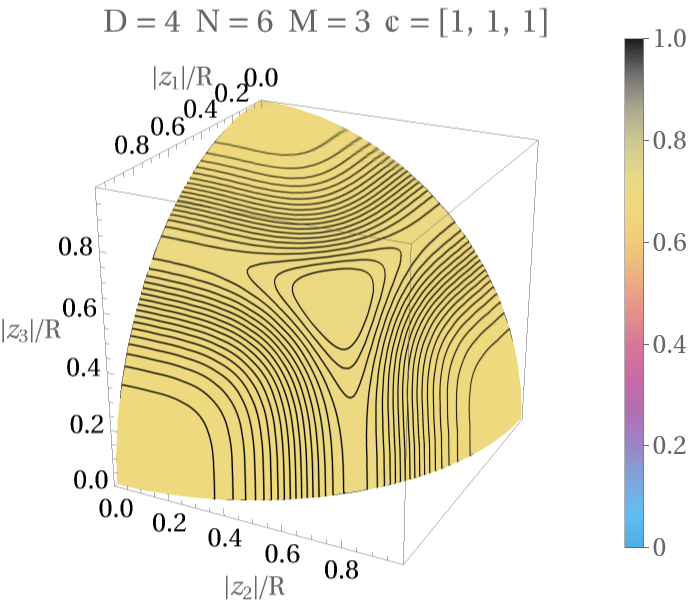}
\caption{Information diagram (left) for $|z_i|\in[0,2]$ and contour plot for the angular dependence (right) for $\|\zb\|=R=10$,  for $D=4, \quad N=6, \quad  M=3, \quad \mathbbm{c}=[1,1,1]$.}
\label{NTL-4-6-3-111}
\end{figure}
\end{center}

In Figures \ref{NTL-4-6-1-000}-\ref{NTL-4-6-3-111} plots of Information Diagrams for $|z_i|\in [0,2]$, for  $D=4$, $N=6$, $M=1,2,3$ and parities $\mathbbm{c}=[0,0,0],[1,0,0],[1,1,0],[1,1,1]$ (for the remaining parities the information diagrams are the same as the ones with the same number of non-zero entries). In this case we also use von Neumann entropy as colormap for the points. Also, contour plots of the von Neumann entropy are shown for $\|\zb\|=R=10$, showing the angular dependence of the isentropic curves in the limit $\|\zb\|\rightarrow \infty$.


 \subsection{Single thermodynamic limit}

\begin{center}
\begin{figure}[h!]
\includegraphics[width=\graphwidth]{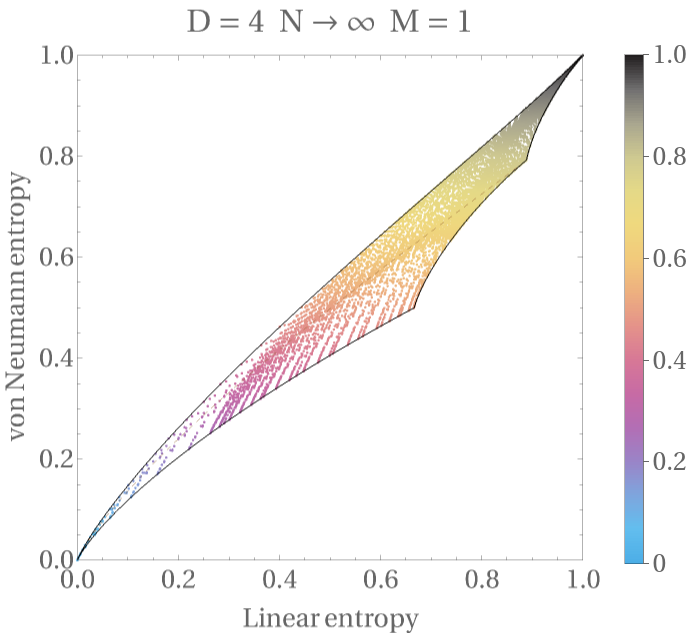}\hspace{\graphsep}
\includegraphics[width=\graphwidth]{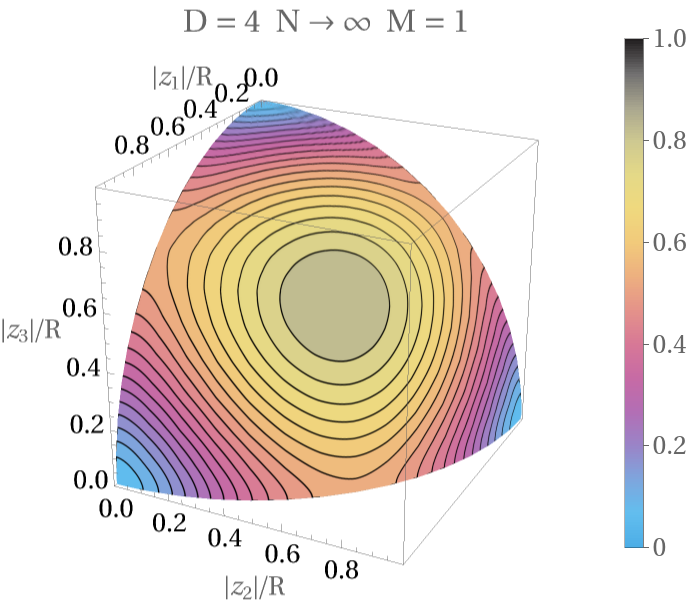}
\caption{Information diagram (left) for $|z_i|\in [0,2]$ and contour plot for the angular dependence (right) for $\|\zb\|_2=R=10$,   for $D=4, \quad N\rightarrow \infty, \quad  M=1$.}
\label{D4-M1-TL}
\end{figure}
\end{center}

\begin{center}
\begin{figure}[h!]
\includegraphics[width=\graphwidth]{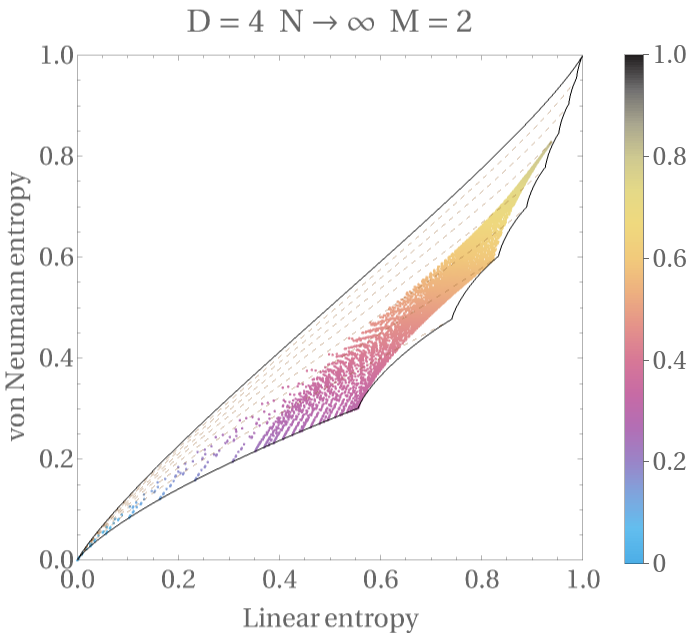}\hspace{\graphsep}
\includegraphics[width=\graphwidth]{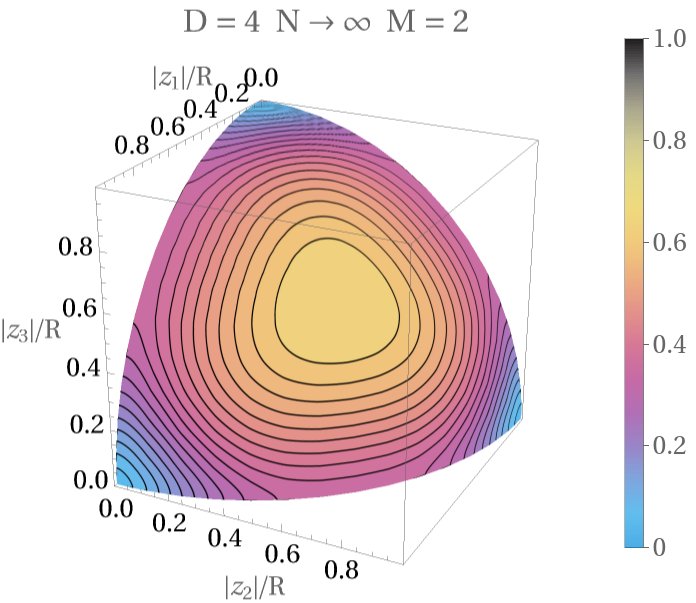}
\caption{Information diagram (left) for $|z_i|\in [0,2]$ and contour plot for the angular dependence (right) for $\|\zb\|_2=R=10$,   for $D=4, \quad N\rightarrow \infty, \quad  M=2$.}
\label{D4-M2-TL}
\end{figure}
\end{center}

\begin{center}
\begin{figure}[h!]
\includegraphics[width=\graphwidth]{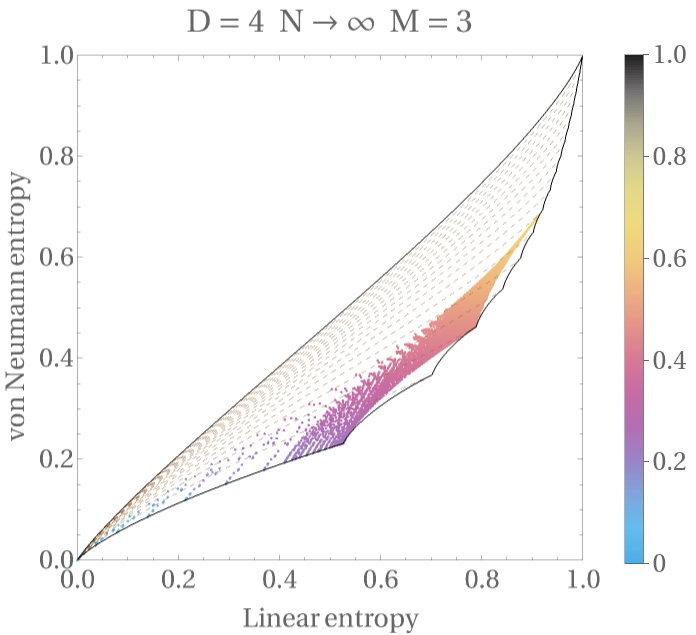}\hspace{\graphsep}
\includegraphics[width=\graphwidth]{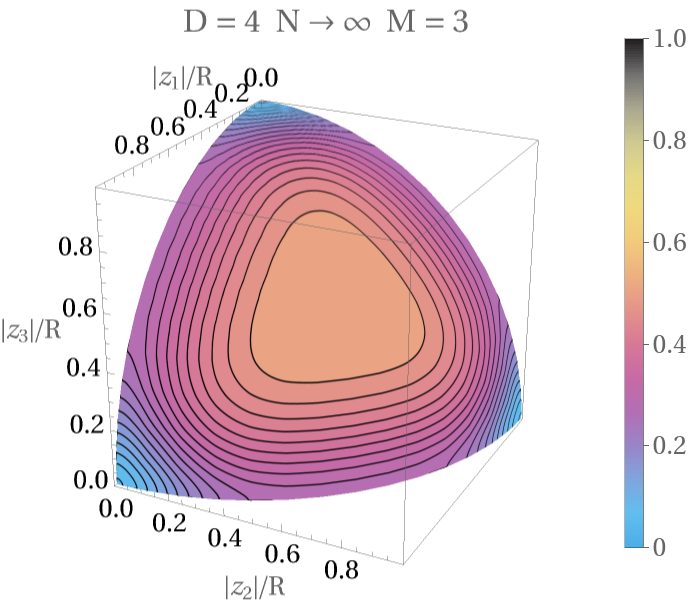}
\caption{Information diagram (left) for $|z_i|\in [0,2]$ and contour plot for the angular dependence (right) for $\|\zb\|_2=R=10$,   for $D=4, \quad N\rightarrow \infty, \quad  M=3$.}
\label{D4-M3-TL}
\end{figure}
\end{center}

 In the case $N\rightarrow\infty$ (the thermodynamic limit), we plot the information diagrams for $|z_i|\in [0,2], i=1,2,3$ and contour plots of the angular dependence of von Neuman entropy for a large fixed value of $\|\zb\|_2=R=10$ for each value of $M=1,2,3$.  They are qualitatively the same as in the finite $N$ case, with the difference that they do not depend on the original parity $\mathbb{c}$ of the state, and the appearance is that of the completely even case. See Figures \ref{D4-M1-TL}, \ref{D4-M2-TL}, \ref{D4-M3-TL}.


\subsection{Rescaled Double thermodynamic limit}

As in the previous case, the double limit $N,M\rightarrow\infty$ (Double thermodynamic Limit) will not be considered since in this case the RDMs correspond to that of a maximally mixed state of dimension $2^k$, with $k=\|\zb\|_0$.

 For the \textit{directional} limit
 $N,M\rightarrow\infty$ with $M=(1-\eta)N$ and $\eta\in[\frac{1}{2},1)$, rescaling $\zb=\frac{\alphab}{\sqrt{N}}$, we have that, as shown in eq. (65) of the paper, the Schmitd eigenvalues factorize as a product of Schmitd eigenvalues for the decomposition of one-dimensional  Schr\"odinger cat states of the harmonic oscillator.

\begin{center}
\begin{figure}[h!]
\includegraphics[width=\graphwidth]{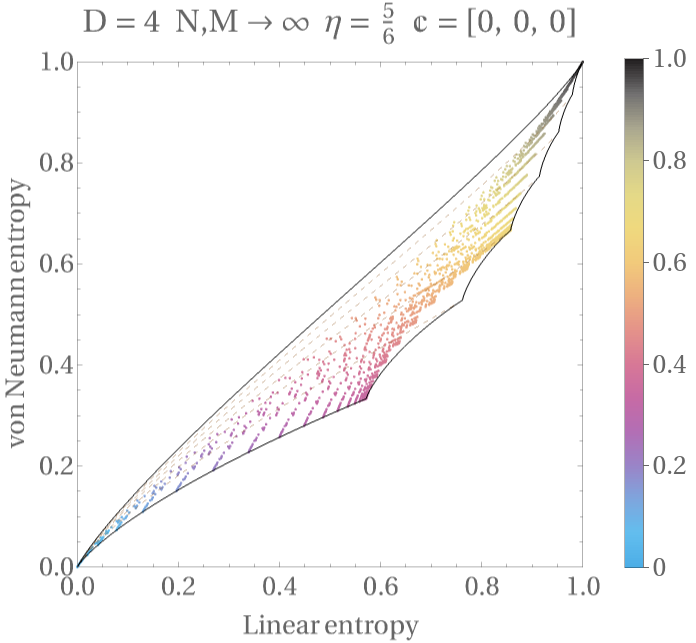}\hspace{\graphsep}
\includegraphics[width=\graphwidth]{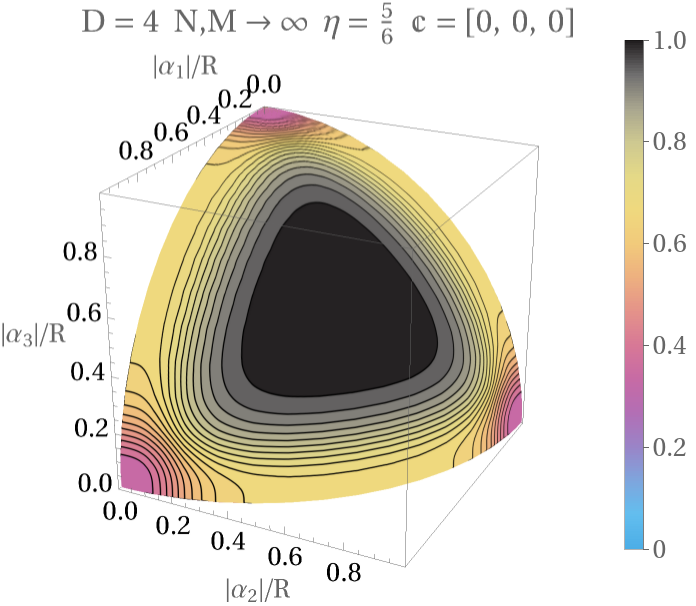}
\caption{Information diagram (left) for $|\alpha_i|\in [0,4]$ and contour plot of angular dependence of von Neumann entropy (right) for $\|\alphab\|=R=5$,  for $D=4, \quad N,M\rightarrow \infty,\quad \eta=\frac{5}{6}, \quad \mathbbm{c}=[0,0,0]$.}
\label{RTL-4-5d6-000}
\end{figure}
\end{center}

\begin{center}
\begin{figure}[h!]
\includegraphics[width=\graphwidth]{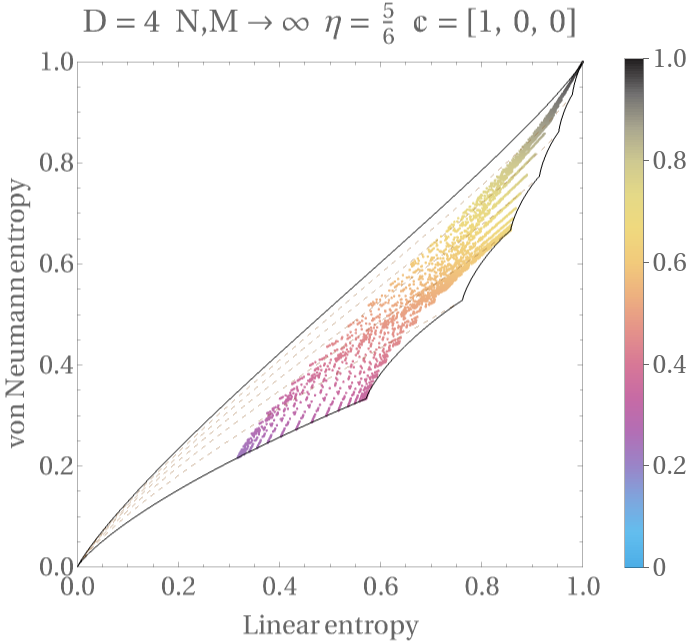}\hspace{\graphsep}
\includegraphics[width=\graphwidth]{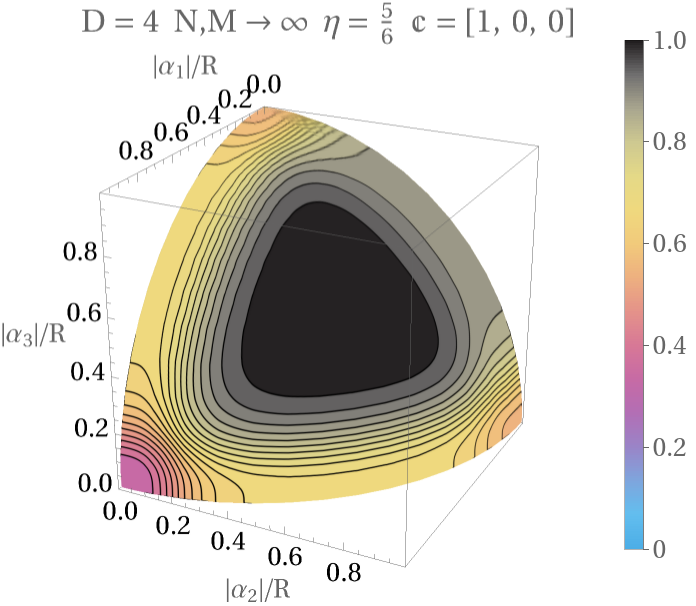}
\caption{Information diagram (left) for $|\alpha_i|\in [0,4]$ and contour plot of angular dependence of von Neumann entropy (right) for $\|\alphab\|=R=5$,  for $D=4, \quad N,M\rightarrow \infty,\quad \eta=\frac{5}{6}, \quad \mathbbm{c}=[1,0,0]$.}
\label{RTL-4-5d6-100}
\end{figure}
\end{center}

\begin{center}
\begin{figure}[h!]
\includegraphics[width=\graphwidth]{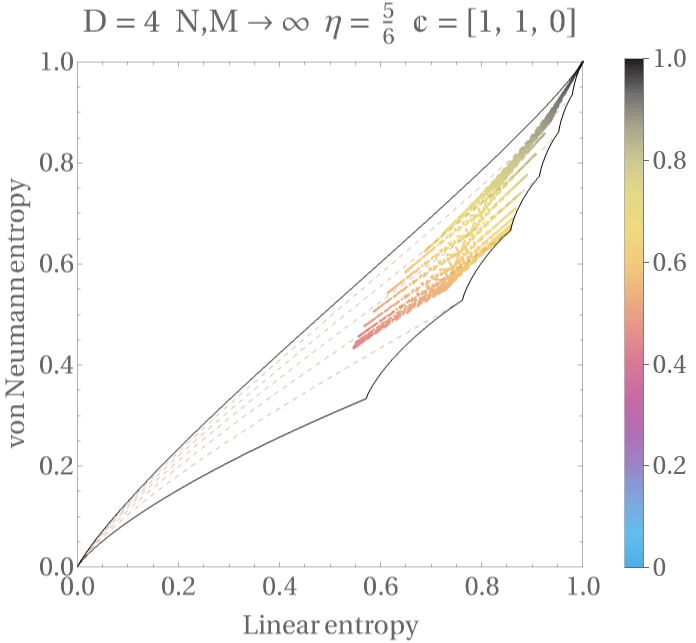}\hspace{\graphsep}
\includegraphics[width=\graphwidth]{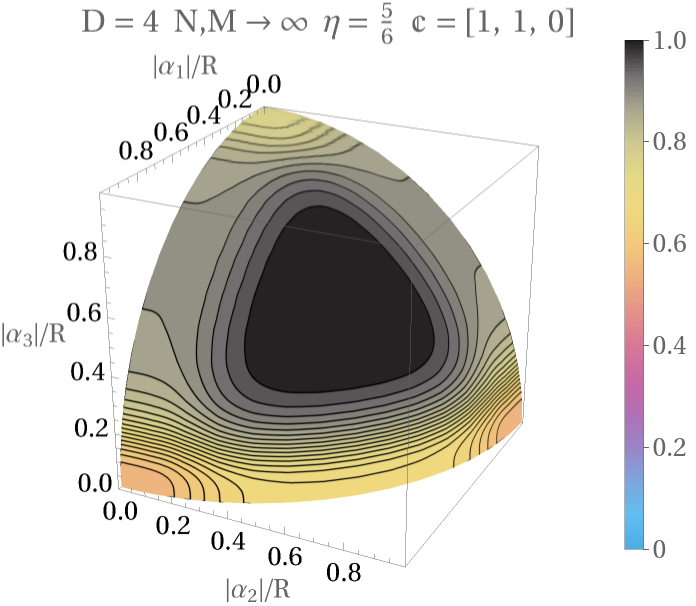}
\caption{Information diagram (left) for $|\alpha_i|\in [0,4]$ and contour plot of angular dependence of von Neumann entropy (right) for $\|\alphab\|=R=5$,  for $D=4, \quad N,M\rightarrow \infty,\quad \eta=\frac{5}{6}, \quad \mathbbm{c}=[1,1,0]$.}
\label{RTL-4-5d6-110}
\end{figure}
\end{center}

\begin{center}
\begin{figure}[h!]
\includegraphics[width=\graphwidth]{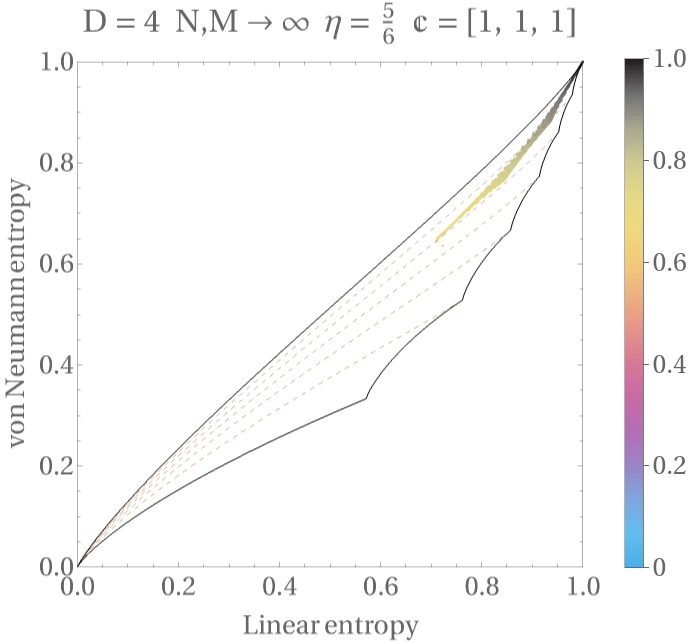}\hspace{\graphsep}
\includegraphics[width=\graphwidth]{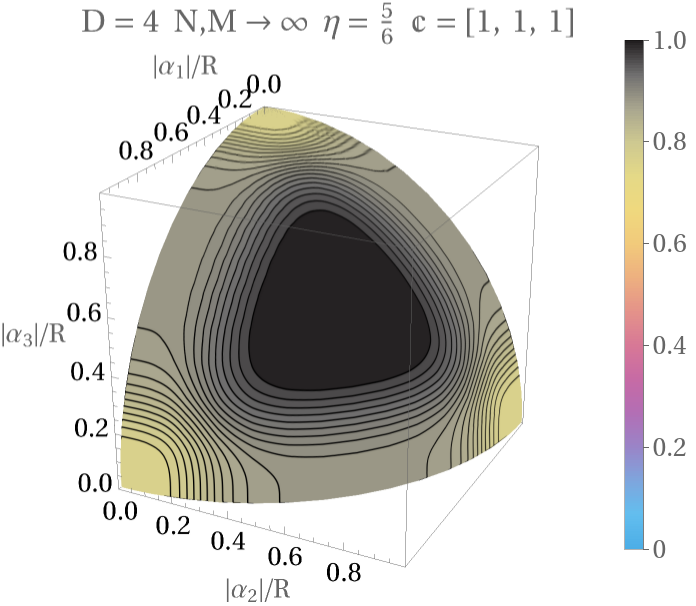}
\caption{Information diagram (left) for $|\alpha_i|\in [0,4]$ and contour plot of angular dependence of von Neumann entropy (right) for $\|\alphab\|=R=5$,  for $D=4, \quad N,M\rightarrow \infty,\quad \eta=\frac{5}{6}, \quad \mathbbm{c}=[1,1,1]$.}
\label{RTL-4-5d6-111}
\end{figure}
\end{center}

\begin{center}
\begin{figure}[h!]
\includegraphics[width=\graphwidth]{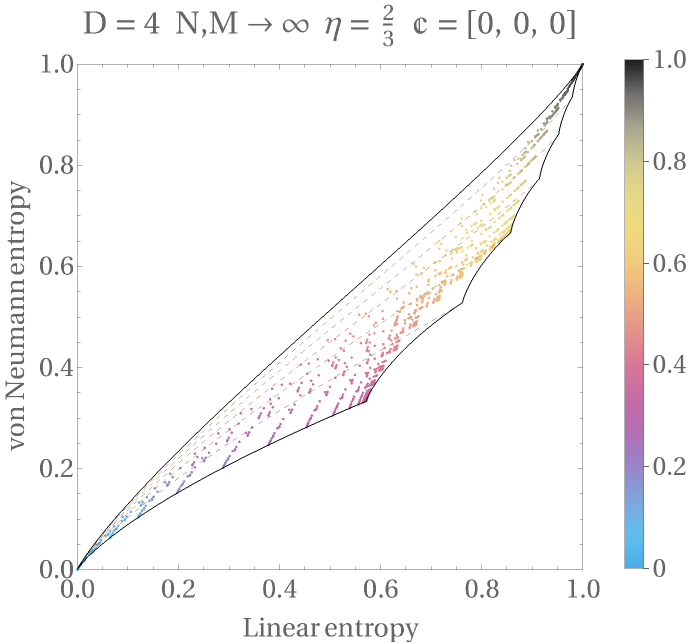}\hspace{\graphsep}
\includegraphics[width=\graphwidth]{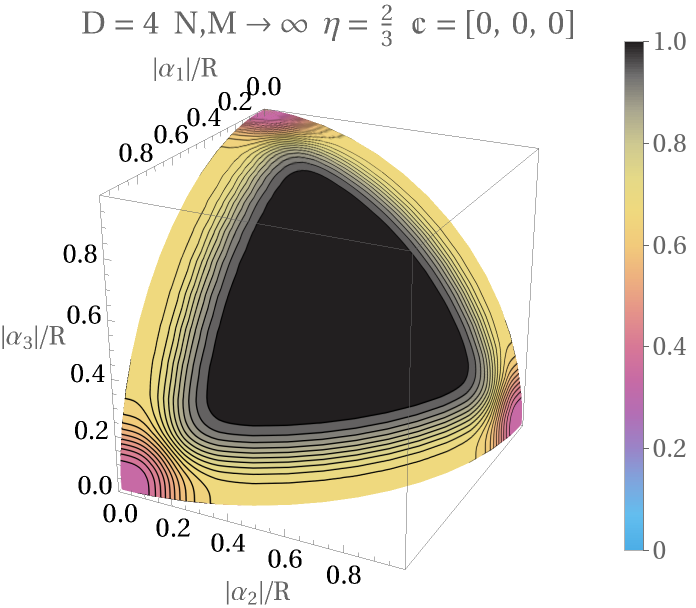}
\caption{Information diagram (left) for $|\alpha_i|\in [0,4]$ and contour plot of angular dependence of von Neumann entropy (right) for $\|\alphab\|=R=5$,  for $D=4, \quad N,M\rightarrow \infty,\quad \eta=\frac{2}{3}, \quad \mathbbm{c}=[0,0,0]$.}
\label{RTL-4-2d3-000}
\end{figure}
\end{center}

\begin{center}
\begin{figure}[h!]
\includegraphics[width=\graphwidth]{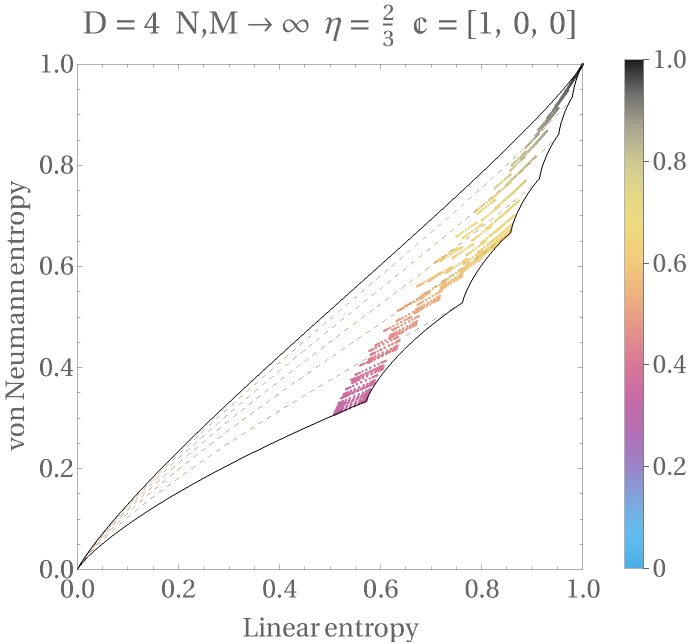}\hspace{\graphsep}
\includegraphics[width=\graphwidth]{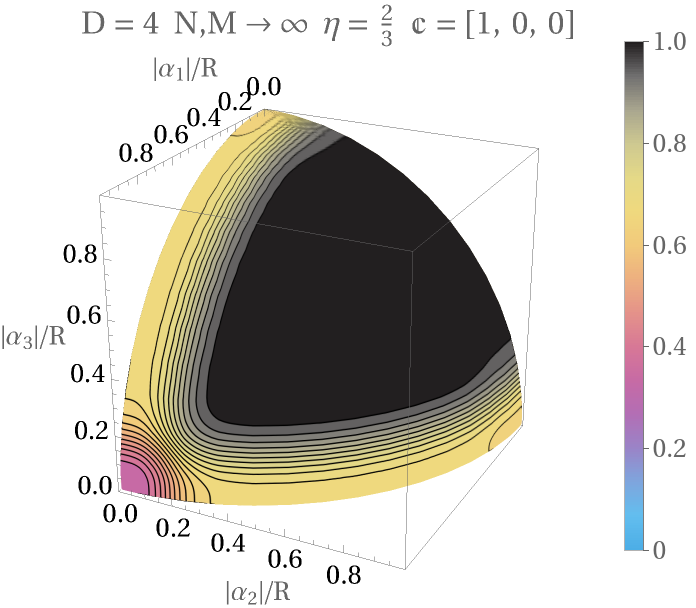}
\caption{Information diagram (left) for $|\alpha_i|\in [0,4]$ and contour plot of angular dependence of von Neumann entropy (right) for $\|\alphab\|=R=5$,  for $D=4, \quad N,M\rightarrow \infty,\quad \eta=\frac{2}{3}, \quad \mathbbm{c}=[1,0,0]$.}
\label{RTL-4-2d3-100}
\end{figure}
\end{center}

\begin{center}
\begin{figure}[h!]
\includegraphics[width=\graphwidth]{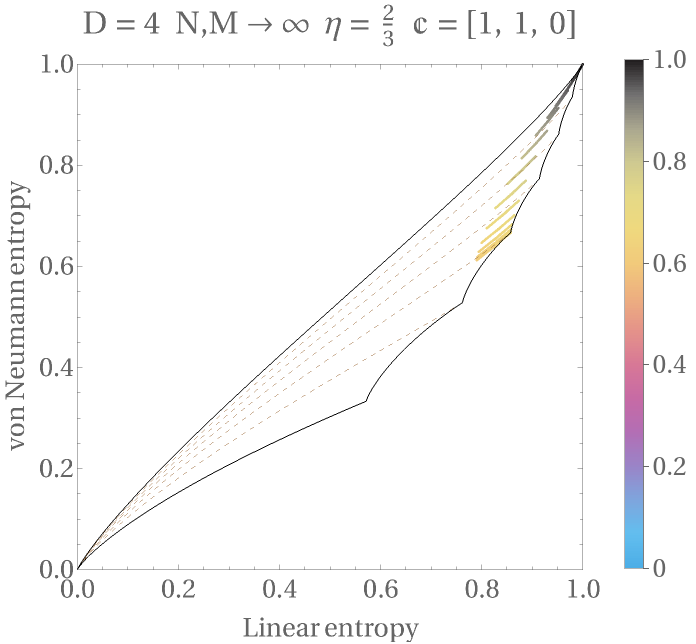}\hspace{\graphsep}
\includegraphics[width=\graphwidth]{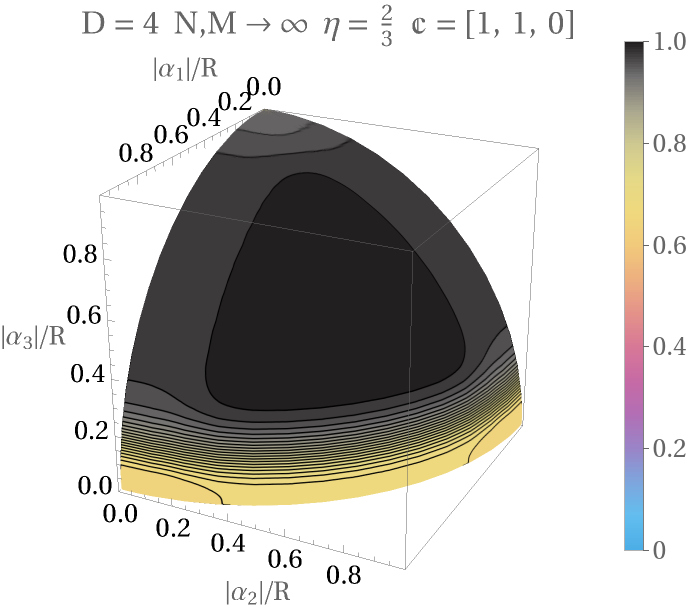}
\caption{Information diagram (left) for $|\alpha_i|\in [0,4]$ and contour plot of angular dependence of von Neumann entropy (right) for $\|\alphab\|=R=5$,  for $D=4, \quad N,M\rightarrow \infty,\quad \eta=\frac{2}{3}, \quad \mathbbm{c}=[1,1,0]$.}
\label{RTL-4-2d3-110}
\end{figure}
\end{center}

\begin{center}
\begin{figure}[h!]
\includegraphics[width=\graphwidth]{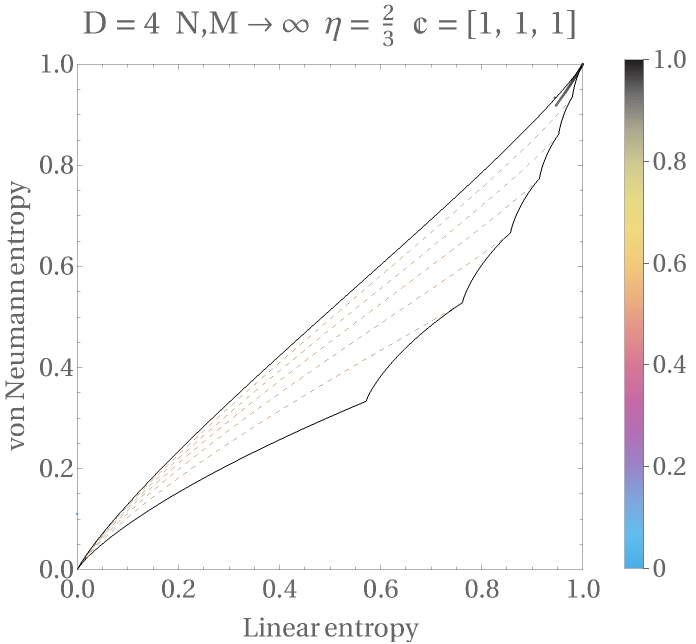}\hspace{\graphsep}
\includegraphics[width=\graphwidth]{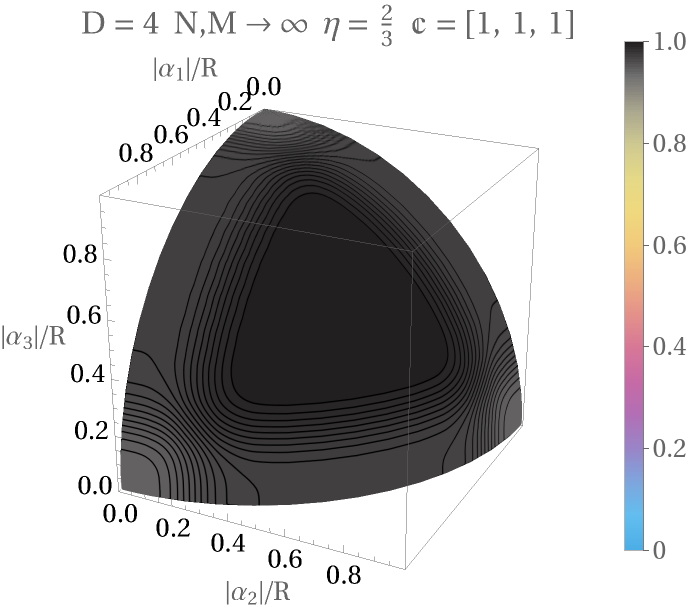}
\caption{Information diagram (left) for $|\alpha_i|\in [0,4]$ and contour plot of angular dependence of von Neumann entropy (right) for $\|\alphab\|=R=5$,  for $D=4, \quad N,M\rightarrow \infty,\quad \eta=\frac{2}{3}, \quad \mathbbm{c}=[1,1,1]$.}
\label{RTL-4-2d3-111}
\end{figure}
\end{center}

\begin{center}
\begin{figure}[h!]
\includegraphics[width=\graphwidth]{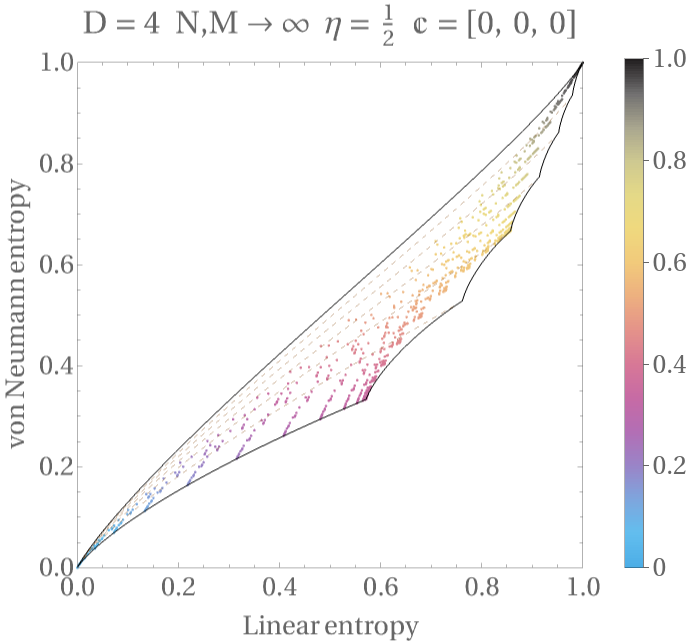}\hspace{\graphsep}
\includegraphics[width=\graphwidth]{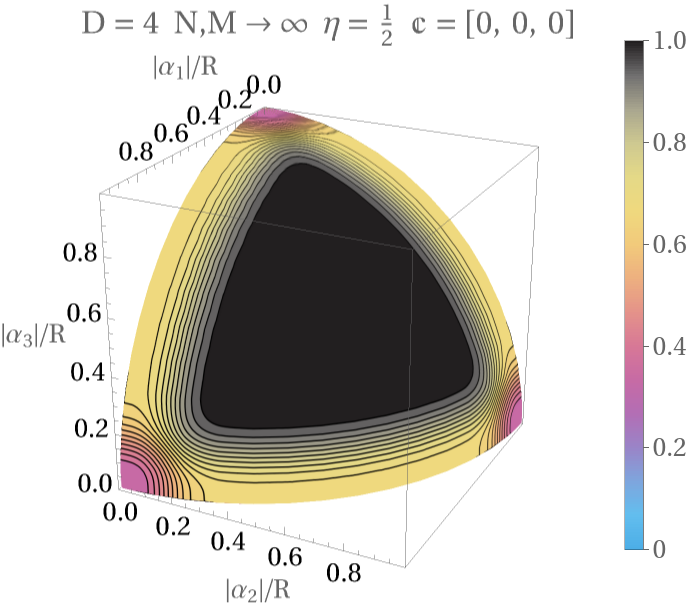}
\caption{Information diagram (left) for $|\alpha_i|\in [0,4]$ and contour plot of angular dependence of von Neumann entropy (right) for $\|\alphab\|=R=5$,  for $D=4, \quad N,M\rightarrow \infty,\quad \eta=\frac{1}{2}, \quad \mathbbm{c}=[0,0,0]$.}
\label{RTL-4-1d2-000}
\end{figure}
\end{center}

\begin{center}
\begin{figure}[h!]
\includegraphics[width=\graphwidth]{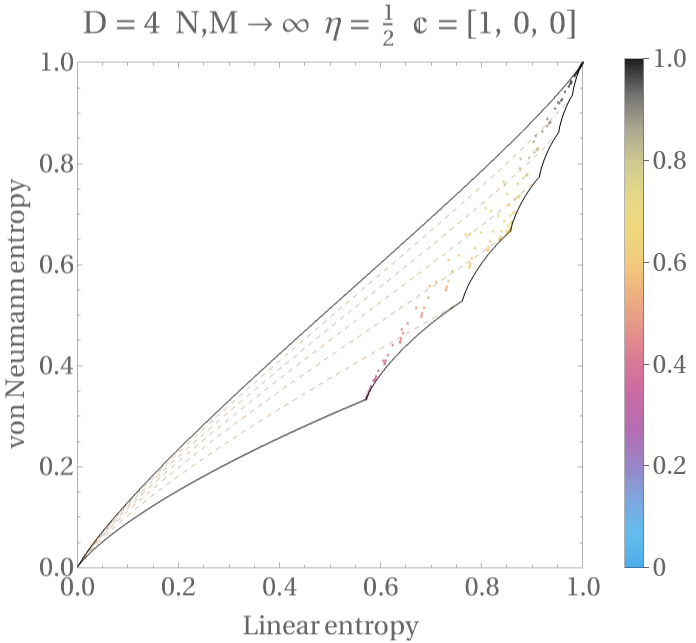}\hspace{\graphsep}
\includegraphics[width=\graphwidth]{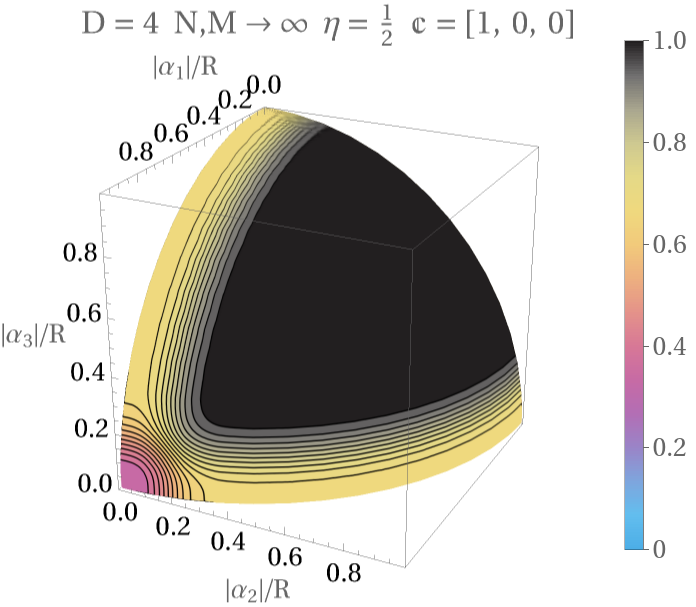}
\caption{Information diagram (left) for $|\alpha_i|\in [0,4]$ and contour plot of angular dependence of von Neumann entropy (right) for $\|\alphab\|=R=5$,  for $D=4, \quad N,M\rightarrow \infty,\quad \eta=\frac{1}{2}, \quad \mathbbm{c}=[1,0,0]$.}
\label{RTL-4-1d2-100}
\end{figure}
\end{center}

\begin{center}
\begin{figure}[h!]
\includegraphics[width=\graphwidth]{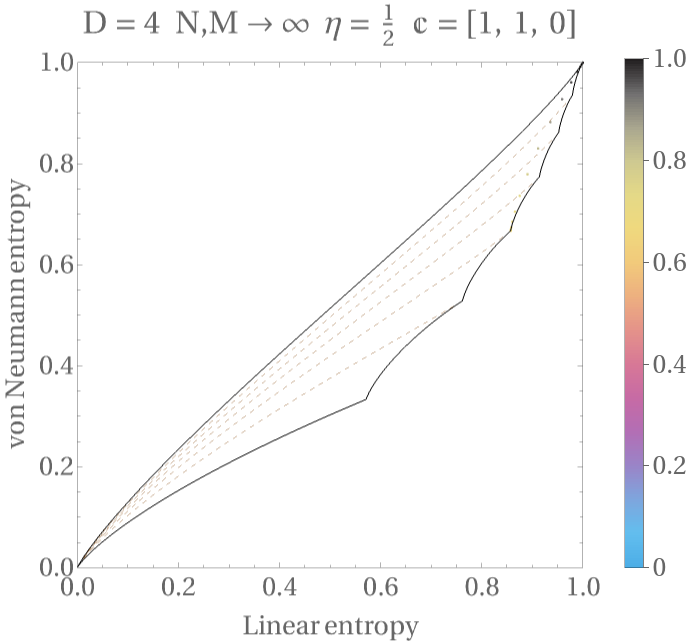}\hspace{\graphsep}
\includegraphics[width=\graphwidth]{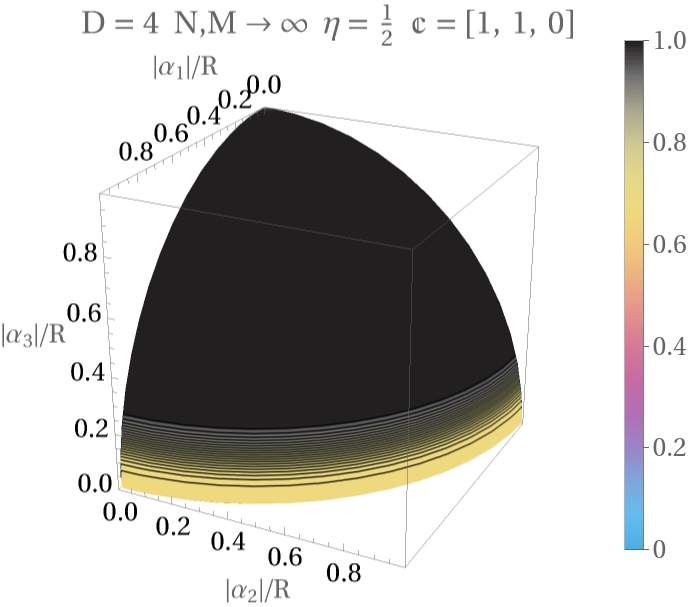}
\caption{Information diagram (left) for $|\alpha_i|\in [0,4]$ and contour plot of angular dependence of von Neumann entropy (right) for $\|\alphab\|=R=5$,  for $D=4, \quad N,M\rightarrow \infty,\quad \eta=\frac{1}{2}, \quad \mathbbm{c}=[1,1,0]$.}
\label{RTL-4-1d2-110}
\end{figure}
\end{center}

\begin{center}
\begin{figure}[h!]
\includegraphics[width=\graphwidth]{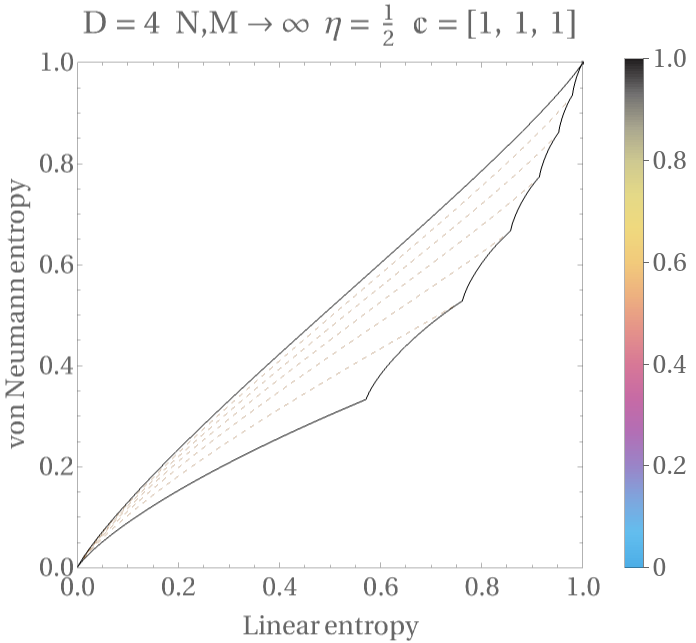}\hspace{\graphsep}
\includegraphics[width=\graphwidth]{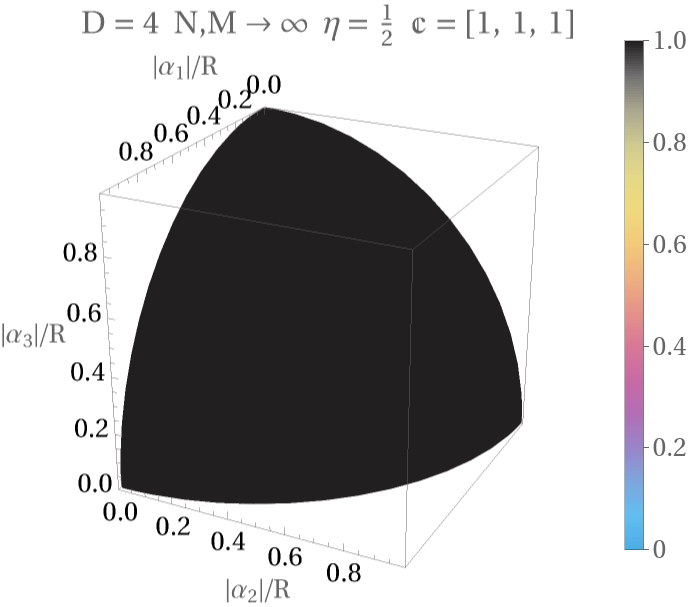}
\caption{Information diagram (left) for $|\alpha_i|\in [0,4]$ and contour plot of angular dependence of von Neumann entropy (right) for $\|\alphab\|=R=5$,  for $D=4, \quad N,M\rightarrow \infty,\quad \eta=\frac{1}{2}, \quad \mathbbm{c}=[1,1,1]$.}
\label{RTL-4-1d2-111}
\end{figure}
\end{center}

In  Figures \ref{RTL-4-5d6-000}-\ref{RTL-4-1d2-111} we show information diagrams with $|\alpha_i|\in [0,4]$, and contour plots of the angular dependence of von Neumann entropy for a large fixed value of $\|\alphab\|_2=R=5$ for each value of $\eta=5/6,2/3,1/2$, and the different parities $\mathbbm{c}=[0,0,0], [1,0,0],[1,1,0],[1,1,1]$. The same comments as in the case $D=3$ apply here.



\section{Figures for a large number of atom levels (qupentits)}
\label{FigD5}

For higher values of $D$, for instance $D=5$, it is more difficult to represent graphically the dependence of the entropy of the RDM in terms of the values of $\zb$, since now we have four variales, $(|z_1|, |z_2|,|z_3|,|z_4|)$.  We can not even use angular plots for a fixed $\|\zb\|=R$, since they lie in a hypersphere $S^3$. This problem is even worse with larger values of $D$. Thus we only have at our disposal information diagrams, which are always bidimensional, helping us with the colors of the points to codify the relevant extra information.

Until now we have codified in the color the value of the von Neumann entropy, which is redundant since it corresponds to the vertical axis. This was done to compare with the contour plots in the case $D=3$, and it was also used for $D=4$.

We could continue to use the value of von Neumann entropy as a colormap for larger values of $D$, but it could be useful to codify another relevant information (since von Neumann entropy is just the vertical axis). It would be interesting to codify the values of $\zb$ to introduce ``spatial'' information in the information diagrams (as it is done in contour plots). The complexity of information diagrams does not allow for a direct codification of the spatial information (for instance, for $D=3$ the map $(|z_1|,|z_2|)\rightarrow ({\cal L},{\cal S})$ is not area preserving, it is not even injective, and clearly is not surjective). However, restricting the region in the $\zb$ domain it is possible to codify, in a qualitative way, the spatial information.

\begin{center}
\begin{figure}[h!]
\includegraphics[width=\graphwidth]{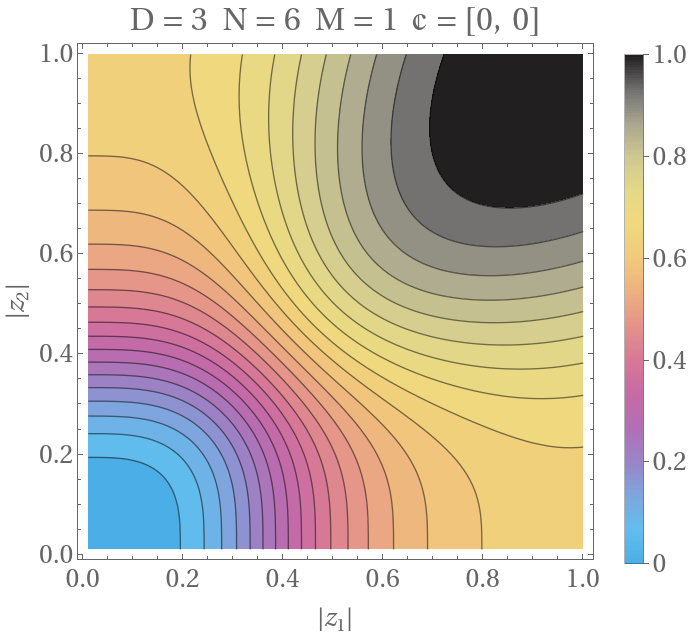}\hspace{\graphsep}
\includegraphics[width=\graphwidth]{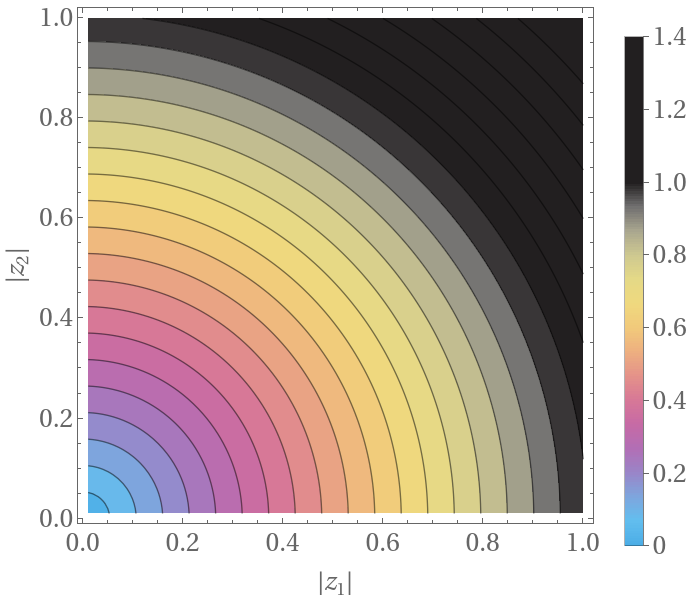}\hspace{\graphsep}
\includegraphics[width=\graphwidth]{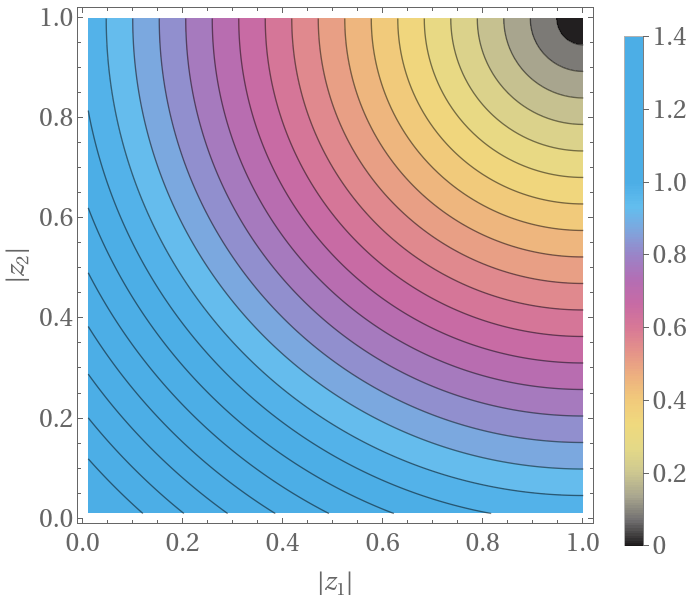}
\caption{Contour plot of von Neumann entropy (left) for $|z_i|\in[0,1]$ for $D=3\,, \quad N=6\,, \quad M=1\,, \quad \mathbbm{c}=[0,0]$, compared with  contour plots of the distance (with the 2-norm) to the point $(0,0)$ (center) and  to the point $(1,1)$ (right). In the case of the distance to the point $(1,1)$ the colormap has been reversed (darker colors correspond to smaller distances) to compare with the values of von Neumann entropies.}
\label{Contour-norm}
\end{figure}
\end{center}

\begin{center}
\begin{figure}[h!]
\includegraphics[width=\graphwidth]{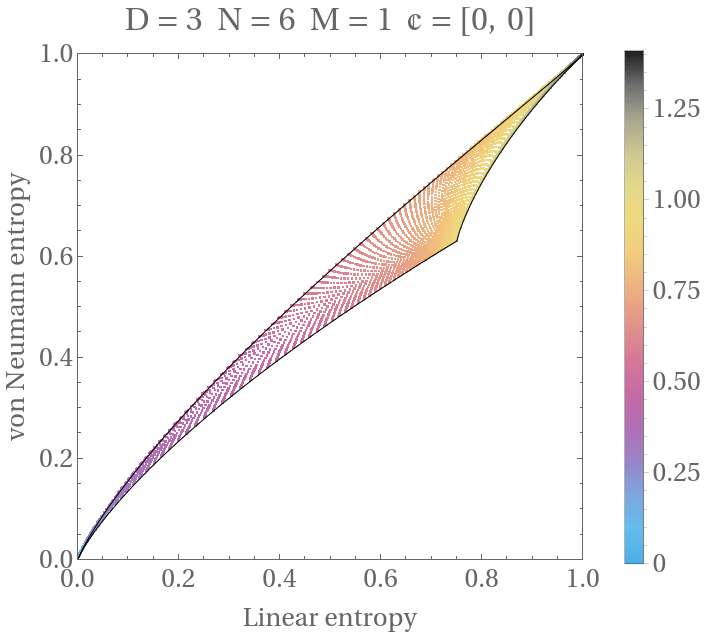}\hspace{\graphsep}
\includegraphics[width=\graphwidth]{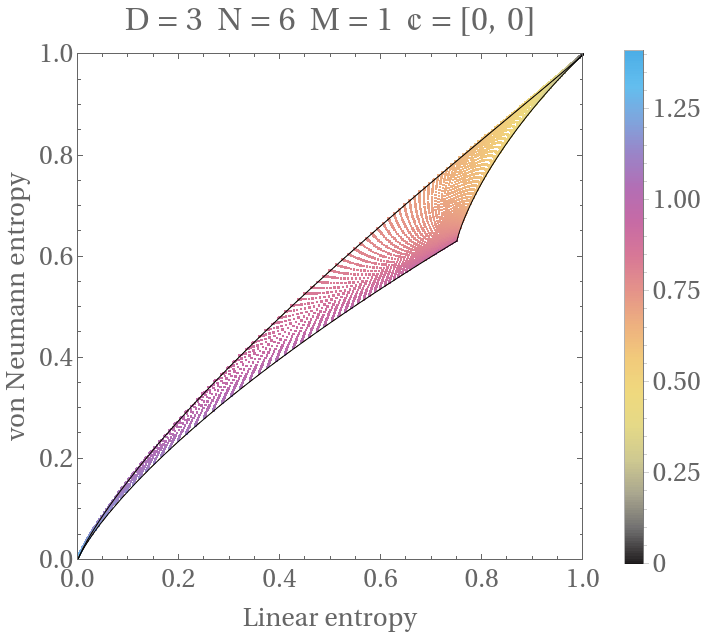}
\caption{Information diagrams for $|z_i|\in[0,1]$ (for $D=3\,, \quad N=6\,, \quad M=1\,, \quad \mathbbm{c}=[0,0]$) using as colormaps the distance to the point $(0,0)$ (left) and $(1,1)$ (right). In the last case the colormap has been reversed to keep darker colors with larger values of entropy.}
\label{InfDiag-Norm-D3}
\end{figure}
\end{center}

As it can be seen in Figure \ref{Contour-norm}, near the point $(0,0)$ there is a good qualitative agreement between the contour plot of the von Neumann entropy and the contour plot of the Euclidean distance to the point $(0,0)$. At the same time, near the point $(1,1)$, there is a good qualitative agreement between the contour plot of the von Neumann entropy and the contour plot of the Euclidean distance to the point $(1,1)$ (the colormap has been reversed in this case, therefore darker colors corresponds to smaller distances).

Using this, in Fig. \ref{InfDiag-Norm-D3} we plot information diagrams for $D=3$ (and $N=6$, $M=1$, $\mathbbm{c}=[0,0]$) using as colormap the distance to the point $(0,0)$ (left) and the distance to the point $(1,1)$ (right), with reversed colormap. Both information diagrams are qualitatively very similar, and indicates that the Euclieand distance is a good indicator to provide spatial information in information diarams. This continues to be true for other values of $M$, and it is a good approximation with different values of $\mathbbm{c}$ (although the approximation is not as good as in the case $\mathbbm{c}=[0,0]$ since the symmetry of the contour lines is lost as the number of ones in $\mathbbm{c}$ increases).

\begin{center}
\begin{figure}[h!]
\includegraphics[width=\graphwidth]{Angular-D4/NTL/Angular-D4-N6-M1-c000.png}\hspace{\graphsep}
\includegraphics[width=\graphwidth]{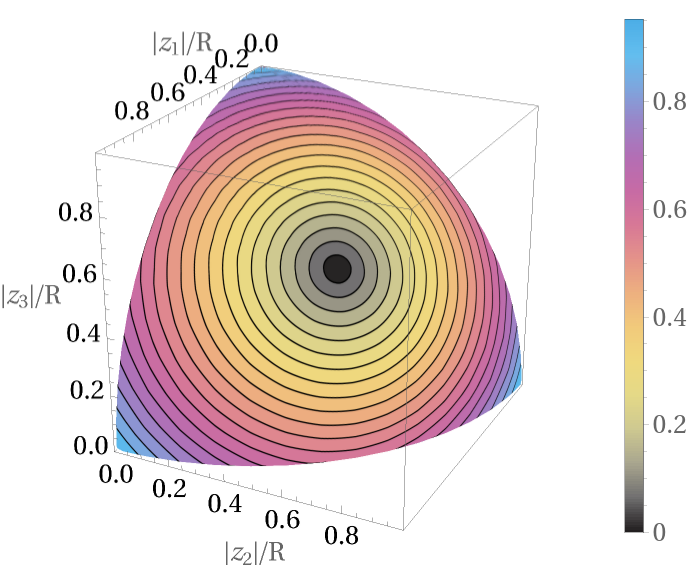}\hspace{\graphsep}
\includegraphics[width=\graphwidth]{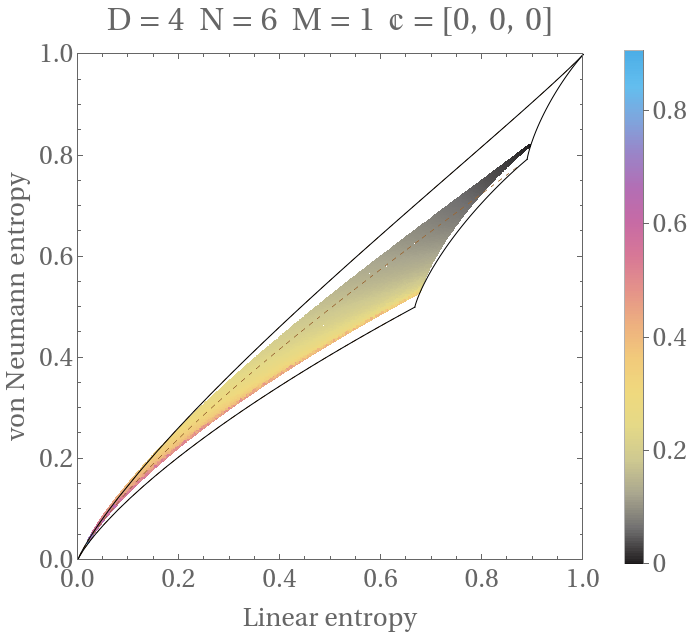}
\caption{(left) Contour plot of the angular dependence (left) of von Neumann entropy for $\|\zb\|=R=5$ (for $D=4\,, \quad N=6\,, \quad M=1\,, \quad \mathbbm{c}=[0,0]$), compared with (center) contour plots of the angular distance to the direction of the vector $(1,1,1)$ (in this case  the colormap has been reversed (darker colors corresponds to smaller distances, to compare with the values of von Neumann entropies).}
\label{Angular-Distance-D4}
\end{figure}
\end{center}

Since in the Eucliedean distance the angular information is lost, and this is important when  the large $\zb$ behaviour is analysed (for instance, for $D=4$ we used contour plots of von Neumann entropy on the first octant of the sphere for fixed $\|\zb\|=R$), in order to generalize this for larger values of $D$ we shall try to incorporate the angular information as a colormap in the information diagrams. In Fig. \ref{Angular-Distance-D4} we compare the contour plot (left) of von Neumann entropy as a function of the angle for $\|\zb\|=R=5$ and $D=4$ (with $N=6$, $M=1$ and $\mathbbm{c}=[0,0,0]$), with a contour plot (centre) of the angular distance to the direction of the vector $(1,1,1)$. Clearly there is a good agreement, which continues for other values of $M$ and that it is not as good when the number of ones in $\mathbbm{c}$ increases (in the sense that the value of the entropy at the diferent axes can be different). In Fig. \ref{Angular-Distance-D4} (right) we plot the information diagram for $\|\zb\|=R=5$ and $D=4$ (with $N=6$, $M=1$ and $\mathbbm{c}=[0,0,0]$) using as colormap the angular distance to the direction of the vector $(1,1,1)$, and we see that there is a clear distinction in the information diagram between the different angular separations.

In the following subsections, information diagrams with both distance and angular dependence are shown for the case $D=5$. In the case of the distance, the range of variation of $\zb$ is $|z_i|\in [0,1],\, i=1,2,3,4$, the distance is measured with respect to the point $(1,1,1,1)$ and the colormap scale has been reversed. For the angular dependence, the value of $\|\zb\|=R=5$ has been chosen, the angular distance is measured with respect to the direction of the vector $(1,1,1,1)$, and the colormap  scale has been also reversed.


\subsection{Finite number $N$ of qupentits}


\begin{center}
\begin{figure}[h!]
\includegraphics[width=\graphwidth]{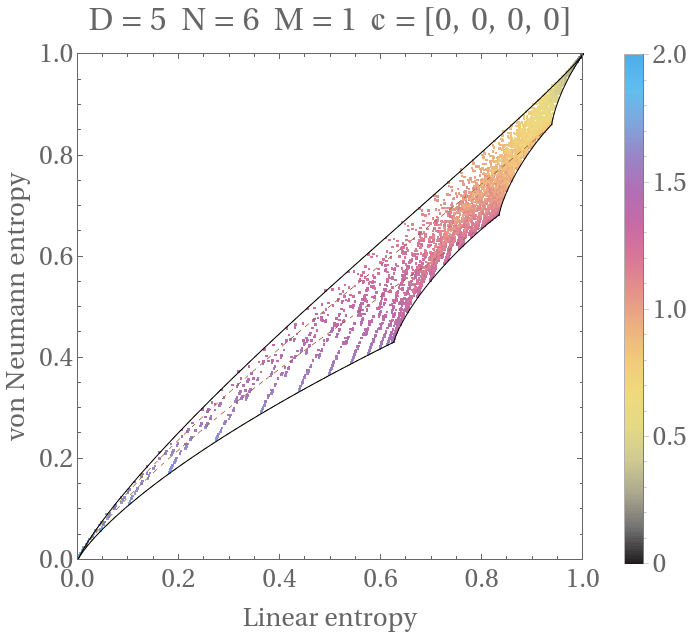}\hspace{\graphsep}
\includegraphics[width=\graphwidth]{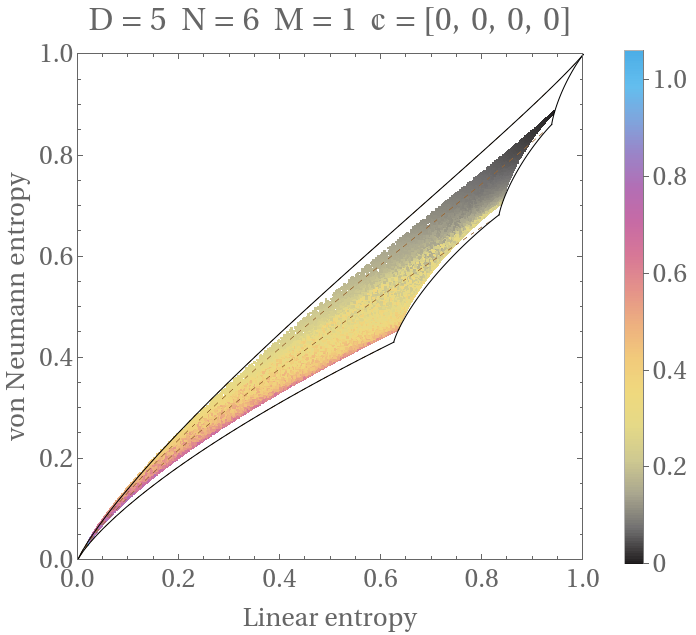}
\caption{Information diagrams using as colormap the distance (left) for $|z_i|\in[0,1]$ and $D=5$, $N=6$, $M=1$ and $\mathbbm{c}=[0,0,0,0]$. The same using as colormap the angular dependence (right) for $\|\zb\|=R=5$.}
\label{NTL-5-6-1-0000}
\end{figure}
\end{center}

\begin{center}
\begin{figure}[h!]
\includegraphics[width=\graphwidth]{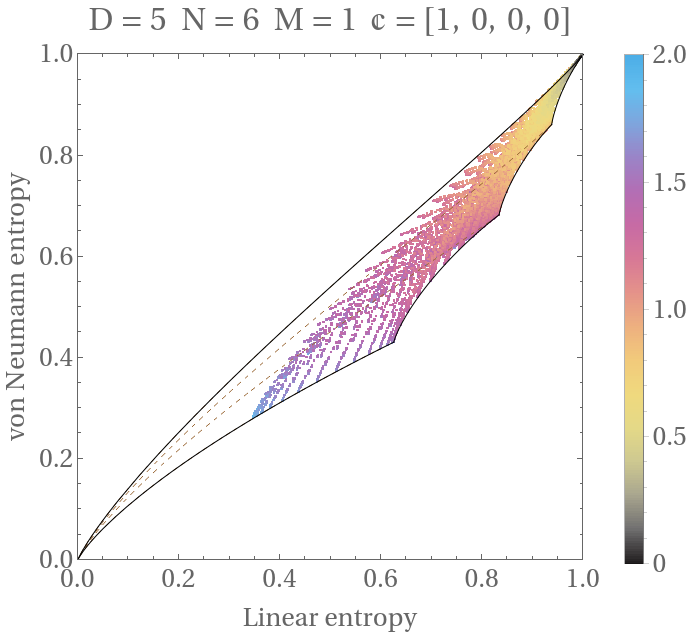}\hspace{\graphsep}
\includegraphics[width=\graphwidth]{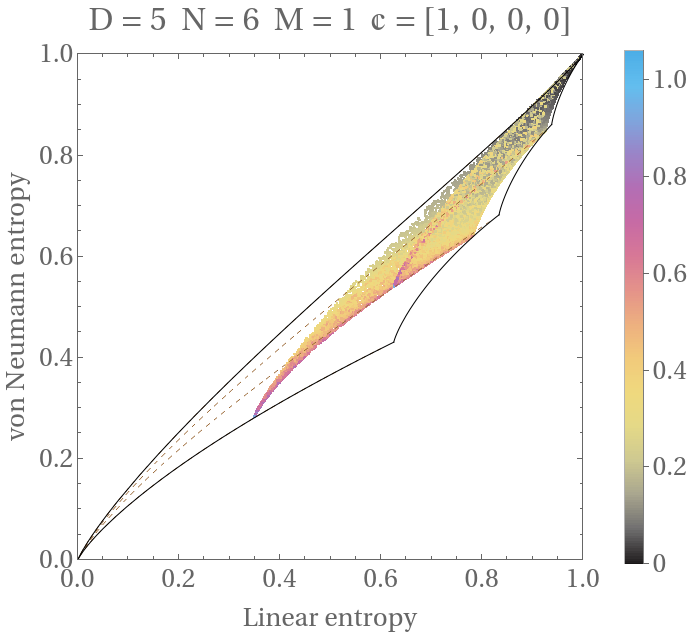}
\caption{Information diagrams using as colormap the distance (left) for $|z_i|\in[0,1]$ and $D=5$, $N=6$, $M=1$ and $\mathbbm{c}=[1,0,0,0]$. The same using as colormap the angular dependence (right) for $\|\zb\|=R=5$.}
\label{NTL-5-6-1-1000}
\end{figure}
\end{center}

\begin{center}
\begin{figure}[h!]
\includegraphics[width=\graphwidth]{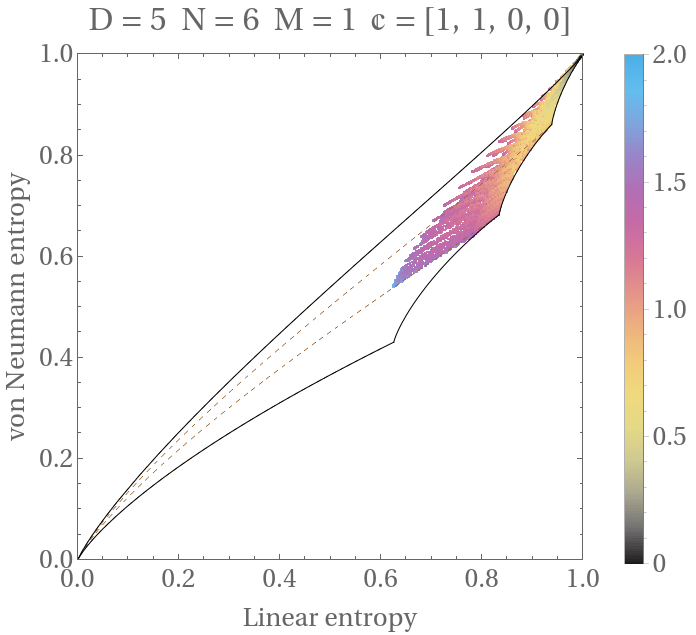}\hspace{\graphsep}
\includegraphics[width=\graphwidth]{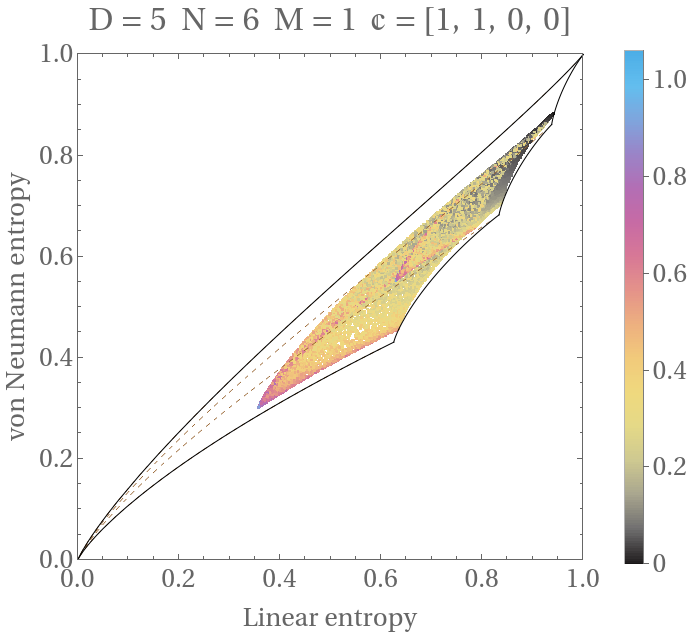}
\caption{Information diagrams using as colormap the distance (left) for $|z_i|\in[0,1]$ and $D=5$, $N=6$, $M=1$ and $\mathbbm{c}=[1,1,0,0]$. The same using as colormap the angular dependence (right) for $\|\zb\|=R=5$.}
\label{NTL-5-6-1-1100}
\end{figure}
\end{center}

\begin{center}
\begin{figure}[h!]
\includegraphics[width=\graphwidth]{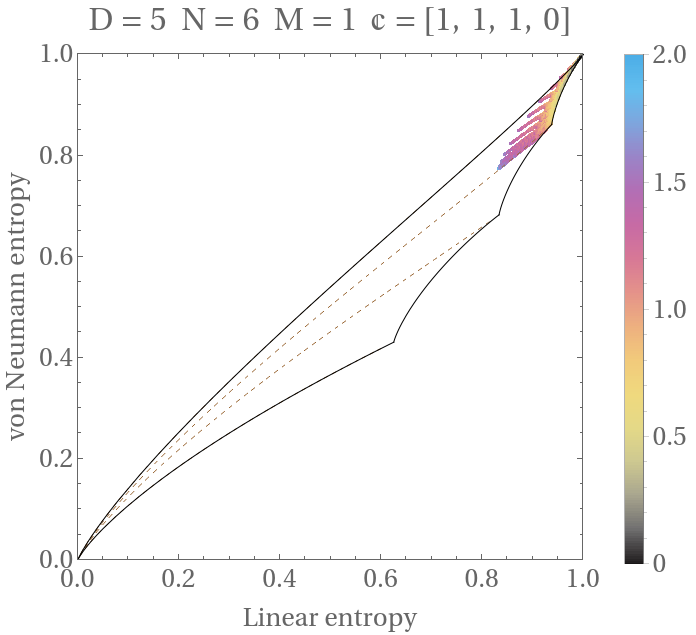}\hspace{\graphsep}
\includegraphics[width=\graphwidth]{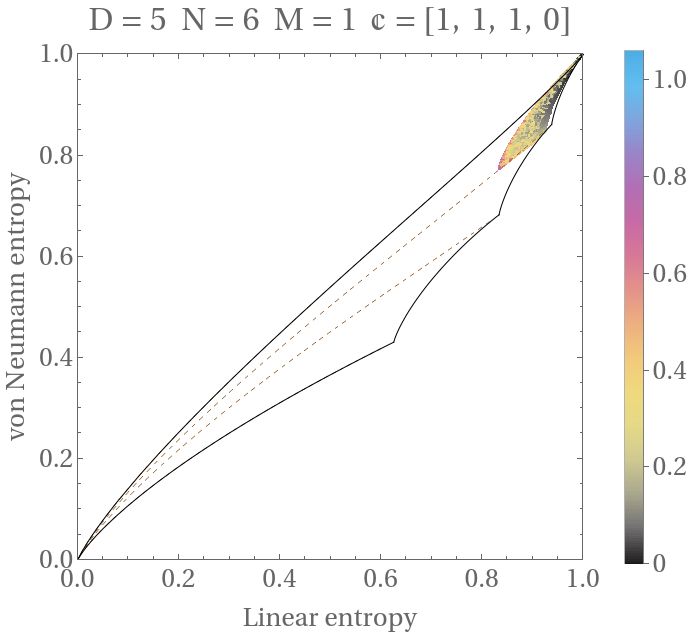}
\caption{Information diagrams using as colormap the distance (left) for $|z_i|\in[0,1]$ and $D=5$, $N=6$, $M=1$ and $\mathbbm{c}=[1,1,1,0]$. The same using as colormap the angular dependence (right) for $\|\zb\|=R=5$.}
\label{NTL-5-6-1-1110}
\end{figure}
\end{center}

\begin{center}
\begin{figure}[h!]
\includegraphics[width=\graphwidth]{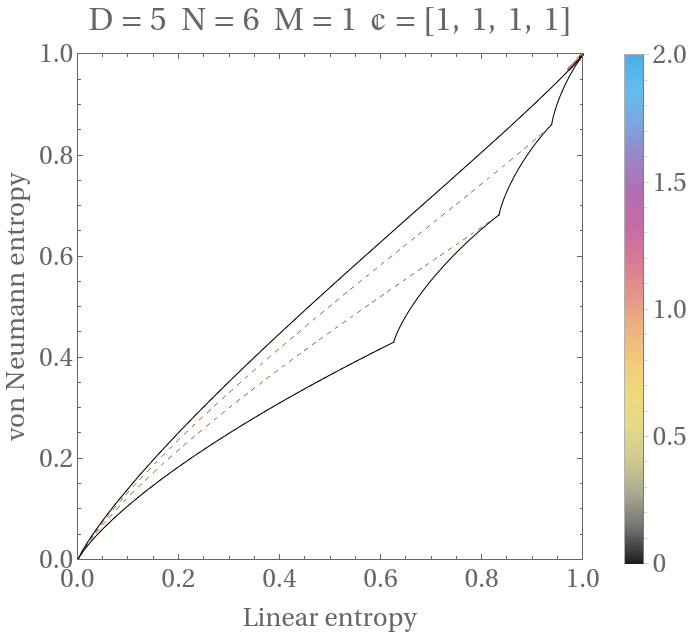}\hspace{\graphsep}
\includegraphics[width=\graphwidth]{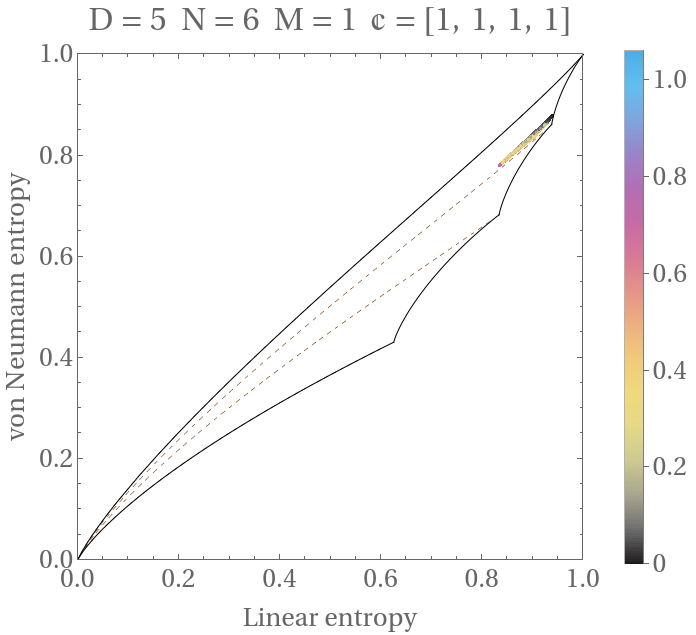}
\caption{Information diagrams using as colormap the distance (left) for $|z_i|\in[0,1]$ and $D=5$, $N=6$, $M=1$ and $\mathbbm{c}=[1,1,1,1]$. The same using as colormap the angular dependence (right) for $\|\zb\|=R=5$.}
\label{NTL-5-6-1-1111}
\end{figure}
\end{center}


\begin{center}
\begin{figure}[h!]
\includegraphics[width=\graphwidth]{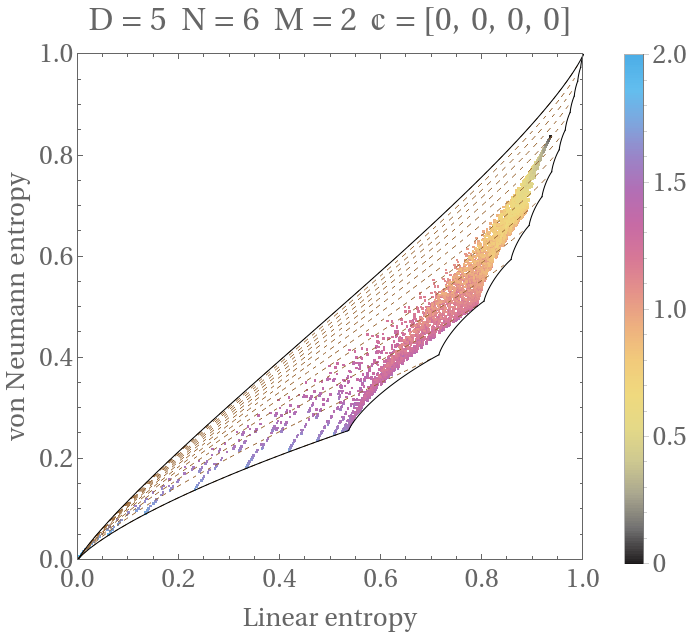}\hspace{\graphsep}
\includegraphics[width=\graphwidth]{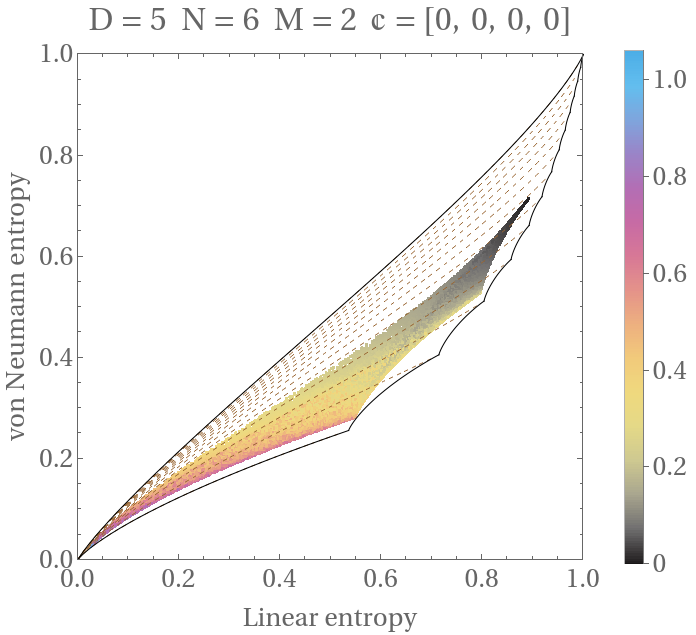}
\caption{Information diagrams using as colormap the distance (left) for $|z_i|\in[0,1]$ and $D=5$, $N=6$, $M=2$ and $\mathbbm{c}=[0,0,0,0]$. The same using as colormap the angular dependence (right) for $\|\zb\|=R=5$.}
\label{NTL-5-6-2-0000}
\end{figure}
\end{center}

\begin{center}
\begin{figure}[h!]
\includegraphics[width=\graphwidth]{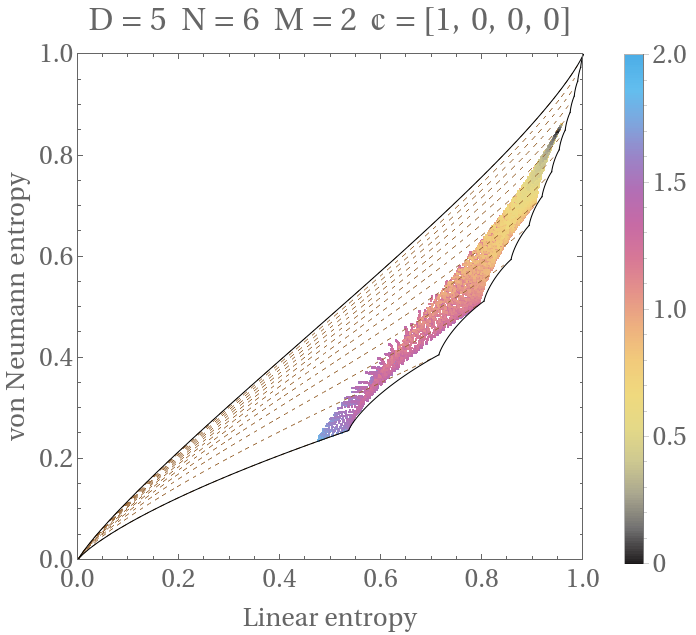}\hspace{\graphsep}
\includegraphics[width=\graphwidth]{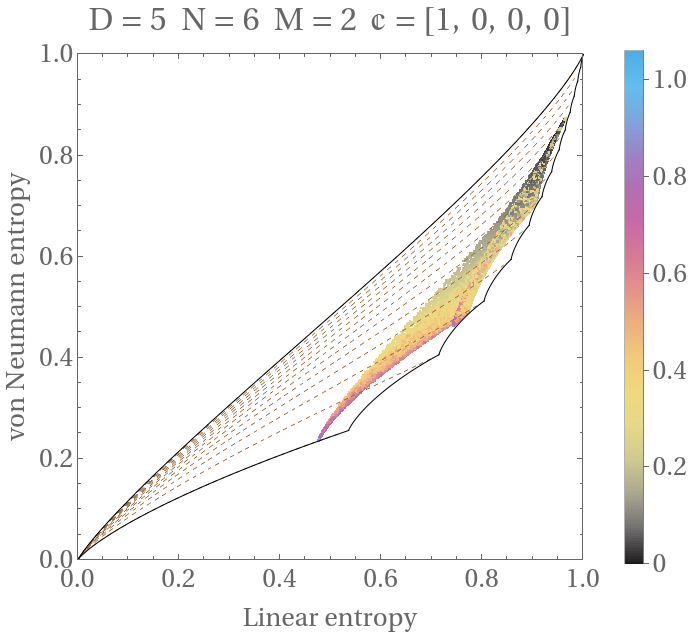}
\caption{Information diagrams using as colormap the distance (left) for $|z_i|\in[0,1]$ and $D=5$, $N=6$, $M=2$ and $\mathbbm{c}=[1,0,0,0]$. The same using as colormap the angular dependence (right) for $\|\zb\|=R=5$.}
\label{NTL-5-6-2-1000}
\end{figure}
\end{center}

\begin{center}
\begin{figure}[h!]
\includegraphics[width=\graphwidth]{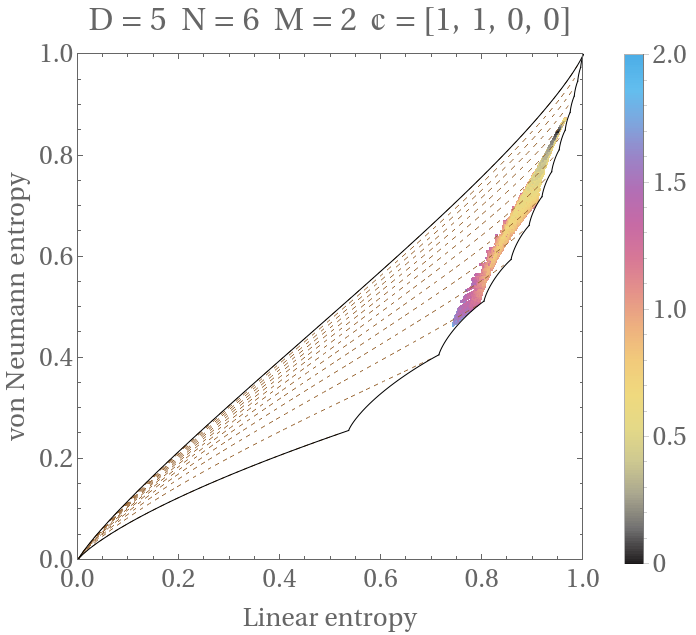}\hspace{\graphsep}
\includegraphics[width=\graphwidth]{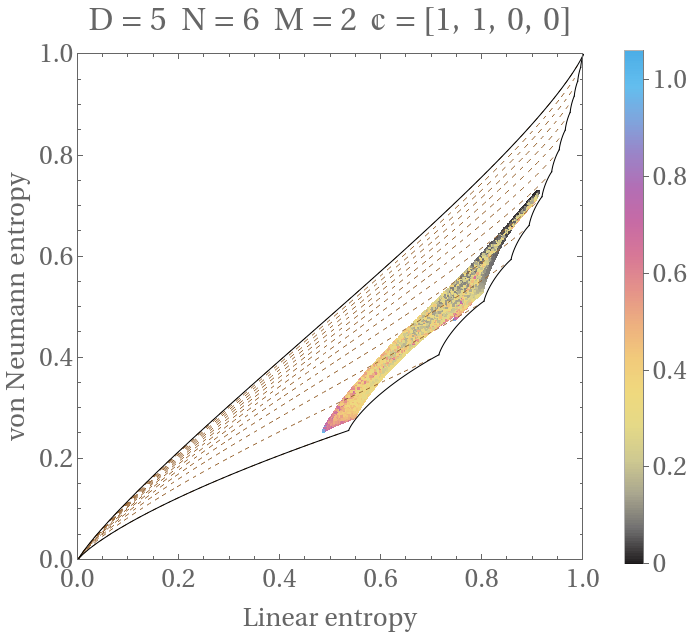}
\caption{Information diagrams using as colormap the distance (left) for $|z_i|\in[0,1]$ and $D=5$, $N=6$, $M=2$ and $\mathbbm{c}=[1,1,0,0]$. The same using as colormap the angular dependence (right) for $\|\zb\|=R=5$.}
\label{NTL-5-6-2-1100}
\end{figure}
\end{center}

\begin{center}
\begin{figure}[h!]
\includegraphics[width=\graphwidth]{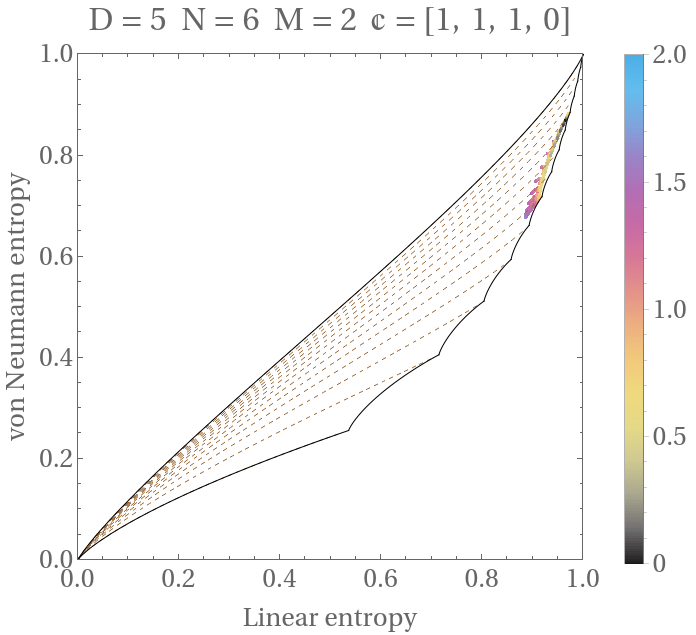}\hspace{\graphsep}
\includegraphics[width=\graphwidth]{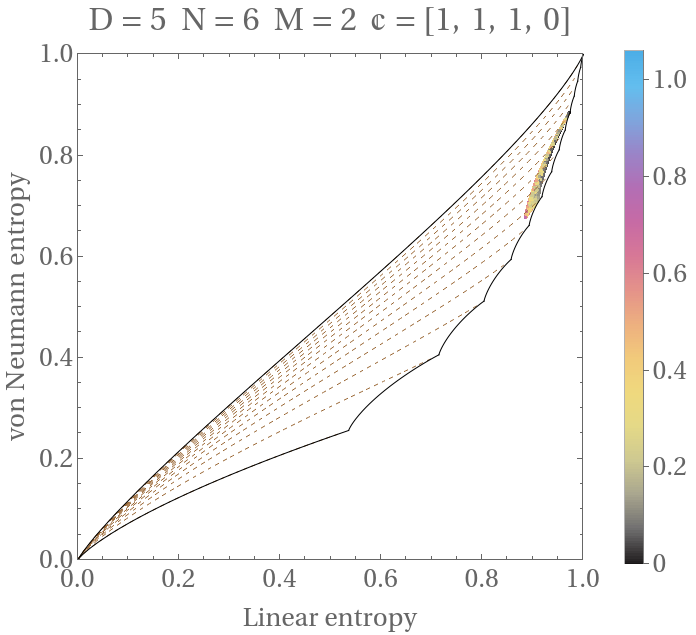}
\caption{Information diagrams using as colormap the distance (left) for $|z_i|\in[0,1]$ and $D=5$, $N=6$, $M=2$ and $\mathbbm{c}=[1,1,1,0]$. The same using as colormap the angular dependence (right) for $\|\zb\|=R=5$.}
\label{NTL-5-6-2-1110}
\end{figure}
\end{center}

\begin{center}
\begin{figure}[h!]
\includegraphics[width=\graphwidth]{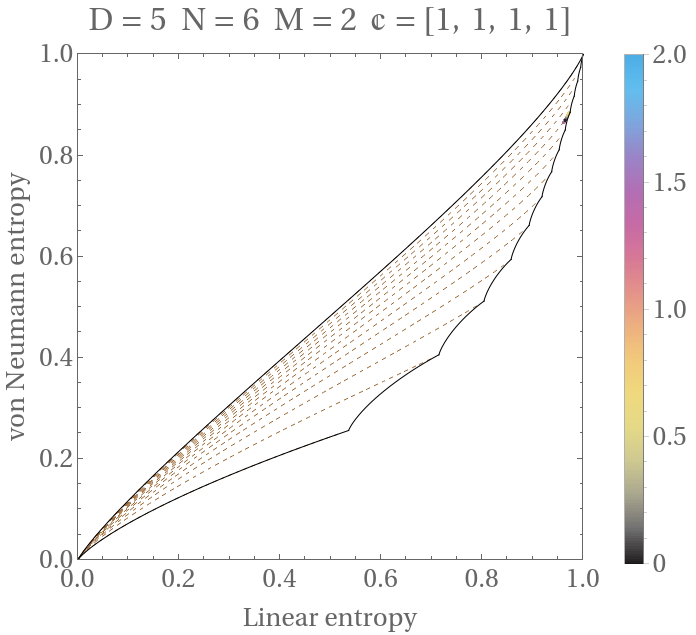}\hspace{\graphsep}
\includegraphics[width=\graphwidth]{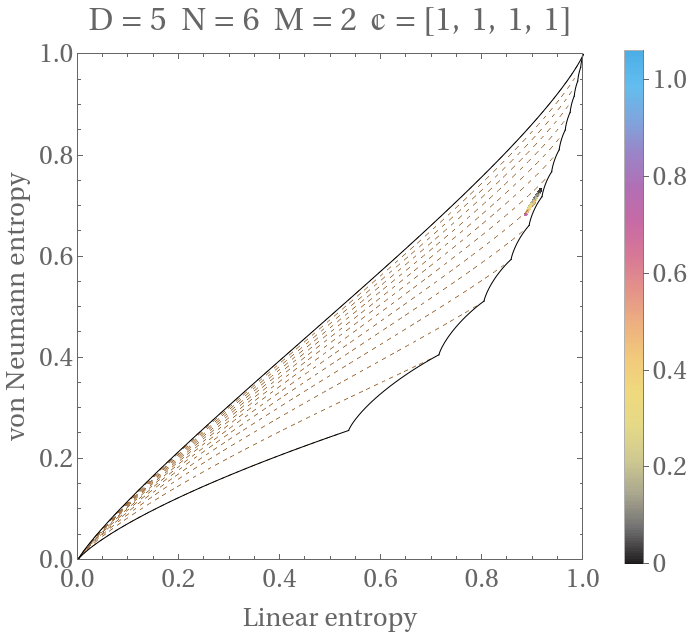}
\caption{Information diagrams using as colormap the distance (left) for $|z_i|\in[0,1]$ and $D=5$, $N=6$, $M=2$ and $\mathbbm{c}=[1,1,1,1]$. The same using as colormap the angular dependence (right) for $\|\zb\|=R=5$.}
\label{NTL-5-6-2-1111}
\end{figure}
\end{center}


\begin{center}
\begin{figure}[h!]
\includegraphics[width=\graphwidth]{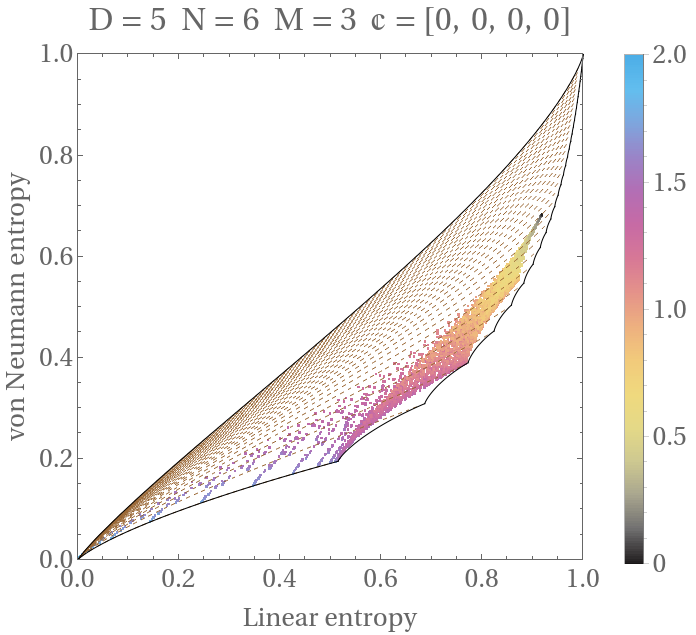}\hspace{\graphsep}
\includegraphics[width=\graphwidth]{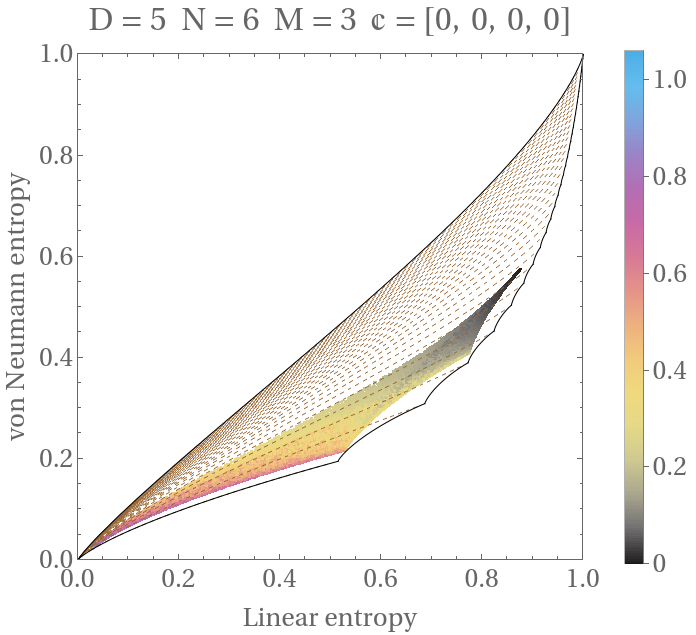}
\caption{Information diagrams using as colormap the distance (left) for $|z_i|\in[0,1]$ and $D=5$, $N=6$, $M=3$ and $\mathbbm{c}=[0,0,0,0]$. The same using as colormap the angular dependence (right) for $\|\zb\|=R=5$.}
\label{NTL-5-6-3-0000}
\end{figure}
\end{center}

\begin{center}
\begin{figure}[h!]
\includegraphics[width=\graphwidth]{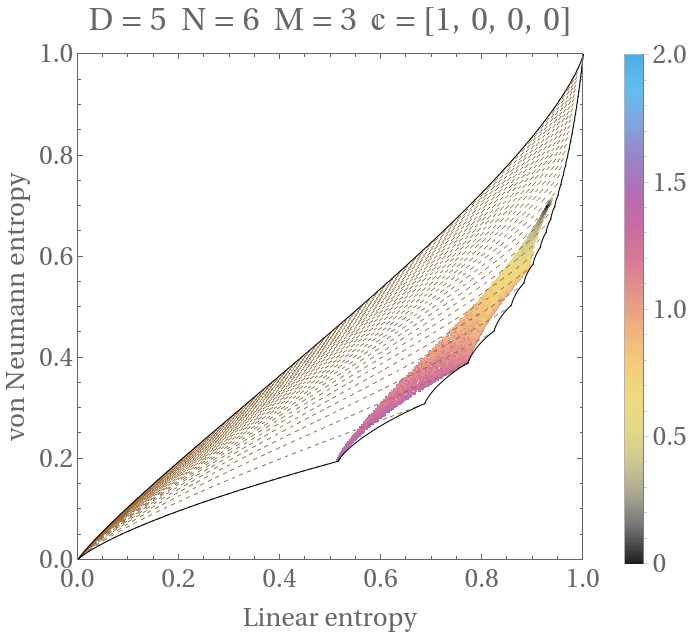}\hspace{\graphsep}
\includegraphics[width=\graphwidth]{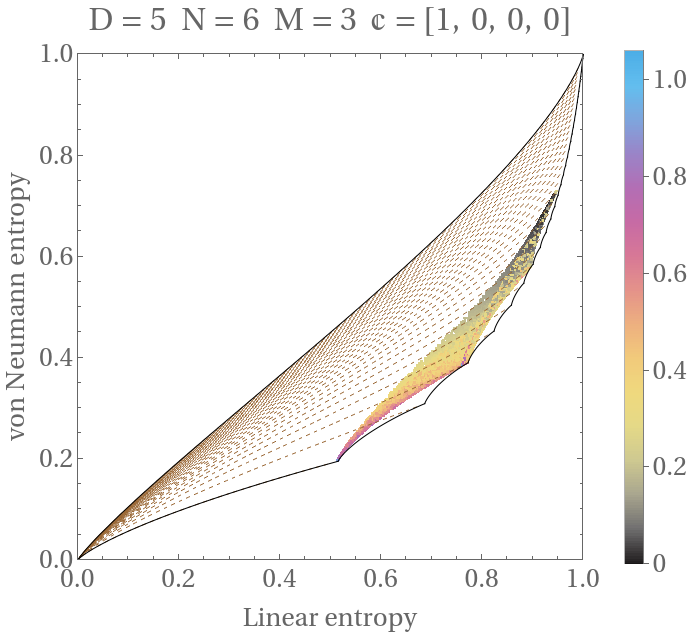}
\caption{Information diagrams using as colormap the distance (left) for $|z_i|\in[0,1]$ and $D=5$, $N=6$, $M=3$ and $\mathbbm{c}=[1,0,0,0]$. The same using as colormap the angular dependence (right) for $\|\zb\|=R=5$.}
\label{NTL-5-6-3-1000}
\end{figure}
\end{center}

\begin{center}
\begin{figure}[h!]
\includegraphics[width=\graphwidth]{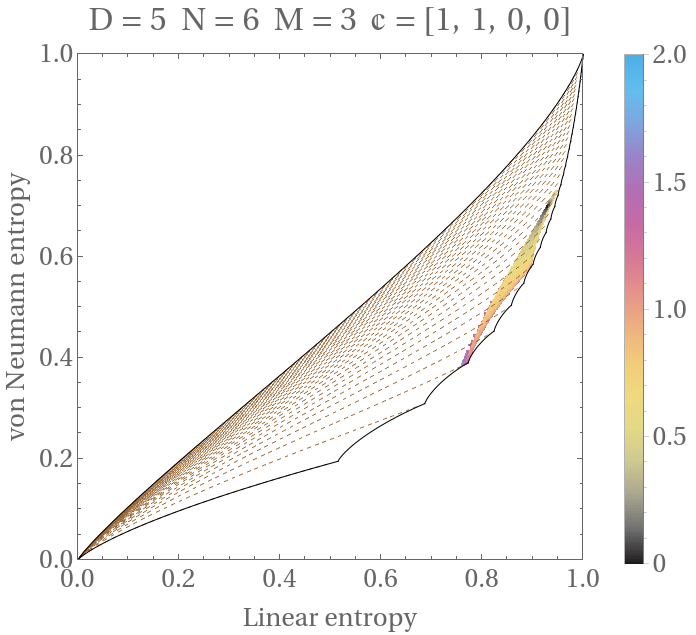}\hspace{\graphsep}
\includegraphics[width=\graphwidth]{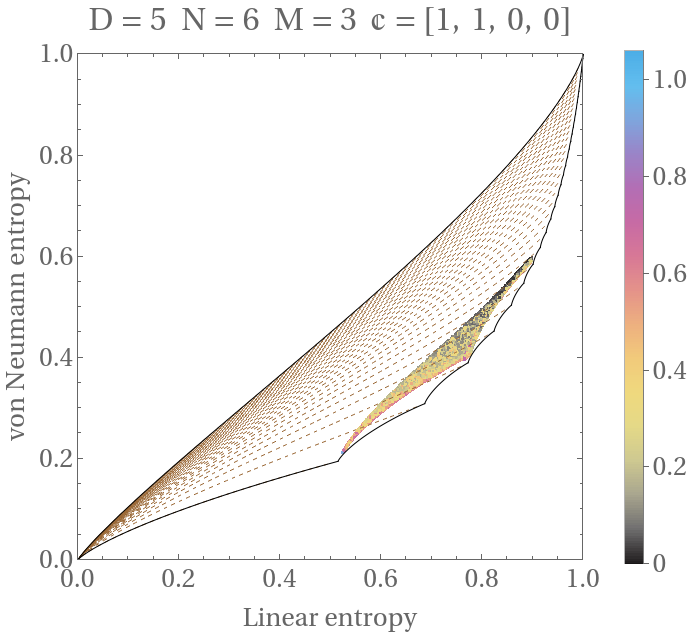}
\caption{Information diagrams using as colormap the distance (left) for $|z_i|\in[0,1]$ and $D=5$, $N=6$, $M=3$ and $\mathbbm{c}=[1,1,0,0]$. The same using as colormap the angular dependence (right) for $\|\zb\|=R=5$.}
\label{NTL-5-6-3-1100}
\end{figure}
\end{center}

\begin{center}
\begin{figure}[h!]
\includegraphics[width=\graphwidth]{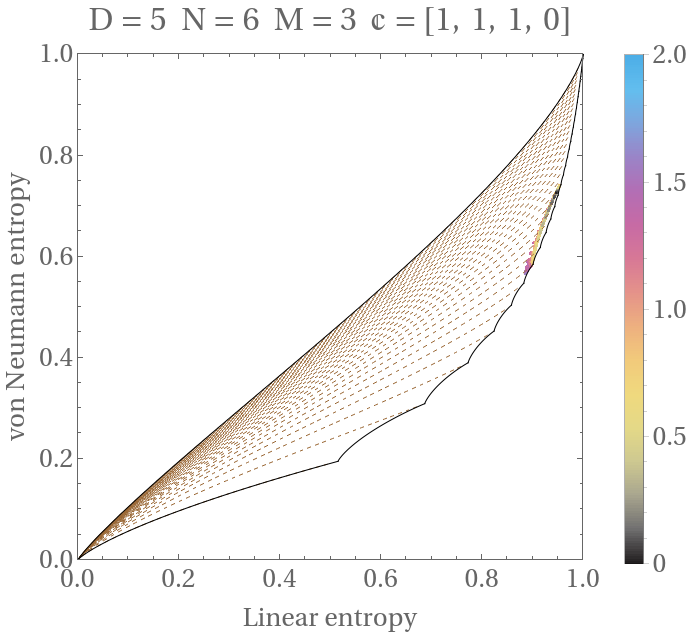}\hspace{\graphsep}
\includegraphics[width=\graphwidth]{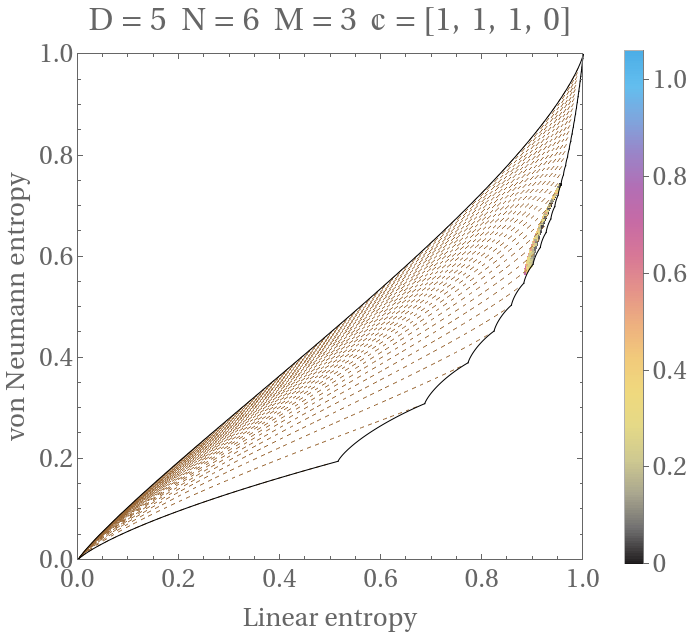}
\caption{Information diagrams using as colormap the distance (left) for $|z_i|\in[0,1]$ and $D=5$, $N=6$, $M=3$ and $\mathbbm{c}=[1,1,1,0]$. The same using as colormap the angular dependence (right) for $\|\zb\|=R=5$.}
\label{NTL-5-6-3-1110}
\end{figure}
\end{center}

\begin{center}
\begin{figure}[h!]
\includegraphics[width=\graphwidth]{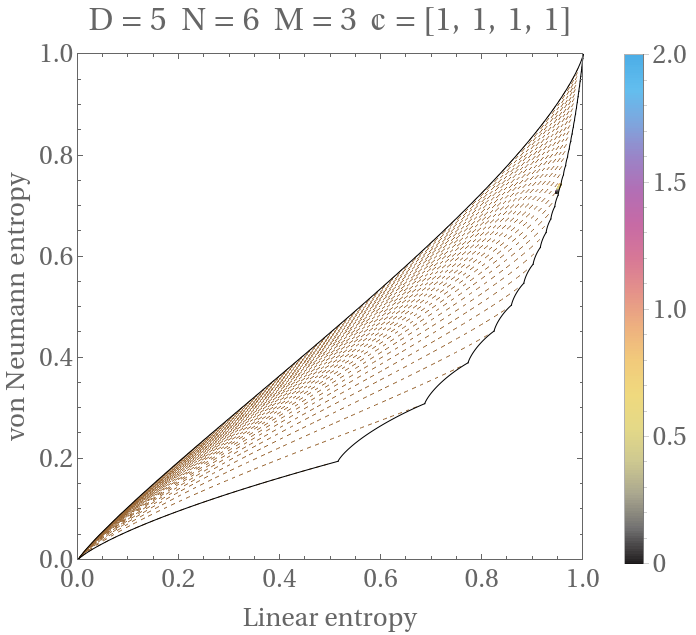}\hspace{\graphsep}
\includegraphics[width=\graphwidth]{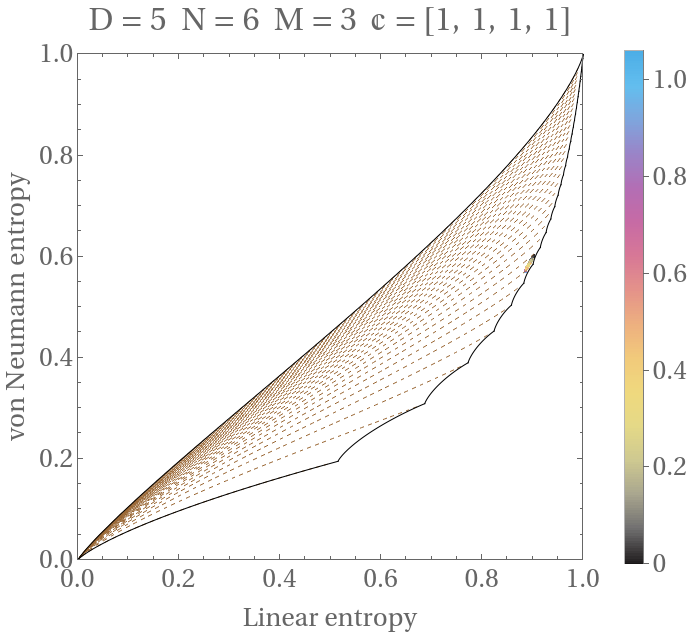}
\caption{Information diagrams using as colormap the distance (left) for $|z_i|\in[0,1]$ and $D=5$, $N=6$, $M=3$ and $\mathbbm{c}=[1,1,1,1]$. The same using as colormap the angular dependence (right) for $\|\zb\|=R=5$.}
\label{NTL-5-6-3-1111}
\end{figure}
\end{center}


 \subsection{Single thermodynamic limit}

\begin{center}
\begin{figure}[h!]
\includegraphics[width=\graphwidth]{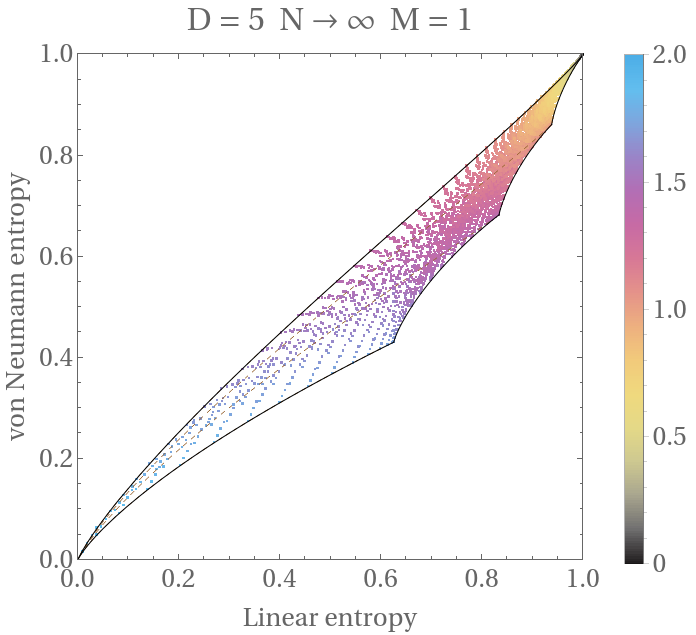}\hspace{\graphsep}
\includegraphics[width=\graphwidth]{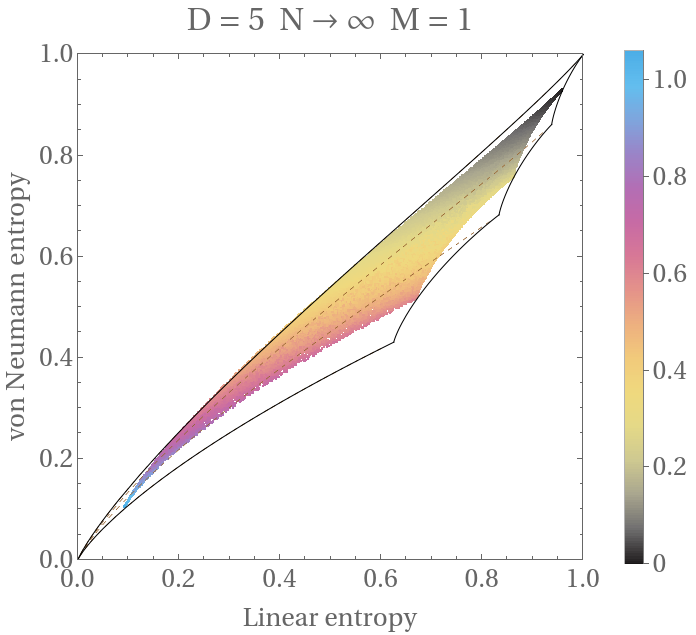}
\caption{Information diagrams using as colormap the distance (left) for $|z_i|\in[0,1]$ and $D=5$, $N\rightarrow\infty$ and $M=1$. The same using as colormap the angular dependence (right) for $\|\zb\|=R=5$.}
\label{D5-M1-TL}
\end{figure}
\end{center}

\begin{center}
\begin{figure}[h!]
\includegraphics[width=\graphwidth]{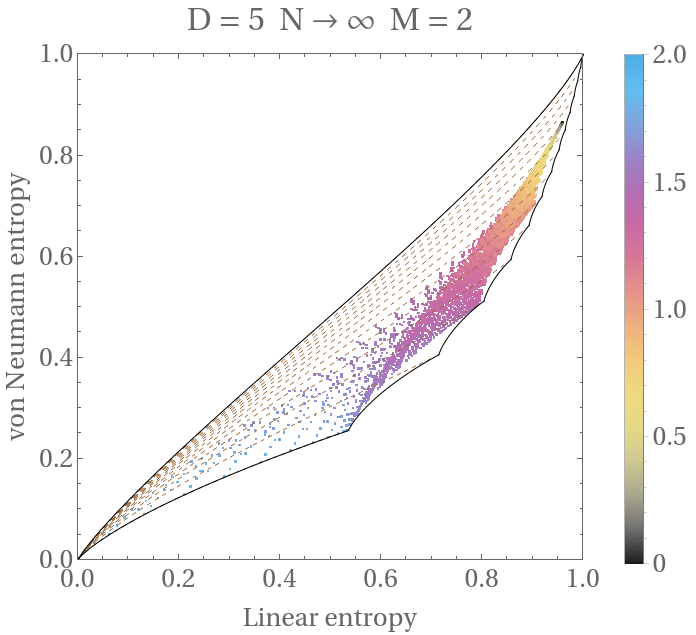}\hspace{\graphsep}
\includegraphics[width=\graphwidth]{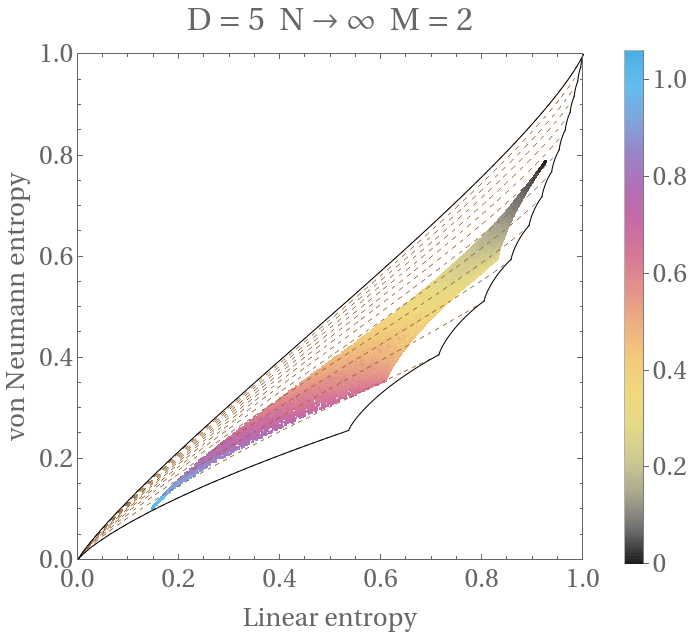}
\caption{Information diagrams using as colormap the distance (left) for $|z_i|\in[0,1]$ and $D=5$, $N\rightarrow\infty$ and $M=2$. The same using as colormap the angular dependence (right) for $\|\zb\|=R=5$.}
\label{D5-M2-TL}
\end{figure}
\end{center}

\begin{center}
\begin{figure}[h!]
\includegraphics[width=\graphwidth]{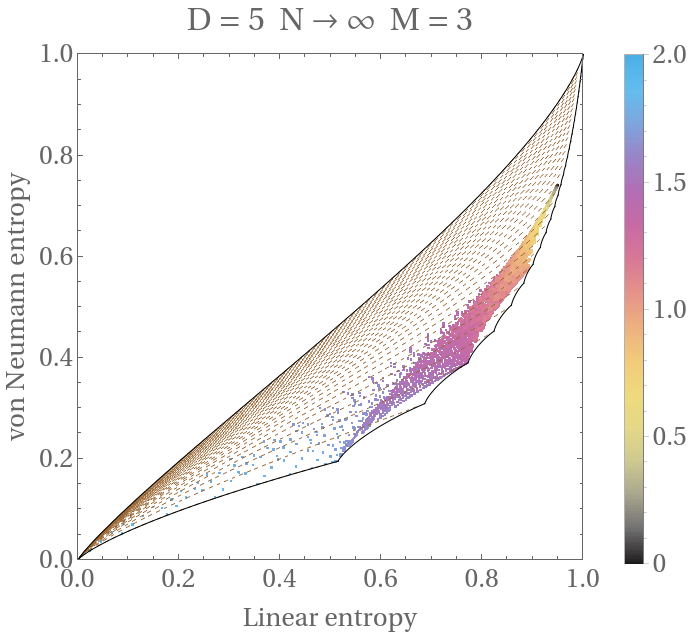}\hspace{\graphsep}
\includegraphics[width=\graphwidth]{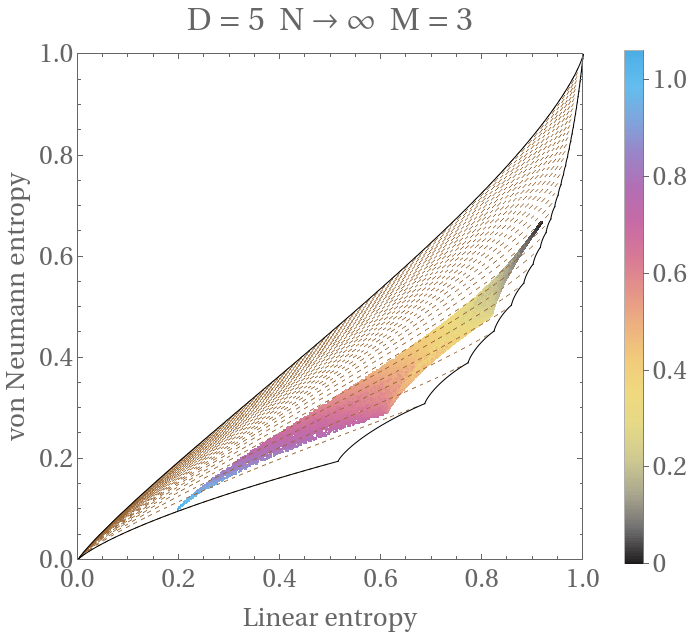}
\caption{Information diagrams using as colormap the distance (left) for $|z_i|\in[0,1]$ and $D=5$, $N\rightarrow\infty$ and $M=3$. The same using as colormap the angular dependence (right) for $\|\zb\|=R=5$.}
\label{D5-M3-TL}
\end{figure}
\end{center}

 In the case $N\rightarrow\infty$ (the thermodynamic limit), we plot the information diagrams for $|z_i|\in [0,1], i=1,2,3$ using as colormap the distance to the point $(1,1,1)$, and the  information diagrams for $\|\zb\|=R=5$ using as colormap the  angular distance to the vector $(1,1,1)$, for each value of $M=1,2,3$.  They are qualitatively the same as in the finite $N$ case, with the difference that they do not depend on the original parity $\mathbb{c}$ of the state, and the appearance is that of the completely even case. See Figures \ref{D5-M1-TL}, \ref{D5-M2-TL}, \ref{D5-M3-TL}.


\subsection{Rescaled Double thermodynamic limit}

As in the previous cases, the double limit $N,M\rightarrow\infty$ (double thermodynamic limit) will not be considered since in this case the RDM correspond to that of a maximally mixed state of dimension $2^k$, with $k=\|\zb\|_0$.

 For the \textit{directional} limit
 $N,M\rightarrow\infty$ with $M=(1-\eta)N$ and $\eta\in[\frac{1}{2},1)$, rescaling $\zb=\frac{\alphab}{\sqrt{N}}$, we have, as shown in eq. (65) of the paper, the Schmitd eigenvalues factorize as a product of Schmitd eigenvalues for the decomposition of one-dimensional Schr\"odinger cat states of the harmonic oscillator.

\begin{center}
\begin{figure}[h!]
\includegraphics[width=\graphwidth]{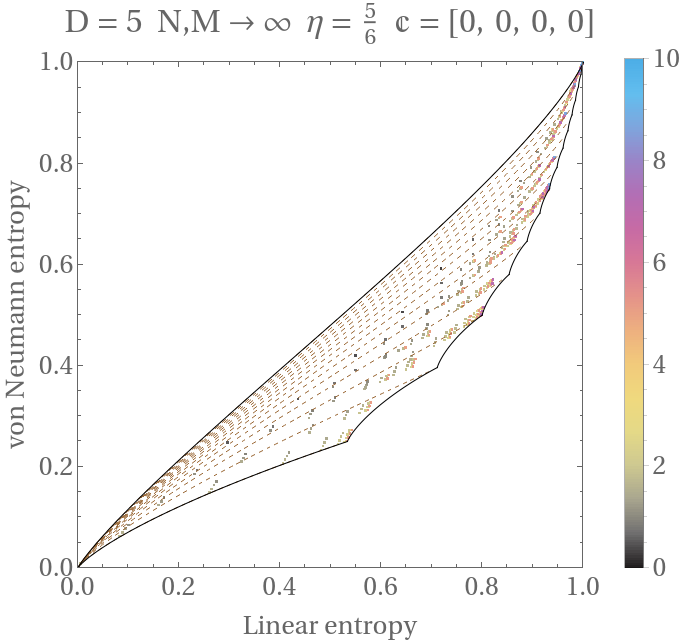}\hspace{\graphsep}
\includegraphics[width=\graphwidth]{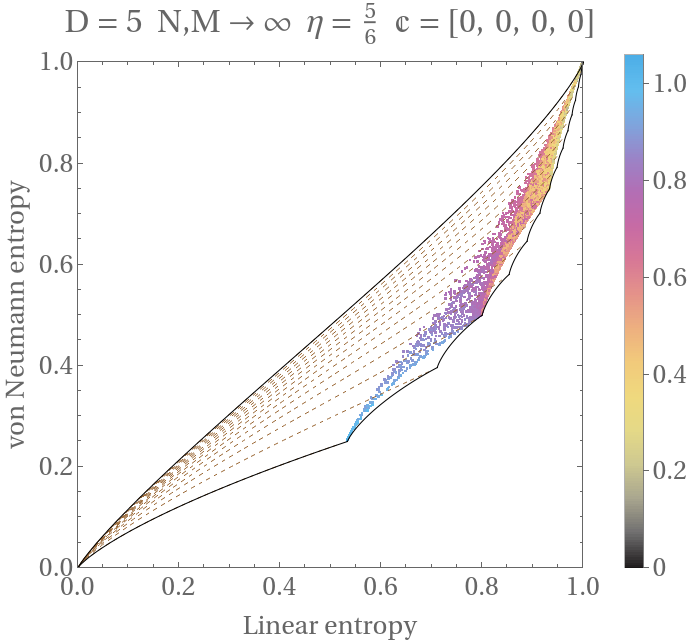}
\caption{Information diagrams using as colormap the distance (left) for $|\alpha_i|\in[0,1]$ and $D=5$, $N,M\rightarrow\infty$ and $\mathbbm{c}=[0,0,0,0]$. The same using as colormap the angular dependence (right) for $\|\alphab\|=R=10$.}
\label{RTL-5-5d6-0000}
\end{figure}
\end{center}

\begin{center}
\begin{figure}[h!]
\includegraphics[width=\graphwidth]{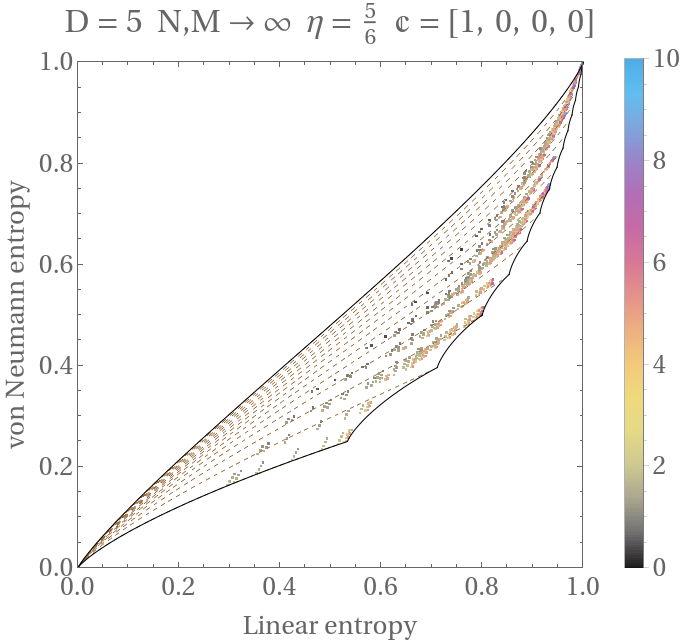}\hspace{\graphsep}
\includegraphics[width=\graphwidth]{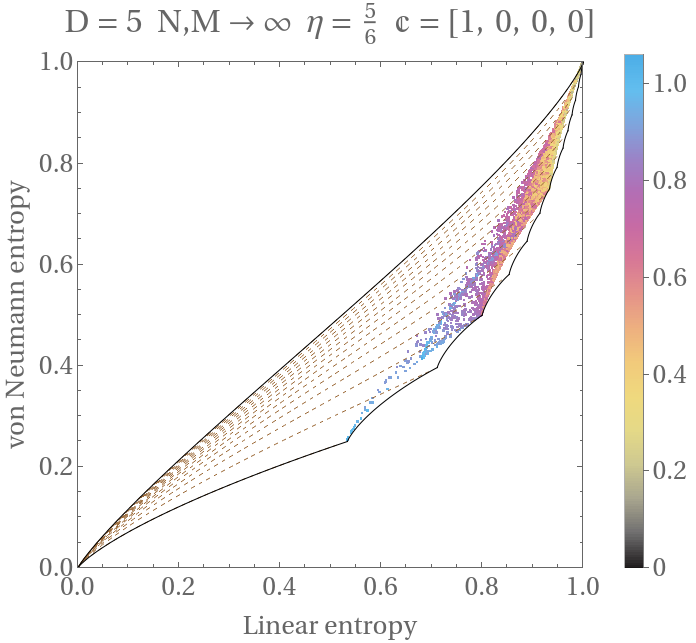}
\caption{Information diagrams using as colormap the distance (left) for $|\alpha_i|\in[0,1]$ and $D=5$, $N,M\rightarrow\infty$ and $\mathbbm{c}=[1,0,0,0]$. The same using as colormap the angular dependence (right) for $\|\alphab\|=R=10$.}
\label{RTL-5-5d6-1000}
\end{figure}
\end{center}

\begin{center}
\begin{figure}[h!]
\includegraphics[width=\graphwidth]{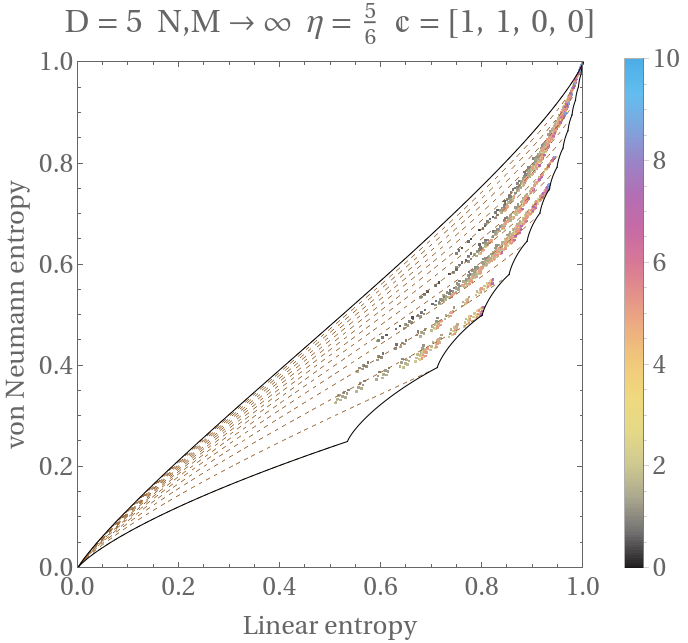}\hspace{\graphsep}
\includegraphics[width=\graphwidth]{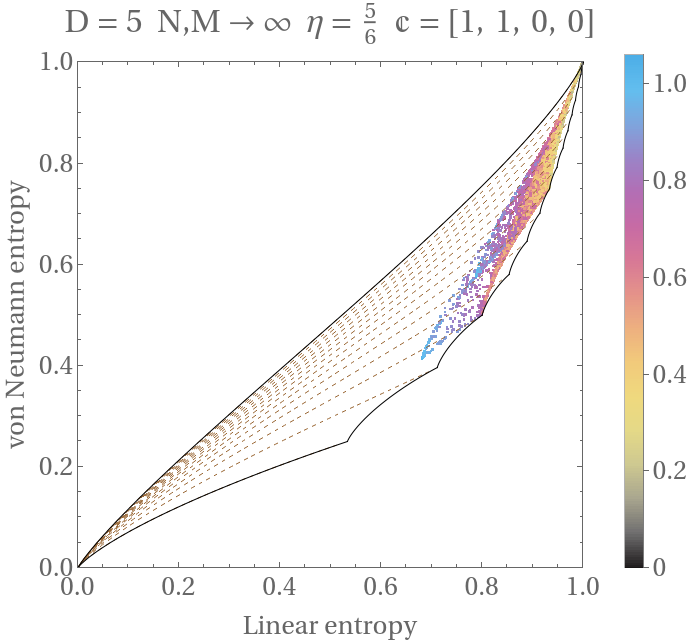}
\caption{Information diagrams using as colormap the distance (left) for $|\alpha_i|\in[0,1]$ and $D=5$, $N,M\rightarrow\infty$ and $\mathbbm{c}=[1,1,0,0]$. The same using as colormap the angular dependence (right) for $\|\alphab\|=R=10$.}
\label{RTL-5-5d6-1100}
\end{figure}
\end{center}

\begin{center}
\begin{figure}[h!]
\includegraphics[width=\graphwidth]{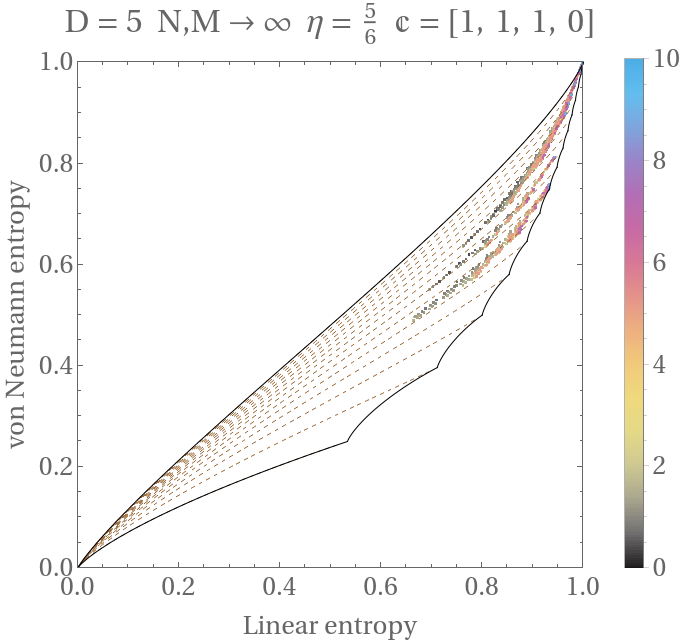}\hspace{\graphsep}
\includegraphics[width=\graphwidth]{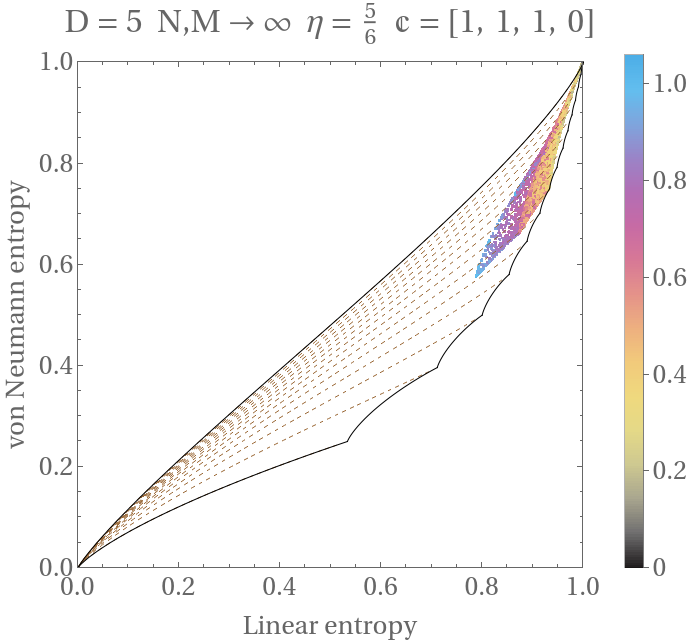}
\caption{Information diagrams using as colormap the distance (left) for $|\alpha_i|\in[0,1]$ and $D=5$, $N,M\rightarrow\infty$ and $\mathbbm{c}=[1,1,1,0]$. The same using as colormap the angular dependence (right) for $\|\alphab\|=R=10$.}
\label{RTL-5-5d6-1110}
\end{figure}
\end{center}

\begin{center}
\begin{figure}[h!]
\includegraphics[width=\graphwidth]{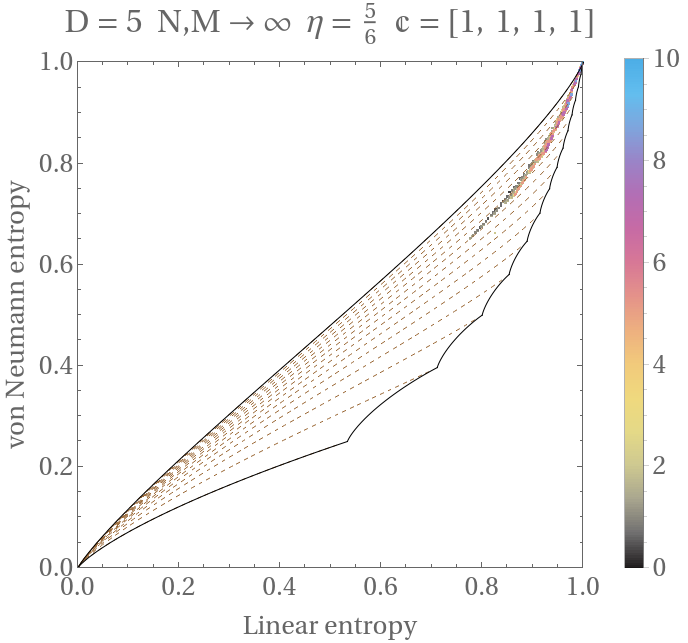}\hspace{\graphsep}
\includegraphics[width=\graphwidth]{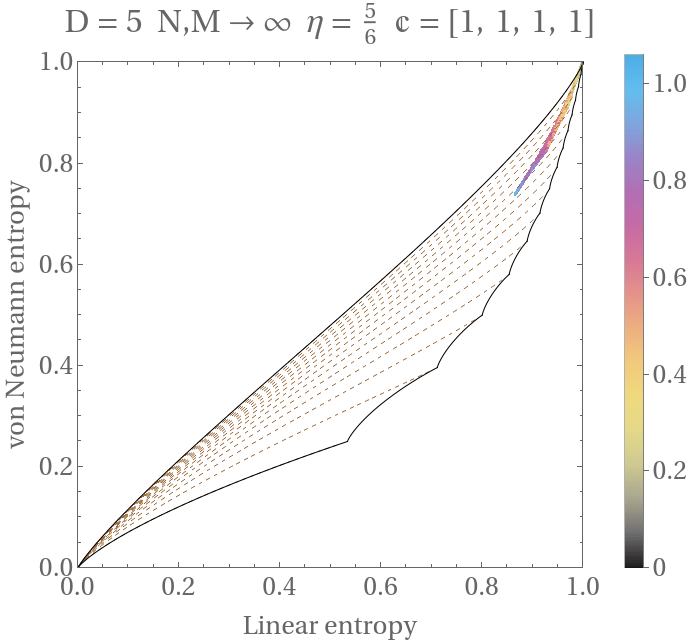}
\caption{Information diagrams using as colormap the distance (left) for $|\alpha_i|\in[0,1]$ and $D=5$, $N,M\rightarrow\infty$ and $\mathbbm{c}=[1,1,1,1]$. The same using as colormap the angular dependence (right) for $\|\alphab\|=R=10$.}
\label{RTL-5-5d6-1111}
\end{figure}
\end{center}

\begin{center}
\begin{figure}[h!]
\includegraphics[width=\graphwidth]{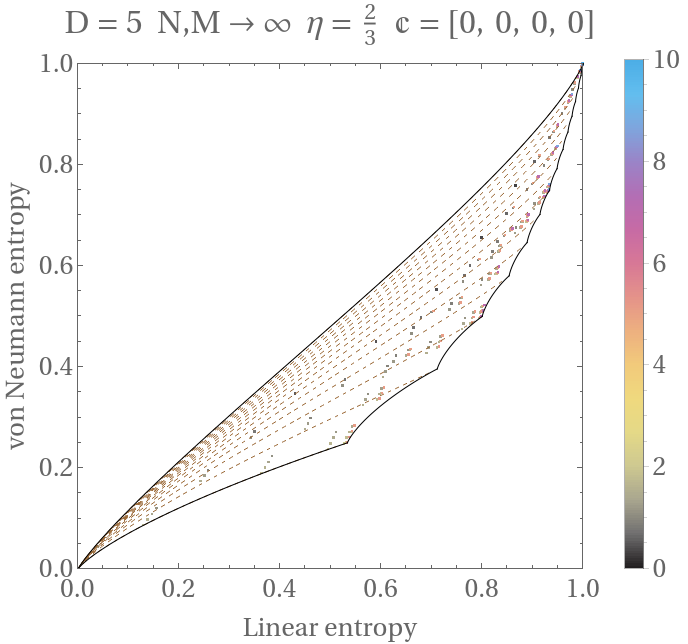}\hspace{\graphsep}
\includegraphics[width=\graphwidth]{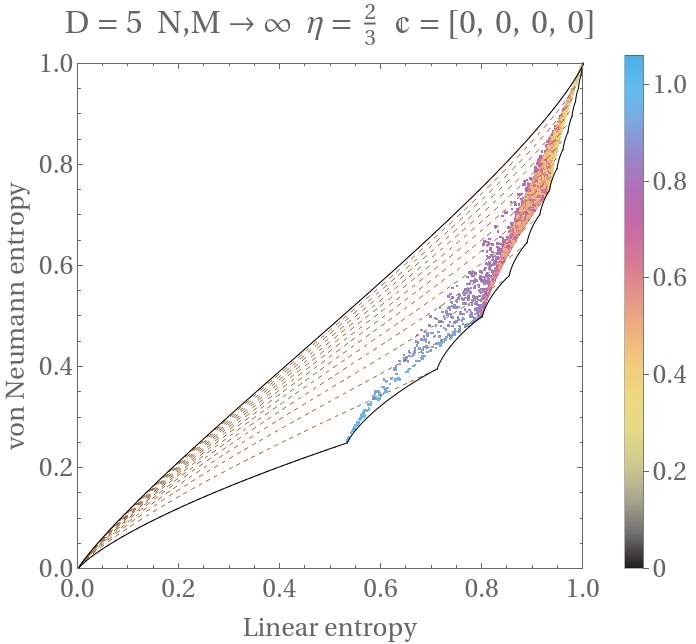}
\caption{Information diagrams using as colormap the distance (left) for $|\alpha_i|\in[0,1]$ and $D=5$, $N,M\rightarrow\infty$ and $\mathbbm{c}=[0,0,0,0]$. The same using as colormap the angular dependence (right) for $\|\alphab\|=R=10$.}
\label{RTL-5-2d3-0000}
\end{figure}
\end{center}

\begin{center}
\begin{figure}[h!]
\includegraphics[width=\graphwidth]{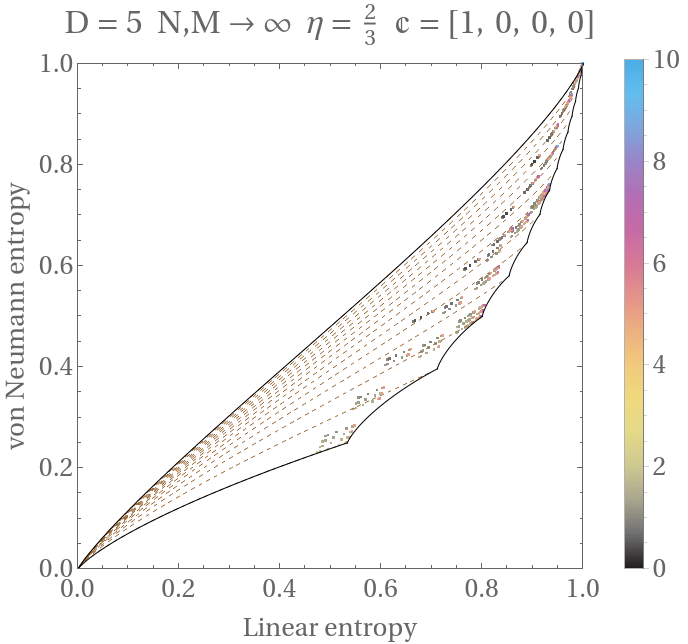}\hspace{\graphsep}
\includegraphics[width=\graphwidth]{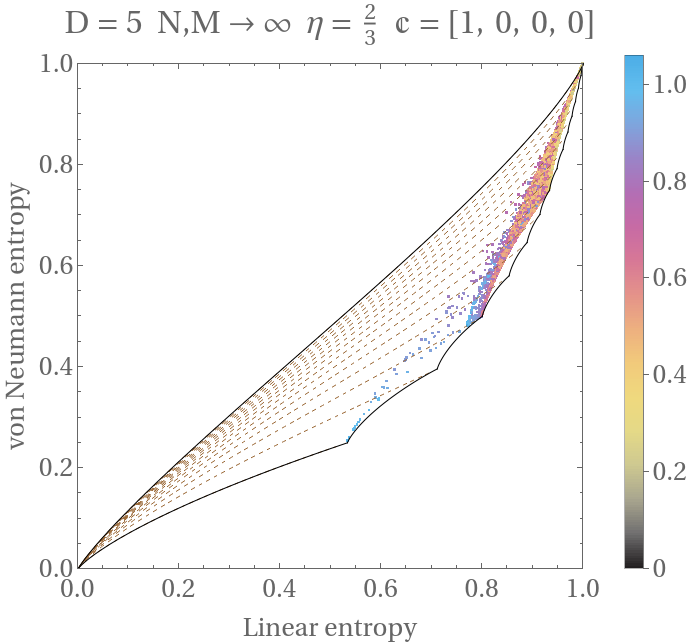}
\caption{Information diagrams using as colormap the distance (left) for $|\alpha_i|\in[0,1]$ and $D=5$, $N,M\rightarrow\infty$ and $\mathbbm{c}=[1,0,0,0]$. The same using as colormap the angular dependence (right) for $\|\alphab\|=R=10$.}
\label{RTL-5-2d3-1000}
\end{figure}
\end{center}

\begin{center}
\begin{figure}[h!]
\includegraphics[width=\graphwidth]{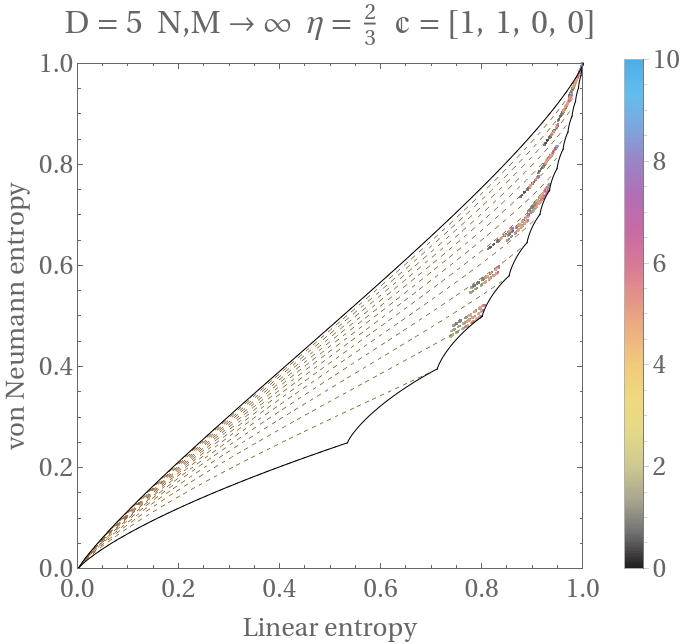}\hspace{\graphsep}
\includegraphics[width=\graphwidth]{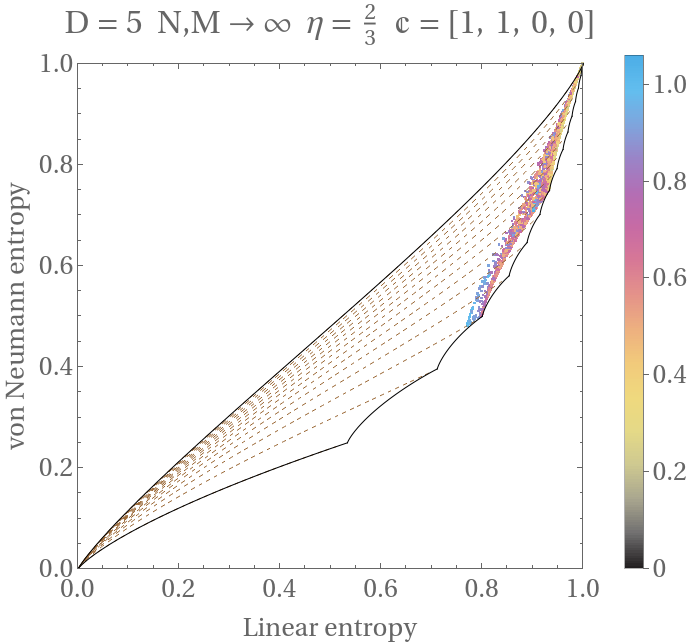}
\caption{Information diagrams using as colormap the distance (left) for $|\alpha_i|\in[0,1]$ and $D=5$, $N,M\rightarrow\infty$ and $\mathbbm{c}=[1,1,0,0]$. The same using as colormap the angular dependence (right) for $\|\alphab\|=R=10$.}
\label{RTL-5-2d3-1100}
\end{figure}
\end{center}

\begin{center}
\begin{figure}[h!]
\includegraphics[width=\graphwidth]{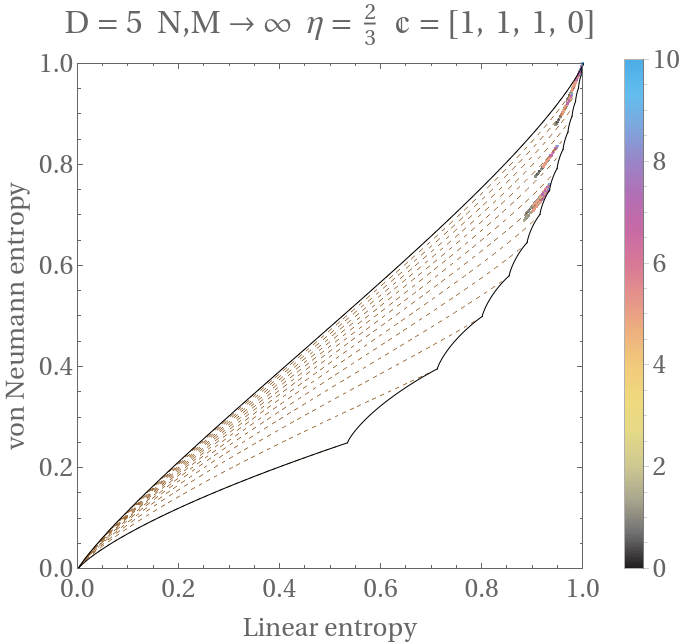}\hspace{\graphsep}
\includegraphics[width=\graphwidth]{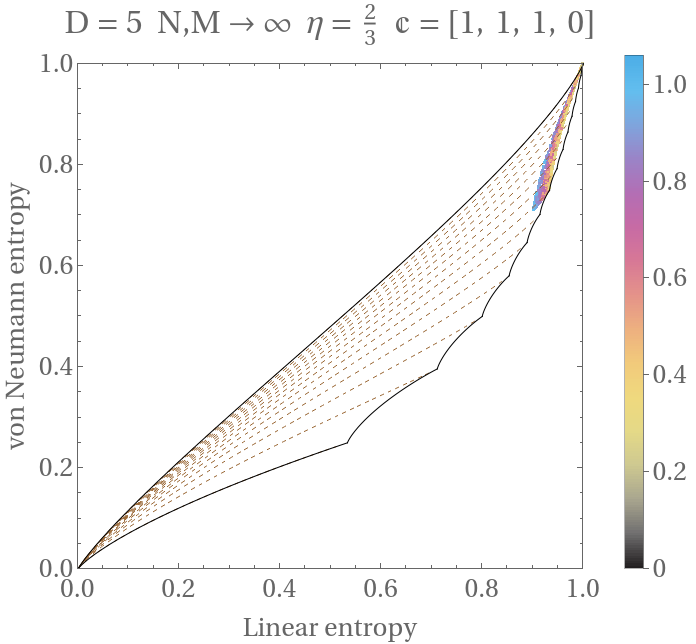}
\caption{Information diagrams using as colormap the distance (left) for $|\alpha_i|\in[0,1]$ and $D=5$, $N,M\rightarrow\infty$ and $\mathbbm{c}=[1,1,1,0]$. The same using as colormap the angular dependence (right) for $\|\alphab\|=R=10$.}
\label{RTL-5-2d3-1110}
\end{figure}
\end{center}

\begin{center}
\begin{figure}[h!]
\includegraphics[width=\graphwidth]{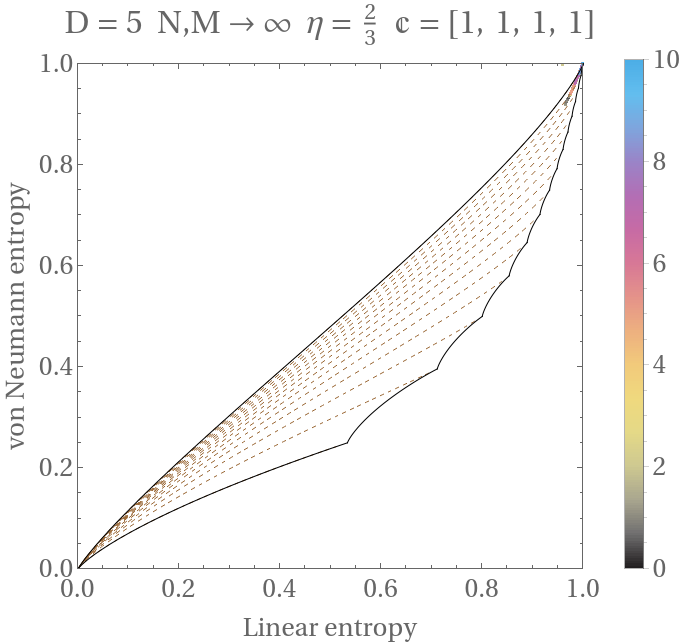}\hspace{\graphsep}
\includegraphics[width=\graphwidth]{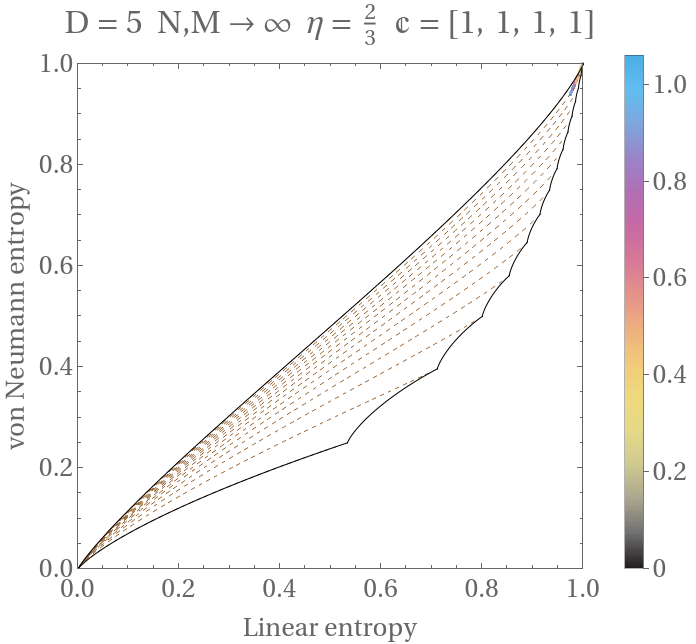}
\caption{Information diagrams using as colormap the distance (left) for $|\alpha_i|\in[0,1]$ and $D=5$, $N,M\rightarrow\infty$ and $\mathbbm{c}=[1,1,1,1]$. The same using as colormap the angular dependence (right) for $\|\alphab\|=R=10$.}
\label{RTL-5-2d3-1111}
\end{figure}
\end{center}

\begin{center}
\begin{figure}[h!]
\includegraphics[width=\graphwidth]{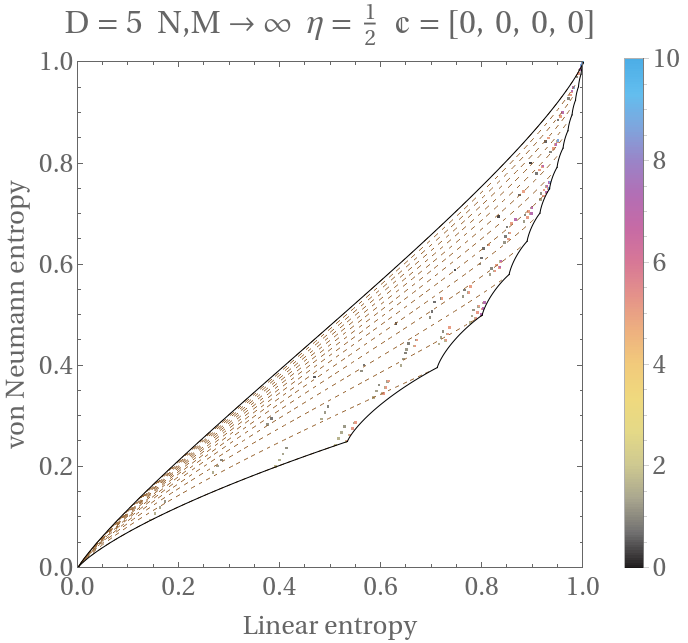}\hspace{\graphsep}
\includegraphics[width=\graphwidth]{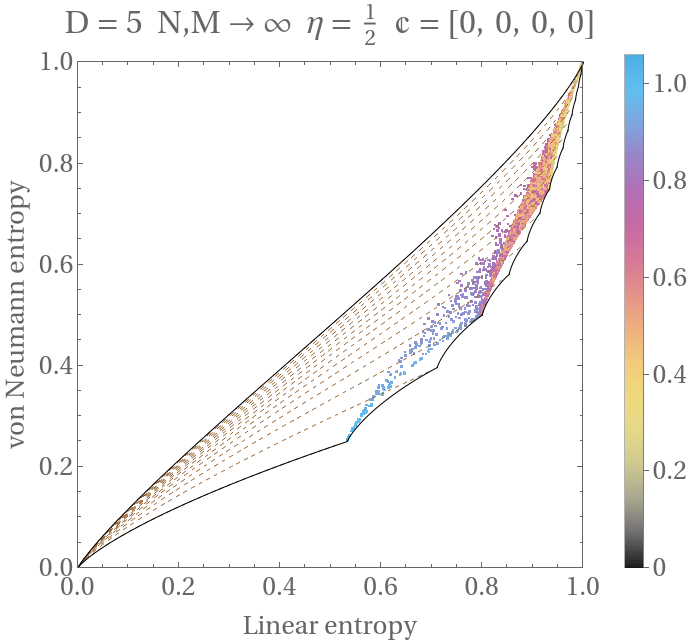}
\caption{Information diagrams using as colormap the distance (left) for $|\alpha_i|\in[0,1]$ and $D=5$, $N,M\rightarrow\infty$ and $\mathbbm{c}=[0,0,0,0]$. The same using as colormap the angular dependence (right) for $\|\alphab\|=R=10$.}
\label{RTL-5-1d2-0000}
\end{figure}
\end{center}

\begin{center}
\begin{figure}[h!]
\includegraphics[width=\graphwidth]{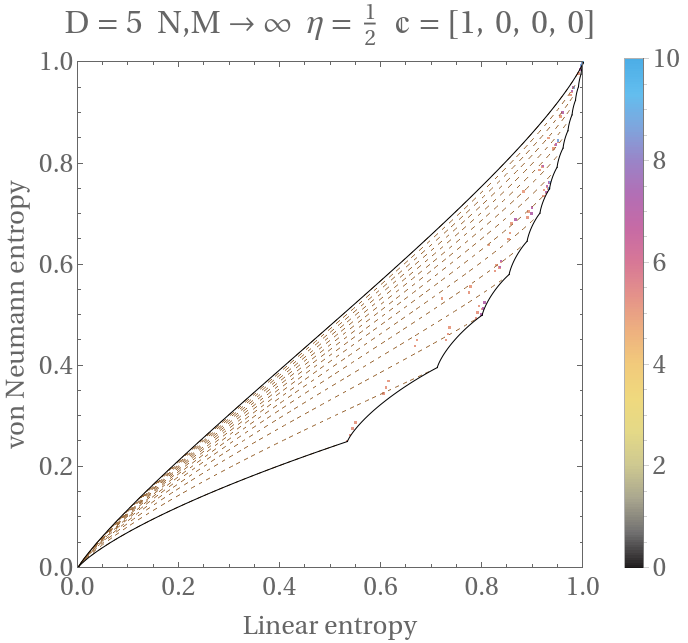}\hspace{\graphsep}
\includegraphics[width=\graphwidth]{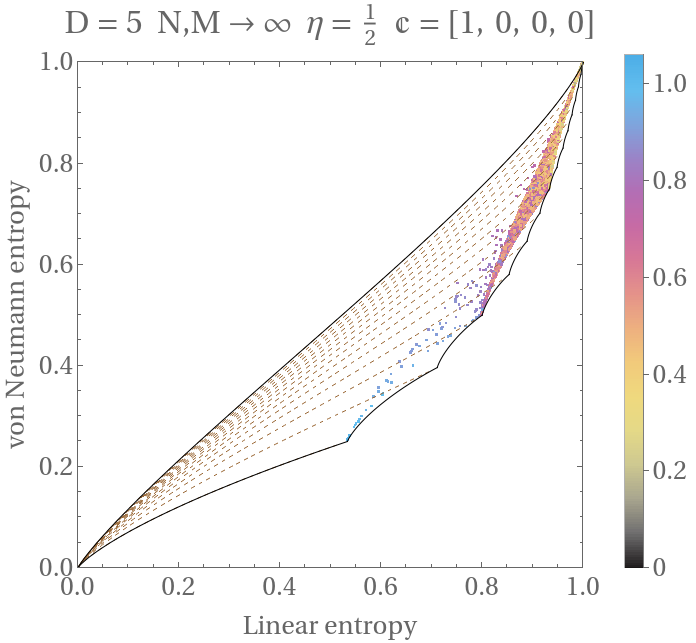}
\caption{Information diagrams using as colormap the distance (left) for $|\alpha_i|\in[0,1]$ and $D=5$, $N,M\rightarrow\infty$ and $\mathbbm{c}=[1,0,0,0]$. The same using as colormap the angular dependence (right) for $\|\alphab\|=R=10$.}
\label{RTL-5-1d2-1000}
\end{figure}
\end{center}

\begin{center}
\begin{figure}[h!]
\includegraphics[width=\graphwidth]{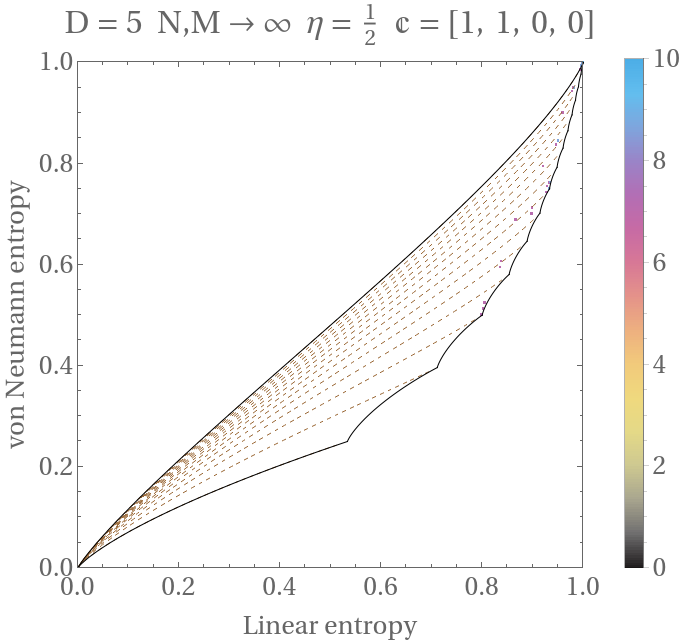}\hspace{\graphsep}
\includegraphics[width=\graphwidth]{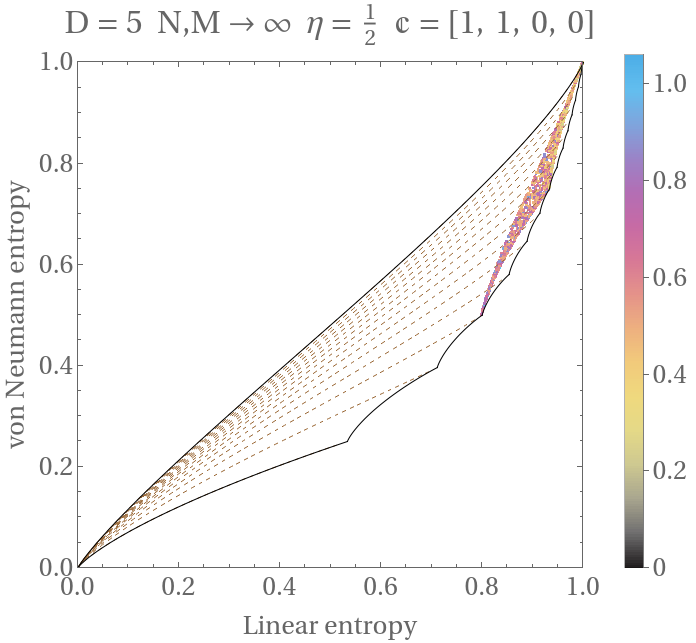}
\caption{Information diagrams using as colormap the distance (left) for $|\alpha_i|\in[0,1]$ and $D=5$, $N,M\rightarrow\infty$ and $\mathbbm{c}=[1,1,0,0]$. The same using as colormap the angular dependence (right) for $\|\alphab\|=R=10$.}
\label{RTL-5-1d2-1100}
\end{figure}
\end{center}

\begin{center}
\begin{figure}[h!]
\includegraphics[width=\graphwidth]{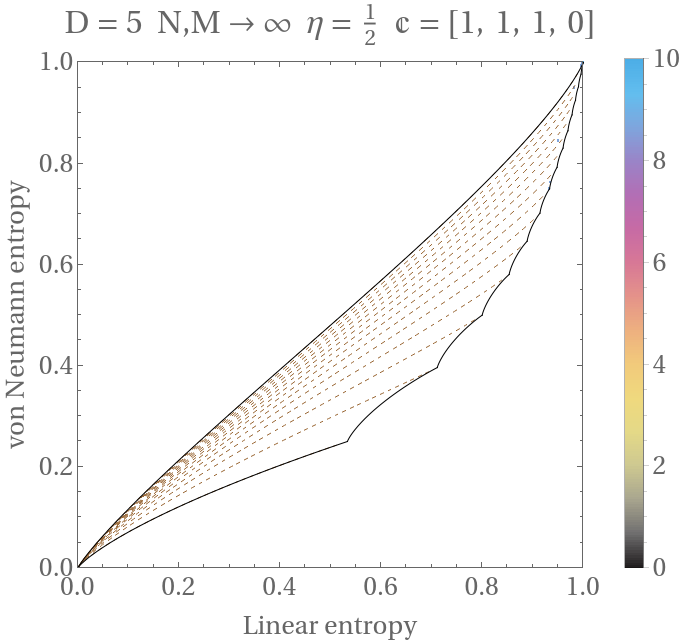}\hspace{\graphsep}
\includegraphics[width=\graphwidth]{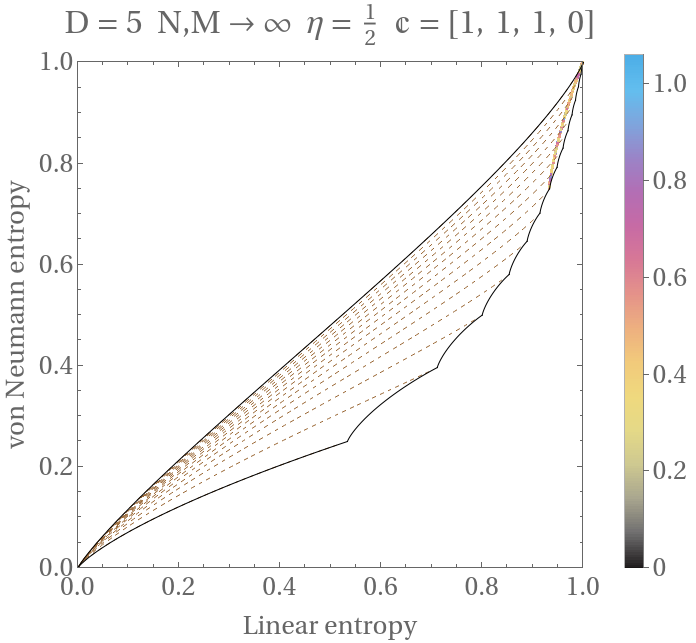}
\caption{Information diagrams using as colormap the distance (left) for $|\alpha_i|\in[0,1]$ and $D=5$, $N,M\rightarrow\infty$ and $\mathbbm{c}=[1,1,1,0]$. The same using as colormap the angular dependence (right) for $\|\alphab\|=R=10$.}
\label{RTL-5-1d2-1110}
\end{figure}
\end{center}

\begin{center}
\begin{figure}[h!]
\includegraphics[width=\graphwidth]{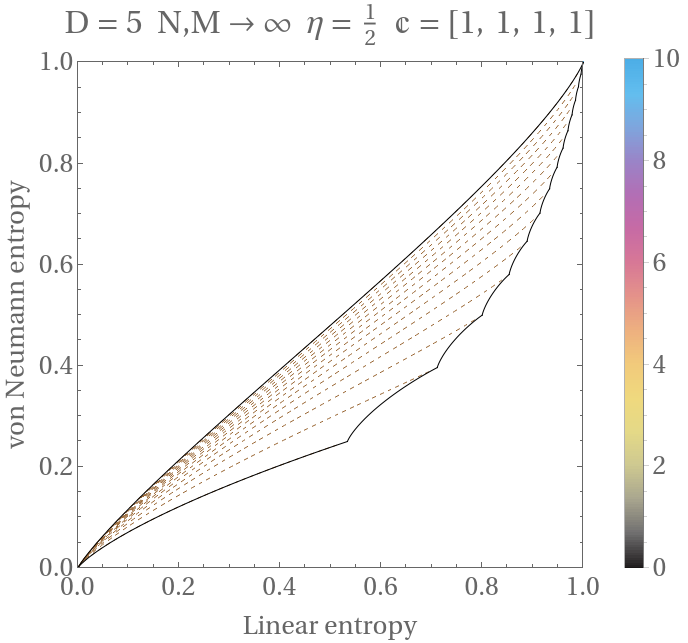}\hspace{\graphsep}
\includegraphics[width=\graphwidth]{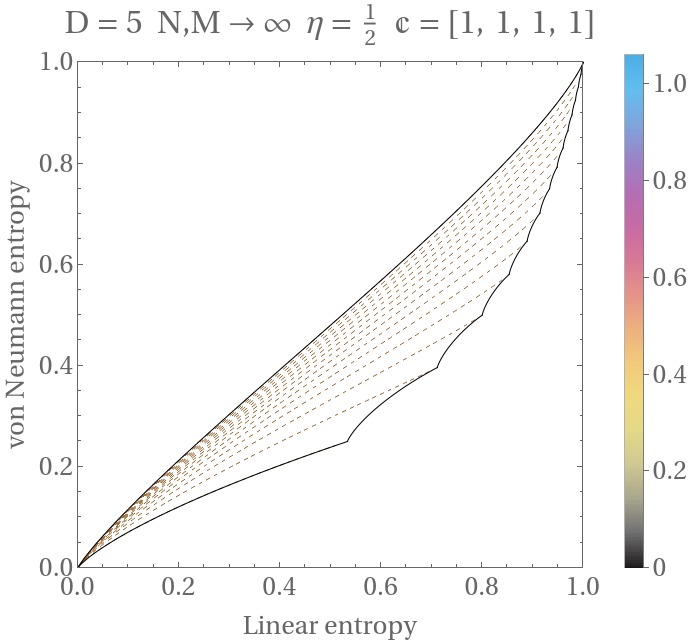}
\caption{Information diagrams using as colormap the distance (left) for $|\alpha_i|\in[0,1]$ and $D=5$, $N,M\rightarrow\infty$ and $\mathbbm{c}=[1,1,1,1]$. The same using as colormap the angular dependence (right) for $\|\alphab\|=R=10$.}
\label{RTL-5-1d2-1111}
\end{figure}
\end{center}

In  Figures \ref{RTL-5-5d6-0000}-\ref{RTL-5-1d2-1111} we show information diagrams with $|\alpha_i|\in [0,4]$, and contour plots of the angular dependence of von Neumann entropy for a large fixed value of $\|\alphab\|_2=R=10$ for each value of $\eta=\frac{5}{6},\frac{2}{3},\frac{1}{2}$. The same comments as in the other values of $D$ apply here.



\bibliography{/home/guerrero/MEGA/Genfimat/Bibliografia/bibliografia.bib}


%

%
%
%
%
%
%
%
%
%
%